\documentclass[10pt]{article}
\usepackage{graphicx}
\usepackage{graphics, cancel}
\usepackage{color}
\usepackage{amsmath, amsthm, slashed,bbold,upgreek}
\usepackage{rotating}
\usepackage{epsfig}
\usepackage{verbatim}
\usepackage{rotating}
\usepackage[affil-it]{authblk}
\usepackage{mathrsfs}
\usepackage{latexsym}
\usepackage{amsmath,amsthm}
\usepackage[greek,english]{babel}
\usepackage{teubner}
\usepackage{MnSymbol}
\usepackage{graphicx}
\usepackage{marvosym}
\usepackage{eufrak}
\usepackage[margin=1in,dvips]{geometry}
\usepackage{color}
\usepackage{hyperref}
\usepackage{comment}

\newtheorem{definition}{Definition}[subsection]

\newtheorem{remark}{Remark}[subsection]

\newtheorem{lemma}{Lemma}[subsection]
\newtheorem{theorem}{Theorem}[subsection]
\newtheorem{bigthe}{Theorem}
\newtheorem*{theorem*}{Theorem}
\newtheorem{moderthem}{Theorem}[section]
\newtheorem*{corollary*}{Corollary}
\newtheorem*{proposition*}{Proposition}
\newtheorem{proposition}{Proposition}[subsection]

\newtheorem{biggerprop}{Proposition}

\newtheorem*{conjecture*}{Conjecture}

\newtheorem{corollary}{Corollary}[subsection]
\newtheorem{modercor}{Corollary}[section]

\def\f12{\frac 1 2}

\def\f12{\frac 1 2}

\newcommand{\nabb}{\mbox{$\nabla \mkern-13mu /$\,}}

\newcommand{\slashg}{\mbox{$g \mkern-9mu /$\,}}

\title{A scattering theory for the wave equation\\ on Kerr black hole exteriors}

\author[1,2]{Mihalis Dafermos}
\author[1]{Igor Rodnianski}
\author[3]{Yakov Shlapentokh-Rothman}

\affil[1]{\small Princeton University, Department of Mathematics, Fine~Hall,~Washington~Road,~Princeton,~NJ~08544,~United~States\vskip.2pc \ }

\affil[2]{\small University of Cambridge, Department of Pure Mathematics and Mathematical
Statistics, Wilberforce~Road,~Cambridge~CB3~0WA,~United~Kingdom\vskip.2pc \ }

\affil[3]{\small Massachusetts Institute of Technology, Department of Mathematics, 77~Massachusetts~Avenue,~Cambridge,~MA~02139, United States\vskip.2pc \ }

\date{December 29, 2014}

\begin{document}

\maketitle

\begin{abstract}
We develop a definitive physical-space scattering theory for the scalar wave equation $\Box_g\psi=0$ on Kerr exterior backgrounds in the general subextremal case $|a|<M$.
In particular, we prove  results corresponding to  ``existence and uniqueness of scattering states'' and ``asymptotic completeness'' and we show moreover that
the resulting ``scattering matrix''  mapping radiation fields on the past horizon $\mathcal{H}^-$
and past
null infinity $\mathcal{I}^-$
to radiation fields on $\mathcal{H}^+$ and $\mathcal{I}^+$
is a bounded operator.
The latter allows us to give a time-domain theory of superradiant
reflection. The boundedness of the scattering matrix shows in particular
that the maximal amplification of
solutions associated to ingoing finite-energy wave packets on past null infinity
$\mathcal{I}^-$ is bounded. On the frequency side, this corresponds to the
novel statement that the suitably normalised reflection and transmission coefficients
are uniformly bounded independently of the frequency parameters.
We further complement this with a demonstration that superradiant reflection
indeed amplifies the energy radiated to future null infinity $\mathcal{I}^+$ of
suitable wave-packets as above.
The results make essential use of a refinement of our recent proof
[M.~Dafermos, I.~Rodnianski and Y.~Shlapentokh-Rothman, \emph{Decay for solutions of the wave equation on Kerr exterior spacetimes III: the full subextremal case $|a| <M$},
arXiv:1402.6034]
of boundedness and
decay for solutions of the Cauchy problem so as to apply in the class
of solutions where only a degenerate energy is assumed finite.
We show in contrast that the analogous scattering maps cannot
be defined for the class of finite non-degenerate energy solutions.
This is due to the fact that the celebrated horizon red-shift effect acts as a blue-shift instability
when solving the wave equation backwards.
\end{abstract}

  \tableofcontents

\section{Introduction}

Black holes play a central role in our present general relativistic picture of the universe.
At the same time, however, they are perhaps the example \emph{par excellence} of
a physical object which cannot be observed ``directly''. An effective approach  to infer both the very presence but also the finer properties of black holes proceeds through the study of the scattering of waves on their exterior. Hence, a theoretical understanding of scattering theory in this context is of  paramount importance.

The bulk of  the now classical black hole scattering-theory
literature concerns only the \emph{fixed-frequency}
study of solutions $u_{(\omega, m, \ell)}(r^*)$ to the radial o.d.e.
\begin{equation}
\label{radodehere}
u''+\omega^2u=Vu,
\end{equation}
where $V=V_{(\omega, m,\ell)}(r^*)$,
resulting from Carter's remarkable  separation~\cite{cartersep} of
the linear scalar wave
equation
\begin{equation}
\label{WAVE}
\Box_g\psi =0
\end{equation}
on Kerr black hole backgrounds $(\mathcal{M},g_{a, M})$.
One can also consider
 more complicated systems like the Maxwell equations or the equations of linearised
gravity.
See Chandrasekhar's monumental~\cite{chandrasekhar} and
the monograph~\cite{physXX}.

Beyond formal fixed-frequency statements concerning $(\ref{radodehere})$, true
scattering results in the ``time-domain'',
describing actual
\emph{finite-energy} solutions of $(\ref{WAVE})$ and related equations,
have only been
obtained in various special cases. Let us already mention the pioneering
results of
Dimock and Kay~\cite{dimock, DK1, DK2}
in the Schwarzschild $a=0$ case. See also~\cite{bachelot, bachelot2}.
In the case of rotating Kerr black holes with $a\ne 0$, on the other hand,
despite recent progress on the Cauchy problem, first for the $|a|\ll M$ case~\cite{partsIandII,
andblue, tattoh}
and then, for the full subextremal range $|a|<M$ in~\cite{partIII},
the most basic questions of scattering theory
for $(\ref{WAVE})$ have
remained to this day unanswered.
In particular:
\begin{enumerate}
\item[(a)]
Can one associate a finite-energy solution
of $(\ref{WAVE})$  to every suitable
finite-energy past/future asymptotic state?
\emph{(Existence of scattering states)}
\item[(b)]
Is the above association unique, i.e.~do two finite-energy solutions
having the same asymptotic state necessarily coincide?
\emph{(Uniqueness of scattering states)}
\item[(c)]
Do the above solutions parametrised by finite-energy
past/future asymptotic states describe the totality of finite-energy
solutions $\psi$ to $(\ref{WAVE})$? \emph{(Asymptotic completeness)}
\end{enumerate}
See the classic~\cite{reedsimon} for a general introduction to the
scattering theory framework
in physics.

At the conceptual level, one
of the most interesting new phenomena of black hole scattering which
arises when passing from the Schwarzschild $a=0$ to the rotating $a\ne 0$ Kerr case
is that of \emph{superradiance}. This already can be seen at the level of
the fixed-frequency o.d.e.~$(\ref{radodehere})$.
We review this very quickly for the benefit of the reader familiar with
the classical physics literature~\cite{chandrasekhar}.\footnote{All notations here will be
explained in detail in the paper. The reader for which this is unfamiliar can skip directly
to the next paragraph!}
For each fixed frequency triple
$(\omega, m, \ell)$ with $\omega\in \mathbb R$,
one can define two complex-valued solutions
$U_{\rm hor}(r^*)$ and $U_{\rm inf}(r^*)$ of $(\ref{radodehere})$ so that
\[
U_{\rm hor}\sim e^{-i(\omega-\upomega_+m )r^*}{\rm\ as\ }r^*\to-\infty,
\qquad\qquad
U_{\rm inf}\sim e^{i\omega r^*}{\rm\ as\ }r^*\to \infty,
\]
corresponding to the asymptotic behaviour of the potential $V$, which is itself real.
Here $\upomega_+$ is related to the Kerr parameters $a, M$ by the formula
$2M \upomega_+(M+\sqrt{M^2-a^2})=a$.
 The linear independence
of $U_{\rm hor}$ and $U_{\rm inf}$
is the statement of    \emph{mode stability} on the real axis
and was proven recently by one of us~\cite{realmodestability}, extending the transformation
theory of~\cite{whiting}. By dimensional considerations, this linear independence
at one go
answers the ``fixed frequency'' analogue of questions (a)--(c) in the affirmative.
It follows that since $\overline{U_{\rm inf}}$ also solves $(\ref{radodehere})$, we may write
\begin{equation}
\label{intheintroform}
\frac{\mathfrak{T}}{-i(\omega-\upomega_+m)}U_{\rm hor} = \frac{\mathfrak{R}}{i\omega}U_{\rm inf} + \frac{\overline{U_{\rm inf}}}{i\omega},
\end{equation}
where $\mathfrak{T}=\mathfrak{T}(\omega, m, \ell)$ and $\mathfrak{R}=\mathfrak{R}(\omega, m, \ell)$
are known as the \emph{transmission} and \emph{reflexion} coefficients.
Formally, these coefficients
describe the proportion of ``energy'' at fixed frequency $(\omega, m, \ell)$
transmitted to the horizon and reflected to infinity, respectively, of purely incoming wave
from past infinity.
With the precise normalisation of $(\ref{intheintroform})$,
which will be in fact motivated by the considerations of this paper,
the energy identity associated to $(\ref{radodehere})$ yields
\begin{equation}
\label{formula1}
|\mathfrak{R}|^2 + \frac{\omega}{\omega-\upomega_+m}|\mathfrak{T}|^2 = 1.
\end{equation}
Superradiance, first discussed by Zeldovich~\cite{zeldovich}, corresponds
to the fact that, for the frequency range
\begin{equation}
\label{supcndin}
\omega(\omega-\upomega_+m)^{-1}<0,
\end{equation}
the transmission coefficient $\mathfrak{T}$
is weighted with a negative sign in $(\ref{formula1})$ allowing thus the
reflection coefficient $\mathfrak{R}$ to have norm strictly greater than $1$
\begin{equation}
\label{softst}
|\mathfrak{R}(\omega, m, \ell)|>1.
\end{equation}
That is to say, there is a nontrivial energy amplification factor at fixed frequency.
The first estimates for
the maximum reflection coefficient in various frequency regimes go back to pioneering
work of Starobinskii~\cite{staro} (see also~\cite{teukpress}), but even the
statement of the uniform boundedness of $\mathfrak{R}(\omega, m, \ell)$ over all superradiant
frequencies $(\ref{supcndin})$
has remained an open problem.

In passing from a fixed-frequency scattering theory to a true time-domain scattering theory,
the
absence of an obvious quantitative frequency-independent control of the coefficient
$\mathfrak{R}(\omega, m, \ell)$ presents itself as a fundamental difficulty.
Moreover, an additional difficulty is identifying the correct notion of ``energy'' with respect
to which solutions should
be defined.
In particular, one requires a notion of energy which
controls
solutions of $(\ref{WAVE})$ not only in the forward but also
in the \emph{backward} direction, i.e.~an energy
\underline{not} subject to the local red-shift effect associated to the event horizon, which when
solving backwards appears as a \emph{blue-shift} instability.

The purpose of this paper is to overcome these difficulties and
develop a definitive finite-energy scattering theory
for $(\ref{WAVE})$ on general subextremal Kerr exteriors $(\mathcal{M},g_{a,M})$ with $|a|<M$, showing in particular:
\begin{quotation} \noindent {\bf \emph{The answer to (a), (b) and (c) is yes.}}
\emph{Existence and uniqueness of
scattering states as well as asymptotic completeness indeed hold
for the space of solutions to $(\ref{WAVE})$ and scattering states
defined by the finiteness of a natural energy flux.}
\end{quotation}

We will understand scattering states in the sense of Friedlander~\cite{friedlander} (for the
Schwarzschild case in this context,
see~\cite{nicolas2}),
and our approach to both constructing and estimating the
scattering
maps can be thought of as a combination of what in the traditional literature are known as ``stationary''
and ``time-dependent'' methods~\cite{kato}.
We will depend heavily on
our recent boundedness results~\cite{partIII}
for the Cauchy problem for $(\ref{WAVE})$, as well as certain decay results of~\cite{partIII},
which indeed succeeded in giving
a first version of quantitative physical-space control
over superradiance, independent of frequency, and also showed that a suitable class of
solutions of $(\ref{WAVE})$
can be indeed understood as superpositions of solutions of $(\ref{radodehere})$
over real frequencies $\omega$.
We will in fact, however, here
require a certain refinement of the estimates of~\cite{partIII}
so as to apply to a degenerate energy not subject to the backwards blue-shift
instability.
This notion of energy lies behind the particular choice of normalisation of
the reflection coefficient
$\mathfrak{R}$ in~$(\ref{intheintroform})$.
Along the way, we shall in particular provide the missing frequency-independent bound on
$\mathfrak{R}$ over all superradiant frequencies $(\ref{supcndin})$:
\begin{equation}
\label{introubo}
\sup_{(\omega, m, \ell)} |\mathfrak{R}(\omega, m, \ell)| = S(a,M)<\infty
\end{equation}
by a finite constant $S(a,M)$ depending only on the Kerr parameters, with $S(a,M)>1$
if $a\ne 0$.

Our asymptotic completeness
results will allow us to define (in the language of Wheeler~\cite{wheeler})
 an $S$-matrix $\mathscr{S}$ whose boundedness in the
operator norm
replaces the usual unitarity property.
A suitable restriction of $\mathscr{S}$ will be related
to a generalisation of the
inverse-Fourier transform applied to multiplication by the coefficients $\mathfrak{R}$ and $\mathfrak{T}$
 defined by $(\ref{intheintroform})$.
Through this, we will give a definitive
physical space (i.e.~time-domain) interpretation of
 superradiant reflection, in particular, showing:
\begin{quotation}
\noindent
\emph{Superradiant reflection indeed} {\bf \emph{strictly amplifies}} \emph{the energy radiated to infinity of suitably constructed
purely ingoing finite-energy wave packets. The maximum
amplification factor, however,  is} {\bf
\emph{bounded}}
\emph{precisely by the constant $S(a,M)$ of $(\ref{introubo})$.}
 \end{quotation}

Our results leave open the extremal case $g_{a,M}$ for $a=M$
(see~\cite{aretakiskerr}). In particular, it is not known whether the limit
$\lim_{|a|\to M} S(a,M)$ is finite.

\subsection{Brief overview of the main theorems}
We introduce briefly  the main theorems of the paper in what follows below.
(We will give a more detailed overview together with
precise statements of all theorems listed here in boldface type
in Section~\ref{DOsec}.)

\subsubsection{Fron Schwarzschild to Kerr: the $T$-energy theory and superradiance}
The first difficulty in constructing  a physical-space scattering theory
is identifying what constitutes the ``correct'' class
of finite energy solutions and asymptotic states.
In the Schwarzschild case, as admissible solutions to $(\ref{WAVE})$
it is natural to consider the class of $\psi$
which
have finite \emph{conserved}
energy (i.e.~finite energy corresponding to the stationary
Killing vector field $T$) on a Cauchy hypersurface.
This in turn suggests a corresponding notion of asymptotic states defined
in terms of the completion (with respect to the natural
$T$-energy flux) of the set of \emph{Friedlander radiation fields}
$r\psi$ on $\mathcal{I}^+$ (see~\cite{friedlander}),
complemented by the analogous completion of the
set of traces of $\psi$ on the event horizon $\mathcal{H}^+$.
See Nicolas~\cite{nicolas2} for a recent formulation of Schwarzschild scattering theory in precisely
these terms. This theory can be constructed entirely in the time-domain, i.e.~using
``time-dependent'' methods. (We will
in fact give
our own self-contained version of the Schwarzschild theory in Section~\ref{veryselfcontained}.)

Turning to the Kerr case, the above conserved energy  corresponding to $T$
is clearly unsuitable for  a scattering theory, because the inner product it
defines is now indefinite,
in view of the existence of the well-known \emph{ergoregion} where $T$ is
spacelike.\footnote{Let us note that, in contrast to the wave equation $(\ref{WAVE})$,
for the Dirac equation, one still has
a coercive $L^2$-conservation law despite the absence of a globally  timelike Killing field.
Using this, H\"afner and Nicolas~\cite{hafnernicolas} have constructed a scattering
theory for the Dirac equation on Kerr backgrounds,  generalising~\cite{nicolas1}.
This has been extended to Kerr--Newman--de Sitter backgrounds by
Daud\'e and Nicoleau~\cite{daudenic}.
In this context, see also H\"afner~\cite{hafnerthesis} for scattering results concerning a non-superradiant class of solutions
of the Klein--Gordon equation for fixed azimuthal mode $m$.}
 This is the physical-space origin of the phenomenon of superradiance discussed with
 respect to $(\ref{formula1})$.
Recent progress on understanding the Cauchy problem for $(\ref{WAVE})$
on Schwarzschild and
Kerr has rested in part on the realisation (see~\cite{redshift, boundedn, lectnotes}) that a more natural
energy quantity for understanding forward evolution
is that defined by a $T$-invariant everywhere-timelike vector field $N$.
Even though this $N$-energy is not conserved,  it remains, as
proven in our recent work~\cite{partIII} (for
the full sub-extremal range of Kerr parameters $|a|<M$),
uniformly bounded through a suitable spacelike
foliation $\Sigma_s^*$ of the exterior region
and controls in fact a spacetime integral quantity. The good divergence properties
of the vector field $N$ are related to the celebrated red-shift effect associated
to the horizon $\mathcal{H}^+$.

\subsubsection{The $N$-energy theory and the backwards blue-shift instability}
Despite its success in the context of the Cauchy problem on Kerr, the above $N$-energy
is again unsuitable for defining a scattering
theory, because the helpful red-shift transforms into a lethal blue-shift when trying to associate
admissible solutions to their natural asymptotic states, which requires solving the
wave equation backwards.
See the discussion in~\cite{BHscatter, sbierski} and also
the more recent comments in~\cite{nicolas2}.
The first two results of our paper are dedicated
to making explicit this obstruction. Our {\bf Theorems~\ref{THEOREM1}
and~\ref{failsurj}}
together show that while one can naturally associate
 (using our results of~\cite{partIII})  asymptotic states to finite $N$-energy
solutions, this map is not surjective, and thus, one cannot define a one-sided inverse
map embodying the existence of scattering states (cf.~(a)).

\subsubsection{The $V$-energy theory}\label{NEWSE}
The correct setting for a scattering theory on Kerr would then appear to be an energy quantity
defined by a vector field $V$ which (like $T$ in Schwarzschild) is null on the horizon
and timelike outside. With the help of the additional axisymmetric Killing field $\Phi$,
one can in fact construct such a vector field $V$ which can be chosen
moreover
 Killing
in  a neighbourhood of both $\mathcal{H}^+$ and $\mathcal{I}^+$ (though not globally
Killing!).
Even the question of uniform boundedness of solutions assumed to lie
only in the energy space defined by $V$, however, has
not previously been answered.  (See however, the very related
higher-order weighted estimates of Andersson--Blue~\cite{andblue}
in the very slowly rotating $|a|\ll M$ case.)

The main results of the present paper ({\bf Theorems~\ref{toinvert} and \ref{existofsc}}) succeed
in constructing a bounded invertible map $\mathscr{F}_+$ associating a unique
future
asymptotic state
to each solution with initially bounded $V$-energy, with two-sided inverse $\mathscr{B}_-$
satisfying
\begin{equation}
\label{areinversesh}
\mathscr{B}_-\circ \mathscr{F}_+ =Id, \qquad  \mathscr{F}_+\circ  \mathscr{B}_-=Id.
\end{equation}
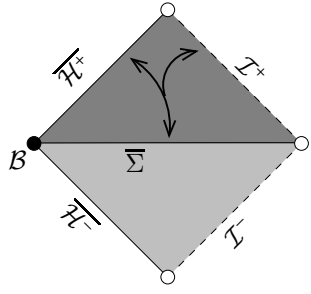
\[
\input{forscater2b.pstex_t}
\]
The boundedness of the map
$\mathscr{F}_+$
requires
a refinement of our previous boundedness results on the Cauchy
problem (see~\cite{partIII}) so as to apply
for admissible solutions defined by the finiteness of a  suitable $V$-energy as above.
This will require us to revisit the fixed frequency o.d.e.~estimates on $(\ref{radodehere})$
proven in~\cite{partIII}.
What will be the inverse map $\mathscr{B}_-$ is constructed explicitly via
the frequency domain by an appropriate superposition
of solutions to the fixed frequency o.d.e~$(\ref{radodehere})$.
Again, to infer the boundedness of $\mathscr{B}_-$  one needs to exploit
quantitative estimates on~$(\ref{radodehere})$ adapted from~\cite{partIII}, again
referring only to the $V$-energy flux. One may define similar maps
$\mathscr{F}_-$, $\mathscr{B}_+$ associating solutions to \emph{past} asymptotic states.

In the traditional language of scattering theory, let us note that
existence of scattering states (cf.~(a)) corresponds to the existence
of  $\mathscr{B}_\mp$, uniqueness of scattering states (cf.~(b)) to the
injectivity of
$\mathscr{F}_\pm$, and asymptotic completeness (cf. (c)) to the surjectivity
of $\mathscr{B}_\mp$. These three statements of course all follow from $(\ref{areinversesh})$.

\subsubsection{The scattering map $\mathscr{S}$, superradiant reflection $\mathscr{R}$
and applications}
 The asymptotic completeness results allow us in particular to define a scattering map
 ($S$-matrix)
 \[
 \mathscr{S} =\mathscr{F}_+\circ \mathscr{B}_+
 \]
 taking asymptotic past states to asymptotic future states
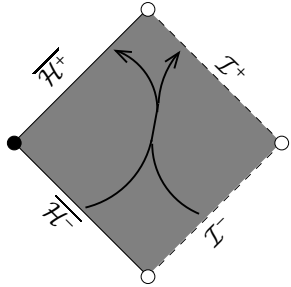
\[
\input{forscater3.pstex_t}
\]
 which is moreover bounded in the operator norm with respect to
 the spaces defined by the flux of the $V$-energy (see {\bf Theorem~\ref{SMatrix}}).

 To connect with the usual discussion of superradiant scattering, we may also
 define a reflection map $\mathscr{R}$ and a transmission map $\mathscr{T}$
 which restricts $\mathscr{S}$ to past asymptotic states with no trace on the past event horizon
 $\mathcal{H}^-$ and
 returns only the radiation to future null infinity $\mathcal{I}^+$ or the future event horizon $\mathcal{H}^+$ respectively.
It follows
in particular that $\mathscr{R}$ and $\mathscr{T}$ are also bounded (see {\bf Theorem~\ref{isbound}}).
On the other hand, we show that the operator norm of $\mathscr{R}$
satisfies $\|\mathscr{R}\|>1$ (see {\bf Theorem~\ref{lowbn}}),
and thus there exist wave packets
corresponding to past asymptotic states supported only on $\mathcal{I}^-$
such that the energy radiated to $\mathcal{I}^+$ is strictly greater than
the energy flux on $\mathcal{I}^-$. As discussed above, this gives a physical
space interpretation of superradiance (cf.~the numerical \cite{numeric}). Next, we will show that $\mathscr{T}\oplus\mathscr{R}$
is pseudo-unitary in that it preserves an indefinite inner product associated to the $T$-energy ({\bf Theorem~\ref{bb}}). Upon restricting to ``non-superradiant'' data along $\mathcal{H}^-$ and $\mathcal{I}^-$ the map $\mathscr{S}$ becomes unitary in the standard sense ({\bf Theorem~\ref{bbb}}).

We finally give a ``unique continuation'' result that finite $V$-energy solutions
are uniquely characterized by their scattering data on any of the ``ill-posed'' pairs
$\mathcal{H}^-\cup\mathcal{H}^+$,  $\mathcal{I}^+\cup\mathcal{I}^-$,
$\mathcal{H}^-\cup\mathcal{I}^+$ or $\mathcal{H}^+\cup\mathcal{I}^-$
(see {\bf Theorem~\ref{introuC}}). This has the interpretation that for this improper
notion of asymptotic states, uniqueness
of scattering states (b) holds without existence (a).

\subsubsection{Back to the fixed-frequency theory}\label{fixedfreqsec}
We have already noted that our results will require revisiting the
estimates proven in~\cite{partIII}
for the radial o.d.e.~$(\ref{radodehere})$ appearing in Carter's classical
separation of $(\ref{WAVE})$.
In this sense, our work makes
contact back with    the formal  scattering theory
literature~\cite{physXX}  concerning $(\ref{radodehere})$ at fixed frequency.
In particular, our o.d.e.~results will yield the uniform
boundedness of the reflection and transmission coefficients
({\bf Theorem~\ref{isbound2}}), in particular,
giving $(\ref{introubo})$.
 This complements  the work of Starobinskii and others (see~\cite{staro, teukpress})
aimed at  numerically estimating the maximum of these for low
fixed values of $m$, $\ell$.
 Our transmission
and reflection maps $\mathscr{T}$ and $\mathscr{R}$ can in fact be represented as a generalised
inverse Fourier transform of multiplication by $\mathfrak{T}$ and
$\mathfrak{R}$ ({\bf Theorem~\ref{PSpREPint}}).
In particular, a posteriori, the boundedness statements of
Theorems~\ref{isbound} and~\ref{isbound2} are equivalent.
This connects the fixed frequency and physical space scattering theories in a very explicit way.

\subsection{Related work and further reading}
Let us specifically
mention here
a related recent important advance by Georgescu, G\'erard and H\"afner~\cite{GGH}
which proves scattering results for fixed-azimuthal mode (i.e.~fixed $m$) solutions of
the Klein-Gordon equation in the
very slowly rotating Kerr-de Sitter case $|a|\ll M, \Lambda$. This is in part
based on work on the Cauchy problem due to Dyatlov~\cite{dyatlov}.
For additional background on the Cauchy problem on
other black hole spacetimes, besides references mentioned previously,
we refer the reader
to the lecture notes~\cite{lectnotes}.

\subsection{Acknowledgements}
MD acknowledges support through NSF grant
DMS-1405291.
IR acknowledges support through NSF grant DMS-1001500 and DMS-1065710.
YS acknowledges support through NSF grants DMS-0943787 and DMS-1065710
as well as the hospitality of Princeton University during the period when
this research was carried out.

\section{Detailed overview and statements of the main theorems}
\label{DOsec}
 In this section, we will give a more detailed overview of the main  results
 of this paper.  We begin in Section~\ref{setupint} with the basic setup
 for our ``time-domain'' scattering theory. We shall then briefly turn in
Section~\ref{TeneSc} to
 a discussion first of the Schwarzschild $a=0$ case based on spaces defined by the conserved
 $T$-energy, and then of  the problem of superradiance in Kerr for $a\ne 0$ which makes this
 approach impossible. With these preliminaries, we  present
 in Section~\ref{MThSec}  the statements of the main theorems
 of our scattering theory in the time domain for Kerr in the general subextremal
 range
 $|a|<M$. We shall relate this back to
 the fixed-frequency theory in Section~\ref{appliestofixed}, stating two additional
 theorems. In Section~\ref{Nonlinearps}, we make a brief comparison with
non-linear scattering problems involving
 black holes, in particular referring to a recent
 scattering construction of solutions to the Einstein equations themselves which
 asymptote in time to the Kerr family~\cite{BHscatter}.
 Finally, we shall give in Section~\ref{OUTLhere}
 a section by section outline of the remainder of the paper, identifying
 in particular where each of the main theorems is proven.

\subsection{The setup for scattering theory in the time-domain}
\label{setupint}
We begin with the basic setup describing our ``time-domain'' scattering theory
in the Kerr black-hole context.

\subsubsection{The exterior region of Kerr}
We will fix subextremal Kerr parameters $|a|<M$ and consider the Kerr metric $g_{a, M}$
defined on a ``domain of outer communication'' $\mathcal{D}$. See Section~\ref{diffCoord}
for an explicit
representation of this manifold with stratified boundary.
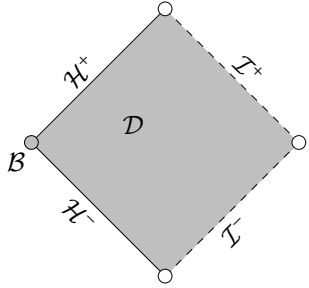
\[
\input{bare.pstex_t}
\]
The boundary components $\mathcal{H}^\pm$ correspond to past and future
event horizons and meet in the so-called bifurcation sphere $
\mathcal{B}$. (Our convention will be that $\mathcal{H}^\pm$ do \underline{not}
contain $\mathcal{B}$.)
Moreover, one can define the two ``asymptotic'' boundary components
future and past null infinity $\mathcal{I}^\pm$, which, in an auxiliary
topology, can indeed  be attached to $\mathcal{D}$ as boundary.
See Section~\ref{formdefine}.

\subsubsection{Hypersurfaces and forward evolution of smooth data}
\label{hypforev}
We begin by considering smooth solutions $\psi$ of $(\ref{WAVE})$
arising from compactly supported
initial data on a suitable hypersurface.
We will in fact consider three distinct classes of such data.

When we are only interested in future scattering, it is more natural to focus
on  solutions parametrised by compactly supported
data $(\uppsi_{\Sigma_0^*}, \uppsi'_{\Sigma_0^*})$
on a hypersurface
\[
\Sigma_0^*=\{t^*=0\},
\]
defined as the level set
of a future-horizon penetrating
$t^*$-coordinate. See Section~\ref{diffCoord}.
Here $\Sigma_0^*$ is understood as a manifold-with-boundary,
so the support of the data can in principle contain the boundary 
$\Sigma_0^*\cap {\mathcal{H}}^+$.
By general theory, such data give rise to a unique smooth  solution $\psi$
of $(\ref{WAVE})$ on
$\mathcal{R}_{\ge 0}= D^+(\Sigma_0)$.
We shall call the map
from smooth initial data to solution \emph{forward evolution}:
\begin{equation}
\label{MAPSTOhere}
(\uppsi,\uppsi')\mapsto\psi.
\end{equation}
See Proposition~\ref{WPProp}.

When we are interested in defining the $S$-matrix,
we need  to parameterise solutions $\psi$ by data which determine $\psi$
globally on $\mathcal{D}$.
It is in fact natural to
distinguish between two cases.
Defining
\[
\mathring\Sigma=\{t=0\}, \qquad \overline\Sigma =\mathring\Sigma\cup\mathcal{B},
\]
where $t$ is the usual Boyer-Lindquist coordinate defined only on the interior of
$\mathcal{D}$,
we can consider smooth compactly supported data
$(\uppsi_{\overline\Sigma}, \uppsi'_{\overline\Sigma})$ on $\overline\Sigma$,
or the more restrictive class of smooth compactly supported data
$(\uppsi_{\mathring\Sigma}, \uppsi'_{\mathring\Sigma})$
on $\mathring\Sigma$. (The latter, thought of as a special case of the former,
must vanish in a neighbourhood of $\mathcal{B}$.)
We can now associate in either case a global smooth
solution $\psi$ on $\mathcal{D}$. See Proposition~\ref{t0wellposed}.
We will again refer to the map $(\ref{MAPSTOhere})$ as forward
evolution.

The significance of considering the restricted data (i.e.~data whose support
as a subset of $\mathring\Sigma$ is compact) is that the support
of the resulting $\psi$ in $\mathcal{D}$ is disjoint from an open neighbourhood
of $\mathcal{B}$. This will be useful technically in defining the backwards map
in the frequency domain. It will also facilitate comparison with other results
where it has often been this scattering theory that has been implicitly or
explicitly considered.

\subsubsection{Radiation fields and horizon traces}\label{radfieldsec}
The most natural formulation of a scattering theory from the point of view
of the present problem describes asymptotic states by an appropriate
Hilbert space completion (see below) of the future and past radiation fields
on $\mathcal{I}^\pm$
augmented by
radiation fields on the horizons.

The notion of radiation field along $\mathcal{I}^+$ is due to Friedlander~\cite{friedlander}
and in our context is given by the following Proposition:
\begin{biggerprop}
\label{PROP1}
If data $(\uppsi, \uppsi')$ are smooth of compact support
on $\Sigma_0^*$, $\mathring\Sigma$ or $\overline{\Sigma}$,
then the solution
$r\psi$ extends to a smooth function $\upphi$ defined on $\mathcal{I}^+$.
\end{biggerprop}
We shall
infer the above as an essentially trivial consequence of
the $r^p$ estimates of~\cite{icmp}.  See Proposition~\ref{rp} and
Corollary~\ref{radinfwelldef}.

The radiation field on the horizon is just the usual
restriction of $\psi$ as  a smooth function.
Let us introduce the notation $\mathcal{H}^+_{\ge 0} = \mathcal{R}_{\ge 0}$,
and $\overline{\mathcal{H}^+}=\mathcal{H}^+\cup \mathcal{B}$.
Since $\psi$ arising from compactly supported
data $(\uppsi_{\Sigma_0^*}, \uppsi_{\Sigma_0^*}')$ is only
defined on $\mathcal{R}_{\ge 0}$, we may define in this case only
$\uppsi|_{\mathcal{H}^+_{\ge 0}}\doteq \psi|_{\mathcal{H}^+_{\ge 0}}$.
In the case of solutions arising from compactly supported data on $\mathring{\Sigma}$
and $\overline{\Sigma}$, respectively, $\psi$ is of course defined on all
of $\overline{\mathcal{H}^+}$; nonetheless, we shall refer to
$\uppsi_{\mathcal{H}^+}\doteq \psi_{\mathcal{H}^+}$ in the former case
and $\uppsi_{\overline{\mathcal{H}^+}}\doteq \psi_{\overline{\mathcal{H}^+}}$
in the latter case.
This notation reminds us  (cf.~the remark at the end of Section~\ref{hypforev} above)
that in the former case,
the support of $\uppsi_{\mathcal{H}^+}$ is disjoint from a neighborhood of $\mathcal{B}$
in $\mathcal{D}$, whereas, in the latter case,
 the support of $\uppsi_{\overline{\mathcal{H}^+}}$ may contain
$\mathcal{B}$.

To summarise, forward evolution $(\ref{MAPSTOhere})$
gives rise to a map on smooth compactly supported
initial data
\begin{equation}
\label{completof}
(\uppsi|_{\Sigma_0^*, \, \mathring\Sigma, {\rm\ or\ }\overline{\Sigma}},\, \uppsi'|_{\Sigma_0^*,\, \mathring\Sigma, {\rm\ or\ }\overline{\Sigma}})\mapsto \psi \mapsto (\uppsi|_{\mathcal{H}^+_{\ge 0}, \, \mathcal{H}^+,{\rm\ or\ } \overline{\mathcal{H}^+}} \doteq \psi|_{\mathcal{H}^+_{\ge 0}, \, \mathcal{H}^+,{\rm\ or\ } \overline{\mathcal{H}^+}},\,\upphi|_{\mathcal{I}^+}\doteq  r\psi|_{\mathcal{I}^+})
\end{equation}
defined by solving the initial value problem for $(\ref{WAVE})$ and restricting to the radiation
fields.
The forward maps of our scattering theory will be constructed by completing
the above map with respect to suitably defined energies.

\subsubsection{Vector fields, energies and asymptotic states}\label{FROMTHEIN}
The states defining scattering theory are associated to energies
which are in turn defined by vector fields.

Recall that
a general vector field $X$ defines an energy current ${\bf J}^X[\psi]$
and an energy flux
\begin{equation}
\label{generalflux}
\int_{\mathcal{S}} {\bf J}^X[\psi]
\end{equation}
through an arbitrary hypersurface $\mathcal{S}$. (See Section~\ref{currentssec}.)

For appropriate vector fields $X$ for which $(\ref{generalflux})$ is nonnegative,
the square root of the expression $(\ref{generalflux})$ can in turn be used as a norm
to define a space
\begin{equation}
\label{statespace}
\mathcal{E}_{\Sigma_0^*}^X, \qquad
\mathcal{E}_{\mathring{\Sigma}}^X, \qquad
\mathcal{E}_{\overline{\Sigma}}^X
\end{equation}
by  completion of the set of smooth compactly supported data
$(\uppsi, \uppsi')$
on $\Sigma_0^*$, $\mathring\Sigma$, $\overline\Sigma$, respectively.
(See Section~\ref{FSsec}.)
Recall that ``compactly supported on $\mathring\Sigma$'' is a
more restrictive assumption than ``compactly supported on $\overline\Sigma$''
and thus
$\mathcal{E}_{\mathring{\Sigma}}^X\subset \mathcal{E}_{\overline{\Sigma}}^X$.

Similarly,  the flux $(\ref{generalflux})$ defines asymptotic spaces
\begin{equation}
\label{asymptoticspaces}
\mathcal{E}_{\mathcal{H}^+_{\geq 0}}^X\oplus \mathcal{E}_{\mathcal{I}^+}^X
\qquad
\mathcal{E}_{\mathcal{H}^+}^X\oplus \mathcal{E}_{\mathcal{I}^+}^X,
\qquad
\mathcal{E}_{\overline{\mathcal{H}^+}}^X\oplus \mathcal{E}_{\mathcal{I}^+}^X,
\end{equation}
via completion of the space of radiation fields arising from $(\ref{completof})$.
Here we have that $\mathcal{E}^X_{\mathcal{H}^+_{\ge 0}}$ embeds (non-uniquely) into $\mathcal{E}^X_{\mathcal{H}^+}$, and also, $\mathcal{E}^X_{\mathcal{H}^+}\subset \mathcal{E}^X_{\overline{\mathcal{H}^+}}$.

\emph{In this picture,
the problems (a)--(c) of scattering theory translate into finding bijective maps between
$(\ref{statespace})$ and $(\ref{asymptoticspaces})$ induced by the
completion of forward evolution $(\ref{completof})$
of smooth data, for a suitable choice of the vector field $X$.} We will not discuss the construction of wave operators
in the spirit of~\cite{friedrichs, moller} as there is no compelling global ``reference dynamics''
with which to compare; see~\cite{nicolas2} for a nice discussion of how to construct
the latter if desired.

\subsection{The $T$-energy theory and its limitations}
\label{TeneSc}
Before turning to our main theorems, we briefly review the Schwarzschild $a=0$ case,
as well as
the physical space manifestation of the difficulty of superradiance, discussed previously,
which arises upon passing to rotating Kerr with $a\ne 0$.

\subsubsection{The Schwarzschild $a=0$ case}
In the Schwarzschild case $a=0$, the stationary Killing field $T$
is timelike in the interior of $\mathcal{D}$
becoming null on $\mathcal{H}^+\cup\mathcal{H}^-$ and vanishing
on $\mathcal{B}$. Thus the energy defined by $T$ degenerates pointwise.
Nonetheless, the completions $\mathcal{E}_{\mathring{\Sigma}}^T$,
$\mathcal{E}_{\mathcal{H}^+}^T$ and $\mathcal{E}_{\mathcal{I}^+}^T$
define Hilbert spaces and
one can obtain a {\bf unitary} isomorphism
\begin{equation}
\label{inSchwcase}
\mathcal{E}^T_{\mathring{\Sigma}} \cong \mathcal{E}^T_{\mathcal{H}^+} \oplus \mathcal{E}^T_{\mathcal{I}^+}.
\end{equation}
In our notation, this is the content of the previously known Schwarzschild
scattering theory~\cite{dimock, DK1, nicolas2}.

We will give our own self-contained treatment in Section~\ref{veryselfcontained}. One obtains
with no additional difficulty the alternative
unitary isomorphisms
$\mathcal{E}^T_{\Sigma_0^*} \cong \mathcal{E}^T_{{\mathcal{H}^+_{\ge 0}}} \oplus \mathcal{E}^T_{\mathcal{I}^+}$ and
$\mathcal{E}^T_{\overline{\Sigma}} \cong \mathcal{E}^T_{\overline{\mathcal{H}^+}} \oplus \mathcal{E}^T_{\mathcal{I}^+}$.

\subsubsection{The case $a\ne 0$ and the ergoregion}
Turning to the Kerr case
$a\ne 0$, there is now a non-empty subset $\mathcal{S}$ of
$\mathcal{D}$ known as the \emph{ergoregion} where
$T$ is spacelike. In particular, the energy-fluxes $\int_{\mathring{\Sigma}} {\bf J}^T[\psi]$,
$\int_{\mathcal{H}^+} {\bf J}^T[\psi]$
defined by $T$ fail to be positive definite. This is the physical space origin
of the phenomenon of superradiance,  discussed in the fixed-frequency
theory in the context of  $(\ref{formula1})$ and $(\ref{supcndin})$.

Part of the conceptual difficulty of formulating a scattering theory in the Kerr case
is thus to find the correct notion of asymptotic states which replaces
those based on $\mathcal{E}^T$. At the same time, one must understand
what property replaces the  notion of unitarity in $(\ref{inSchwcase})$ as a means of
quantifying the good properties of the scattering map.
We turn now to the statements of the main results of this paper
that give a definitive resolution of this problem.

\subsection{A scattering theory for Kerr: the main theorems}
\label{MThSec}

In this section, we will present in detail
the main theorems of our paper concerning physical-space (time-domain)
scattering theory for the wave equation $(\ref{WAVE})$ on Kerr in the general subextremal case
$|a|<M$.

\subsubsection{The $N$-energy forward map}\label{Nenergyforsec}
The first candidate replacement for the (degenerate) Schwarzschild $T$-energy
is the so-called $N$-energy.
Here, $N$ is a globally timelike vector field which is $T$-invariant outside a
neighbourhood of the bifurcation sphere $\mathcal{B}$ and moreover such that
$N=T$ in a neighbourhood
of $\mathcal{I}^+$.
The energies $(\ref{statespace})$
associated to this vector field are indeed manifestly positive-definite
and pointwise non-degenerate.

The first main theorem defines asymptotic states for all solutions
arising from finite $N$-energy data on the hypersurface $\Sigma_0^*$, i.e.,~in the notation $(\ref{statespace})$, for all solutions
parametrised by $\mathcal{E}_{\Sigma_0^*}^N$.

\begin{bigthe}
\label{THEOREM1}
Forward evolution $(\ref{completof})$ with data on $\Sigma_0^*$
extends to a bounded map
$\mathscr{F}_+: \mathcal{E}^N_{\Sigma_0^*}\to \mathcal{E}_{\mathcal{H}^+_{\geq 0}}^N\oplus
\mathcal{E}^T_{\mathcal{I}^+}$.
\end{bigthe}
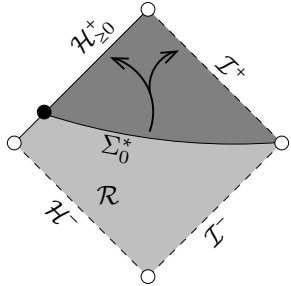
\[
\input{forscater0.pstex_t}
\]

See Theorem~\ref{boundNONDeg}. (Note that $\mathcal{E}^N_{\mathcal{I}^+}=
\mathcal{E}^T_{\mathcal{I}^+}$.)
For the hard analysis behind the above,
the proof relies in particular on a uniform boundedness statement
for the energy $\int_{\Sigma_s^*} {\bf J}^N$ through a foliation $\Sigma_s^*$ defined
by future-translating $\Sigma_0^*$ by the flow of $T$,
as well as a weak decay statement, both of which follow from the
results of~\cite{partIII} mentioned previously, here quoted as
Theorem~\ref{theResult}.

\subsubsection{A blue-shift instability and the \underline{non}-existence of an $N$-energy backwards map}
\label{failsurjdisc}

Satisfactory though the forward theory may be, it turns out that the
above $N$-energy is ill-suited for defining the asymptotic states of a scattering theory.
The fundamental origin of this is the red-shift effect on the horizon (so favourable
for controlling forward evolution!), which for backwards evolution is now seen
as a blue-shift. See~\cite{BHscatter} and Section~3.1.2 of Sbierski~\cite{sbierski}.
It turns out that one can show explicitly that the map of Theorem~\ref{THEOREM1} fails
to be surjective:
\begin{bigthe}
\label{failsurj}
Already in the Schwarzschild $a=0$ case,
the map $\mathscr{F}_+$ of Theorem~\ref{THEOREM1}  fails to  be surjective.
\end{bigthe}
It follows that
 there does not exist even a \emph{one-sided} inverse $\mathscr{B}_-$
satisfying $\mathscr{F}_+\circ \mathscr{B}_-=Id$; thus, \emph{existence
of scattering states} (cf.~(a)) does \underline{not} hold in the $N$-theory.
(As we shall see in Section~\ref{intr-V-}, the above map $\mathscr{F}_+$ is however injective.)

Our proof of the above theorem exploits monotonicity satisfied by the spherical mean under
spatial evolution. Though essentially independent of the rest of the paper, the precise statement proven (Theorem~\ref{noScatter}) is
deferred to the end (Section~\ref{correlate}), so that it can be interpreted
both as a non-surjectivity result with
respect to  our $N$-energy scattering theory (Corollary~\ref{thenonsurcor})
and also
constructively (Corollary~\ref{awonderfulcor}) using
Theorem~\ref{existofsc} of our $V$-energy scattering theory
to be discussed below.  Let us already remark, however, that
the non-surjectivity statement we obtain in  Corollary~\ref{thenonsurcor} is
more precise than what we have just stated above.
 We elaborate briefly below.

First let us note that with the notations of the present paper,
the considerations of Section~1.1.6.1 of~\cite{BHscatter} show
that by introducing sufficiently
high exponential weights in the spaces defining the scattering data,
i.e.~considering the spaces
$\mathcal{E}^{e^{\alpha v} N}_{\mathcal{H}^+}$
and
$\mathcal{E}^{e^{\alpha u} T}_{\mathcal{I}^+}$, then
there indeed exists a bounded one-sided inverse
\begin{equation}
\label{indeeDEX}
\mathscr{B}_- : \mathcal{E}^{e^{\alpha v} N}_{\mathcal{H}^+_{\geq 0}}\oplus
\mathcal{E}^{e^{\alpha u} T}_{\mathcal{I}^+} \to \mathcal{E}^N_{\Sigma_0^*}
\end{equation}
such that $\mathscr{F}_+\circ \mathscr{B}_-= id$.
Thus, we do have existence of a \emph{restricted class} of future scattering states.

With this setting, our Theorem~\ref{noScatter} in fact shows (see Corollary~\ref{thenonsurcor}) that $e^{\alpha v}$ above \underline{cannot}
be replaced by $|v|^p$ no matter how large $p$ is taken, i.e.~the map
$\mathscr{F}_+$ of Theorem~\ref{THEOREM1}
is not surjective as a map $\mathscr{F}_+^{-1}(\mathcal{E}^{|v|^pN}_{\mathcal{H}^+_{\geq 0}}
\oplus \{0\})\to \mathcal{E}^{|v|^pN}_{\mathcal{H}^+_{\geq 0}}
\oplus \{0\}$.
The question of precise characterization of the range of $\mathscr{F}_+$
remains open.
We shall return to this issue in Section~\ref{Nonlinearps}.

\subsubsection{The $V$-energy forward map}\label{venergyforwardsec}
To define a forward map which one can indeed
hope to show is invertible, we must pass to a degenerate energy class
which does not see the red-shift at the horizon.

Recall that $g_{M,a}$ admits an additional Killing vector field $\Phi$
corresponding to axisymmetry.
Although for $a\ne 0$, the vector field
$T$ fails to be globally timelike in the interior
of $\mathcal{D}$, the span
of $T$ and $\Phi$ does form a timelike plane,
and  the Killing combination $K=  T+\upomega_+ \Phi$
is timelike in  a neighbourhood of $\mathcal{H}^+$, becoming null
on $\mathcal{H}^+$ itself.
(Note that if $a=0$, then $K=T$, but if $a\ne 0$, then
$K$ is spacelike away from the axis of symmetry near $\mathcal{I}^+$.)
We define a $T$-invariant
vector field $V$ with the property
that $V=K$ near $\mathcal{H}^+$ and $V=T$ near $\mathcal{I}^+$ and $V$ is
timelike in the interior of $\mathcal{D}$. The energy associated to this vector field
is manifestly non-negative definite, though degenerate analogous to the $T$-energy
in the Schwarzschild case. In the case $a\ne0$, there is necessarily a region where
$V$ fails to be Killing.

Our third main theorem is  a degenerate $V$-energy analogue of
Theorem~\ref{THEOREM1} given by
\begin{bigthe}
\label{toinvert}
Forward evolution $(\ref{completof})$ extends to bounded maps
\[
\mathscr{F}_+:\mathcal{E}^V_{\Sigma_0^*}\to
\mathcal{E}_{\mathcal{H}^+_{\geq 0}}^K \oplus\mathcal{E}_{\mathcal{I}^+}^T, \qquad
\mathscr{F}_+:\mathcal{E}^V_{\mathring\Sigma}\to
\mathcal{E}_{\mathcal{H}^+}^K \oplus\mathcal{E}_{\mathcal{I}^+}^T, \qquad
\mathscr{F}_+:\mathcal{E}^V_{\overline\Sigma}\to
\mathcal{E}_{\overline{\mathcal{H}^+}}^K \oplus\mathcal{E}_{\mathcal{I}^+}^T.
\]
\end{bigthe}
See Theorems~\ref{boundNONDeg},~\ref{forwardt00} and~\ref{forwardt00bif}.
The above theorem requires a new version of the  boundedness part of
Theorem~\ref{theResult}
of~\cite{partIII},
depending only on   the degenerate energy. This result, which
is of independent interest, is stated as Theorem~\ref{boundDegen}
and proven in Section~\ref{secDegenBound}.
The reader can compare with
the higher-order weighted boundedness result  of Andersson and Blue~\cite{andblue} for the $|a|\ll M$ case,
whose degenerate horizon
weights are similar to the $V$-energy.

Let us note that the proof of Theorem~\ref{boundDegen} will require us to revisit the quantitative
study of the o.d.e.~$(\ref{radodehere})$
at fixed frequency,  on which the original results of~\cite{partIII} were
based, in particular
in the form of Theorem~\ref{odeEstimates}, and a new result,
Theorem~\ref{scatterEst2}, which we will prove here by adapting the proof
of~\cite{partIII}.
In particular, from these statements, one can
already infer novel results on the fixed-frequency scattering;
we defer specific discussion of these till Section~\ref{appliestofixed}.

\subsubsection{The $V$-energy backwards map}\label{vbackwardssec}
\label{intr-V-}
Our degenerate-energy class is indeed
suitable to construct a bounded inverse
of the map of Theorem~\ref{toinvert} and thus infer the existence of a satisfactory
scattering theory satisfying (a)--(c).
\begin{bigthe}
\label{existofsc}
There exist  bounded maps
\begin{equation}
\label{boundedinverse}
\mathscr{B}_-:\mathcal{E}_{\mathcal{H}^+_{\geq 0}}^K\oplus \mathcal{E}_{\mathcal{I}^+}^T
\to \mathcal{E}^V_{\Sigma_0^*},
\qquad
\mathscr{B}_-:\mathcal{E}_{\mathcal{H}^+}^K\oplus \mathcal{E}_{\mathcal{I}^+}^T
\to \mathcal{E}^V_{\mathring\Sigma}, \qquad
\mathscr{B}_-:\mathcal{E}_{\overline{\mathcal{H}^+}}^K\oplus \mathcal{E}_{\mathcal{I}^+}^T
\to \mathcal{E}^V_{\overline\Sigma},
\end{equation}
which  are two-sided inverses to the maps of Theorem~\ref{toinvert}, i.e.~$\mathscr{B}_-\circ\mathscr{F}_+=Id$
and $\mathscr{F}_+\circ \mathscr{B}_-=Id$.
\end{bigthe}
\[
\input{forscater1.pstex_t}\qquad
\input{forscater2.pstex_t}\qquad
\input{forscater2b.pstex_t}
\]

See Theorems~\ref{defScat},~\ref{isot0} and~\ref{torevisitornottorevisit}.
As explained in Section~\ref{NEWSE},
it is the existence of the map $\mathscr{B}_-$ which gives
the \emph{existence} of future scattering states (a),
the injectivity of $\mathscr{F}_+$ which gives the \emph{uniqueness}
of future scattering spaces (b), and the surjectivity of $\mathscr{B}_-$
corresponds to \emph{asymptotic completeness} of
future scattering states (c).
We note that the map of Theorem~\ref{THEOREM1} is in fact the restriction of
the first map of Theorem~\ref{toinvert}. Thus a corollary of
the above is that the map $\mathscr{F}$ of Theorem~\ref{THEOREM1} is
injective.
In this sense, for
the $N$-energy theory, one
still has uniqueness (b)--but \underline{not} existence (a)!--of scattering states.
Cf.~the discussion of the ill-posed problems of Section~\ref{theillpos}.

Let us note that in our proof, we construct $\mathscr{B}_-$ with the help of the frequency
domain,  again using our o.d.e.~result Theorem~\ref{scatterEst2}, together with
a decomposition first given in~\cite{partIII} and which exploits the fact that the span of $T$
and $\Phi$ is timelike
(See Section~\ref{someboundedargument}),  to give
us the quantitative statement of boundedness. Due to this use of the frequency
domain, it is in fact the map
$\mathscr{B}_-:\mathcal{E}_{\mathcal{H}^+}^K\oplus \mathcal{E}_{\mathcal{I}^+}^T
\to \mathcal{E}^V_{\mathring\Sigma}$
which is most natural to construct first.

It is perhaps worth explicity noting that even to show the existence of  $\mathscr{B}_-$,
we require appeal to an o.d.e.~result which in essence already embodies the totality
of the quantitative decay statement for the Cauchy problem $(\ref{WAVE})$.
This should emphasise how intricately tied
 in the Kerr case the problem of boundedness is
to the problem of quantitative decay.
This is in contrast to many usual problems in scattering theory where
``existence of scattering states'' (cf.~(a))
is a relatively soft result, which can be proven
independently of the structure necessary to obtain asymptotic completeness-type statements.

\subsubsection{Existence and boundedness of the $S$-matrix}
\label{EBSmat}
We will base our discussion here on the scattering theory associated to
$\mathring\Sigma$ or $\overline{\Sigma}$.
First, note that applying a discrete isometry of $\mathcal{D}$ which interchanges the future and past of $\mathring{\Sigma}$,
we
infer analogously to Theorems~\ref{toinvert} and~\ref{existofsc} the existence
of  bounded \underline{past} forward maps,
\[
\mathscr{F}_- :\mathcal{E}_{\mathring{\Sigma}}^V \to  \mathcal{E}_{\mathcal{H}^-}^K\oplus
\mathcal{E}_{\mathcal{I}^-}^T,
\qquad
\mathscr{F}_- :\mathcal{E}_{\overline{\Sigma}}^V \to  \mathcal{E}_{\overline{\mathcal{H}^-}}^K\oplus
\mathcal{E}_{\mathcal{I}^-}^T,
\]
and the corresponding bounded two-sided inverses
\[
\mathscr{B}_+:
\mathcal{E}_{\mathcal{H}^-}^K\oplus
\mathcal{E}_{\mathcal{I}^-}^T \to \mathcal{E}_{\mathring{\Sigma}}^V,
\qquad
\mathscr{B}_+:
\mathcal{E}_{\overline{\mathcal{H}^-}}^K\oplus
\mathcal{E}_{\mathcal{I}^-}^T \to \mathcal{E}_{\overline{\Sigma}}^V.
\]
We thus have both existence and uniqueness for \emph{past} scattering states as well
as \emph{past} asymptotic completeness.

The following is then an immediate corollary
\begin{bigthe}
\label{SMatrix}
The composition of $\mathscr{S}= \mathscr{F}_+ \circ \mathscr{B}_+$ defines bounded
invertible maps
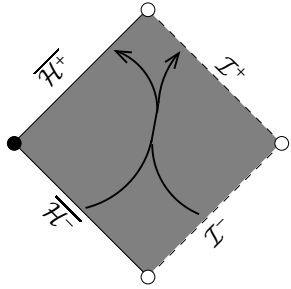
\begin{equation}
\label{heretheS}
\mathscr{S}:\mathcal{E}_{\mathcal{H}^-}^K\oplus\mathcal{E}_{\mathcal{I}^-}^T \to
\mathcal{E}_{\mathcal{H}^+}^K\oplus\mathcal{E}_{\mathcal{I}^+}^T,\qquad \mathscr{S}:\mathcal{E}_{\overline{\mathcal{H}^-}}^K\oplus\mathcal{E}_{\mathcal{I}^-}^T \to
\mathcal{E}_{\overline{\mathcal{H}^+}}^K\oplus\mathcal{E}_{\mathcal{I}^+}^T.
\end{equation}
\end{bigthe}
\[
\input{forscater3.pstex_t}
\]

The boundedness $\|\mathscr{S}\|\le C$ of the map $\mathscr{S}$
in the operator norm
should be viewed as the quantitative replacement for the usual unitarity
property.

\subsubsection{A physical space theory of superradiant reflection}
\label{pST}
Given the scattering map $\mathscr{S}$, we can now give an account of
superradiant reflection in physical space, i.e.~in the ``time domain''.

Recall the standard physical set-up: One wishes to study the scattering of waves
with no ingoing contribution from the past event horizon $\mathcal{H}^-$ and we are interested
only in the part of the wave reflected to future null infinity $\mathcal{I}^+$.
We thus
 pass from  $\mathscr{S}$
to the transmission map $\mathscr{T}$ and reflection map $\mathscr{R}$ defined by
\begin{equation}
\label{wepasstohere}
\mathscr{T}=\pi_{\mathcal{E}^K_{\mathcal{H}^+}}\circ
\mathscr{S}|_{\{0\}\oplus \mathcal{E}^T_{\mathcal{I}^-}}, \qquad
\mathscr{R}=\pi_{\mathcal{E}^T_{\mathcal{I}^+}}\circ
\mathscr{S}|_{\{0\}\oplus \mathcal{E}^T_{\mathcal{I}^-}}
\end{equation}
where
\[
\pi_{\mathcal{E}^T_{\mathcal{I}^+}} : \mathcal{E}^K_{\mathcal{H}^+} \oplus
\mathcal{E}^T_{\mathcal{I}^+} \to \mathcal{E}^T_{\mathcal{I}^+},
\qquad
\pi_{\mathcal{E}^K_{\mathcal{H}^+}} : \mathcal{E}^K_{\mathcal{H}^+} \oplus
\mathcal{E}^T_{\mathcal{I}^+} \to \mathcal{E}^K_{\mathcal{H}^+}
\]
are the natural projections.
Note that this map does not depend on whether we consider the domain
of $\mathscr{S}$ to be either of the choices in $(\ref{heretheS})$.
The map
\[
\mathscr{R}: \mathcal{E}^T_{\mathcal{I}^-} \to \mathcal{E}^T_{\mathcal{I}^+}
\]
takes an asymptotic state corresponding to an incoming wave packet supported solely on past infinity
$\mathcal{I}^-$ (i.e.~with no incoming radiation from $\mathcal{H}^-$) and maps it
to the part of the asymptotic state which is reflected to future null infinity $\mathcal{I}^+$ (i.e.~projecting out the
part transmitted to the future horizon $\mathcal{H}^+$). Similarly,
the map
\[
\mathscr{T}: \mathcal{E}^T_{\mathcal{I}^-} \to \mathcal{E}^K_{\mathcal{H}^+}
\]
takes an asymptotic state corresponding to an incoming wave packet supported solely on past infinity
$\mathcal{I}^-$ and maps it to the part of the asymptotic state which is transmitted to
the future event horizon $\mathcal{H}^+$.

Since $\mathscr{S}|_{\{0\}\oplus \mathcal{E}^T_{\mathcal{I}^-}}=
\mathscr{T}\oplus\mathscr{R}$, the boundedness of $\mathscr{S}$ above immediately yields
the strictly weaker statement
\begin{bigthe}
\label{isbound}
The reflection and transmission maps $\mathscr{R}$ and $\mathscr{T}$ are bounded,
i.e.~$\| \mathscr{R} \| , \| \mathscr{T} \| \le C$.
\end{bigthe}
See the first statement of Theorem~\ref{notreallysparta}.
In view of the relation with the fixed-frequency theory to be discussed in Section~\ref{appliestofixed} below,
we have
\begin{equation}
\label{inparticularwh}
\sup_{(\omega, m, \ell)} |\mathfrak{R}(\omega, m, \ell)|=\|\mathscr{R}\|,
\end{equation}
and thus, a posteriori,
Theorem~\ref{isbound}
gives in particular $(\ref{introubo})$. We note however that in the logic of the proof,
we  will have essentially
already used  $(\ref{introubo})$ in proving the boundedness of both
the maps $\mathcal{F}_+$ and $\mathcal{B}_-$.

Let
us here already mention a further application of the relationship $(\ref{inparticularwh})$
to our physical-space scattering theory. First, note that
general soft o.d.e.~theory
is sufficient to show that the reflection coefficient satisfies
$|\mathfrak{R}(\omega, m, \ell)|>1$
for any superradiant frequency triple (see
Corollary~\ref{superradiantAmp}).
Thus, one immediately obtains from $(\ref{inparticularwh})$ the statement
\begin{bigthe}
\label{lowbn}
For $a\ne0$, the reflection map $\mathscr{R}$ has norm strictly greater than
1, i.e.~$\| \mathscr{R} \|>1$.
\end{bigthe}
See the second statement of Theorem~\ref{notreallysparta}.
The above theorem can be viewed as
the definitive physical-space interpretation of the phenomenon
of superradiant reflection.
To connect with the numerical setting often studied~(e.g.~\cite{numeric, LazRacz})
in which it is difficult
to implement past scattering data on $\mathcal{I}^-$,
we will extract in addition the following somewhat
less natural statement concerning Cauchy data on $\mathring\Sigma$ via a density argument
(see Theorem~\ref{makeitsuperr}):
\emph{There exists a smooth solution $\psi$ with the property that its $T$-energy
flux through $\mathcal{I}^+$ is greater than its $T$-energy flux through $\mathring{\Sigma}$
and moreover,
the support of the solution on $\mathring{\Sigma}$ is compact and can be made arbitrarily close to spatial
infinity.} Cf.~\cite{FKSY}.
This addresses in particular some questions raised in~\cite{LazRacz}.

\subsubsection{Pseudo-unitarity and non-superradiant unitarity}\label{pseunitsec}
As we have already discussed, when $a \neq 0$ one does not have a unitary
scattering theory; however, one still expects to recover the conservation of the indefinite inner product associated to the $T$-energy, provided this inner product is finite.

The $T$-energy is \underline{not} finite on the full  domain of the scattering matrix
$\mathscr{S}$ of
$(\ref{heretheS})$.
It is, however, finite if one for instance restricts to past scattering
data supported only on $\mathcal{I}^-$.
Recalling the notation~$(\ref{wepasstohere})$, one
statement of ``pseudo-unitarity'' is then captured by the following theorem.

\begin{bigthe}\label{bb}The map $\mathscr{T}\oplus\mathscr{R}$ preserves the $T$-energy:
\[\int_{\mathcal{H}^+}\mathbf{J}^T_{\mu}\left[\mathscr{T}\upphi\right]n^{\mu}_{\mathcal{H}^+} + \int_{\mathcal{I}^+}\mathbf{J}^T_{\mu}\left[\mathscr{R}\upphi\right]n^{\mu}_{\mathcal{I}^+} = \int_{\mathcal{I}^-}\mathbf{J}^T_{\mu}\left[\upphi\right]n^{\mu}_{\mathcal{I}^-}.\]
\end{bigthe}
In particular, if the right hand side above is bounded, then the first term on the left hand
side, which is unsigned, is integrable.
See Theorem~\ref{aunitscatter}.

If we restrict to past scattering data on $\mathcal{H}^-\cup \mathcal{I}^-$ that
are \underline{non-superradiant},
i.e.~supported in frequency space outside the superradiant range,
then our scattering map $\mathscr{S}$ will indeed be unitary in the usual sense.
For this we define
Hilbert spaces
$\mathcal{E}^{T,\natural}_{\mathcal{H}^{\pm}}\oplus \mathcal{E}^{T,\natural}_{\mathcal{I}^{\pm}}$ by the completion under the inner product
\begin{equation}\label{theunitprod}
\langle \left(\uppsi_1,\upphi_1\right),\left(\uppsi_2,\upphi_2\right)\rangle = \int_{-\infty}^{\infty}\sum_{m\ell}\left[\omega\left(\omega-\upomega_+m\right)\text{Re}\left(\hat{\uppsi}_1\overline{\hat{\uppsi}}_2\right) + \omega^2\text{Re}\left(\hat{\upphi}_1\overline{\hat{\upphi}_2}\right)\right]
\end{equation}
of scattering data whose Fourier transforms are supported in the
\emph{non-superradiant range}
\[
\{(\omega,m,\ell) : \omega(\omega-\upomega_+m) > 0\}.
\]

We then have
\begin{bigthe}\label{bbb}The restriction of the first map of $(\ref{heretheS})$ extends
to a unitarity isomorphism
 $\mathscr{S} : \mathcal{E}^{T,\natural}_{\mathcal{H}^-} \oplus \mathcal{E}^{T,\natural}_{\mathcal{I}^-} \to \mathcal{E}^{T,\natural}_{\mathcal{H}^-} \oplus \mathcal{E}^{T,\natural}_{\mathcal{I}^-}$ with respect to the positive definite inner product~(\ref{theunitprod}).
\end{bigthe}
See Theorem~\ref{unitNonsup}.
Note that the above theorem retrieves in particular the unitarity of
the first map of $(\ref{heretheS})$ in the Schwarzschild case $a=0$ (which we in fact
provide an independent treatment of; see
Theorem~\ref{schscatunit})
as well as the unitarity of $\mathscr{S}$ restricted to
axisymmetric data in the full $|a|<M$ case.

\subsubsection{Uniqueness of scattering states for ill-posed scattering data}
\label{theillpos}
Finally, we note that our scattering theory allows us to make the following
injectivity statements which can be understood as
statements just of \emph{uniqueness} of
scattering states (cf.~(b)) for
scattering data determined on any of the four ``ill-posed'' pairs
of asymptotic boundaries
$\overline{\mathcal{H}^+}\cup{\overline{\mathcal{H}^-}}$,
$\mathcal{I}^+\cup\mathcal{I}^-$,  $\overline{\mathcal{H}^+}\cup \mathcal{I}^-$
and
$\overline{\mathcal{H}^-}\cup \mathcal{I}^+$.
\begin{bigthe}
\label{introuC}
The maps
\[
\mathscr{F}:\mathcal{E}_{\overline{\Sigma}}^{V}\to\mathcal{E}^K_{\overline{\mathcal{H}^+}}\oplus\mathcal{E}^K_{\overline{\mathcal{H}^-}}, \qquad
\mathscr{F}:\mathcal{E}_{\overline{\Sigma}}^{V}\to \mathcal{E}^T_{\mathcal{I}^+}\oplus\mathcal{E}^T_{\mathcal{I}^-},
\qquad
\mathscr{F}:\mathcal{E}_{\overline{\Sigma}}^{V}\to \mathcal{E}^K_{\overline{\mathcal{H}^+}}\oplus\mathcal{E}^T_{\mathcal{I}^-},
\qquad
\mathscr{F}:\mathcal{E}_{\overline{\Sigma}}^{V}\to \mathcal{E}^K_{\overline{\mathcal{H}^-}}\oplus\mathcal{E}^T_{\mathcal{I}^+}
\]
are all injective.
\end{bigthe}
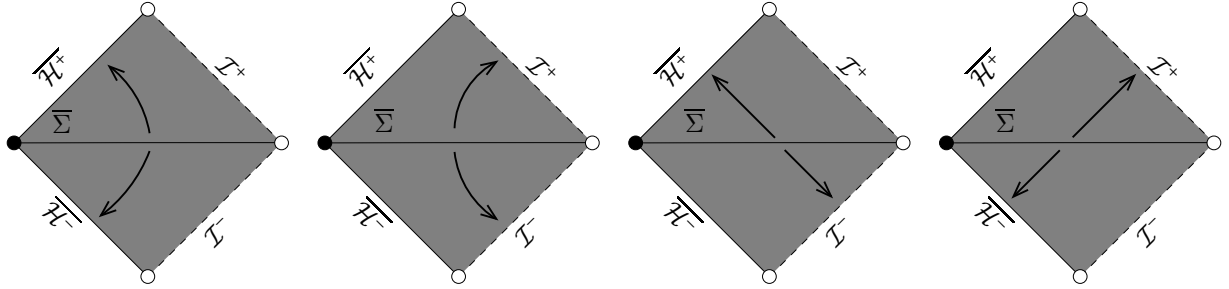
\[
\input{forscater4a.pstex_t}\,\,\,\,\,\,\,
\input{forscater4b.pstex_t}\,\,\,\,\,\,\,
\input{forscater4c.pstex_t}\,\, \,\,\,\,\,
\input{forscater4d.pstex_t}
\]
See Corollary~\ref{uniquecont}.
Together with the previous results, the above implies that finite $V$-energy solutions are uniquely determined by their
fluxes to any pair of the set $\{\mathcal{H}^+, \mathcal{H}^-, \mathcal{I}^+, \mathcal{I}^-\}$.
In contrast, however, to the forward maps of Theorem~\ref{toinvert}, it follows
already from general local ill-posedness type results  for the wave equation
(see e.g.~the classic textbook~\cite{hadamard}) that \underline{the above maps
$\mathscr{F}$ are not
surjective}.
Thus, one does not have the analogue of ``existence of scattering states'' (cf.~(a))
for scattering states parameterized as above.\footnote{\label{footnote3}This is of
course in sharp distinction to the fixed-frequency
theory, for which ``existence of scattering states'' associated to $\mathcal{H}^+\cup\
\mathcal{H}^-$ and $\mathcal{I}^+\cup\mathcal{I}^-$, respectively, corresponds precisely to the existence and
linear independence of
the pairs $U_{\rm hor}$, $\overline{U}_{\rm hor}$ or
alternatively $U_{\rm inf}$, $\overline{U}_{\rm inf})$
described in the beginning of this introduction, on which the whole theory is based.}

\subsection{Applications to fixed frequency scattering theory}
\label{appliestofixed}
As we have discussed, the proofs of our theorems of physical
space scattering theory required us to revisit  our quantitative
fixed frequency study of the o.d.e.~$(\ref{radodehere})$  conducted
in~\cite{partIII}.
Thus, along the way, we have in fact obtained new results for the
fixed-frequency scattering theory initiated by Chandrasekhar~\cite{chandrasekhar},
as well as a precise connection
of the two through the scattering matrix $\mathscr{S}$.
We collect these statements in this section.

\subsubsection{Uniform boundedness of the coefficients $\mathfrak{R}$ and $\mathfrak{T}$}\label{uniformboundsec}
We begin with the statement of the uniform boundedness of the transmission
and reflection coefficients.
\begin{bigthe}
\label{isbound2}
The reflection and transmission coefficients
as normalised in~$(\ref{intheintroform})$
are uniformly
bounded over all frequencies:
\begin{equation}
\label{uniformlybounded}
\sup_{(\omega, m, \ell)} |\mathfrak{R}(\omega, m, \ell)| \le C, \qquad
\sup_{(\omega, m, \ell)} |\mathfrak{T}(\omega, m, \ell)| \le C.
\end{equation}
\end{bigthe}
We in fact have a statement for the complete
set of coefficients where we also allow for waves normalised to the past
horizon. See Theorem~\ref{refltransBound}.

We will infer the above theorem as an immediate corollary of our
o.d.e.~estimate Theorem~\ref{scatterEst2}, which itself is an easy adaptation of
an estimate of our previous~\cite{partIII}. We emphasise
again that this result requires in particular appeal to the
real-mode stability theorem of~\cite{realmodestability}.

To connect with the pioneering heuristic work of Starobinski~\cite{staro}, we may define the following constant
depending only on the Kerr parameters
\[
S(a,M) \doteq \sup_{(\omega, m, \ell)} |\mathfrak{R}(\omega, m, \ell)|,
\]
and by Theorem~\ref{isbound2}, together with the soft statement
Corollary~\ref{superradiantAmp}
(mentioned already in the context of Theorem~\ref{lowbn}),
we have
\[
1< S(a,M)<\infty, \qquad 0<|a|<M.
\]
(For $a=0$ we have of course $S(0,M)\le 1$ and in fact, by an easy high
angular frequency $\ell$ estimate given by Corollary~\ref{largelR},
$S(0,M)=1$.)
It would be very interesting from the point of view
of applications, following~\cite{staro}, to find effective
upper and
lower bounds for $S(a,M)$, and to understand in particular
the limit
\begin{equation}
\label{toextremal}
\lim_{|a|\to M} S(a,M).
\end{equation}

\subsubsection{Connection with physical-space theory}
\label{connectPST}
The full scattering map $\mathscr{S}$ defined in Section~\ref{EBSmat}
can be represented as a generalised Fourier
transform involving the
transmission and reflection coefficients $\mathfrak{T}$ and $\mathfrak{R}$
defined via $(\ref{intheintroform})$, together with coefficients $\tilde{\mathfrak{T}}$ and $\tilde{\mathfrak{R}}$
associated to analogously defined solutions $U$ of $(\ref{radodehere})$
normalised to the past event horizon
$\mathcal{H}^-$. So as not to define the latter here,
for convenience, let us simply state the relations for the physical space transmission and
reflection maps
$\mathscr{T}$ and $\mathscr{R}$ defined in Section~\ref{pST}:

\begin{bigthe}
\label{PSpREPint}
We may represent
\[
\mathscr{R}\left[\upphi\right]= \frac{1}{\sqrt{2\pi}}\int_{-\infty}^{\infty}\sum_{m\ell}a_{\mathcal{I}^-}\mathfrak{R}\,e^{-it\omega}e^{im\phi}S_{m\ell}(a\omega, \cos\theta)\ d\omega
\]
and
\[
\mathscr{T}\left[\upphi\right]= -\frac{1}{\sqrt{4M\pi r_+}}\int_{-\infty}^{\infty}\sum_{m\ell}\left(\frac{\omega}{\omega-\upomega_+m}\right)a_{\mathcal{I}^-}\mathfrak{T}\,e^{-it\omega}e^{im\phi}S_{m\ell}(a\omega,
\cos\theta)\ d\omega.
\]
Here
\[-i\omega a_{\mathcal{I}^-} \doteq \frac{1}{\sqrt{2\pi}}\int_{-\infty}^{\infty}\int_{\mathbb{S}^2}\partial_t\upphi \,e^{it\omega}e^{-im\phi}S_{m\ell}(a\omega, \cos \theta)\, \sin\theta\ dt\, d\theta\, d\phi.\]
In particular, $(\ref{inparticularwh})$ holds.
\end{bigthe}
See Theorem~\ref{constructaway} for the full statement concerning $\mathscr{S}$.

In fact, a posteriori, in view
of Theorem~\ref{constructaway}, the
statement of Theorem~\ref{refltransBound} is
equivalent to the
boundedness of the map $\mathscr{S}$ of Theorem~\ref{isbound}.
We note in contrast that the  boundedness of
the maps $\mathscr{F}_+$ and $\mathscr{B}_+$ individually (already asserted)
does \underline{not} have an obvious natural interpretation purely in terms of the formal fixed frequency scattering theory. Similarly, the boundedness statement of
Theorem~\ref{THEOREM1} (and the boundedness statement of~\cite{partIII} quoted
here as Theorem~\ref{theResult} which concerns
boundedness through a spacelike foliation) cannot be directly interpreted
purely in terms of the formal fixed frequency scattering theory.
These are all distinct manifestations
of ways that the phenomenon of ``superradiance'' allowed by the presence of
an ergoregion can be quantified.
As with the question of the finiteness of $(\ref{toextremal})$, it
is a completely open question
which if any of these boundedness statements survives in the extremal case
$|a|=M$.
See~\cite{aretakiskerr}.

\subsection{Nonlinear problems and scattering constructions of dynamical black holes}
\label{Nonlinearps}
We make a few comments on
scattering theory for non-linear generalisations of $(\ref{WAVE})$.
Perhaps the ultimate nonlinear such
 generalisation  is provided by the Einstein vacuum equations
\begin{equation}
\label{EVac}
{\rm Ric}(g)=0
\end{equation}
themselves, where the background geometry is now itself unknown.

The problem of characterizing \emph{all} ``admissible'' solutions by appropriate asymptotic states
may turn out to be too ambitious for equations as nonlinear as $(\ref{EVac})$.
The mere constructing of some, however, in the spirit of
the map $(\ref{indeeDEX})$,
can serve as an important way of obtaining non-trivial examples
of  solution spacetimes which cannot otherwise  easily be inferred to exist. A result in that direction
has recently been provided by
\begin{theorem*}\cite{BHscatter}
Consider asymptotic data on $\mathcal{H}^+\cup \mathcal{I}^+$
for the Einstein vacuum equations~(\ref{EVac}), decaying towards Kerr data corresponding
to $g_{a,M}$ with $|a| \leq M$ at a sufficiently fast \underline{exponential} rate.  Then
there exists a vacuum spacetime $(\mathcal{M},g)$ attaining the data.
\end{theorem*}
The spacetimes $(\mathcal{M},g)$ constructed in the above are in fact the first known
examples of dynamical vacuum black holes settling down to Kerr.

The above
theorem can be thought of as a non-linear analogue
of the map~$(\ref{indeeDEX})$ (for energies which have additional weights in $r$ however!).
In fact, proving the above requires capturing a complicated $r^p$-hierarchy of decay of
various components of curvature which in turn allows one to identify a null condition in
the implicit non-linearities in $(\ref{EVac})$. (We note in contrast that without additional special
structure, the analogue of the above theorem
does \underline{not} hold even say for the general scalar semilinear equation
of the form $\Box_g\psi = Q(\nabla \psi,\nabla\psi)$.)
 We refer the reader to~\cite{BHscatter}.

In the context of our present paper, let us simply remark that
the degenerate $V$-scattering theory developed here, together with the blow-up result
Theorem~\ref{noScatter} and the upcoming~\cite{DafLuk}, gives further support to the
 following
conjecture of~\cite{BHscatter}:

\begin{conjecture*}\cite{BHscatter}
\label{BHscatterconji}
Consider asymptotic data on $\mathcal{H}^+\cup \mathcal{I}^+$
as above but which decay
to $g_{a,M}$ only at a sufficiently fast inverse \underline{polynomial} rate. Then
there exists a vacuum spacetime $(\mathcal{M},g)$ attaining the data.
For generic such data, $\mathcal{H}^+$ is a ``weak null singularity'' across
which the metric extends continuously but with Christoffel symbols which
fail to be locally square integrable.
\end{conjecture*}

See the discussion of Section~\ref{constructi} and~\cite{jluk}.

\subsection{Outline}
\label{OUTLhere}
The logic of the paper will depart slightly from the order we have presented the main
results above. We thus close this introduction with a brief section by section outline of the contents of the remainder of the paper, highlighting in bold where each of the main theorems above
are actually proven. (More detailed outlines
will be given in the body of the paper at the beginning of
each individual section.)

In Section~\ref{prelim}, we briefly review the structure of the Kerr spacetime, introduce various conventions, and quote some previous results on forward evolution
which will be important, in particular, we will state  general well-posedness
results (Propositions~\ref{WPProp}--\ref{phybackwp}),
precise versions of our previous boundedness and integrated local decay
results of~\cite{partIII} (quoted as Theorem~\ref{theResult}
and the higher order Theorem~\ref{h.o.s.}) as well as a general
$r^p$ weighted estimate (Proposition~\ref{rp}) which we will derive from~\cite{icmp}.

In Section~\ref{radenergy}, we define and establish some basic properties of the radiation fields and energy fluxes along $\mathcal{H}^+_{\geq 0}$ (or $\overline{\mathcal{H}^+}$) and $\mathcal{I}^+$ for solutions $\psi$ to~(\ref{WAVE}) arising from smooth initial data along $\Sigma_0^*$
(or $\overline{\Sigma}$)
which are compactly supported. The main result is Proposition~\ref{radinfwelldef}, which is the precise
statement of {\bf Proposition~\ref{PROP1}} above.

In Section~\ref{cartsecmicrorad}, we first review Carter's separation of variables for the wave equation and then define the \emph{radial o.d.e.}, recalling also some results from its basic asymptotic analysis.  This will allow
us to define the reflection and transmission coefficients (Definition~\ref{reftransDef}),
deduce fixed-frequency superradiant amplification in the form of Corollary~\ref{superradiantAmp},
and define the so-called \emph{microlocal radiation fields} (Definition~\ref{microRad}) and
\emph{fluxes} (Definition~\ref{microFluxDef}).

In Section~\ref{radodesec} we establish various estimates for the radial o.d.e.~and give some useful applications. We start by proving Theorem~\ref{scatterEst2}, an estimate for general solutions to the homogeneous radial o.d.e. The proof of Theorem~\ref{scatterEst2} will heavily rely on our o.d.e.~estimates from~\cite{partIII}.
Next, in Section~\ref{herethesuperproof} we establish an important estimate for $U_{\rm hor}$
in the superradiant regime
(Proposition~\ref{asymExpandGood}).  We then use related ideas in Section~\ref{ellisquitelarge} to prove Proposition~\ref{largelT} which states that for fixed $\omega$ and $m$, $\mathfrak{T}$ vanishes in the large-$\ell$ limit. In Section~\ref{isovanish} we show that for each $m$ and $\ell$, the reflection coefficient $\mathfrak{R}$ is not identically $0$ as a function of $\omega$. In Section~\ref{MICROLrp} we prove Proposition~\ref{microrp}, which is the microlocal version of the $r^p$ estimates of~\cite{icmp} (cf. Proposition~\ref{rp}). The goal of Section~\ref{Qestie} is to prove Proposition~\ref{microradconverge} which establishes uniform estimates, over all frequency parameters, for the rate of convergence of solutions to the radial o.d.e.~to their microlocal radiation fields.
Finally, in Section~\ref{relatephy} we prove Propositions~\ref{microEqualInf} and~\ref{microequalhor} which establish that for suitable solutions $\psi$ to the wave equation, the microlocal radiation fields are essentially the Fourier transform of the physical space radiation fields defined in
Section~\ref{radenergy}.

In Section~\ref{secDegenBound}, we prove Theorem~\ref{boundDegen},
the statement that the total flux to null infinity $\mathcal{I}^+$ and the degenerate $K$-flux to the
horizon $\mathcal{H}^+$ of a solution $\psi$ to the wave equation may be
controlled by the $V$-energy of $\psi$ along $\Sigma_0^*$.
The theorem is stated in Section~\ref{MTandC},
after which the reader  impatient to proceed to the construction of our scattering theory
may skip to Section~\ref{forward} below.
The proof of Theorem~\ref{boundDegen},
which occupies Sections~\ref{outline}
is a modification of the proof of Theorem~\ref{theResult} quoted from~\cite{partIII}.
In a brief aside in Section~\ref{yetanotheraside}, we shall state
Theorem~\ref{degentheResult},
which is the full
degenerate-energy analogue of our results of~\cite{partIII}, quoted
as Theorem~\ref{theResult} above. We emphasise that Theorem~\ref{degentheResult}
is not in fact necessary for
the rest of the paper, and we defer its proof to Section~\ref{asideforthepr},
where we can make
use of the backwards maps of our scattering theory.

 In Section~\ref{forward}, we introduce
 the $\mathcal{E}_{\mathring{\Sigma}}^V$ and $\mathcal{E}_{\mathring{\Sigma}}^N$ spaces, etc.,
 and define the various ``forward'' maps and establish their boundedness.
 Theorem~\ref{boundNONDeg},  the precise version of {\bf Theorem~\ref{THEOREM1}},
 is independent of Section~\ref{secDegenBound}, as it relies directly on Theorem~\ref{theResult} of~\cite{partIII}.
 Theorem~\ref{forwarddegen}, on the other hand, which together
 with its corollaries Theorems~\ref{forwardt00} and~\ref{forwardt00bif} embodies the precise version of
 {\bf Theorem~\ref{toinvert}}, uses in  a fundamental way Theorem~\ref{boundDegen}.

  In Section~\ref{secScatter}, we prove
  first Theorem~\ref{isot0}, then  Theorem~\ref{defScat},
  then Theorem~\ref{torevisitornottorevisit}.
This obtains all statements in
{\bf Theorem~\ref{existofsc}}.
As an aside in Section~\ref{asideforthepr}, we obtain the proof of Theorem~\ref{degentheResult}
referred to above.
 Next, we construct the ``scattering'' map $\mathscr{S}$ and show that it is a bounded
 invertible map
 from data along $\mathcal{H}^-\cup\mathcal{I}^-$ to data along $\mathcal{H}^+\cup\mathcal{I}^+$
  (Theorem~\ref{scatAll}, the precise version of {\bf Theorem~\ref{SMatrix}}).
  We then prove Theorem~\ref{constructaway} which establishes a formula for the scattering map $\mathscr{S}$ explicitly exhibiting the roles of the reflexion and transmission coefficients. This formula will in particular establish the relationship between physical space and fixed frequency
  scattering theories embodied by {\bf Theorem~\ref{PSpREPint}}. Finally, as an additional aside
  in Section~\ref{veryselfcontained},
  we give an alternative self-contained treatment for the Schwarzschild case where it
  is possible to exploit purely physical space arguments.

  In Section~\ref{somanyapplications}, we begin by interpreting our scattering results for the  reflection operator
$\mathscr{R}$. Theorem~\ref{notreallysparta} combines the statements of
{\bf Theorems~\ref{isbound} and~\ref{lowbn}}.
We also infer the related Theorem~\ref{makeitsuperr}.  Following this, we study the pseudo-unitarity properties of $\mathscr{S}$ and prove the corresponding Theorems~\ref{aunitscatter} and~\ref{unitNonsup} (cf.~{\bf Theorems~\ref{bb} and~\ref{bbb}}). Finally, our ``uniqueness
of improper scattering states'' results are stated   as Theorem~\ref{uniquecont}, giving
{\bf Theorem~\ref{introuC}}.

  In Section~\ref{correlate}, we prove  Theorem~\ref{noScatter}, the statement that solutions
  $\psi$ of $(\ref{WAVE})$ on Schwarzschild
  whose radiation fields on the horizon $\mathcal{H}^+$ have
  a precise polynomial tail and whose
  radiation fields  on $\mathcal{I}^+$ vanish
   must necessarily have infinite $N$-energy on the hypersurface $\Sigma_0^*$.
  This statement can be understood independently of the results concerning
  our scattering maps, and indeed, Sections~\ref{metricandwave} and~\ref{proofofithere}
  can be read independently of   the rest of the paper.
  In Sections~\ref{constructi} and~\ref{nonsurji}
  we will then return to the scattering framework of our paper. We first use the
 backwards map of our $V$-scattering theory to infer the
 existence (See Corollary~\ref{awonderfulcor}) of
 solutions satisfying the
  assumptions of Theorem~\ref{noScatter}.
 Finally, we  infer {\bf Theorem~\ref{failsurj}} as Corollary~\ref{thenonsurcor}.

\section{Preliminaries}\label{prelim}
We begin in this section with various preliminaries.

After reviewing our notations for energy currents associated to vector
fields in Section~\ref{currentssec},
we will define  carefully in Section~\ref{diffCoord}
the ambient spacetime $\mathcal{D}$
(and related subsets) on which we will consider the Kerr metric
$g_{a,M}$ for subextremal values $|a|<M$. Our conventions for constants depending only
on the Kerr parameters will be reviewed in Section~\ref{gencon}. These follow
our conventions from~\cite{partIII}.
Some auxilliary useful vector fields will be presented in Section~\ref{useful?}.

It will be useful to define a hyperboloidal-type foliation $S_\tau$
 of $\mathcal{R}$    and we shall do this in Section~\ref{hyperboloid}.
 The form of the $T$ energy-flux through such a foliation is recorded in Lemma~\ref{iftediou}.
 Section~\ref{wellposedness} states general well posedness
results (Proposition~\ref{WPProp}--\ref{phybackwp})
 for the wave equation $(\ref{WAVE})$ on the Kerr exterior. We shall then quote
 our boundedness and integrated decay statement from~\cite{partIII} in Section~\ref{edweivaiola},
  as Theorem~\ref{theResult} and the higher order Theorem~\ref{h.o.s.}.
The foliation of Section~\ref{hyperboloid} will then
 allow us in Section~\ref{rpSec} to easily quote the $r^p$ hierarchy of estimates (introduced in~\cite{icmp})
 in the form of Proposition~\ref{rp} and the higher order Proposition~\ref{rpCommute}.

\subsection{Currents}
\label{currentssec}
Given a general Lorentzian manifold $(\mathcal{M},g)$,
let $\Psi$
be a sufficiently regular complex function.
We define
\[
{\bf T}_{\mu\nu}[\Psi]
\doteq \text{Re}\left(\partial_\mu\Psi\overline{\partial_\nu\Psi}\right) -\frac12 g_{\mu\nu}g^{\alpha\beta}
\text{Re}\left(\partial_\alpha\Psi \overline{\partial_\beta\Psi}\right).
\]
Given a sufficiently regular vector field $X$ on $\mathcal{M}$,
we define the currents
\[
{\bf J}^X_\mu[\Psi] = {\bf T}_{\mu\nu}[\Psi] X^\nu,
\]
\[
{\bf K}^X[\Psi] ={\bf T}_{\mu\nu}[\Psi]\nabla^\mu X^\nu = \frac{1}{2}{\bf T}_{\mu\nu}\text{}^{(X)}\pi^{\mu\nu},
\]
\[
\mathcal{E}^X[\Psi] = -\text{Re}\left((\Box_g\Psi)\overline{X^\nu \Psi_{,\nu}}\right).
\]
Here $\text{}^{(X)}\pi^{\mu\nu} \doteq \nabla^{\mu}X^{\nu} + \nabla^{\nu}X^{\mu}$ is the deformation tensor of $X$. In particular, ${\bf K}^X=0$ where $X$ is Killing.

Recall the fundamental identity:
\[\nabla^{\mu} {\bf J}^X_\mu[\Psi] = {\bf K}^X[\Psi] - \mathcal{E}^X[\Psi].\]
Then the divergence identity between two homologous spacelike
hypersurfaces $S^-$, $S^+$, bounding a region $\mathcal{C}$,
with $S^+$ in the future of $S^-$, yields
\begin{equation}
\label{ingeneralform}
\int_{S^+}{\bf J}^X_\mu [\Psi]n^\mu_{S^+}+
\int_{\mathcal{C}} ({\bf K}^X[\Psi] - \mathcal{E}^X[\Psi])
=\int_{S^-}{\bf J}^X_\mu [\Psi]n^\mu_{S^-},
\end{equation}
where $n_{S_{\pm}}$ denotes the future directed timelike unit normal, and the induced volume forms are to be understood.
\begin{remark}\label{volumeforms}In general, in integrals we will either write explicitly a volume form or it is to be understood that the integration is with respect to the induced volume form. In the case of a null hypersurface, the volume element depends on the choice of a null generator and is defined so that the divergence theorem holds.
\end{remark}

We direct the reader unfamiliar with the use of energy currents to the concise introductory book~\cite{alinhacbook}. See~\cite{christbook} for a systematic discussion.

\subsection{The ambient differentiable structure and the Kerr metric}
\label{diffCoord}
In this section we will briefly review the background differentiable structure and various convenient coordinate systems for the Kerr spacetime. We direct the reader to~\cite{partsIandII} and~\cite{partIII} for a more thorough discussion of our conventions and to the books~\cite{hawkingellis} and~\cite{oneill} for a proper introduction to Kerr.

As is well known, the Kerr spacetimes $(\mathcal{M},g_{a,M})$ are a $2$-parameter family of spacetimes which in the parameter range $|a| < M$ may be thought of as the maximal Cauchy development of a Cauchy hypersurface with two asymptotically flat ends.
The spacetime
$\mathcal{M}$ possesses a bifurcate Killing horizon separating two asymptotically flat exterior regions from a \emph{black hole} and a \emph{white hole} region (in the case $a\ne 0$, then
$\mathcal{M}$ is further extendible beyond a smooth Cauchy horizon to a larger spacetime which fails however to be globally hyperbolic and is thus not uniquely determined by initial data). In this paper, we will work on the subregion $\mathcal{D}$ which is the closure of one of the exterior regions
$\mathcal{M}$. The boundary of the region $\mathcal{D}$ consists of the union of two null hypersurfaces $\mathcal{H}^+$ and $\mathcal{H}^-$, the \emph{future event horizon} and the \emph{past event horizon}, along with $\mathcal{B}$, the \emph{bifurcate sphere}.
Our convention will be that $\mathcal{B}$ is not included in $\mathcal{H}^\pm$ and
$\mathcal{H}^+\cup\mathcal{B}\cup\mathcal{H}^-$ is a bifurcate null hypersurface.

We proceed to describe explicitly the underlying structure and metric.
We  start with the smooth manifold with boundary
\begin{equation}
\label{diffStruct}
\mathcal{R}=\mathbb R_{\geq 0}\times\mathbb R\times  \mathbb S^ 2,
\end{equation}
parameterized by $y^*\in\mathbb R_{\geq 0}$, $t^*\in\mathbb R$ and a choice of standard spherical coordinates $(\theta^*,\phi^*)\in\mathbb S^2$.
This coordinate system will be known as ``Kerr-star coordinates''.
Let us denote the coordinate vector field $T=\partial_{t^*}$ and $\Phi=\partial_{\phi^*}$
and let us denote by $\Omega_1$, $\Omega_2$, $\Omega_3$ a basis of standard angular momentum operators corresponding to the $\mathbb S^2$ factor of $(\ref{diffStruct})$.\footnote{We may take $\Omega_1=\Phi$
for instance.}
In particular, the $\Omega_i$
span the tangent space of $\mathbb S^2$.

We define what shall be the future event horizon $\mathcal{H}^+$ by $\mathcal{H}^+=\partial\mathcal{R}=\{y^*=0\}$. It will be useful to adopt the conventions:
\[\mathcal{H}^+_{\geq s} \doteq \mathcal{H}^+\cap \{t^* \geq s\},\]
\[\mathcal{H}^+(s_1,s_2)\doteq \mathcal{H}^+ \cap \{t^* \in [s_1,s_2]\},\]
\[\mathcal{R}_{\geq s} \doteq \{t^* \geq s\},\]
\[\mathring{\mathcal{R}} \doteq {\rm int}(\mathcal{R})=\mathcal{R}\setminus \mathcal{H}^+, \]
\[\mathring{\mathcal{R}}_{\geq s} \doteq \mathcal{R}_{\geq s}\setminus \mathcal{H}^+_{\geq s}.\]

Next, given a choice of parameters $(a,M)$ satisfying $|a|<M$, we define a new coordinate function
$r=r(y^*)$ on $\mathcal{R}$ (with $\infty>C>\frac{d r}{d y^*}\ge c>0$)
 so that $r|_{\mathcal{H}^+} =r_+(a,M)$
 where $r_\pm\doteq M \pm \sqrt{M^2-a^2}$. It is often convenient to replace $r$ with yet another rescaled version, $r^*=r^*(r)$,
defined in $\mathring{\mathcal{R}}$, by
\begin{equation}
\label{r*def}
\frac{dr^*}{dr}=\frac{r^2+a^2}{\Delta}, \qquad r^*(3M)=0,
\end{equation}
where
\begin{equation}\label{delta}
\Delta= r^2-2Mr+a^2 = (r-r_+)(r-r_-).
\end{equation}
Since $r_-<r_+$, it follows that $\Delta$ vanishes to first order on $\mathcal{H}^+$, and thus the coordinate range $\infty> r>r_+$ covering $\mathring{\mathcal{R}}$
corresponds to the range $\infty>r^*>-\infty$. It will also be useful to sometimes employ what will be
an ``approximately null'' coordinate system $(\tilde u,\tilde v,\theta,\phi)$ defined by
\[
\tilde u = \frac{1}{2}(t-r^*),\qquad \tilde v = \frac{1}{2}(t+r^*).
\]

Next, we introduce the new coordinates
\begin{equation}\label{newcoord}
t(t^*,r)\doteq t^* -  \bar t(r), \qquad
\phi(\phi^*,r)\doteq \phi^*- \bar \phi(r) \mod 2\pi, \qquad
\theta \doteq \theta^*
\end{equation}
where $\bar{t}(r)$ and $\bar{\phi}(r)$
are appropriately chosen smooth functions depending
on $a$ and $M$ (see~\cite{partsIandII} and~\cite{partIII} for details) which vanish for sufficiently large $r$.

In these ``Boyer-Lindquist coordinates'' $(t,r,\theta,\phi)$, we finally define the Kerr metric by
\begin{align}
\label{eleme}
g_{a,M}=
-\frac{\Delta}{\rho^2}\left(dt-a\sin^2\theta d\phi\right)^2
+\frac{\rho^2}{\Delta}dr^2+\rho^2d\theta^2 +\frac{\sin^2\theta}{\rho^2}
\left(a\,dt-(r^2+a^2)d\phi\right)^2,
\end{align}
where $\rho^2 \doteq r^2+a^2\cos^2\theta$. Though \emph{a priori} $(\ref{eleme})$ is only
defined in $\mathring{\mathcal{R}}$,
by examining the expression of the metric in Kerr-star coordinates $y^*$ and $t^*$ (see~\cite{partsIandII} for the computation), one checks easily that $g_{a, M}$ extends
smoothly to $\mathcal{H}^+$ making  $(\mathcal{R},g_{a,M})$  a smooth Lorentzian manifold-with-boundary. Let us note moreover that $T$ and $\Phi$
 defined previously can be expressed  again as
coordinate vector fields
$T=\partial_t$ and $\Phi=\partial_{\phi}$,
whence it follows from $(\ref{eleme})$
that $T$ and $\Phi$ are Killing on $\mathcal{R}$. These are the so-called stationary and axisymmetric Killing vector fields.

Recall that when $a \neq 0$ the vector field $T$ is \emph{not} everywhere timelike. The region $\mathcal{S}$ where $T$ is spacelike is known as the ``ergoregion''. Explicitly, we have
\begin{equation}\label{erg}
\mathcal{S} = \{\Delta - a^2\sin^2\theta < 0\}.
\end{equation}
Note that
\[\mathcal{S} \subset \{r < 2M\}.\]

Let us also recall that in~\cite{partsIandII} and~\cite{partIII} we chose the function $\bar{t}$ of~(\ref{newcoord}) so that the hypersurfaces $t^*= s$, denoted by $\Sigma_s^*$, are spacelike with respect to the Kerr
metric as just defined. Furthermore, we will have $\mathcal{R}_{\geq s} = D^+(\Sigma^*_s)$.
Let us introduce the notation
\begin{equation}\label{ringsigma}
\mathring{\Sigma} \doteq \{t = 0\}.
\end{equation}
We have that $\mathring\Sigma$ is also spacelike and a Cauchy hypersurface for
$\mathring{\mathcal{R}}$.

Some additional notation from~\cite{partIII}:
Note that the definition $\partial_r$ is ambiguous since it depends on the choice of coordinate system. Thus, we define
\begin{equation}\label{zstar}
Z^* \doteq \partial_r \qquad \text{with\ respect\ to\ coordinates}\qquad (t^*,r,\theta^*,\phi^*),
\end{equation}
\begin{equation}\label{znostar}
Z \doteq \partial_r \qquad \text{with\ respect\ to\ coordinates}\qquad (t,r,\theta,\phi).
\end{equation}
Note that $Z^*$ is well defined in $\mathcal{R}$ and is
transversal to $\mathcal{H}^+$ while $Z$ is only well defined in $\mathring{\mathcal{R}}$.
Finally, we will use $\nabb$ to denote the induced covariant derivative on the $\mathbb{S}^2$ factor of $\mathcal{R}$.

Though all explicit computations will take place on the manifold-with-boundary $\mathcal{R}$ defined above, it is of fundamental importance to understand the existence
and properties of a  further smooth extension to $\mathcal{D}=\mathcal{R}\cup
\mathcal{H}^-\cup\mathcal{B}$, which will represent precisely the $\mathcal{D}$ described
at the beginning of this section. We will be brief in our presentation; we direct the reader to~\cite{oneill} for a very careful and detailed exposition.

We begin by attaching $\mathcal{H}^-$. Starting with Boyer-Lindquist coordinates $(t,r,\theta,\phi)$, one defines a new coordinate system $({}^*t,{}^*\phi,r,{}^*\theta)$ by
\[
{}^*t(t,r) \doteq  t  -  \bar t(r), \qquad
{}^*\phi(t,r) \doteq \phi - \bar \phi(r) \mod 2\pi, \qquad
{}^*\theta\doteq \theta,
\]
where $\bar{t}$ and $\bar{\phi}$ are as above.
A straightforward computation shows that the metric naturally extends smoothly so as
to be defined also at $r = r_+$ in this chart.
We may thus use this coordinate chart to extend $\mathcal{R}$ to a
larger manifold-with-boundary $\mathcal{R}\cup\mathcal{H}^-$ where $\mathcal{H}^-$
corresponds to the hypersurface $r=r_+$ of this new chart.
We shall refer to $\mathcal{H}^-$ as the \emph{past event horizon}.

 One may easily check that the Boyer-Lindquist coordinate defined map
\begin{equation}
\label{discreteIso}
(t,\phi) \mapsto (-t,-\phi)
\end{equation}
 is an isometry of $\mathring{\mathcal{R}}$ which smoothly extends to an isometry of $\mathcal{R}\cup\mathcal{H}^-$ and furthermore sends $\mathcal{H}^+$ to $\mathcal{H}^-$.

Finally, one may even further extend $\mathcal{R}\cup\mathcal{H}^-$ to a larger
Lorentzian manifold $\widetilde{\mathcal{M}}$ so that
the boundary of $\mathcal{R}$ (as a subset) in $\widetilde{\mathcal{M}}$ consists of a bifurcate null hypersurface
$\mathcal{B}\cup\mathcal{H}^-\cup\mathcal{H}^+$, with
$\mathcal{B}\subset \widetilde{\mathcal{M}}$
a sphere. Our region of interest $\mathcal{D}$
described at the beginning of this section
is simply then the manifold-with-stratified boundary $\mathcal{D}=\mathcal{R}\cup\mathcal{H}^-\cup
\mathcal{B}$. We remark that
$\mathcal{D}$ admits a globally regular coordinate system\footnote{It is globally regular up to the usual degeneration of spherical coordinates.} $(U^+,V^+,\theta,\phi) \in [0,-\infty) \times [0,\infty) \times \mathbb{S}^2$ so that $\mathcal{H}^+ = \{U^+ = 0,V^+ \in (0,\infty)\}$, $\mathcal{H}^- = \{V^+ = 0, U^+ \in (-\infty,0)\}$ and $\mathcal{B}= \{(U^+,V^+) = (0,0)\}$.
Moreover, along $\mathcal{B}$ we have
\begin{equation}\label{metricbifurcates}
g_{U^+U^+} = g_{U^+\theta} = g_{U^+\phi} = g_{V^+V^+} = g_{V^+\theta} = g_{V^+\phi} = 0.
\end{equation}
We shall not here
require the form of the explicit coordinate transformations defining
$U^+$ and $V^+$ in terms of our previously described charts on $\mathcal{R}\cup\mathcal{H}^-$ but we remark that
\begin{equation}
\label{Sigmabardefin}
\overline{\Sigma} \doteq \mathcal{B}\cup \mathring{\Sigma}
\end{equation}
is a smooth manifold-with-boundary (with boundary $\mathcal{B}$) and interior $\mathring{\Sigma}$.
Note that smooth functions $(\uppsi, \uppsi')$
``compactly supported on $\mathring{\Sigma}$'' extend to smooth compact supported
functions on
$\overline{\Sigma}$ which moreover vanish
in a neighbourhood of $\mathcal{B}$. On the other hand, smooth functions
compactly supported on $\overline{\Sigma}$ do not restrict to compactly
supported functions on $\mathring{\Sigma}$.

It will be convenient to introduce the notation
\begin{equation}\label{barH}
\overline{\mathcal{H}^{\pm}} \doteq \mathcal{H}^{\pm}\cup \mathcal{B},
\end{equation}
\begin{equation}\label{barHless}
\overline{\mathcal{H}^+_{\leq \tau}} \doteq \overline{\mathcal{H}^+} \cap (\{t^* \leq \tau\})\cup\mathcal{B}).
\end{equation}
These are again smooth hypersurfaces-with-boundary, with
boundary $\mathcal{B}$ for $\overline{\mathcal{H}^{\pm}}$ and boundary $\mathcal{B}\cup \left(\Sigma_0^* \cap \mathcal{H}^+\right)$ for $\overline{\mathcal{H}^+_{\leq 0}}$.
The reader should in particular again contrast the distinct notions of ``compactly
supported'' on $\mathcal{H}^+$ and $\overline{\mathcal{H}^\pm}$.

We have already noted that the vector fields $T$ and $\Phi$ are Killing. The event horizon $\mathcal{H}^+$ is
also a \emph{Killing horizon}: the Killing field given by the linear combination
\begin{equation}\label{thisisK}
K \doteq T+  \upomega_+\Phi,
\end{equation}
where $\upomega_+ \doteq \frac{a}{2Mr_+}$ is the ``angular velocity'' of the event horizon. The vector field $K$ is null and normal to $\mathcal{H}^+$; thus, $\mathcal{H}^+$ is in particular a null hypersurface. In integrals associated to energy currents we will denote $K$ by $n^{\mu}_{\mathcal{H}^+}$. It will be useful to recall that the vectorfield $K$ restricted to $\mathcal{H}^+$ coincides with the
smooth extension of the coordinate vector field $\partial_{r^*}$ of the $(r^*,t,\theta, \phi)$ coordinate system.

The past event horizon $\mathcal{H}^-$ is also a Killing horizon with a Killing field also given by $K$. Note however that the restriction of $K$ to $\mathcal{H}^-$ coincides with the smooth extensions of the coordinate vector field $-\partial_{r^*}$ of the $(r^*,t,\theta, \phi)$ coordinate system.

Note finally that the vector fields $T$ and $\Phi$ and the  discrete isometry $(\ref{discreteIso})$ both extend smoothly to all of $\mathcal{D}$ and
\begin{equation}\label{vanish}
K|_{\mathcal{B}} = 0.
\end{equation}

\subsection{Dependence on $a$ and $M$ and conventions on constants}
\label{gencon}
{\it In all propositions to follow, unless otherwise stated, $|a|<M$ are fixed
parameters and everything refers to the Kerr metric $g_{M,a}$ on $\mathcal{D}$
as described in the previous section.}

Let us briefly review our conventions from~\cite{partsIandII} and~\cite{partIII} regarding constants
depending on the parameters $a$ and $M$. Large positive constants will be denoted by $B$,
and small positive constants by $b$. Both constants $B$ and $b$ depend only on $M$ and a lower bound for $M - |a|$, and this dependence is always to be understood even when not mentioned explicitly. Often these constants will blow up $B\to\infty$, $b^{-1}\to\infty$ in
the extremal limit $|a|\to M$.

We recall the usual arithmetic properties of $b$ and $B$:
\[
b+b=b,  \, B+B=B, \, B\cdot B=B, \, B^{-1}=b, \ldots
\]
The statement $f \sim g$ will mean
\[
bg \leq f \leq Bg.
\]
The statement ``for $R$ sufficiently large'', etc., without further qualification, will mean
``there exists a constant $R_0(a,M)$ such that for $R\ge R_0$''.

Lastly, if the constant $B$ or $b$ depends on the value of a yet to be fixed parameter, then that dependence will be explicitly noted. For example, if $B$ depends on a parameter $c$ which has not been fixed, we shall denote it by $B(c)$. Once the constant $c$ is fixed, we then write $B$.

\subsection{Useful vector fields}
\label{useful?}
We recall the following two lemmas proved in~\cite{partIII}.

\begin{lemma}\label{timelikeVector}
The vector field
\[
T + \frac{2Mar}{\left(r^2+a^2\right)^2}\Phi
\]
is a smooth vector field in $\mathcal{D}$, is timelike  in $\mathring{\mathcal{R}}$
and  null on $\mathcal{H}^{\pm}$.
\end{lemma}
\begin{lemma}\label{horizonTimelike}
There exists a constant $\epsilon_0 = \epsilon_0(a,M) > 0$ such that the vector field $K$~(\ref{thisisK})
is timelike for $r \in (r_+,r_++\epsilon_0)$.
\end{lemma}

These lemmas allow us to make the following definition.
\begin{definition}\label{Vvect}
Let $\epsilon_0 > 0$ be from Lemma~\ref{horizonTimelike}. Let $\alpha(r)$ be a function such that $V \doteq T + \alpha(r)\Phi$ is a smooth vector field in $\mathcal{D}$, timelike in $\mathring{\mathcal{R}}$ and which satisfies
\begin{align*}
V& = K, \text{ when } r \in [r_+,r_+ + \epsilon_0/2],\\
V& = T + \frac{2Mar}{\left(r^2+a^2\right)^2}\Phi, \text{ when }r \in \left[r_++\epsilon_0,\frac{M\left(7+\sqrt{2}\right)}{4}\right],\\
V& = T, \text{ when } r \geq \frac{M\left(3+\sqrt{2}\right)}{2}.
\end{align*}
\end{definition}
\begin{remark}This vector field will be useful because it is manifestly $T$-invariant, it is timelike (hence the associated energy fluxes $\mathbf{J}^V$ are positive definite) and because it is Killing
for $r\le r_++\epsilon_0/2$ and $r\ge M(3+\sqrt{2})/2$ (hence the error terms $\mathbf{K}^V$ from the energy identity~(\ref{ingeneralform}) are supported in $r_++\epsilon_0/2\le r\le M(3+\sqrt{2})/2$).
\end{remark}

It will be useful to observe the following immediate corollary of Lemmas~\ref{timelikeVector} and~\ref{horizonTimelike}.
\begin{corollary}\label{tempKill}For every $\epsilon > 0$ sufficiently small and any
$r_0 \in (r_+,\infty)$, there exists a vector field $\tilde V=\tilde V(r_0, \epsilon)$
of the form $T + \tilde\alpha (r)\Phi$ for an appropriate function $\tilde \alpha(r)$ such that $\tilde V$ is a smooth vector field on $\mathcal{D}$, is timelike in $\mathring{\mathcal{R}}$, is Killing in the region
\[
r \in\left({\rm min}\left(r_+,r_0-\epsilon\right),r_0+\epsilon\right],
\]
and is equal to $V$ for $r$ sufficiently close to $r_+$ and $r$ sufficiently large.
\end{corollary}
We shall apply the above Corollary, for finitely many distinct choices of $r_0$,
in the context of Section~\ref{someboundedargument}.

In order for our non-degenerate energies to have a fixed meaning, it is useful to fix once and for all a choice of a globally defined smooth timelike vector field on $\mathcal{D}$.

\begin{definition}\label{Nvect}Let $N$ denote any fixed choice of a smooth timelike vector field on $\mathcal{D}$ which is invariant under the flow of $T$ on the complement of a compact set containing the bifurcate sphere $\mathcal{B}$,\footnote{Note that in view of the vanishing of
$T$ on $\mathcal{B}$, one cannot define such a timelike $N$ which is
invariant on \underline{all} of $\mathcal{D}$.} and satisfies $N = T$ for sufficiently large $r$.
\end{definition}

Finally, we note the following easy calculations.
\begin{remark}\label{sobol1}
Fix an open set $\mathcal{U}\subset \mathcal{D}$ containing the bifurcate sphere
$\mathcal{B}\subset \mathcal{U}$. Then, for $s$ such that $\Sigma_s^* \cap U = \emptyset$ we have
\begin{equation}
\int_{\Sigma_s^*} {\bf J}_\mu^N[\psi]n^\mu_{\Sigma_s^*}\sim
\| \psi\|^2_{\dot{H}^1(\Sigma_s^*)} +\|n_{\Sigma_s^*}\psi\|^2_{L^2(\Sigma_s^*)}
\sim \int_{\theta,\phi^*} \int_{r_+}^\infty
\left( |\partial_{t^*}\psi|^2 + |\partial_{r}\psi|^2 +|\nabb\psi|^2_{\slashg}\right)\, dr\, dV_{\slashg}
\end{equation}
with respect to coordinates $(t^*, r, \theta^*, \phi^*)$.
\end{remark}
\begin{remark}\label{sobol2}
We have
\begin{equation}
\int_{\Sigma_s^*} {\bf J}_\mu^V[\psi]n^\mu_{\Sigma_s^*}
\sim \int_{\theta,\phi^*} \int_{r_+}^\infty
\left( |\partial_{t^*}\psi|^2 + \left(1-\frac{r_+}{r}\right)|\partial_{r}\psi|^2 +|\nabb\psi|^2_{\slashg}\right)\, dr\, dV_{\slashg}
\end{equation}
with respect to coordinates $(t^*, r, \theta^*, \phi^*)$.
\end{remark}

\subsection{A foliation by hyperboloidal hypersurfaces}\label{hyperboloid}
It will be convenient to have the following explicit foliation of $\mathcal{R}$ by a family of hyperboloidal hypersurfaces.
\begin{definition}For every $\tau \in \mathbb{R}$ we set
\[S_{\tau} \doteq \left\{
        \begin{array}{ll}
            t^* = \tau & r \leq 5M \\
            t^* - r^* + \frac{10M}{r} = \tau - (5M)^* + 2 & \quad r > 5M.
        \end{array}
    \right.\]
\end{definition}
(This hypersurface could be smoothed out, but this does not in fact make a difference.)

Some straightforward, if tedious, calculations yield the following lemma.
\begin{lemma}\label{iftediou}
For every $\tau \in \mathbb{R}$, $S_{\tau}$ is a spacelike hypersurface, $\mathcal{R} = \cup_{\tau \in \mathbb{R}}S_{\tau}$, and, for sufficiently large $R$,
\[\int_{S_{\tau} \cap \{r \geq R\}}\mathbf{J}^T_{\mu}[\psi]n^{\mu}_{S_{\tau}} \sim \int_{S_{\tau} \cap \{r \geq R\}}\left[\left|\partial_{\tilde v}\psi\right|^2 + r^{-2}\left|\partial_{\tilde u}\psi\right|^2 + \left|\nabb\psi\right|^2\right]\, r^2\sin\theta\, dv\, d\theta\, d\phi.\]
\end{lemma}

In comparing the way these two integrals are written, we recall our convention that if no volume form is written explicitly (as on the left hand side of the above), the integration is with respect to the induced volume form.

Later, when $r$ is sufficiently large we will often work in the coordinate system $(\tau,r,\theta,\phi)$ associated to the
foliation $\{S_{\tau}\}_{\tau \in \mathbb{R}}$. We will in fact use this coordinate system to define
our notion of null infinity $\mathcal{I}^+$ in Section~\ref{formdefine}.

\subsection{Well-posedness}\label{wellposedness}
Let us briefly recall some basic well-posedness statements.

First we consider the case of initial data prescribed on $\Sigma_0^*$.
Recall that $\mathcal{R}_{\geq 0} = \{t^* \geq 0\} = D^+(\Sigma_0^*)$. In the propositions below the $H^s$ and $C^k$ spaces will refer to complex valued functions.

\begin{proposition}
\label{WPProp}
Let $(\uppsi,\uppsi') \in H^s_{\rm loc}(\Sigma_0^*) \times H^{s-1}_{\rm loc}(\Sigma_0^*)$. Then there exists a unique solution $\psi$ to the wave equation~$(\ref{WAVE})$ on $\mathcal{R}_{\geq 0}$
such that
\[\psi \in C^0_{\tau \in [0,\infty)}(H^s_{\rm loc}(\Sigma^*_{\tau})) \cap C^1_{\tau \in [0,\infty)}(H^{s-1}_{\rm loc}(\Sigma^*_{\tau})) \cap H^s_{\rm loc}(\mathcal{H}^+_{\geq 0}),\]
$\psi|_{\Sigma_0^*} = \uppsi$, and $n_{\Sigma_0^*}\psi|_{\Sigma_0^*} = \uppsi'$. Furthermore, the solution map depends continuously on the initial data. Finally, we note that if the initial data $(\uppsi,\uppsi')$ are smooth, then the solution $\psi$ will be smooth.
\end{proposition}

Next we consider the case of initial data along $\overline{\Sigma}$. Let us define $\tilde{\Sigma}_{\tau}$ to be the image of $\overline{\Sigma}$ at time $\tau$ of the flow map associated to the vector field $N$ from Definition~\ref{Nvect}.

\begin{proposition}\label{t0wellposed}
Let $(\uppsi,\uppsi') \in H^s_{\rm loc}(\overline{\Sigma}) \times H^{s-1}_{\rm loc}(\overline{\Sigma})$. Then there exists a unique solution $\psi$ to the wave equation~$(\ref{WAVE})$ in $\mathcal{D}$ such that
\[\psi \in C^0_{\tau \in (-\infty,\infty)}(H^s_{\rm loc}(\tilde\Sigma_{\tau})) \cap C^1_{\tau \in (-\infty,\infty)}(H^{s-1}_{\rm loc}(\tilde\Sigma_{\tau})) \cap H^s_{\rm loc}(\mathcal{H}^+) \cap H^s_{\rm loc}(\mathcal{H}^-),\]
$\psi|_{\overline{\Sigma}} = \uppsi$, and $n_{\overline{\Sigma}}\psi|_{\overline{\Sigma}} = \uppsi'$. Furthermore, the solution map depends continuously on the initial data. Finally, we note that if the initial data $(\uppsi,\uppsi')$ are smooth, then the solution $\psi$ will be smooth.
\end{proposition}
\begin{remark}
\label{giasthrig}
In the case when the initial data $(\uppsi,\uppsi')$ are compactly supported along $\mathring{\Sigma}$, then a $\mathbf{J}^K$ energy estimate immediately implies that bifurcate sphere $\mathcal{B}$
lies outside the support of the solution $\psi$ produced by Proposition~\ref{t0wellposed}.
Also note that if $(\uppsi,\uppsi')$ are compactly supported on $\overline\Sigma$,
then $(\psi|_{\Sigma_0^*}, n_{\Sigma_0^*}\psi|_{\Sigma_0^*})$
are compactly supported on $\Sigma_0^*$.
\end{remark}

It will also be useful to consider the following two mixed characteristic-spacelike initial value problems. For convenience these will both be stated in the smooth category. First we have

\begin{proposition}\label{mixedwellposed}Let $\uppsi_{\overline{\mathcal{H}^+_{\leq 0}}}$ be a smooth function on $\overline{\mathcal{H}^+_{\leq 0}}$ and $(\uppsi_{\Sigma_0^*},\uppsi'_{\Sigma_0^*})$ be a pair of smooth functions on $\Sigma_0^*$ such that there exists a smooth function $\tilde\Psi$ on $\mathcal{D}$ satisfying
\[\tilde\Psi|_{\overline{\mathcal{H}^+_{\leq 0}}} = \uppsi_{\overline{\mathcal{H}^+_{\leq 0}}},\qquad (\tilde\Psi|_{\Sigma_0^*},n_{\Sigma_0^*}\tilde\Psi|_{\Sigma_0^*}) = (\uppsi_{\Sigma_0^*},\uppsi'_{\Sigma_0^*}).\]
Then there exists a unique smooth solution $\psi$ to the wave equation~(\ref{WAVE}) in the past of $\Sigma_0^*$ such that
\[\psi|_{\overline{\mathcal{H}^+_{\leq 0}}} = \uppsi_{\overline{\mathcal{H}^+_{\leq 0}}},\qquad (\psi|_{\Sigma_0^*},n_{\Sigma_0^*}\psi|_{\Sigma_0^*}) = (\uppsi_{\Sigma_0^*},\uppsi'_{\Sigma_0^*}).\]
\end{proposition}
See
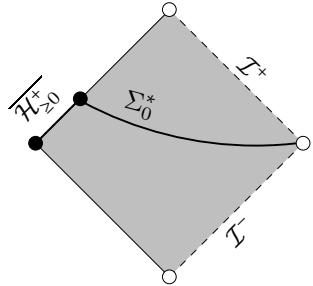
\[
\input{initkan.pstex_t}
\]

Before giving the next proposition, it is useful to define a function $r(\tau,s)$: Let $\tau > -\infty$. Then, for each $s > 0$ sufficiently large we define the value $r(\tau,s)$ to be the largest solution to
\[s - r^*\left(\tau,s\right) + \frac{10M}{r\left(\tau,s\right)} \doteq \tau - (5M)^* + 2.\]
Observe that the hypersurface $\{t = s\}$ will intersect the hypersurface $S_{\tau}$ along the surface where $(t,r) = (s,r(\tau,s))$.
Refer to
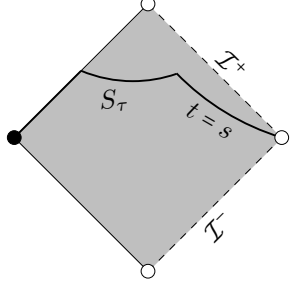
\[
\input{initper.pstex_t}
\]
We have
\begin{proposition}\label{phybackwp}Let $\tau < \infty$,
let $\uppsi_{\overline{\mathcal{H}^+_{\leq \tau}}}$ be a smooth function on $\overline{\mathcal{H}^+_{\leq \tau}}$ which vanishes in a neighborhood of $S_{\tau} \cap \mathcal{H}^+$ and let $\Phi_{\{t = s\} \cap \{r \geq r(\tau,s)\}}$ be a smooth compactly supported function on $\{t = s\} \cap \{r \geq r(\tau,s)\}$ which vanishes in a neighborhood of $\{t = s\} \cap \{r = r(\tau,s)\}$. Then there exists a unique smooth solution $\psi$ to the wave equation~(\ref{WAVE}) in the past of $\overline{\mathcal{H}^+_{\leq \tau}} \cup \left(S_{\tau} \cap \{r \leq r(\tau,s)\}\right) \cup \left(\{t = s\} \cap \{r \geq r(\tau,s)\}\right)$ such that
\[
\psi|_{\overline{\mathcal{H}^+_{\leq \tau}}} = \uppsi_{\overline{\mathcal{H}^+_{\leq \tau}}},
\]
\[
(\psi|_{S_{\tau} \cap \{r \leq r(\tau,s)\}},n_{S_{\tau}}\psi|_{S_{\tau} \cap \{r \leq r(\tau,s)\}}) =(0,0),
\]
\[
r\psi|_{\{t = s\} \cap \{r \geq r(\tau,s)\}} = \Phi_{\{t = s\} \cap \{r \geq r(\tau,s)\}}.
\]
\end{proposition}

In accordance with our conventions (recall Section~\ref{gencon}),
the above propositions refer always to the Kerr metric with fixed parameters $|a|<M$.
Let us remark that we have defined the differentiable structure in~\cite{partIII}
so that we can assert also the smooth dependence of $\psi$ on our parameters $a$ and
$M$;
this, however, shall play no role in the current paper.

\subsection{The non-degenerate boundedness and integrated energy decay statements}
\label{edweivaiola}
In this section, we shall recall the precise boundedness and integrated energy decay statements proved in~\cite{partIII}.

First we recall a few additional notations from~\cite{partIII}:
\begin{definition}Given $s^-$
satisfying $r_+ < 3M-s^- < \infty$, let us define a cutoff function $\chi(r)$ such that
$\chi=1$ for $r\ge 3M-s^{-}$ and $\chi=0$ for $r\le (r_++3M-s^-)/2$.
We then set
\[\tilde{Z}^*=\chi Z+(1-\chi)Z^*.\]
\end{definition}
\begin{definition}Given $s^-$ and $s^+$ satisfying $r_+ < 3M-s^- < 3M + s^+ < \infty$ we define
\begin{equation}
\label{degenerationfunc}
\zeta(r)\doteq (1-3M/r)^2(1- \eta_{[3M-s^-,3M+s^+]}(r)),
\end{equation}
where $\eta$ is the indicator function.
\end{definition}

The main result of~\cite{partIII} was
\begin{theorem}
\label{theResult}\cite{partIII}
There exist parameters $s^-(a,M)$ and $s^+(a,M)$ satisfying $r_+ < 3M-s^- < 3M+s^+ < \infty$
such that for all
$\delta>0$, all
 sufficiently regular solutions $\psi$ to~(\ref{WAVE}) on $\mathcal{R}_{ \geq 0}$ satisfy the following estimates:
\begin{equation}
\label{protasn1b}
\int_{\mathcal{R}_{\geq 0}}\Big(r^{-1}\zeta |\nabb\psi|^2+r^{-1-\delta}\zeta \left|T\psi\right|^2+r^{-1-\delta}\left|\tilde Z^*\psi\right|^2+ r^{-3-\delta} \left|\psi-\psi_\infty\right|^2\Big) \le B(\delta)
\int_{\Sigma_0^*} {\bf J}_\mu^N[\psi]n^\mu_{\Sigma_0^*},
\end{equation}
\begin{equation}
\label{fluxho...}
\int_{\mathcal{H}^+_{\geq 0}}\left( {\bf J}^N_\mu[\psi]n^\mu_{\mathcal{H}^+}
+\left|\psi-\psi_{\infty}\right|^2\right)
\le B \int_{\Sigma_0^*} {\bf J}_\mu^N[\psi]n^\mu_{\Sigma_0^*},
\end{equation}
\begin{equation}
\label{bndts1b}
 \int_{\Sigma_s^*} {\bf J}_\mu^N[\psi]
n^\mu_{\Sigma_s^*}\le B \int_{\Sigma_0^*} {\bf J}_\mu^N[\psi]n^\mu_{\Sigma_0^*}, \qquad
\forall s\ge 0,
\end{equation}
where $4\pi\psi_\infty^2= \lim_{r'\to\infty}\int_{\Sigma_0^*\cap\{r=r'\}} r^{-2}\left|\psi\right|^2$.
\end{theorem}

We also proved the following higher order version of Theorem~\ref{theResult}:
\begin{theorem}
\label{h.o.s.}\cite{partIII}
With $s^\pm(a,M)$ as above, then
for all
$\delta>0$, $j\ge 1$,
all sufficiently regular solutions $\psi$ to~(\ref{WAVE}) on $\mathcal{R}_{\geq 0}$ satisfy the following estimates:
\begin{align}
\nonumber
\label{protasn1}
\int_{\mathcal{R}_{\geq 0}} &
r^{-1-\delta}\zeta \sum_{1\le i_1+i_2+i_3\le j}
|\nabb^{i_1}T^{i_2}(\tilde Z^*)^{i_3}\psi|^2+r^{-1-\delta}\sum_{1\le i_1+i_2+i_3\le j-1}
\left(|\nabb^{i_1}T^{i_2}(\tilde Z^*)^{i_3+1}\psi|^2+|\nabb^{i_1}T^{i_2}(Z^*)^{i_3}\psi|^2\right)\\
&\le B(\delta,j)\int_{\Sigma_0^*} \sum_{0\le i \le j-1}
{\bf J}^N_\mu[N^{i}\psi]n^\mu_{\Sigma_0^*},
\end{align}
\begin{equation}
\label{bndts2}
\int_{\mathcal{H}^+_{\geq 0}} \sum_{0\le i \le j-1}{\bf J}^N_\mu[N^{i}\psi]n^\mu_{\mathcal{H}^+}
\le  B(j)\int_{\Sigma_0^*} {\sum_{0\le i \le j-1}
{\bf J}^N_\mu[N^{i}\psi]n^\mu_{\Sigma_0^*}},
\end{equation}
\begin{equation}
\label{bndts1}
\int_{\Sigma_s^* } \sum_{0\le i \le j-1}{\bf J}^N_\mu[N^{i}\psi]n^\mu_{\Sigma_s^*}
\le  B(j)\int_{\Sigma_0^*} {\sum_{0\le i \le j-1}
{\bf J}^N_\mu[N^{i}\psi]n^\mu_{\Sigma_0^*}}, \qquad
\forall s\ge 0.
\end{equation}
\end{theorem}
\begin{remark}Sufficiently regular may be taken to mean that the initial data lies in $H^s_{\rm loc}\left(\Sigma_0^*\right)$ for $s$ suitably large and that the right hand sides of each inequality are finite.
\end{remark}
\begin{remark}Recall that a straightforward elliptic estimate would yield
\begin{equation}
\label{inasyflatcas}
\int_{\Sigma_s^* } \sum_{0\le i \le j-1}{\bf J}^N_\mu[N^{i}\psi]n^\mu_{\Sigma_s^*}
\sim \sum_{1\le i\le j} \|\psi\|^2_{\dot{H}^i(\Sigma_s^*)} +\|n_{\Sigma_s^*}\psi\|^2_{\dot{H}^{i-1}(\Sigma_s^*)}.
\end{equation}
\end{remark}
\begin{remark}\label{theResultflipped}In view of the discrete isometry~(\ref{discreteIso}), one immediately obtains versions of Theorem~\ref{theResult} and Theorem~\ref{h.o.s.} for solutions defined in the past of the hypersurface ${}^*t = 0$.
\end{remark}

\subsection{The $r^p$ estimates}\label{rpSec}
It will be useful to exploit the hierarchy
of
``$r^p$ estimates'' from~\cite{icmp}. For our purposes, it is convenient to apply these estimates in the following form.

\begin{proposition}\label{rp}
Let $R$ be sufficiently large. Then for all $\tau_1<\tau_2$, $p\in[0,2]$, and
 all $\psi$  sufficiently regular solutions to~(\ref{WAVE}) on $\mathcal{D}(\tau_1,\tau_2)\doteq
J^+(S_{\tau_1})\cap J^-(S_{\tau_2})$, then
 setting $\varphi \doteq (r^2+a^2)^{1/2}\psi$ and keeping Remark~\ref{volumeforms} in mind, we have
\begin{align*}
&\int_{S_{\tau_2}\cap \{r \geq R\}}\left[r^p\left|\partial_{\tilde v}\varphi\right|^2 + r^{p-2}\left|\nabb\varphi\right|^2 + r^{-2}\left|\partial_{\tilde u}\varphi\right|^2\right]\sin\theta\, dv\, d\theta\, d\phi\\
&\ \ \ \ \ \ +\int_{\mathcal{D}(\tau_1,\tau_2)\cap \{r \geq R\}}\left[pr^{p-1}\left|\partial_{\tilde v}\varphi\right|^2 + \left((2-p)r^{p-1} + r^{-1}\right)\left|\nabb\varphi\right|^2 + r^{p-4}\left|\varphi\right|^2 + r^{-2}\left|\partial_{\tilde u}\varphi\right|^2\right]\sin\theta\, du\, dv\, d\theta\, d\phi \\
&\ \ \ \ \ \leq B\int_{\mathcal{D}(\tau_1,\tau_2)\cap \{R \leq r \leq R+1\}}r^p\left[\left|T\varphi\right|^2 + \left|Z\varphi\right|^2 + \left|\nabb\varphi\right|^2\right]\sin\theta\, du\, dv\, d\theta\, d\phi
\\ &\ \ \ \ \ \ \ \ +B\int_{S_{\tau_1}\cap \{r \geq R\}}\left[r^p\left|\partial_{\tilde v}\varphi\right|^2 + r^{p-2}\left|\nabb\varphi\right|^2 + r^{-2}\left|\partial_{\tilde u}\varphi\right|^2\right]\sin\theta\, dv\, d\theta\, d\phi.
\end{align*}
\end{proposition}
\begin{proof}One combines the estimates of~\cite{icmp} with an energy estimate, Hardy inequalities, and a Morawetz estimate. This is a special case of a more
general computation done in detail in~\cite{moschidis} for the
general setting of asymptotically flat spacetimes.
\end{proof}

\begin{remark}\label{staudecay}Note that one may easily check that the boundary terms of the $p=1$ estimate relate to the spacetime terms of the $p=2$ estimate in such a way as to allow one to combine Theorem~\ref{h.o.s.} with the iterated pigeon hole argument of~\cite{icmp} in order to conclude for instance that
\[\int_{S_{\tau}}\mathbf{J}^N_{\mu}[\psi]n^{\mu}_{S_{\tau}} \leq B\tau^{-2}E_0\left[\psi\right]\qquad \forall \tau > 0,\]
where $E_0\left[\psi\right]$ denotes a weighted second order energy of $\psi$ along $\Sigma_0^*$.

Note that we will \underline{not} require such quantitative decays results in this paper.
\end{remark}
We will also need to commute with angular momentum operators $\Omega^{(\alpha)}$. We obtain
\begin{proposition}\label{rpCommute}For every multi-index $\alpha$, let $\Omega^{(\alpha)}$ denote an arbitrary product of angular momentum operators as defined in Section~\ref{diffCoord}, and set $\varphi^{(\alpha)} \doteq \Omega^{(\alpha)}(r^2+a^2)^{1/2}\psi$. For all sufficiently large $R$, multi-indices $\alpha$, $\tau_1 < \tau_2$, and $p \in [0,2]$, we obtain the estimate of Proposition~\ref{rp} with $\varphi$ replaced by $\varphi^{(\alpha)}$.
\end{proposition}
\begin{proof}
If $a=0$, this is of course immediate since then $\left[\Omega^{(\alpha)},\Box_g\right] =0$.
Otherwise, one proceeds inductively in $|\alpha|$ and observes that the the error terms
arising
from $\left[\Omega^{(\alpha)},\Box_g\right]$ have sufficiently strong $r$ decay so as to
be either absorbed by good bulk terms on the left hand side of the estimate or controlled by the previous step.
\end{proof}

\section{Radiation fields and energy fluxes}\label{radenergy}
In this section, we will define the radiation fields along $\mathcal{H}^+_{\geq 0}$ or $\overline{\mathcal{H}^+}$ and $\mathcal{I}^+$ for solutions $\psi$ to the wave equation~(\ref{WAVE}) arising from
smooth initial data along $\Sigma_0^*$ or $\overline{\Sigma}$ which are compactly supported. Since $\mathring{\Sigma} \subset \overline{\Sigma}$, this \emph{a fortiori} defines the radiation field for solutions with compactly supported data along $\mathring{\Sigma}$.

The considerations at the horizon are straightforward and will be given in
Section~\ref{heretheHOR}.  The finiteness of both the
non-degenerate and degenerate radiation fluxes
follows as a soft application of Theorem~\ref{theResult} quoted in the previous section.

Null infinity will be handled in Section~\ref{formdefine}.
We will first have to explicitly define
$\mathcal{I}^+$ as an
additional boundary which can be attached to $\mathcal{D}$ (Defintion~\ref{defNullInf}).
The main result is Proposition~\ref{radinfwelldef}, which gives the
statement of {\bf Proposition~\ref{PROP1}}
of Section~\ref{radfieldsec}.
We shall then relate the radiation field as defined
to the limiting energy flux of $\psi$ along $\mathcal{I}^+$. Theorem~\ref{theResult}
immediately implies the latter is finite (see Theorem~\ref{theResultinf}), and according
to Proposition~\ref{efluxnullinf} it can by computed from the radiation field.

\subsection{The horizon}
\label{heretheHOR}
\subsubsection{The radiation field along $\mathcal{H}^+_{\ge 0}$ and $\overline{\mathcal{H}^+}$}
We begin with the radiation field along the horizon.
\begin{definition}\label{radHor}Given a solution $\psi$ to~(\ref{WAVE}) on $\mathcal{R}_{\geq 0}$ arising from smooth initial data along $\Sigma_0^*$ which are compactly supported, the \textbf{radiation field of $\psi$ along $\mathcal{H}_{\geq 0}^+$} is simply defined to be the restriction of $\psi$ to the horizon $\mathcal{H}_{\geq 0}^+$.
\end{definition}
Similarly, we have
\begin{definition}Given a solution $\psi$ to~(\ref{WAVE}) on $\mathcal{R}$ arising from smooth initial data along $\overline{\Sigma}$ which are compactly supported, the \textbf{radiation field of $\psi$ along $\overline{\mathcal{H}^+}$} is simply defined to be the restriction of $\psi$ to the horizon $\overline{\mathcal{H}^+}$.
\end{definition}
\begin{remark}Note that it follows immediately from Proposition~\ref{t0wellposed} that the radiation field is smooth along the horizon.
\end{remark}
\begin{remark}If the initial data for $\psi$ is compactly supported on $\mathring{\Sigma}$, then Remark~\ref{giasthrig} implies that the radiation field for $\psi$ is supported in $\mathcal{H}^+$.
\end{remark}
\begin{remark}Of course, given a solution $\psi$ to~(\ref{WAVE}) defined in the past of $\{{}^*t = 0\}$, one may make an analogous definition for the radiation field along $\mathcal{H}^-_{\ge 0}$. Similarly, one may define the radiation field along $\overline{\mathcal{H}^-}$ for a solution $\psi$ to~(\ref{WAVE}) arising from smooth initial data along $\overline{\Sigma}$.
\end{remark}

\subsubsection{The energy flux through $\mathcal{H}^+_{\ge 0}$ and $\overline{\mathcal{H}^+}$}\label{BALA}
We next define the non-degenerate energy flux along the horizon.
\begin{definition}\label{nonDegenHor}Given a solution $\psi$ to~(\ref{WAVE}) on $\mathcal{R}_{\geq 0}$ arising from smooth initial data along $\Sigma_0^*$ which are compactly supported, the \textbf{non-degenerate $N$-energy flux of $\psi$ through $\mathcal{H}^+_{\geq 0}$} is defined by
\[\int_{\mathcal{H}^+_{\geq 0}}\mathbf{J}^N_{\mu}[\psi]n^{\mu}_{\mathcal{H}^+}.\]
\end{definition}
\begin{remark}Note that Theorem~\ref{theResult} implies that this energy flux is finite.
\end{remark}
Observe that a straightforward computation shows that
\[\int_{\mathcal{H}^+_{\geq 0}}\mathbf{J}^N_{\mu}[\psi]n^{\mu}_{\mathcal{H}^+} \sim \int_{\mathcal{H}^+_{\geq 0}}\left[\left|K\psi\right|^2 + \left|\nabb\psi\right|^2\right].\]
In particular, all of the derivatives are tangent to the horizon; thus one may think of the non-degenerate flux as depending only on the radiation field.

Finally, we define the degenerate flux along the horizon.
\begin{definition}\label{degenHor}Given a solution $\psi$ to~(\ref{WAVE}) on $\mathcal{R}_{\geq 0}$ arising from smooth initial data along $\Sigma_0^*$ which are compactly supported, the \textbf{degenerate $K$-energy flux of $\psi$ through $\mathcal{H}^+_{\geq 0}$} is defined by
\[\int_{\mathcal{H}^+_{\geq 0}}\mathbf{J}^K_{\mu}[\psi]n^{\mu}_{\mathcal{H}^+}.\]
\end{definition}
A straightforward computation shows that
\[\int_{\mathcal{H}^+_{\geq 0}}\mathbf{J}^K_{\mu}[\psi]n^{\mu}_{\mathcal{H}^+} = \int_{\mathcal{H}^+_{\geq 0}}\left|K\psi\right|^2.\]
Similarly,
\begin{definition}Given a solution $\psi$ to~(\ref{WAVE}) on $\mathcal{R}$ arising from smooth initial data along $\overline{\Sigma}$ which are compactly supported, the \textbf{degenerate $K$-energy flux of $\psi$ through $\mathcal{H}^+$} is defined by
\[\int_{\mathcal{H}^+}\mathbf{J}^K_{\mu}[\psi]n^{\mu}_{\mathcal{H}^+}.\]
\end{definition}
A straightforward computation shows that
\[\int_{\mathcal{H}^+}\mathbf{J}^K_{\mu}[\psi]n^{\mu}_{\mathcal{H}^+} = \int_{\mathcal{H}^+}\left|K\psi\right|^2.\]

\subsection{Null infinity}
\label{formdefine}
We first define $\mathcal{I}^+$ as a suitable additional boundary
which can be attached to our spacetime.

\begin{definition}\label{defNullInf}As a differentiable manifold we define
\[\mathcal{I}^+ \doteq \mathbb{R}\times \mathbb{S}^2\]
and parameterize $\mathcal{I}^+$ in the standard fashion by coordinates $(\tau,\theta,\phi)$. Next, we extend our background differentiable structure $\mathcal{R}$ to a manifold with boundary
\[\mathcal{\tilde R} \doteq \mathcal{R} \cup \mathcal{I}^+\]
by declaring that for every sufficiently large $R$ and open set $\mathcal{U} \subset \mathcal{I}^+$, the set
\[\mathcal{U}_R \doteq \{(\tau,r,\theta,\phi) : r > R\text{ and }(\tau,\theta,\phi) \in \mathcal{U}\}\]
is open (where $(\tau,r,\theta,\phi)$ are the coordinates associated to the foliation $\{S_{\tau}\}_{\tau \in \mathbb{R}}$ which we defined in Section~\ref{hyperboloid}), identifying $\mathcal{I}^+$ with the points $(\tau,\infty,\theta,\phi)$, and then covering the sets $\mathcal{U}_R$ by a coordinate chart $\left(\tau,s,\theta,\phi\right) \in \mathbb{R} \times [0,1) \times \mathbb{S}^2$ via the map
\[\left(\tau,s,\theta,\phi\right) \mapsto \left(\tau,Rs^{-1},\theta,\phi\right).\]
\end{definition}
\begin{remark}Note that for every fixed $(\tau,\theta,\phi)$ there exists a unique limit $\lim_{r\to\infty}(\tau,r,\theta,\phi) \in \mathcal{I}^+$, and, if we denote these limits by $(\tau,\infty,\theta,\phi)$, then the map $(\tau,\theta,\phi) \mapsto (\tau,\infty,\theta,\phi)$ is a diffeomorphism from $\mathbb{R}\times\mathbb{S}^2$ to $\mathcal{I}^+$.
\end{remark}
\begin{remark}The above ``pedestrian''
definition of $\mathcal{I}^+$ is completely equivalent to the usual one involving a conformal compactification (see \cite{hawkingellis}).
\end{remark}

\begin{definition}Apply the discrete isometry $(t,\phi)\mapsto (-t,-\phi)$ to the foliation $\{S_{\tau}\}$ to define a new foliation $\{\tilde S_{\tau}\}$:
\[\tilde S_{\tau} \doteq \left\{
        \begin{array}{ll}
            -\text{}^*t = \tau & r \leq 5M \\
            -\text{}^*t - r^* + \frac{10M}{r} = \tau+ \text{}^*(5M) + 2 & \quad r > 5M.
        \end{array}
    \right.\]
Repeating the construction above with respect to this new foliation then defines past null infinity $\mathcal{I}^-$. Proceeding in an analogous fashion to Definition~\ref{defNullInf}, $\mathcal{I}^-$ may be glued to $\mathcal{\tilde R}$ as a suitable boundary.
\end{definition}
Lastly, it will be useful to introduce the notations
\[
\mathcal{I}^+_{\ge s} \doteq \{(\tau,\theta,\phi) : \tau \geq s\},\qquad \mathcal{I}^+_{\le s} \doteq \{(\tau,\theta,\phi) : \tau \leq s\}.
\]

\subsubsection{The radiation field along $\mathcal{I}^+$}
Recall that given a function $\psi$, in Section~\ref{rpSec} we introduced the notation
\[\varphi \doteq (r^2+a^2)^{1/2}\psi,\]
\[\varphi^{(\alpha)} \doteq \Omega^{(\alpha)}\varphi.\]
We now have the following straightforward corollary of Propositions~\ref{rp} and~\ref{rpCommute}.

\begin{proposition}\label{radinfwelldef}For all solutions $\psi$ to~(\ref{WAVE}) on $\mathcal{R}_{\geq 0}$ arising from smooth initial data along $\Sigma_0^*$ which are compactly supported, and each $(\tau,\theta,\phi) \in \mathbb{R}\times \mathbb{S}^2$, the function
\[
\varphi(\tau,\infty,r,\theta) \doteq \lim_{r\to\infty}\varphi(\tau,r,\theta,\phi) = \lim_{r\to\infty}(r^2+a^2)^{1/2}\psi(\tau,r,\theta,\phi)
\]
is well defined, and is in fact a smooth function on $\mathcal{I}^+$.
\end{proposition}
\begin{proof}Let $r_2 > r_1$. The fundamental theorem of calculus, Cauchy-Schwarz, and a Sobolev inequality on $\mathbb{S}^2$ imply

\begin{align*}
\left|\varphi(\tau,r_2,\theta,\phi) - \varphi(\tau,r_1,\theta,\phi)\right|^2 &\leq B\left(\sum_{\left|\alpha\right| \leq 2}\int_{\mathbb{S}^2}\left|\varphi^{(\alpha)}(\tau,r_2,\theta,\phi) - \varphi^{(\alpha)}(\tau,r_1,\theta,\phi)\right|\, \sin\theta\, d\theta\, d\phi\right)^2 \\
&\leq B\sum_{\left|\alpha\right|\leq 2}\left(\int_{S_{\tau}\cap\{r \geq r_1\}}\left[\left|\partial_{\tilde v}\varphi^{(\alpha)}\right| + r^{-2}\left|\partial_{\tilde u}\varphi^{(\alpha)}\right|\right]\, \sin\theta\, dr\, d\theta\, d\phi\right)^2 \\
&\leq Br_1^{-2}\sum_{\left|\alpha\right|\leq 2}\int_{S_{\tau}\cap\{r \geq r_1\}}\left[r^2\left|\partial_{\tilde v}\varphi^{(\alpha)}\right|^2 + r^{-2}\left|\partial_{\tilde u}\varphi^{(\alpha)}\right|^2\right]\, \sin\theta\, dr\, d\theta\, d\phi.
\end{align*}
In the second inequality we have used the fundamental theorem of calculus along $S_{\tau}$ and expressed the resulting derivative in terms of $\partial_{\tilde v}$ and $\partial_{\tilde u}$.

Now we conclude the proof of existence of the function $\varphi(\tau,\infty,\phi,\theta)$ by observing that Proposition~\ref{rpCommute} implies that this last quantity is bounded by $B(\tau)r_1^{-2}$.

Smoothness of $\varphi$ as a function on $\mathcal{I}^+$ follows in a straightforward manner by applying the above argument to $\partial_{\tau}^i\Omega^{(\alpha)}\varphi$, for $i \in \mathbb{Z}_{\geq 0}$ and $\left|\alpha\right| \in \mathbb{Z}_{\geq 0}$.
\end{proof}
\begin{remark}\label{linf}If we combine the proof of Proposition~\ref{radinfwelldef} with Theorem~\ref{theResult} we may easily conclude that for any $\tau_0 \in \mathbb{R}$, $(r^2+a^2)^{1/2}\psi$ converges to its limit $\varphi|_{r=\infty}$ in $L^{\infty}_{\mathbb{R}_{\geq \tau_0}\times\mathbb{S}^2}$.
\end{remark}

Following~\cite{friedlander} and using the previous proposition, we may now define the radiation field along $\mathcal{I}^+$.
\begin{definition}Given a solution $\psi$ to~(\ref{WAVE}) on $\mathcal{R}_{\geq 0}$ arising from smooth initial data along $\Sigma_0^*$ which are compactly supported, the \textbf{radiation field of $\psi$ along $\mathcal{I}^+$} is defined to be the function $\varphi(\tau,\infty,\theta,\phi)$.
\end{definition}

\begin{remark}Note that any solution $\psi$ to~(\ref{WAVE}) on $\mathcal{D}$ arising from smooth initial data along $\overline{\Sigma}$ which are compactly supported is, a fortiori, a solution to~(\ref{WAVE}) on $\mathcal{R}_{\geq 0}$ arising from smooth initial data along $\Sigma_0^*$ which are compactly supported (cf.~Remark~\ref{giasthrig}).
Thus, this definition of the radiation field may be applied to such solutions.
\end{remark}

\begin{remark}Of course, given a solution $\psi$ to~(\ref{WAVE}) defined in the past of $\{{}^*t = 0\}$, one may analogously define the radiation field along $\mathcal{I}^-$.
In particular, smooth compactly supported
data on $\overline{\Sigma}$ give rise to radiation fields along both $\mathcal{I}^+$
and $\mathcal{I}^-$.
\end{remark}
\begin{remark}In passing, we observe that the weighted estimates of Proposition~\ref{rpCommute} would allow us to easily show that the radiation field decays along null infinity:
\[\left|\varphi(\tau,\infty,\theta,\phi)\right| \leq B\tau^{-1/2}\sqrt{E_0\left[\psi\right]}\qquad \forall \tau > 0,\]
where $E_0$ is a weighted higher order energy along $\Sigma_0^*$. Again, we emphasize that we shall \underline{not} need to use such quantitative decay rates in this paper.
\end{remark}

\subsubsection{The energy flux through $\mathcal{I}^+$}\label{YALA}
In this section we will define the energy flux to future null infinity $\mathcal{I}^+$ for solutions to the wave equation~(\ref{WAVE}) arising from smooth initial data along $\Sigma_0^*$ which are compactly supported. Recall that
$\Sigma_s^*$ denotes the hypersurface $\{t^* = s\}$. We begin with the following lemma:

\begin{lemma}\label{fluxWellDef}Given a solution
$\psi$ to~(\ref{WAVE}) on $\mathcal{R}_{\geq 0}$ arising from smooth initial data along $\Sigma_0^*$ which are compactly supported, then for
every $\tau > 0$, the following limit exists:
\begin{equation}\label{isWellDef}
\lim_{s\to\infty}\int_{\Sigma_s^*\cap J^-(S_{\tau})}\mathbf{J}^T_{\mu}[\psi]n^{\mu}_{\Sigma_s^*}.
\end{equation}
\end{lemma}
\begin{proof}First of all, observe that for sufficiently large $s$, depending on $\tau$, the integration in~(\ref{isWellDef}) occurs far outside the ergoregion~(\ref{erg}), so that in particular, $T$ is a timelike Killing vector field in the region under consideration. With this in mind, a $\mathbf{J}^T$ energy estimate implies that
\[\int_{S_{\tau}}\mathbf{J}^T_{\mu}[\psi]n^{\mu}_{S_{\tau}} < \infty.\]
Consequently,
\[\lim_{s \to \infty}\int_{S_{\tau}\cap J^+(\Sigma_s^*)}\mathbf{J}^T_{\mu}[\psi]n^{\mu}_{S_{\tau}} = 0.\]
Let $s_1< s_2$ both be sufficiently large.
Refer to the figure below:
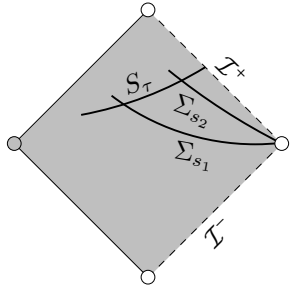
\[
\input{forflux.pstex_t}
\]
 It now suffices to observe the following immediate consequence of a $\mathbf{J}^T$ energy estimate:
\[\left|\int_{\Sigma_{s_2}\cap J^-(S_{\tau})}\mathbf{J}^T_{\mu}[\psi]n^{\mu}_{\Sigma_{s_2}} - \int_{\Sigma_{s_1}\cap J^-(S_{\tau})}\mathbf{J}^T_{\mu}[\psi]n^{\mu}_{\Sigma_{s_1}}\right| \leq \int_{S_{\tau}\cap J^+(\Sigma_{s_1})}\mathbf{J}^T_{\mu}[\psi]n^{\mu}_{S_{\tau}}.\]
\end{proof}

\begin{remark}Observe that this lemma holds for essentially any asymptotically flat spacetime possessing a suitable notion of future null infinity; in particular, we do not appeal to Theorem~\ref{theResult}.
\end{remark}

\begin{remark}\label{finiteTonly}We observe that one may easily check that if one considers smooth solutions which satisfy $\int_{\overline{\Sigma} \cap \{r \geq R\}}\mathbf{J}^T_{\mu}[\psi]n^{\mu}_{\overline{\Sigma}} < \infty$ for all sufficiently large $R$, but are not necessarily compactly supported, then an easy modification of the proof of Lemma~\ref{fluxWellDef} shows that for all $\tau_0 < \tau_1$, the limit
\[\lim_{s\to\infty}\int_{\Sigma_s^*\cap J^-(S_{\tau_1})\cap J^+(S_{\tau_0})}\mathbf{J}^T_{\mu}[\psi]n^{\mu}_{\Sigma_s^*}\]
exists.
\end{remark}

Lemma~\ref{fluxWellDef} allows us to make the following definitions.
\begin{definition}\label{defNullFlux}Given a solution $\psi$ to~(\ref{WAVE}) on $\mathcal{R}_{\geq 0}$ arising from smooth initial data along $\Sigma_0^*$ which are compactly supported, and $\tau > -\infty$, \textbf{the energy flux of $\psi$ through $\mathcal{I}^+_{\leq \tau}$} is defined by
\begin{equation}\label{limNullInf}
\int_{\mathcal{I}^+_{\leq \tau}}\mathbf{J}^T_{\mu}[\psi]n^{\mu}_{\mathcal{I}^+} \doteq \lim_{s\to\infty}\int_{\Sigma_s^*\cap J^-(S_{\tau})}\mathbf{J}^T_{\mu}[\psi]n^{\mu}_{\Sigma_s^*}.
\end{equation}
\end{definition}
\begin{remark}There is, of course, great flexibility in the choice of the hypersurfaces $S_0$ and $\Sigma_0^*$, but we will forgo a systematic treatment of which choice of hypersurfaces leaves the limit~(\ref{limNullInf}) unchanged.
\end{remark}

Since $\tau_1 < \tau_2$ implies that $J^-(S_{\tau_1}) \subset J^-(S_{\tau_2})$, it immediately follows that $\int_{\mathcal{I}^+_{\leq \tau}}\mathbf{J}^T_{\mu}[\psi]n^{\mu}_{\mathcal{I}^+}$ is an increasing function of $\tau$. Thus, we can make the following definition.

\begin{definition}\label{totalNullFlux}Given a solution $\psi$ to~(\ref{WAVE}) on $\mathcal{R}_{\geq 0}$ arising from smooth initial data along $\Sigma_0^*$ which are compactly supported, the (total) \textbf{flux of $\psi$ through null infinity $\mathcal{I}^+$} is defined by
\begin{equation}\label{totEngFlux}
\int_{\mathcal{I}^+}\mathbf{J}^T_{\mu}[\psi]n^{\mu}_{\mathcal{I}^+} \doteq \lim_{\tau\to\infty}\lim_{s\to\infty}\int_{\Sigma_s^*\cap J^-(S_{\tau})}\mathbf{J}^T_{\mu}[\psi]n^{\mu}_{\Sigma_s^*} \in \mathbb{R}_{\geq 0}\cup\{\infty\}.
\end{equation}
\end{definition}
\begin{remark}As with Definition~\ref{defNullFlux}, we note that this definition also makes sense for essentially any spacetime possessing a suitable notion of future null infinity.
\end{remark}

Now we observe the following immediate consequence of Definition~\ref{totalNullFlux} and Theorem~\ref{theResult}.
\begin{theorem}\label{theResultinf}All sufficiently regular solutions $\psi$ to~(\ref{WAVE}) on $\mathcal{R}_{\geq 0}$ satisfy
\[\int_{\mathcal{I}^+}\mathbf{J}^T_{\mu}[\psi]n^{\mu}_{\mathcal{I}^+} \leq B\int_{\Sigma_0^*}\mathbf{J}^N_{\mu}[\psi]n^{\mu}_{\Sigma_0^*}.\]
In particular, in the case of smooth compactly supported initial data of $\Sigma_0^*$ or $\overline{\Sigma}$, the total flux to null infinity~(\ref{totEngFlux}) is finite.
\end{theorem}

Finally, the next proposition establishes the expected connection between the radiation field along null infinity with the energy flux to null infinity
\begin{proposition}\label{efluxnullinf}Given a solution $\psi$ to~(\ref{WAVE}) on $\mathcal{R}_{\geq 0}$ arising from smooth initial data along $\Sigma_0^*$ which are compactly supported, we have
\[\int_{(-\infty,\tau)\times\mathbb{S}^2}\left|\partial_{\tau}\varphi(\infty, \tau, \theta, \phi)
\right|^2\, \sin\theta\, d\tau\, d\theta\, d\phi  = \int_{\mathcal{I}^+_{\leq \tau}}\mathbf{J}^T_{\mu}[\psi]n^{\mu}_{\mathcal{I}^+}\qquad \forall \tau \in (-\infty,\infty].\]
\end{proposition}
\begin{proof}First of all, a straightforward computation gives
\begin{align*}
\int_{\Sigma_s^*\cap J^-(S_{\tau})}\mathbf{J}^T_{\mu}[\psi]n^{\mu}_{\Sigma_s^*} =&
\int_{\Sigma_s^*\cap J^-(S_{\tau})}\left|\partial_{\tau}\varphi\right|^2\, \sin\theta\, dv\, d\theta\, d\phi
\\& \qquad +
O\left(\int_{\Sigma_s^*\cap J^-(S_{\tau})}\left[|\partial_{\tilde v}\varphi|^2 + |\nabb\varphi|^2\right]\, \sin\theta\, dv\, d\theta\, d\phi\right) \text{ as }s\to\infty.
\end{align*}

Now we simply observe that Proposition~\ref{rp} (with any choice of $p \in (0,2]$) implies that we can find a (dyadic) sequence\footnote{The point being that $\int_1^{\infty}\frac{|f(x)|}{x}dx < \infty$ implies that there exists a sequence $\{x_i\}_{i=1}^{\infty}$ with $x_i \in [2^i,2^{i+1}]$ such that $\lim_{i\to\infty}f(x_i) = 0$.} $\{s_i\}_{i=1}^{\infty}$ such that $\lim_{i\to\infty}s_i = \infty$ and
\[
\lim_{i\to\infty}\int_{\Sigma_{s_i}\cap J^-(S_{\tau})}\left[|\partial_{\tilde v}\varphi|^2 + |\nabb\varphi|^2\right]\, \sin\theta\, dv\, d\theta\, d\phi = 0.
\]
\end{proof}

\section{Carter's separation and the microlocal radiation fields}\label{cartsecmicrorad}
As in our previous work~\cite{partIII}, estimates obtained by exploiting
Carter's separation of the wave equation $(\ref{WAVE})$
will play a fundamental role in our analysis. In this section, we quote a number of
results from~\cite{realmodestability}
and~\cite{partIII} concerning the theory of the radial o.d.e~$(\ref{radodehere})$
and its relation to $(\ref{WAVE})$.
(In Section~\ref{radodesec} to follow, we will then obtain various refinements of the quantitative o.d.e.~estimates
of~\cite{partIII} which will be fundamental for our arguments.)

We begin in Section~\ref{separingh} by reviewing
our relevant  formalism based on the Fourier transform of ``sufficiently integrable'' solutions
(Definition~\ref{sufficient}); the reader should consult~\cite{partIII} for more details.

We shall then quote in Section~\ref{asymAnalysis2}
some results from~\cite{realmodestability} concerning
the asymptotics of solutions of $(\ref{radodehere})$, which in particular allow
us to define the special solutions $U_{\rm hor}$, $U_{\rm inf}$ referred to in the introduction.
We state Proposition~\ref{microEnergyEst}, the microlocal version of the energy identity (we will consider more general currents in Section~\ref{sepcurr} below).

The Wronskian $\mathfrak{W}$, as well as
the reflection $\mathfrak{R}$ and transmission coefficients $\mathfrak{T}$
referred to already
(together with their dual coefficients $\tilde{\mathfrak{R}}$ and $\tilde{\mathfrak{T}}$),
are all defined in Section~\ref{wronskianetal},
appealing to the real-mode stability theorem of~\cite{realmodestability}.
We then obtain Corollary~\ref{superradiantAmp} which gives that the strict inequality
$(\ref{softst})$ indeed holds for any superradiant frequency and establish a fundamental solution formula for the radial o.d.e.~in Proposition~\ref{fundamental}.

Finally, our separation will allow us to define the ``microlocal'' radiation fields
and fluxes in Section~\ref{secMicroRad}. (Later, in Section~\ref{relatephy}, these will be related to the radiation fields and degenerate-energy
fluxes defined in physical space.)

\subsection{Separating the wave equation}
\label{separingh}
We begin by recalling the following definition.
\begin{definition}\label{sufficient}
We say that a smooth function
$\Psi:\mathring{\mathcal{R}}\to \mathbb C$ is ``sufficiently integrable'' if
for every $j \geq 1$ and $r_0 > r_+$, we have
            \[\sum_{0\leq j_1 + j_2 + \left|\alpha\right| \leq j}\int_{-\infty}^{\infty}\int_{\mathbb{S}^2}\left|\nabb^{\alpha}\partial_{r^*}^{j_1}T^{j_2}\Psi\right|^2|_{r = r_0}\sin\theta\, dt\, d\theta\, d\phi < \infty.\]
\end{definition}
\begin{remark}We note that this definition is in fact weaker than that given in~\cite{partIII}.
\end{remark}
\begin{remark}\label{obvsuffint}Observe that it follows immediately from Proposition~\ref{t0wellposed}, Theorem~\ref{theResult} and Remark~\ref{theResultflipped} that any solution to the wave equation arising from smooth compactly supported initial data along $\overline{\Sigma}$ is sufficiently integrable in the sense of Definition~\ref{sufficient} (cf. Remark~\ref{giasthrig}).
\end{remark}
Next, we recall the oblate spheroidal harmonics
\[
\{S_{m\ell}(\nu,\cos \theta)e^{im\phi}\}_{m\ell},\, \nu \in \mathbb{R},
\]
which are the eigenfunctions of the
self-adjoint operator
\[
P(\nu)\, f= -\frac 1{\sin\theta} \frac{\partial}{\partial\theta} \left (\sin\theta \frac{\partial}{\partial\theta}f\right)
-\frac{\partial^2 f}{\partial\phi^2}\frac{1}{\sin^2\theta}
- \nu^2 \cos^2\theta f
\]
on $L^2(\sin\theta\, d\theta\, d\phi)$. We denote the corresponding eigenvalues by
$\lambda^{(\nu)}_{m\ell}\in \mathbb{R}$ where $m \in \mathbb{Z}$ and $l \geq \left| m\right|$. The labeling is uniquely determined by requiring that $\lambda^{(\nu)}_{m\ell}$ depends smoothly on $\nu$ and setting $\lambda^{(0)}_{m\ell} = \ell\left(\ell+1\right)$.\footnote{See Proposition B.1 of~\cite{growingmodes} for a proof that this does indeed uniquely determine $\{\lambda^{(\nu)}_{m\ell}\}$.} These satisfy
\begin{equation}\label{eq:lam}
\lambda_{m\ell}^{(\nu)}+\nu^2\ge |m|(|m|+1),
\end{equation}
\begin{equation}\label{lamBound}
\lambda_{m\ell}^{(\nu)} + \nu^2 \geq 2\left|m\nu\right|.
\end{equation}
Because of the above relations, it is often convenient to work with
\[\Lambda_{m\ell}(\nu) \doteq \lambda_{m\ell}(\nu) + \nu^2.\]

Let $\Psi$ be sufficiently integrable in the sense of Definition~\ref{sufficient}. Then, setting $\nu = a\omega$, where $a$ is the Kerr parameter, for each $\omega\in \mathbb R$, we decompose
\[
\Psi(t,r,\theta,\phi)=\frac{1}{\sqrt{2\pi}}\int_{-\infty}^\infty\sum_{m\ell} e^{-i\omega t} \Psi^{(a\omega)}_{m\ell}(r)S_{m\ell}(a\omega,\cos\theta)e^{im\phi} d\omega.
\]
The sufficiently integrable assumption implies that for each fixed $r$, this equality may be interpreted in $L^2_tL^2_{\mathbb{S}^2}$. Now define
\begin{equation}
\label{homogdefeq}
F=\Box_g\Psi.
\end{equation}

The sufficiently integrable assumption implies that we may  define  the coefficients $\left(\rho^2F\right)_{m\ell}^{(a\omega)}(r)$ as above (recall that $\rho^2 = r^2+a^2\cos^2\theta$).

Carter's formal separation~\cite{cartersep} of the wave operator yields:
\begin{proposition}
Let $\Psi$ be sufficiently integrable in the sense of Definition~\ref{sufficient}, and let $F$ be defined by $(\ref{homogdefeq})$. Then
\begin{align}
\label{CartersODE}
\Delta \frac{d}{dr} \left (\Delta \frac{d\Psi_{m\ell}^{(a\omega)}}{dr}\right)& + \left (a^2m^2 + (r^2+a^2)^2\omega^2-4Mra\omega m - \Delta\Lambda_{m\ell} \right) \Psi_{m\ell}^{(a\omega)}=\Delta\,
\left(\rho^2F\right)_{m\ell}^{(a\omega)}.
\end{align}
Note that the sufficiently integrable assumption allows us to interpret this equality for each $r$ in $L^2_{\omega}l^2_{m\ell}$.
\end{proposition}
\begin{remark}
\label{ONLysmooth}
It will turn out to suffice that we study smooth \underline{smooth} solutions to the o.d.e.~(\ref{CartersODE}). See the discussion in Definition~\ref{microRad}.
\end{remark}
Using the definition $(\ref{r*def})$ of $r^*$ and setting
\begin{equation}\label{uDef}
u^{(a\omega)}_{m\ell}(r)=(r^2+a^2)^{1/2}
 \Psi^{(a\omega)}_{m\ell} (r),
\end{equation}
\begin{equation}\label{hDef}
 H^{(a\omega)}_{m\ell}(r)=\frac{\Delta \left(\rho^2F\right)^{(a\omega)}_{m\ell}(r)}{(r^2+a^2)^{3/2}},
\end{equation}
we obtain
\begin{equation}
\label{e3iswsntouu}
\frac{d^2}{(dr^*)^2}u^{(a\omega)}_{m\ell}+(\omega^2 - V^{(a\omega)}_{m\ell }(r))u =
H^{(a\omega)}_{m\ell},
\end{equation}
where
\begin{equation}
\label{defofV}
V^{(a\omega)}_{m \ell}(r)= \frac{4Mram\omega-a^2m^2+\Delta\Lambda_{m\ell}}{(r^2+a^2)^2}
+\frac{\Delta(3r^2-4Mr+a^2)}{(r^2+a^2)^3}
-\frac{3\Delta^2 r^2}{(r^2+a^2)^4}.
\end{equation}
We will often refer to~(\ref{e3iswsntouu}) as the ``radial o.d.e.''

As in~\cite{partIII}, we shall often suppress the dependence of
$u$, $H$ and $V$ on $a\omega$, $m$, $\ell$ in our notation.
We will also use the notation
\begin{equation}
\label{primenotation}
'=\frac{d}{dr^*}.
\end{equation}
Note that
\[
r' = \frac{\Delta}{r^2+a^2}.
\]

\subsection{Asymptotic analysis of the radial o.d.e.}\label{asymAnalysis2}
In this section we will collect various facts concerning the asymptotic analysis of the radial o.d.e.~(\ref{e3iswsntouu}).
In view of our applications and Remark~\ref{ONLysmooth}, all results stated will concern smooth
solutions.
We will omit proofs as the material is standard (see, e.g., \cite{olver}).

\begin{proposition}\label{asymAnalysis}
Fix parameters $(\omega,m,\ell) \in \mathbb{R}\times \mathbb{Z}\times \mathbb{Z}_{\geq |m|}$ with $\omega \neq 0$ and $\omega \neq \upomega_+m$, and let $u$ be a smooth solution of the radial o.d.e.~(\ref{e3iswsntouu})
\[u'' + (\omega^2-V)u = H,\]
where $H(r)$ smoothly extends to $r = r_+$ and vanishes for large $r$ (of course, by using the relation $\frac{d}{dr} = \frac{r^2+a^2}{\Delta}\frac{d}{dr^*}$, the smoothness condition at $r = r_+$ can be translated to a condition on the limits of $\frac{d^kH}{d(r^*)^k}$ as $r^* \to -\infty$). Then there exist unique complex numbers $a_{\mathcal{H}^+}$, $a_{\mathcal{H}^-}$, $a_{\mathcal{I}^+}$, and $a_{\mathcal{I}^-}$, depending on $u$,
such that
\begin{equation}\label{expansion1}
u = a_{\mathcal{I}^+}e^{i\omega r^*} + a_{\mathcal{I}^-}e^{-i\omega r^*} + O(r^{-1})\text{ as }r\to\infty,
\end{equation}
\begin{equation}\label{expansion2}
u = a_{\mathcal{H}^+}e^{-i(\omega -\upomega_+m)r^*} + a_{\mathcal{H}^-}e^{i(\omega -\upomega_+m)r^*} + O(r-r_+)\text{ as }r\to r_+.
\end{equation}
Here the $O(r^{-1})$ and $O(r-r_+)$ are both preserved upon differentiation in $r^*$.
\end{proposition}

Next, we turn to the ``microlocal energy identity''.
\begin{proposition}\label{microEnergyEst}Fix parameters $(\omega,m,\ell) \in \mathbb{R}\times \mathbb{Z}\times \mathbb{Z}_{\geq |m|}$ with $\omega \neq 0$ and $\omega \neq \upomega_+m$, and let $u$ be a smooth solution of the radial o.d.e.~(\ref{e3iswsntouu}) with $H(r^*)$ compactly supported in $r^*$. Then, we have
\[\omega^2|a_{\mathcal{I}^+}|^2 - \omega^2|a_{\mathcal{I}^-}|^2 + \omega(\omega - \upomega_+m)|a_{\mathcal{H}^+}|^2 - \omega(\omega-\upomega_+m)|a_{\mathcal{H}^-}|^2 = \omega\int_{-\infty}^{\infty}\text{Im}\left(H\overline{u}\right)dr^*.\]
\end{proposition}
\begin{proof}We recall the microlocal energy current from~\cite{partIII}:
\[
Q^T \doteq \omega\text{Im}(u'\overline{u}),
\]
which satisfies
\[
(Q^T)' = \omega\text{Im}\left(H\overline{u}\right).
\]
(The above is of course the most basic energy current associated to $(\ref{e3iswsntouu})$.
We will discuss this and several other currents  in Section~\ref{sepcurr}). The proposition then follows immediately from the fundamental theorem of calculus and the expansions~(\ref{expansion1}) and~(\ref{expansion2}).
\end{proof}

It will be useful to introduce the following definitions.
\begin{definition}\label{hor}Let $(\omega,m,\ell) \in \mathbb{R} \times \mathbb{Z} \times \mathbb{Z}_{\geq |m|}$. Then we define $U_{\rm hor}(r^*,\omega,m,l)$ to be the unique function satisfying
    \begin{enumerate}
        \item $U_{\rm hor}'' + \left(\omega^2-V\right)U_{\rm hor} = 0$.
        \item $U_{\rm hor} \sim e^{-i\left(\omega-\upomega_+m\right)r^*}\text{ near }r^* = -\infty$.\footnote{More precisely, the requirement is that $U_{\rm hor}e^{i\left(\omega-\upomega_+m\right)r^*}$ extends to $r = r_+$ as a smooth function of $r$.}
        \item $\left|U_{\rm hor}\left(-\infty\right)\right|^2 = 1$.
    \end{enumerate}
\end{definition}

\begin{remark}Note that this definition makes sense even when $\omega - \upomega_+ m = 0$ or $\omega = 0$; see, e.g., the discussion in Appendix C.1 of~\cite{growingmodes}.
\end{remark}

\begin{remark}The physical space interpretation of $U_{\rm hor}$ is that $e^{-it\omega}e^{im\phi}S_{m\ell}\left(\theta\right)U_{\rm hor}(r^*)$ corresponds to an amplitude normalised solution of the wave equation ``frequency localised'' to $(\omega,m,\ell)$, with a vanishing energy flux along $\mathcal{H}^-$ and a finite energy flux on any compact subset of $\mathcal{H}^+$.
\end{remark}

\begin{definition}\label{out}For $\omega \neq 0$, define $U_{\rm inf}(r^*,\omega,m,l)$ to be the unique function satisfying
    \begin{enumerate}
        \item $U_{\rm inf}'' + \left(\omega^2-V\right)U_{\rm inf} = 0$.
        \item $U_{\rm inf} \sim e^{i\omega r^*}\text{ near }r^* = \infty$.\footnote{More precisely, this means that $U_{\rm inf}$ exhibits a (generally divergent) asymptotic expansion $U_{\rm inf} = e^{i\omega r^*}\sum_{i=0}^{\infty}\frac{A_i}{r^i}$ as $r\to \infty$.}
        \item $\left|U_{\rm inf}\left(\infty\right)\right|^2 = 1$.
    \end{enumerate}
\end{definition}

\begin{remark}The physical space interpretation of $U_{\rm inf}$ is that $e^{-it\omega}e^{im\phi}S_{m\ell}\left(\theta\right)U_{\rm inf}(r^*)$ corresponds to a an amplitude normalised solution of the wave equation ``frequency localised'' to $(\omega,m,\ell)$, with a vanishing energy flux along $\mathcal{I}^-$ and a finite energy flux on any compact subset of $\mathcal{I}^+$.
\end{remark}

\begin{remark}When $H = 0$, by exploiting the linear independence of the pairs $\{U_{\rm hor}, \overline{U_{\rm hor}}\}$ and $\{U_{\rm inf},\overline{U_{\rm inf}}\}$, one may easily check that expansions~(\ref{expansion1}) and~(\ref{expansion2}) may be written as the identities
\[
u = a_{\mathcal{I}^+}U_{\rm inf} + a_{\mathcal{I}^-}\overline{U_{\rm inf}},
\]
\[
u = a_{\mathcal{H}^+}U_{\rm hor} + a_{\mathcal{H}^-}\overline{U_{\rm hor}}.
\]
Cf.~footnote~[\ref{footnote3}]
\end{remark}

\begin{proposition}\label{smoothenough}The constructions of $U_{\rm hor}$ and $U_{\rm inf}$ imply that for each $k \geq 0$,
\[\left\vert\left\vert \frac{d^k}{d(r^*)^k}U_{\rm hor}\right\vert\right\vert_{L^{\infty}_{r^*}} \leq B\left(\omega,m,\ell,k\right),\]
and $\frac{d^k}{d(r^*)^k}U_{\rm hor}$ depends analytically on $\omega$. Similarly, if we additionally assume that $\omega \neq 0$, we also have
\[\left\vert\left\vert \frac{d^k}{d(r^*)^k}U_{\rm inf}\right\vert\right\vert_{L^{\infty}_{r^*}} \leq B\left(\omega,m,\ell,k\right),\]
and $\frac{d^k}{d(r^*)^k}U_{\rm inf}$ depends analytically on $\omega \in \mathbb{R}\setminus\{0\}$.
\end{proposition}

\subsection{The Wronskian and the reflection and transmission coefficients}
\label{wronskianetal}

\begin{definition}\label{wronk}For $\omega \neq 0$, we define $\mathfrak{W}(\omega,m,\ell)$ to be the Wronskian of $U_{\rm hor}$ and $U_{\rm inf}$:
\[\mathfrak{W} \doteq U_{\rm inf}'U_{\rm hor} - U_{\rm inf}U_{\rm hor}'.\]
\end{definition}
\begin{remark}\label{linDep}Note that one may easily check that $\mathfrak{W}$ does not depend on $r^*$ and vanishes if and only if $U_{\rm hor}$ and $U_{\rm inf}$ are linearly dependent.
\end{remark}

In~\cite{realmodestability}, the following was  shown:
\begin{theorem}\label{modestability}\cite{realmodestability} For all $(\omega,m,\ell) \in \mathbb{R}\times \mathbb{Z} \times \mathbb{Z}_{\geq |m|}$ with $\omega\ne 0$ we have
\[\mathfrak{W}\left(\omega,m,\ell\right) \neq 0,\]
and thus the functions $U_{\rm hor}$ and $U_{\rm inf}$ are linearly independent.
\end{theorem}

The non-vanishing of the Wronskian will allow us to define the \emph{reflection} and \emph{transmission} coefficients. First we need the following lemma which follows immediately from Remark~\ref{linDep} and the non-vanishing of the Wronskian.

\begin{lemma}\label{lemmReflTrans}For $\omega \neq 0$ and $\omega \neq \upomega_+m$,
there exists a unique set of complex numbers
$\mathfrak{R}(\omega,m,\ell)$, $\tilde{\mathfrak{R}}(\omega,m,\ell)$, $\mathfrak{T}(\omega,m,\ell)$ and $\tilde{\mathfrak{T}}(\omega,m,\ell)$ which satisfy
\begin{align}
\frac{\mathfrak{T}}{-i(\omega-\upomega_+m)}U_{\rm hor} &= \frac{\mathfrak{R}}{i\omega}U_{\rm inf} + \frac{\overline{U_{\rm inf}}}{i\omega},
\\ \frac{\tilde{\mathfrak{T}}}{i\omega}U_{\rm inf} &= \frac{\tilde{\mathfrak{R}}}{-i(\omega-\upomega_+m)}U_{\rm hor} + \frac{\overline{U_{\rm hor}}}{-i(\omega-\upomega_+m)},
\end{align}
\end{lemma}

Now we can define the reflection and transmission coefficients.
\begin{definition}\label{reftransDef}
The complex numbers
$\mathfrak{R}$ and $\tilde{\mathfrak{R}}$ are
called the \textbf{reflection} coefficients, and $\mathfrak{T}$ and $\tilde{\mathfrak{T}}$ are called the \textbf{transmission} coefficients.
\end{definition}
\begin{remark}If one considers a solution to the wave equation which is ``sourced'' with a flux along $\mathcal{I}^-$ equal to $1$ and no energy along $\mathcal{H}^-$ and which is furthermore approximately localised to the frequency $(\omega,m,\ell)$, then $\mathfrak{R}$ measures the amount of energy ``reflected'' back to future null infinity $\mathcal{I}^+$, and $\mathfrak{T}$ measures the energy ``transmitted'' to the future event horizon $\mathcal{H}^+$. There is a similar interpretation for $\tilde{\mathfrak{R}}$ and $\tilde{\mathfrak{T}}$. Our Theorem~\ref{constructaway} will make these interpretations rigorous.
\end{remark}

\begin{remark}
One often sees the reflection and transmission coefficients $\mathfrak{R}$ and $\mathfrak{T}$ defined so that they measure the \underline{amplitude} transmitted to the future event horizon and reflected to future null infinity of a wave of \underline{amplitude} $1$ along $\mathcal{I}^-$, see e.g.~Section
28 of~\cite{chandrasekhar}. However, in the context of scattering theory for finite energy solutions, one does \underline{not} expect to control the radiation fields $\uppsi$ and $\upphi$ in $L^2$ along $\mathcal{H}^{\pm}$ and $\mathcal{I}^{\pm}$, hence an energy normalisation is most natural.
\end{remark}

Applying Proposition~\ref{microEnergyEst} immediately yields
\begin{corollary}
\label{superradiantAmp}
Fix a frequency triple $(\omega,m,\ell)$ which satisfy $\omega \neq 0 $ and $\omega \neq \upomega_+m$. Then
\[
|\mathfrak{R}|^2 + \frac{\omega}{\omega-\upomega_+m}|\mathfrak{T}|^2 = 1.
\]
In particular, if
\begin{equation}\label{thesearesuperradiant}
\omega(\omega-\upomega_+m) < 0,
\end{equation}
i.e. the parameters are \underline{superradiant}, then
\[|\mathfrak{R}|^2 > 1.\]
\end{corollary}

\begin{proof}
For the second statement, it
suffices to note that the basic local existence theory for the radial o.d.e.~implies that $\mathfrak{T} \neq 0$ (see~\cite{olver}).
\end{proof}

Though the reflection and transmission coefficient have a nice interpretation in terms of the scattering of waves coming from $\mathcal{H}^-$ and $\mathcal{I}^-$, for technical reasons they are not always the most convenient way to parameterize solutions to the radial o.d.e. Instead we shall often use the following quantities.
\begin{definition}
\label{reflectransmit}
For $\omega \neq 0$ and $\omega \neq \upomega_+ m$, we define the complex
numbers $\mathfrak{A}_{\mathcal{I}^+}(\omega,m,\ell)$, $\mathfrak{A}_{\mathcal{I}^-}(\omega,m,\ell)$, $\mathfrak{A}_{\mathcal{H}^+}(\omega,m,\ell)$, and $\mathfrak{A}_{\mathcal{H}^-}(\omega,m,\ell)$ by
\[
U_{\rm hor} = \mathfrak{A}_{\mathcal{I}^+}e^{i\omega r^*} + \mathfrak{A}_{\mathcal{I}^-}e^{-i\omega r^*} + O\left(r^{-1}\right)\text{ as }r^*\to\infty,
\]
\[
U_{\rm inf} = \mathfrak{A}_{\mathcal{H}^+}e^{-i(\omega-\upomega_+m) r^*} + \mathfrak{A}_{\mathcal{H}^-}e^{i(\omega-\upomega_+m) r^*} + O\left(r-r_+\right)\text{ as }r \to r_+.
\]
\end{definition}

Observe that
$\mathfrak{A}_{\mathcal{I}^+}(\omega,m,\ell)$, $\mathfrak{A}_{\mathcal{I}^-}(\omega,m,\ell)$, $\mathfrak{A}_{\mathcal{H}^+}(\omega,m,\ell)$, and $\mathfrak{A}_{\mathcal{H}^-}(\omega,m,\ell)$ must obey the following constraints.
\begin{lemma}\label{aconstr}
\[\mathfrak{A}_{\mathcal{I}^+}\mathfrak{A}_{\mathcal{H}^+} + \mathfrak{A}_{\mathcal{I}^-}\overline{\mathfrak{A}}_{\mathcal{H}^-} = 1,\]
\[\mathfrak{A}_{\mathcal{I}^+}\mathfrak{A}_{\mathcal{H}^+} + \overline{\mathfrak{A}}_{\mathcal{I}^-}\mathfrak{A}_{\mathcal{H}^-} = 1,\]
\[\mathfrak{A}_{\mathcal{I}^+}\mathfrak{A}_{\mathcal{H}^-} + \mathfrak{A}_{\mathcal{I}^-}\overline{\mathfrak{A}}_{\mathcal{H}^+} = 0,\]
\[\mathfrak{A}_{\mathcal{I}^-}\mathfrak{A}_{\mathcal{H}^+} + \mathfrak{A}_{\mathcal{H}^-}\overline{\mathfrak{A}}_{\mathcal{I}^+} = 0.\]

\end{lemma}
\begin{proof}We may write
\begin{align}
U_{\rm hor} &= \mathfrak{A}_{\mathcal{I}^+}U_{\rm inf} + \mathfrak{A}_{\mathcal{I}^-}\overline{U}_{\rm inf}
\\ \nonumber &= \mathfrak{A}_{\mathcal{I}^+}\left(\mathfrak{A}_{\mathcal{H}^+}U_{\rm hor} + \mathfrak{A}_{\mathcal{H}^-}\overline{U}_{\rm hor}\right) + \mathfrak{A}_{\mathcal{I}^-}\left(\overline{\mathfrak{A}}_{\mathcal{H}^+}\overline{U}_{\rm hor} + \overline{\mathfrak{A}}_{\mathcal{H}^-}U_{\rm hor}\right)
\\ \nonumber &= \left(\mathfrak{A}_{\mathcal{I}^+}\mathfrak{A}_{\mathcal{H}^+} + \mathfrak{A}_{\mathcal{I}^-}\overline{\mathfrak{A}}_{\mathcal{H}^-}\right)U_{\rm hor} + \left(\mathfrak{A}_{\mathcal{I}^+}\mathfrak{A}_{\mathcal{H}^-} + \mathfrak{A}_{\mathcal{I}^-}\overline{\mathfrak{A}}_{\mathcal{H}^+}\right)\overline{U}_{\rm hor}.
\end{align}
Similarly,
\[U_{\rm inf} = \left(\mathfrak{A}_{\mathcal{H}^+}\mathfrak{A}_{\mathcal{I}^+} + \overline{\mathfrak{A}}_{\mathcal{I}^-}\mathfrak{A}_{\mathcal{H}^-}\right)U_{\rm inf} + \left(\mathfrak{A}_{\mathcal{H}^+}\mathfrak{A}_{\mathcal{I}^-} + \overline{\mathfrak{A}}_{\mathcal{I}^+}\mathfrak{A}_{\mathcal{H}^-}\right)\overline{U}_{\rm inf}.\]
The lemma follows immediately.
\end{proof}

The following relationships may be easily verified in a similar fashion to Lemma~\ref{aconstr}.
\begin{lemma}\label{quitenicerelations}
\[\mathfrak{W} = 2i\omega \mathfrak{A}_{\mathcal{I}^-}, \qquad
\mathfrak{W} = 2i\left(\omega- \upomega_+m\right)\mathfrak{A}_{\mathcal{H}^-},\]
\[\mathfrak{R} = -\overline{\mathfrak{A}}_{\mathcal{H}^+}\left(\mathfrak{A}_{\mathcal{H}^-}\right)^{-1}, \qquad
\mathfrak{T} = -\frac{\left(\omega-\upomega_+m\right)}{\omega}\left(\mathfrak{A}_{\mathcal{I}^-}\right)^{-1},\]
\[\tilde{\mathfrak{R}} = -\overline{\mathfrak{A}}_{\mathcal{I}^+}\left(\mathfrak{A}_{\mathcal{I}^-}\right)^{-1},
\qquad
\tilde{\mathfrak{T}} = -\frac{\omega}{\left(\omega-\upomega_+m\right)}\left(\mathfrak{A}_{\mathcal{H}^-}\right)^{-1}.\]

\end{lemma}
We close the section with a final remark:
\begin{remark}By exploiting the underlying analyticity
(cf.~Corollary~\ref{novanishR}), one can in fact define the reflection and transmission coefficients almost everywhere without the mode stability result of~\cite{realmodestability}, quoted here as Theorem~\ref{modestability}. Given this, we see that Theorem~\ref{modestability} is equivalent to the statement that the reflection and transmission coefficients are bounded on any compact set of frequencies, with a bound depending however on the set. The fact the reflection and transmission coefficients are uniformly bounded \underline{over all frequencies} is the content of Theorem~\ref{refltransBound}, to be proven in Section~\ref{microlocaliledsec}.
\end{remark}

We end this section with a final corollary
of Theorem~\ref{modestability} which concerns a fundamental-solution representation
of solutions $u$ of $(\ref{e3iswsntouu})$ with vanishing $a_{\mathcal{H}^-} = a_{\mathcal{I}^-} = 0$.
\begin{proposition}\label{fundamental}Let $u$ be a smooth solution to the radial o.d.e.~(\ref{e3iswsntouu}) with a right hand side $H$ such that $H(r)$ smoothly extends to $r = r_+$ and vanishes for large $r$, and such that $u$ satisfies $a_{\mathcal{H}^-} = a_{\mathcal{I}^-} = 0$.
Then $u$ is given by the following explicit formula:
\begin{align*}
u(r^*) =\ \mathfrak{W}^{-1}\Bigg(U_{\rm inf}(r^*)\int_{-\infty}^{r^*}U_{\rm hor}(x^*)H(x^*)dx^* + U_{\rm hor}(r^*)\int_{r^*}^{\infty}U_{{\rm inf}}(x^*)H(x^*)dx^*\Bigg).
\end{align*}
\end{proposition}
\begin{proof}Given the non-vanishing of the Wronskian (Theorem~\ref{modestability}), this is a trivial computation.
\end{proof}

\subsection{The microlocal radiation fields and fluxes}\label{secMicroRad}
We are now ready to define the microlocal radiation fields. As the name suggests, the definition of the microlocal radiation fields relies on the Fourier transform; hence, we will only be able to define the microlocal radiation fields for a solution $\psi$ if it is defined on all of $\mathring{\mathcal{R}}$, not just $\mathring{\mathcal{R}}_{\geq 0}$.

\begin{definition}\label{microRad}
For all solutions $\psi$ to~(\ref{WAVE}) on $\mathring{\mathcal{R}}$, which are sufficiently integrable in the sense of Definition~\ref{sufficient}, we may apply Carter's separation to $\psi$ and define the corresponding
function $u$. An easy argument (one can slightly modify the proof of Lemma 5.4.1 of~\cite{partIII}) implies that for \underline{almost
every} $\omega$ and every $(m,\ell)$, $u$ will be a \underline{smooth} solution to the radial o.d.e.~(\ref{e3iswsntouu}) with $H = 0$. In particular, we may apply Proposition~\ref{asymAnalysis} and easily show that the corresponding $a_{\mathcal{I}^{\pm}}\left(\omega,m,\ell\right)$ and $a_{\mathcal{H}^{\pm}}\left(\omega,m,\ell\right)$ are measurable functions of $(\omega,m,\ell)$.

The {\bf microlocal radiation field along $\mathcal{I}^{\pm}$} associated to $\psi$ is then defined almost everywhere by the measurable function
\[
a_{\mathcal{I}^{\pm}}(\omega,m,\ell) : \mathbb{R} \times \mathbb{Z}\times\mathbb{Z}_{\geq |m|}
\to \mathbb{C},
\]
and the {\bf microlocal radiation field along $\mathcal{H}^{\pm}$} associated to $\psi$ is defined almost everywhere by the measurable function
\[
a_{\mathcal{H}^{\pm}}(\omega,m,\ell) : \mathbb{R} \times \mathbb{Z}\times\mathbb{Z}_{\geq |m|} \to \mathbb{C}.
\]
\end{definition}

We also have the corresponding total fluxes.
\begin{definition}\label{microFluxDef}
For all solutions $\psi$ to~(\ref{WAVE}) on
$\mathring{\mathcal{R}}$, which are sufficiently integrable in the sense of Definition~\ref{sufficient}, the {\bf total microlocal energy flux through $\mathcal{I}^{\pm}$} associated to $\psi$ is given by
        \[
        \int_{-\infty}^{\infty}\sum_{m\ell}\omega^2|a_{\mathcal{I}^{\pm}}|^2\, d\omega \in \mathbb{R}_{\geq 0} \cup \{\infty\},
        \]
and the {\bf total microlocal (degenerate) energy flux through $\mathcal{H}^{\pm}$} associated to $\psi$ is given by
        \[
        \int_{-\infty}^{\infty}\sum_{m\ell}(\omega-\upomega_+m)^2|a_{\mathcal{H}^{\pm}}|^2\, d\omega \in \mathbb{R}_{\geq 0} \cup \{\infty\}.
        \]
\end{definition}
These latter fluxes will be related in Section~\ref{relatephy} below to the
flux to $\mathcal{I}^+$ and the degenerate $K$-energy flux to $\mathcal{H}^+$
defined previously
in Sections~\ref{YALA} and~\ref{BALA}, respectively.

\section{Estimates for the radial o.d.e. and applications}\label{radodesec}
In this section we will produce estimates for the radial o.d.e.~(\ref{e3iswsntouu})
and given some useful applications. These estimates are refinements of
estimates originally proven in~\cite{partIII}.

In Section~\ref{sepcurr} we review the separated current template from~\cite{stabi} and~\cite{partIII}. These currents form the essential ingredients for all of the o.d.e.~estimates of this section.

In Section~\ref{microlocaliledsec} we start by proving Theorem~\ref{scatterEst2} which is a general estimate for solutions to the radial o.d.e.~with a vanishing right hand side; the proof of Theorem~\ref{scatterEst2} will heavily rely on Theorem 8.1~from~\cite{partIII}. As a corollary
we will obtain the uniform boundedness of all reflexion and transmission coefficients
(Thereom~\ref{refltransBound}).
This gives in particular {\bf Theorem~\ref{isbound2}} of Section~\ref{uniformboundsec}.
We will also obtain a Wronskian bound (Proposition~\ref{theWronkIsReallyBounded})
which will be used in Section~\ref{secDegenBound}.

In Section~\ref{herethesuperproof} we will prove Proposition~\ref{asymExpandGood} which gives asymptotic control of $U_{\rm hor}$ in the superradiant regime as $r \to r_+$ independent of the frequency parameters. Proposition~\ref{asymExpandGood} plays an important role in Section~\ref{secDegenBound}. The proof of Proposition~\ref{asymExpandGood} will require us to quote a special case of Theorem 8.1~from~\cite{partIII} (here
given as Theorem~\ref{odeEstimates}).

Next, using closely related ideas, in Section~\ref{ellisquitelarge} we will prove Proposition~\ref{largelT} which states that for fixed $\omega$ and $m$, the large-$\ell$ limit of $\mathfrak{T}$ must vanish. As a corollary, we deduce that $\lim_{\ell\to \infty}\left|\mathfrak{R}\right| = 1$.

In Section~\ref{isovanish} we will approve Proposition~\ref{notid0} which states that for each fixed $m$ and $\ell$, the reflection coefficient $\mathfrak{R}$ is not identically $0$ as a function of $\omega$. Using analyticity of $\mathfrak{R}$, one corollary will be that $\mathfrak{R}$ can only vanish at isolated points.

In Section~\ref{MICROLrp} we will interpret the weighted $r^p$ hierarchy
of estimates of~\cite{icmp} (given previously as  Proposition~\ref{rp} of Section~\ref{rpSec}) directly
at the level of the o.d.e.~$(\ref{radodehere})$. The main result is Proposition~\ref{microrp}.
We will then use this in Section~\ref{Qestie} to give a quantitative
estimate on the rate of convergence  of the microlocal radiation field
(Proposition~\ref{microradconverge}). Using these results, in Section~\ref{relatephy}, we will
succeed in relating the microlocal radiation fields of Section~\ref{secMicroRad}
with the physical-space definitions given  previously Section~\ref{radenergy}.

\subsection{The separated current templates}\label{sepcurr}

In this section we will recall the separated current template from~\cite{stabi} and~\cite{partIII}. All of our o.d.e.~estimates will be based on suitable combinations of these currents.
\begin{proposition}
\label{microreddef}Fix parameters $(\omega,m,\ell) \in \mathbb{R}\times \mathbb{Z}\times \mathbb{Z}_{\geq |m|}$ with $\omega \neq 0$, and let $u$ be a smooth solution of the radial o.d.e.~(\ref{e3iswsntouu})
\[u'' + (\omega^2-V)u = H.\]
Let $h(r^*)$ be a $C^2$ function, $y(r^*)$ be a $C^1$ function and $z(r)$ a $C^1$ function of $r$. Set $\tilde V = V - V|_{r=r_+}$.
Then we define the {\bf $\text {\fontencoding{LGR}\selectfont \Coppa}^h$ current}
\[\text {\fontencoding{LGR}\selectfont \Coppa}^h[u] \doteq h {\rm Re} (u'\bar u)-\frac12 h' |u|^2,\]
the {\bf $\text {\fontencoding{LGR}\selectfont \koppa}^y$ current}
\[\text{\fontencoding{LGR}\selectfont \koppa}^y[u] \doteq y\left(\left|u'\right|^2 + \left(\omega^2-V\right)\left|u\right|^2\right),\]
the {\bf microlocal redshift current}
\begin{equation}\label{microred}
Q^z_{\rm red}[u] \doteq z\left|u' + i(\omega -\upomega_+ m)u\right|^2 - z\tilde V\left|u\right|^2,
\end{equation}
the {\bf microlocal $r^p$ current}
\begin{equation}\label{microrpcur}
Q^z_{r^p}[u] \doteq z\left|u' - i\omega u\right|^2 - zV\left|u\right|^2,
\end{equation}
the {\bf microlocal $T$-energy current}
\begin{equation}\label{tenergymic}
{\rm Q}^T[u] \doteq \omega\text{Im}\left(u'\overline{u}\right)
\end{equation}
and the {\bf microlocal $K$-energy current}
\begin{equation}
\label{orismostouKcur}
{\rm Q}^K[u] \doteq \left(\omega - \upomega_+m\right)\text{Im}\left(u'\overline{u}\right).
\end{equation}
We have
\begin{equation}\label{eq:Q1for}
(\text {\fontencoding{LGR}\selectfont \Coppa}^h[u])'=h\left(|u'|^2 +(V-\omega^2)|u|^2\right) -\frac12 h'' |u|^2 +h\, {\rm Re} (u\bar H),
\end{equation}
\begin{equation}\label{thekoppa}
\left(\text{\fontencoding{LGR}\selectfont \koppa}^y[u]\right)' = y'\left(\left|u'\right|^2 + \left(\omega^2-V\right)\left|u\right|^2\right) - yV'\left|u\right|^2 + 2y\text{Re}\left(u'\overline{H}\right),
\end{equation}
\begin{equation}
\label{redshiftlocid}
\left(Q^z_{\rm red}[u]\right)' = z'\left|u' + i(\omega -\upomega_+m)u\right|^2 - \left(z\tilde V\right)'\left|u\right|^2 + 2z\text{Re}\left(H\overline{u'+i\left(\omega-\upomega_+m\right)u}\right),
\end{equation}
\begin{equation}\label{derrp}
\left(Q^z_{r^p}[u]\right)' = z'\left|u' - i\omega u\right|^2 - \left(zV\right)'\left|u\right|^2 + 2z\text{Re}\left(H\overline{u'-i\omega u}\right),
\end{equation}
\[\left({\rm Q}^T[u]\right)' = \omega\text{Im}\left(H\overline{u}\right),\]
\begin{equation}\label{Kconscons}
\left({\rm Q}^K[u]\right)' = (\omega-\upomega_+m)\text{Im}\left(H\overline{u}\right).
\end{equation}
\end{proposition}
The identities above follow by direct computation. Note that we have already used the ${\rm Q}^T$ current~(\ref{tenergymic}) in Proposition~\ref{microEnergyEst}.

\begin{remark}Note that the microlocal $r^p$ current appears for the first time in this paper. The reader may find it illuminating to compare~(\ref{microred}) with~(\ref{microrpcur}).
\end{remark}

\subsection{The microlocal ILED estimate and applications}\label{microlocaliledsec}

Recall that the microlocal radiation fields $a_{\mathcal{I}^{\pm}}$ and $a_{\mathcal{H}^{\pm}}$ were defined in Definition~\ref{microRad}. With this notation,
in our previous work~\cite{partIII}, estimates for the radial o.d.e.~(\ref{e3iswsntouu}), in all frequency ranges, with a non-zero right hand side $H$ and $u$ satisfying $a_{\mathcal{H}^-} = a_{\mathcal{I}^-} = 0$ played a fundamental role in the proof of Theorem~\ref{theResult}.
Besides depending on Theorem~\ref{theResult} as stated, in the present paper
we will also require
Theorem~\ref{scatterEst2}, which is a variant
of the o.d.e.~estimates of~\cite{partIII},
concerning now solutions to the \underline{homogeneous} radial o.d.e.~(\ref{e3iswsntouu}) where we do \underline{not} however
assume that $a_{\mathcal{H}^-} = a_{\mathcal{I}^-} = 0$. We will thus prove this latter
theorem in the present section, referring to constructions in our~\cite{partIII}. We will close the section with two corollaries of Theorem~\ref{scatterEst2}: Theorem~\ref{refltransBound} which gives the boundedness of the reflection and transmission coefficients and Proposition~\ref{theWronkIsReallyBounded} which gives a uniform bound on the Wronskian.

\subsubsection{The microlocal ILED estimate for the homogeneous radial o.d.e.}
We will prove here the following variant of Theorem 8.1 of~\cite{partIII} which
applies to solutions of the \emph{homogeneous} o.d.e.~$(\ref{e3iswsntouu})$ (with $H=0$)
but allows
general asymptotics $a_{\mathcal{H}^\pm} \ne 0$, $a_{\mathcal{I}^\pm} \ne 0$.

\begin{theorem}\label{scatterEst2}There exist parameters
$s_-$ and $s_+$ satisfying $r_+ < 3M - s_- < 3M + s_+ < \infty$ such that for all $-\infty < R_-^* < R_+^* < \infty$, the following is true. Given
$(\omega,m,\ell)$ satisfying $\omega \neq 0$ and $\omega \neq \upomega_+m$, there
exists a parameter $r_{\rm trap}\left(\omega,m,\ell\right)$ with
\[
r_{\rm trap} = 0\qquad \text{ or }\qquad r_{\rm trap} \in [3M-s_-,3M+s_+],
\]
such that
for all smooth solutions $u$ to the radial o.d.e.~(\ref{e3iswsntouu}) with vanishing right hand side $H=0$,
\begin{align}\label{theesttobeproved}
(\omega-\upomega_+m)^2|a_{\mathcal{H}^-}|^2 + \omega^2|a_{\mathcal{I}^-}|^2 + &\int_{R^*_-}^{R_+^*}\left[\left| u'\right|^2 + \left(\left(1-r_{\rm trap}r^{-1}\right)^2(\omega^2 + \Lambda) + 1\right)\left|u\right|^2\right]\, dr^* \\ \nonumber &\leq B(R^*_-,R^*_+)\left[(\omega-\upomega_+m)^2| a_{\mathcal{H}^+}|^2 + \omega^2|a_{\mathcal{I}^+}|^2\right].
\end{align}
\end{theorem}
\begin{remark}Recall that the degeneration due to the $(1-r_{\rm trap}r^{-1})^2$ term arises because of trapping. See the discussion in~\cite{partIII}. 

\end{remark}
\begin{remark}Note that applying the theorem to $\overline{u}$ yields the same statement with the roles of $a_{\mathcal{H}^-}$ and $a_{\mathcal{I}^-}$ interchanged with $a_{\mathcal{H}^+}$ and $a_{\mathcal{I}^+}$.
\end{remark}
\begin{remark}Let us emphasise the even though we require $\omega \neq 0$ and $\omega \neq \upomega_+m$ in order to define $a_{\mathcal{H}^{\pm}}$ and $a_{\mathcal{I}^{\pm}}$, the constant
$B(R_-^*, R^*_+)$ is according to our conventions in Section~\ref{gencon} independent of the frequency
parameters and in particular
does \underline{not} blow up in either of the limits $\omega \to 0$ or $\omega \to \upomega_+m$.
\end{remark}
\begin{proof}We recall that in~\cite{partIII} we studied solutions to radial o.d.e.~(\ref{e3iswsntouu}) with a non-zero right hand side $H$ and $u$ satisfying $a_{\mathcal{H}^-} = a_{\mathcal{I}^-} = 0$,
whereas here $H=0$ but all $a_{\mathcal{H}^\pm}$, $a_{\mathcal{I}^\pm}$ are in general
nontrivial.

We begin with the important observation that in (version 2! of)~\cite{partIII} in the proof of Theorem 8.1 we used microlocal currents (see Section~\ref{sepcurr}) where the functions $f$, $h$, etc. were all \emph{bounded} as $r^* \to \pm\infty$. The currents which led to the positive bulk of the microlocal ILED statement produced (1) a term associated to the inhomogeneity $H$ and (2) boundary terms which were proportional to $(\omega-\upomega_+m)^2\left|u(-\infty)\right|^2$ and $\omega^2\left|u(\infty)\right|^2$. These boundary terms were eventually controlled with suitable applications of (cut-off versions of) the ${\rm Q}^T$ and ${\rm Q}^K$ currents. In order for this to work, one key point was that under the assumptions $a_{\mathcal{H}^-} = a_{\mathcal{I}^-} = 0$ we have
\begin{equation}\label{qtsigned}
\left|{\rm Q}^T|_{r=r_+}\right| = \left|\omega(\omega-\upomega_+m)\left|u(-\infty)\right|\right|^2,\qquad  {\rm Q}^T_{r = \infty} = \omega^2\left|u(\infty)\right|^2,
\end{equation}
\begin{equation}\label{qksigned}
{\rm Q}^K_{r = r_+} = \left(\omega-\upomega_+m\right)^2|u(-\infty)|^2,\qquad  \left|{\rm Q}^K_{r=\infty}\right| = \left|\omega(\omega-\upomega_+m)\left|u(\infty)\right|\right|^2.
\end{equation}

Now we consider the case of a solution $u$ to the radial o.d.e.~(\ref{e3iswsntouu}) with a vanishing right hand side $H$ but where we make no assumption about the vanishing or non-vanishing of $a_{\mathcal{H}^{\pm}}$ and $a_{\mathcal{I}^{\pm}}$. Since all of the multipliers discussed above are bounded, we immediately observe that we may apply all the currents from (version 2 of)~\cite{partIII} to $u$.

The term associated to the inhomogeneity in the resulting identity, of course, now vanishes since $H=0$.

However, every application of the microlocal energy currents Q$^K$ and Q$^T$ will yield now various boundary terms each of which will be proportional to one of $(\omega-\upomega_+m)^2|a_{\mathcal{H}^-}|^2$, $(\omega-\upomega_+m)^2|a_{\mathcal{H}^+}|^2$, $\omega^2|a_{\mathcal{I}^+}|^2$, or $\omega^2|a_{\mathcal{I}^-}|^2$. Furthermore, the term proportional to $(\omega-\upomega_+m)^2|a_{\mathcal{H}^-}|^2$ will always enter with the opposite sign of the term proportional to $(\omega-\upomega_+m)^2|a_{\mathcal{H}^+}|^2$. An analogous relation holds for the terms proportional to $\omega^2|a_{\mathcal{I}^+}|^2$ and $\omega^2|a_{\mathcal{I}^-}|^2$. In particular, we do not have~(\ref{qtsigned}) and~(\ref{qksigned}) and we cannot hope to prove an estimate with all of the microlocal radiation fields on the left hand side\footnote{This is not so surprising of course, because if we could prove such an estimate we would deduce that $u$ had to vanish!}. We are thus forced to always put the boundary terms associated to one of the pairs $(a_{\mathcal{H}^-},a_{\mathcal{I}^-})$ and $(a_{\mathcal{H}^+},a_{\mathcal{I}^+})$ on the right hand side.

Given these observations, the following estimate immediately follows from the proof of Theorem 8.1 of~\cite{partIII}:
\begin{align}
(\omega-\upomega_+m)^2|a_{\mathcal{H}^-}|^2 + \omega^2|a_{\mathcal{I}^-}|^2 + &\int_{R^*_-}^{R_+^*}\left[\left| u'\right|^2 + \left(\left(1-r_{\rm trap}r^{-1}\right)^2(\omega^2 + \Lambda) + 1\right)\left|u\right|^2\right]\, dr^* \\ \nonumber &\leq B(R^*_-,R^*_+)\Big[(\omega-\upomega_+m)^2| a_{\mathcal{H}^+}|^2 + \omega^2|a_{\mathcal{I}^+}|^2
\\ \nonumber &\ \ +  1_{\{\omega_{\rm low} \leq \left|\omega\right| \leq \omega_{\rm high}\}\cap \{\Lambda \leq \epsilon_{\rm width}^{-1}\omega_{\rm high}^2\}}\left|a_{\mathcal{H}^-}\right|^2\Big],
\end{align}
for a parameter $r_{\rm trap}(\omega, m, \ell)$ satisfying
\[
r_{\rm trap} = 0\qquad \text{ or }\qquad r_{\rm trap} \in [3M-s_-,3M+s_+],
\]
where $1_{\{\omega_{\rm low} \leq \left|\omega\right| \leq \omega_{\rm high}\}\cap \{\Lambda \leq \epsilon_{\rm width}^{-1}\omega_{\rm high}^2\}}$ denotes the indicator function for the set
\[\mathcal{F}_{\flat} \doteq \{(\omega,m,\ell) : \omega_{\rm low} \leq \left|\omega\right| \leq \omega_{\rm high}\text{ and } \{\Lambda \leq \epsilon_{\rm width}^{-1}\omega_{\rm high}^2\}\}.\]
The $\omega_{\rm low}$, $\omega_{\rm high}$ and $\epsilon_{\rm width}$ are fixed constants which arise during the proof of Theorem 8.1. Thus we have established~(\ref{theesttobeproved}) for frequencies $(\omega,m,\ell) \not\in \mathcal{F}_{\flat}$.

In order to finish the proof we need to show that $(\omega,m,\ell) \in \mathcal{F}_{\flat}$ implies
\[
\left|a_{\mathcal{H}^-}\right|^2 \leq B\left[\left(\omega-\upomega_+m\right)^2\left|a_{\mathcal{H}^+}\right|^2 + \omega^2\left|a_{\mathcal{I}^+}\right|^2\right].
\]
Consider the solution $u^{\dagger}$ to the radial o.d.e.~defined by
\[
 u^{\dagger} = \left(a_{\mathcal{I}^+}\overline{\mathfrak{W}^{-1}(2i\omega)}\right)\overline{U_{\rm hor}} +\left(a_{\mathcal{H}^+}\overline{\mathfrak{W}^{-1}(2i\left(\omega-\upomega_+m\right))}\right)\overline{U_{\rm inf}},
\]
and let $ a^{\dagger}_{\mathcal{H}^+}$ and $a^{\dagger}_{\mathcal{I}^+}$ denote the microlocal radiation fields of $u^{\dagger}$.

Observe that Lemma~\ref{quitenicerelations} implies that
\[
 a^{\dagger}_{\mathcal{H}^+} = a_{\mathcal{H}^+},
\qquad
 a^{\dagger}_{\mathcal{I}^+} = a_{\mathcal{I}^+}.
\]
Thus, applying Theorem~\ref{modestability} and Remark~\ref{linDep} to $u-u^{\dagger}$ implies that $u = u^{\dagger}$. Using the explicit definition of $u^{\dagger}$, appealing to Theorem~\ref{modestability} again and using the compactness of $\mathcal{F}_{\flat}$, we immediately conclude that
\[
\left|a_{\mathcal{H}^-}\right|^2 = \left|a^{\dagger}_{\mathcal{H}^-}\right|^2   \leq B\mathfrak{W}^{-2}\left[\omega^2\left|a_{\mathcal{I}^+}\right|^2 + \left(\omega-\upomega_+m\right)^2\left|a_{\mathcal{H}^+}\right|^2\right] \leq B\left[\omega^2\left|a_{\mathcal{I}^+}\right|^2 + \left(\omega-\upomega_+m\right)^2\left|a_{\mathcal{H}^+}\right|^2\right].
\]
\end{proof}

\subsubsection{Uniform boundedness of $\mathfrak{R}$ and $\mathfrak{T}$}
\label{veoeda}

Applying Theorem~\ref{scatterEst2} to the solutions $U_{\rm hor}$ and $U_{\rm inf}$ immediately implies that the reflection and transmission coefficients are bounded uniformly in $(\omega,m,\ell)$.
\begin{theorem}\label{refltransBound}
The reflection and transmission coefficients are uniformly bounded:
\[
\left|\mathfrak{R}\right|^2 + \left|\tilde{\mathfrak{R}}\right|^2 + \left|\mathfrak{T}\right|^2
+  \left|\tilde{\mathfrak{T}}\right|^2 \leq B.
\]
\end{theorem}
\begin{proof}We simply note that by the definition of $\mathfrak{R}$ and $\mathfrak{T}$, there exists a solution $u$ to the radial o.d.e.~such that
\[
a_{\mathcal{H}^+} = \frac{\mathfrak{T}}{-i\left(\omega-\upomega_+m\right)}, \qquad
a_{\mathcal{H}^-} = 0, \qquad
a_{\mathcal{I}^+} = \frac{\mathfrak{R}}{iw}, \qquad
a_{\mathcal{I}^-} = \frac{1}{i\omega}.
\]
Theorem~\ref{scatterEst2} immediately yields
\[\left|\mathfrak{T}\right|^2 + \left|\mathfrak{R}\right|^2  = \left(\omega-\upomega_+m\right)^2\left|a_{\mathcal{H}^+}\right|^2 + \omega^2\left|a_{\mathcal{I}^+}\right|^2  \leq B\left[\left(\omega-\upomega_+m\right)^2\left|a_{\mathcal{H}^-}\right|^2 + \omega^2\left|a_{\mathcal{I}^-}\right|^2\right] \leq B.\]

An analogous argument applies for $\tilde{\mathfrak{R}}$ and $\tilde{\mathfrak{T}}$.
\end{proof}
The above in particular already yields  {\bf Theorem~\ref{isbound2}} of Section~\ref{uniformboundsec}.

\subsubsection{A bound for the Wronskian $\mathfrak{W}$}
We close the section with a uniform bound on the Wronskian
which will be useful in Section~\ref{secDegenBound}.
\begin{proposition}\label{theWronkIsReallyBounded}For all $(\omega,m,\ell) \in \mathbb{R}\setminus\{0\} \times \mathbb{Z}\times \mathbb{Z}_{\geq |m|}$ we have
\[\frac{\omega^2 + \upomega_+^2m^2}{\left|\mathfrak{W}\right|^2} \leq B.\]
\end{proposition}
\begin{proof}We first apply Theorem~\ref{scatterEst2} with $u = \overline{U}_{\rm hor}$. In this case we have
\[a_{\mathcal{H}^-} = 1,\]
\[a_{\mathcal{H}^+} = 0,\]
\[a_{\mathcal{I}^-} = \overline{\mathfrak{A}_{\mathcal{I}^+}},\]
\[a_{\mathcal{I}^+} = \overline{\mathfrak{A}_{\mathcal{I}^-}} = -\frac{\overline{\mathfrak{W}}}{2i\omega}.\]
In the last equality we have appealed to Lemma~\ref{quitenicerelations}. Theorem~\ref{scatterEst2} then implies
\[\left(\omega - \upomega_+m\right)^2 + \omega^2\left|\mathfrak{A}_{\mathcal{I}^+}\right|^2 \leq B\left|\mathfrak{W}\right|^2.\]
Dividing through by $\mathfrak{W}^2$ implies
\begin{equation}\label{afirststep}
\frac{\left(\omega-\upomega_+m\right)^2}{\left|\mathfrak{W}\right|^2} \leq B.
\end{equation}

Next we apply Theorem~\ref{scatterEst2} with $u = \overline{U}_{\rm inf}$. In this case we have
\[a_{\mathcal{H}^-} = \overline{\mathfrak{A}_{\mathcal{H}^+}},\]
\[a_{\mathcal{H}^+} = \overline{\mathfrak{A}_{\mathcal{H}^-}} = -\frac{\overline{\mathfrak{W}}}{2i\left(\omega-\upomega_+m\right)},\]
\[a_{\mathcal{I}^-} = \overline{\mathfrak{A}_{\mathcal{I}^+}} = 0,\]
\[a_{\mathcal{I}^+} = \overline{\mathfrak{A}_{\mathcal{I}^-}} = 1.\]
Again we have appealed to Lemma~\ref{quitenicerelations}. Theorem~\ref{scatterEst2} then implies
\[\left(\omega-\upomega_+m\right)^2\left|\mathfrak{A}_{\mathcal{H}^+}\right|^2 + \omega^2 \leq B\left|\mathfrak{W}\right|^2.\]
Dividing through by $\mathfrak{W}^2$ yields
\begin{equation}\label{asecondstep}
\frac{\omega^2}{\left|\mathfrak{W}\right|^2} \leq B.
\end{equation}

Since
\[\upomega_+^2m^2 = \left(\omega - \upomega_+m - \omega\right)^2 \leq B\left[\left(\omega-\upomega_+m\right)^2 + \omega^2\right],\]
it is clear that~(\ref{afirststep}) and~(\ref{asecondstep}) conclude the proof.
\end{proof}

\subsection{Superradiant estimates for $U_{\rm hor}$}
\label{herethesuperproof}
The frequency range defined below will play an important role in our arguments.
\begin{definition}\label{Bep}For every $\epsilon > 0$ we define the set $\mathcal{F}^{(\epsilon)}_{\sharp}$ by
\[\mathcal{F}^{(\epsilon)}_{\sharp} \doteq \{(\omega,m,\ell) \in \mathbb{R} \times \mathbb{Z} \times \mathbb{Z}_{\geq |m|} : am\omega > 0\text{ and }\left|\omega\right| - \left|\upomega_+m\right| < \epsilon\left|m\right|.\}\]
\end{definition}
\begin{remark}Observe that if we set $\epsilon = 0$, then $\mathcal{F}^{(\epsilon)}_{\sharp}$ would exactly correspond to the superradiant frequencies~(\ref{thesearesuperradiant}). When $\epsilon > 0$ is small, then $\mathcal{F}^{(\epsilon)}_{\sharp}$ contains all frequencies which are ``close'' to being superradiant. These frequencies will later pose the most serious difficulties in the analysis of Section~\ref{secDegenBound}.
\end{remark}
\begin{remark}Note that for frequencies in $\mathcal{F}^{(\epsilon)}_{\sharp}$ we have $\Lambda \geq b(\epsilon)\left(1+\omega^2\right)$.
\end{remark}

In this section, we shall prove
\begin{proposition}\label{asymExpandGood}Let
$E_{\rm hor}$  be defined by
\begin{equation*}
U_{\rm hor}(r^*) = e^{-i\left(\omega-\upomega_+m\right)r^*} + E_{\rm hor}(r^*).
\end{equation*}

Then $(\omega,m,\ell) \in \mathcal{F}^{(\epsilon)}_{\sharp}$ for a sufficiently small $\epsilon > 0$  and $\Lambda$ sufficiently large imply
\begin{equation}\label{asymExpantoprove}
\left|E_{\rm hor}\right| \leq \frac{B(\epsilon)\left|\mathfrak{W}\right|}{\sqrt{\Lambda}}\sqrt{r-r_+},
\end{equation}
for sufficiently small $r-r_+$.
\end{proposition}
\begin{remark}Note that the $\sqrt{\Lambda}$ factor above represents a ``gain of a derivative'' over what one would expect to prove if we were not restricting to $(\omega,m,\ell) \in \mathcal{F}^{(\epsilon)}_{\sharp}$.
\end{remark}
This proposition will be of fundamental importance in Section~\ref{secDegenBound}.
To prove it, we will need again to return to our o.d.e~theory for $(\ref{e3iswsntouu})$.
We begin with some preliminaries reviewing some additional results
and notation from~\cite{partIII}. The proof proper will be contained in Section~\ref{theproofproper}.

\subsubsection{An inhomogeneous ILED in the superradiant regime}

The following estimate is a special case of Theorem 8.1 from~\cite{partIII}.\footnote{Note that we have strengthened the statement of Theorem 8.1 in version 2 of~\cite{partIII} with Theorems~\ref{odeEstimates} and~\ref{scatterEst2} in mind.}

\begin{theorem}\label{odeEstimates}\cite{partIII}
Let $\epsilon > 0$ be sufficiently small and $-\infty < R^*_- < R^*_+ < \infty$, then there exists a constant $B\left(R^*_-,R^*_+,\epsilon\right)$ such that for all smooth solutions $u$ to the radial o.d.e.~(\ref{e3iswsntouu}) with a smooth compactly supported right hand side $H$, $u$ satisfying $a_{\mathcal{H}^-} = a_{\mathcal{I}^-} = 0$, and frequencies $(\omega,m,\ell) \in \mathcal{F}^{(\epsilon)}_{\sharp}$ with $\Lambda$ sufficiently large, we have
\begin{align}
\label{notrap!}
(\omega-\upomega_+m)^2|a_{\mathcal{H}^+}|^2 &+ \omega^2|a_{\mathcal{I}^+}|^2 + \int_{R^*_-}^{R_+^*}\left[\left|u'\right|^2 + \Lambda\left|u\right|^2\right]\, dr^* \\ \nonumber &\leq B(R^*_-,R^*_+,\epsilon)\int_{-\infty}^{\infty}\left(\left|Hu'\right| + \sqrt{\Lambda}\left|Hu\right|\right)\, dr^*.
\end{align}
\end{theorem}

\begin{remark}
Note that the integrand on the right hand side
of $(\ref{notrap!})$ does not degenerate (cf.~$(\ref{theesttobeproved})$ below).
This is because the ($\epsilon$-enlarged) superradiant frequency
range $\mathcal{F}^{(\epsilon)}_{\sharp}$ is
 \underline{not}  trapped. See the discussion in~\cite{partIII} regarding
 the fortuitous disjointness of the difficulties of superradiance and trapping.
\end{remark}

\subsubsection{Properties of the potential $V$ in the superradiant regime}
We  recall the following two propositions proved in~\cite{partIII}.
\begin{proposition}\label{classicForbid}Let $(\omega,m,\ell) \in \mathcal{F}^{(\epsilon)}_{\sharp}$ for a sufficiently small $\epsilon > 0$. Then there exists a unique $r$ value $r_{\rm max}$ where the potential $V$ of $(\ref{defofV})$
achieves its maximum. Furthermore, there exists $\delta > 0$, independent of the frequency parameters, such that
\[\left(V - \omega^2\right)|_{r \in [r_{\rm max}-\delta,r_{\rm max} + \delta]} \geq b\Lambda.\]
Furthermore, $r_{\rm max}$ is uniformly bounded away from $r_+$ and $\infty$.
\end{proposition}

\begin{proposition}\label{potincrease}Let $(\omega,m,\ell) \in \mathcal{F}^{(\epsilon)}_{\sharp}$ for a sufficiently small $\epsilon > 0$, then there exists $\delta_1 > 0$, independent of the frequency parameters, such that
\[r \in [r_+,r_++\delta_1] \Rightarrow \frac{dV}{dr} \geq b(\epsilon)\Lambda.\]
\end{proposition}

\subsubsection{An improved estimate in the superradiant regime}

We begin by applying Theorem~\ref{scatterEst2} to $U_{\rm hor}$ and refer to Lemma~\ref{quitenicerelations} concerning the Wronskian. We obtain
\begin{corollary}\label{almostathorizon}For all frequencies $(\omega,m,\ell) \in \mathcal{F}^{(\epsilon)}_{\sharp}$ for a sufficiently small $\epsilon > 0$ and sufficiently large $\Lambda$, and for any constants $-\infty < R^*_- < R^*_+ < \infty$, we have
\begin{align}
\left(\omega-\upomega_+ m\right)^2\cdot 1+\omega^2\left|\mathfrak{A}_{\mathcal{I}^+}\right|^2 + &\int_{R^*_-}^{R_+^*}\left[\left|U_{\rm hor}'\right|^2 + \Lambda\left|U_{\rm hor}\right|^2\right]\, dr^*\leq B(R^*_-,R^*_+,\epsilon)\left|\mathfrak{W}\right|^2.
\end{align}
\end{corollary}

Proposition~\ref{classicForbid} allows us to ``gain a derivative''
in comparison with Corollary~\ref{almostathorizon}
in the following lemma.
\begin{lemma}\label{derGain}
There exists $r_1^*  > -\infty$ such that for all frequencies $(\omega,m,\ell) \in \mathcal{F}^{(\epsilon)}_{\sharp}$ with a sufficiently small $\epsilon > 0$ and $\Lambda$ sufficiently large, and $r_0^* < r_1^*$, we have
\begin{align}
\int_{r_0^*}^{r_1^*}\left[\left|U_{\rm hor}'\right|^2 + \Lambda\left|U_{\rm hor}\right|^2\right]\, dr^* \leq B(r_0^*,\epsilon)\frac{\left|\mathfrak{W}\right|^2}{\Lambda}.
\end{align}
\end{lemma}

\begin{proof}Let $u$ be an arbitrary smooth solution to the \underline{homogeneous} radial o.d.e.~(\ref{e3iswsntouu}), with $(\omega,m,\ell) \in \mathcal{F}^{(\epsilon)}_{\sharp}$ for a sufficiently small $\epsilon > 0$ and $\Lambda$ sufficiently large. Let $\tilde h$ be a smooth positive function supported in $[r_{\rm max}-\delta,r_{\rm max}+\delta]$ which is identically $1$ within $[r_{\rm max}-\delta/2,r_{\rm max}+\delta/2]$. Then set $h \doteq \Lambda \tilde h$. Using~(\ref{eq:Q1for}), we obtain
\begin{align}\label{gainthatder}
\Lambda\int_{r_{\rm max}-\delta/2}^{r_{\rm max}+\delta/2}\left[\left|u'\right|^2 + \Lambda\left|u\right|^2\right]\, dr^* &\leq \int_{-\infty}^{\infty}(\text {\fontencoding{LGR}\selectfont \Coppa}^h[u])' + B(\epsilon)\Lambda\int_{r_{\rm max}-\delta}^{r_{\rm max}+\delta}\left|u\right|^2
\\ \nonumber &= B(\epsilon)\Lambda\int_{r_{\rm max}-\delta}^{r_{\rm max}+\delta}\left|u\right|^2
\\ \nonumber &\leq B(\epsilon)\left[(\omega-\upomega_+m)^2|a_{\mathcal{H}^+}|^2 + \omega^2|a_{\mathcal{I}^+}|^2\right].
\end{align}
In the last line we used Theorem~\ref{scatterEst2}.

In particular, applying the estimate~(\ref{gainthatder}) to $U_{\rm hor}$ and then appealing to Corollary~\ref{almostathorizon} implies
\begin{align}\label{nicegain}
\int_{r_{\rm max}-\delta/2}^{r_{\rm max}+\delta/2}\left[\left|U_{\rm hor}'\right|^2 + \Lambda\left|U_{\rm hor}\right|^2\right]\, dr^* \leq \frac{B(\epsilon)\left|\mathfrak{W}\right|^2}{\Lambda}.
\end{align}

Now let $\chi$ be a function which is identically $1$ on $[r_+,r_{\rm max}-\delta/2]$ and identically $0$ on $[r_{\rm max}+\delta/2,\infty)$, and then set $\tilde u \doteq \chi U_{\rm hor}$. We have
\[\tilde u'' + \left(\omega^2 - V\right)\tilde u = \chi''U_{\rm hor} + 2\chi'U_{\rm hor}' \doteq \tilde H,\]
\[\tilde a_{\mathcal{H}^-} = \tilde a_{\mathcal{I}^-} = 0.\]

Thus, taking $r_1^*$ sufficiently negative, applying Theorem~\ref{odeEstimates} to $\tilde u$ yields
\begin{align*}
\int_{r_0^*}^{r_1^*}\left[\left| U_{\rm hor}'\right| + \Lambda\left| U_{\rm hor}\right|^2\right]\, dr^* &\leq B(r_0^*,\epsilon)\int_{r_{\rm max}-\delta/2}^{r_{\rm max}+\delta/2}\left[\left|U_{\rm hor}'\right|^2 + \Lambda\left|U_{\rm hor}\right|^2\right]\, dr^*
\\ \nonumber &\leq \frac{B(r_0^*,\epsilon)\left|\mathfrak{W}\right|^2}{\Lambda}.
\end{align*}
In the last line we used the estimate~(\ref{nicegain}).
\end{proof}

Now we are ready for the following lemma.
\begin{lemma}\label{transverseder}
There exists a constant $\tilde\epsilon > 0$ such that for all frequencies $(\omega,m,\ell) \in \mathcal{F}^{(\epsilon)}_{\sharp}$ with a sufficiently small $\epsilon > 0$ and $\Lambda$ sufficiently large, we have
\begin{align*}
\int_{-\infty}^{\left(r_++\tilde\epsilon\right)^*}&\left|U'_{\rm hor} + i\left(\omega-\upomega_+m\right)U_{\rm hor}\right|^2(r-r_+)^{-1}\, dr^* \leq \frac{B(\epsilon)\left|\mathfrak{W}\right|^2}{\Lambda}.
\end{align*}
\end{lemma}

\begin{proof}
We consider the microlocal redshift current~(\ref{microred}) with
\[
z \doteq -\frac{\Lambda}{\tilde V}\chi(r),
\]
where $\chi$ is a bump function which is identically $1$ for $r \in [r_+,r_++\tilde\epsilon]$ and $0$ for $r \in [2\tilde\epsilon,\infty)$, for a small positive constant $\tilde\epsilon$ to be determined.  We obtain from $(\ref{redshiftlocid})$ the estimate
\begin{align}\label{anEst7}
\int_{-\infty}^{(r_++\tilde\epsilon)^*}&\left[z'\left|U_{\rm hor}' + i(\omega -\upomega_+m)U_{\rm hor}\right|^2\right]\, dr^*  \\
\nonumber &\qquad \le B(\epsilon,\tilde\epsilon)\int_{(r_++\tilde\epsilon)^*}^{(r_++2\tilde\epsilon)^*}\left[\left|U_{\rm hor}'\right|^2 + \Lambda\left|U_{\rm hor}\right|^2\right]\, dr^* - Q^z_{\rm red}|_{r=r_+}.
\end{align}

If $\tilde\epsilon > 0$ is small enough, then via Proposition~\ref{potincrease} we see that $r \in (r_+,r_++\tilde\epsilon]$ implies that $z' \geq b(r-r_+)^{-1}$. In particular, after fixing a small choice of $\tilde\epsilon$, we may combine~(\ref{anEst7}) and Lemma~\ref{derGain} to conclude
\begin{align}\label{anEst9}
\int_{-\infty}^{(r_++\tilde\epsilon)^*}&\left|U_{\rm hor}' + i(\omega -\upomega_+m)U_{\rm hor}\right|^2(r-r_+)^{-1}\, dr^* \leq \frac{B(\epsilon)\left|\mathfrak{W}\right|^2}{\Lambda} - BQ^z_{\rm red}|_{r=r_+}.
\end{align}

We conclude the proof by noting that
\[-Q^z_{\rm red}|_{r=r_+} = -\Lambda \leq 0.\]
\end{proof}

\subsubsection{Proof of Proposition~\ref{asymExpandGood}}
\label{theproofproper}
Finally, Lemma~\ref{transverseder} easily allows us to prove Proposition~\ref{asymExpandGood}.
\begin{proof}[Proof of Proposition~\ref{asymExpandGood}]
Let $\Lambda$ be sufficiently large.
Recall the definition $(\ref{somestuff000})$ of $E_{\rm hor}$.
It follows that
\[
E'_{\rm hor} = U'_{\rm hor}+i(\omega-\upomega_+ m)U_{\rm hor}.
\]
Assuming $r-r_+$ sufficiently small, we then have
\begin{align*}
\left|E_{\rm hor}(r^*)\right| &\leq \int_{-\infty}^{r^*}\left|E_{\rm hor}'\right|\, ds^*
\\ \nonumber &= \int_{-\infty}^{r^*}\left|U_{\rm hor}' + i\left(\omega-\upomega_+m\right)U_{\rm hor}\right|\, ds^*
\\ \nonumber &\leq B\int_{r_+}^r\left|U_{\rm hor}'+i\left(\omega-\upomega_+m\right)U_{\rm hor}\right|(s-r_+)^{-1}\, ds
\\ \nonumber &\leq B\sqrt{r-r_+}\sqrt{\int_{r_+}^r\left|U'_{\rm hor} + i\left(\omega-\upomega_+m\right)U_{\rm hor}\right|^2(s-r_+)^{-2}\, ds}
\\ \nonumber &\leq B\sqrt{r-r_+}\sqrt{\int_{-\infty}^{r^*}\left|U'_{\rm hor} + i\left(\omega-\upomega_+m\right)U_{\rm hor}\right|^2(s-r_+)^{-1}\, ds^*}
\\ \nonumber &\leq \frac{B(\epsilon)\left|\mathfrak{W}\right|}{\sqrt{\Lambda}}\sqrt{r-r_+}.
\end{align*}
\end{proof}

\subsection{The large-$\ell$ limit of $\mathfrak{T}$}\label{ellisquitelarge}
It is useful to observe that $\mathfrak{T}$ must vanish in the large-$\ell$ limit.
\begin{proposition}\label{largelT}For each fixed value of $\omega$ and $m$ satisfying $\omega - \upomega_+m \neq 0$, we have
\[\lim_{\ell\to\infty}\mathfrak{T}\left(\omega,m,\ell\right) = 0,\]
\[\lim_{\ell\to\infty}\mathfrak{\tilde T}\left(\omega,m,\ell\right) = 0.\]
\end{proposition}
\begin{proof}We will only consider the case of $\mathfrak{T}$ as the proof for $\mathfrak{\tilde T}$ is exactly the same.

Fix a pair $\omega$ and $m$ such that $\omega -\upomega_+m \neq 0$. Next, pick and fix some value of $r_0 \in (r_+,\infty)$. Then, for all sufficiently large $\ell$, there will exist a $\delta > 0$ such that
\begin{equation}\label{potBig}
\left(V - \omega^2\right)|_{r \in [r_0-\delta,r_0+\delta]} \geq b\Lambda.
\end{equation}

The basic intuition is that for $\Lambda$ sufficiently large, this large potential barrier will prevent the transmissions of waves to $\mathcal{H}^+$. To make this rigorous, we observe that an examination of the beginning of proof of Lemma~\ref{derGain} shows that~(\ref{potBig}) implies that if $\ell$ is sufficiently large
\begin{equation}\label{bigpotconsequence}
\int_{r_0-\delta/2}^{r_0+\delta/2}\left[\left|U_{\rm hor}'\right|^2 + \Lambda\left|U_{\rm hor}\right|^2\right]\, dr^* \leq \frac{B}{\Lambda}\int_{r_0-\delta}^{r_0+\delta}\left[\left|U_{\rm hor}'\right|^2 + \Lambda\left|U_{\rm hor}\right|^2\right]\, dr^*.
\end{equation}
Keeping in mind that Lemma~\ref{quitenicerelations} implies
\[U_{\rm hor} = -\frac{\mathfrak{R}\left(\omega-\upomega_+m\right)}{\omega \mathfrak{T}}e^{i\omega r^*} - \frac{\left(\omega-\upomega_+m\right)}{\omega\mathfrak{T}}e^{-i\omega r^*} + O\left(r^{-1}\right)\text{ as }r\to\infty,\]
an application of Theorem~\ref{scatterEst2} implies that
\begin{equation}
\int_{r_0-\delta}^{r_0+\delta}\left[\left|U_{\rm hor}'\right|^2 + \Lambda\left|U_{\rm hor}\right|^2\right]\, dr^* \leq B\left[1 + \frac{(\omega-\upomega_+m)^2\left|\mathfrak{R}\right|^2}{\left|\mathfrak{T}\right|^2}\right].
\end{equation}
Combining this with~(\ref{bigpotconsequence}) implies
\begin{equation}\label{amostatfrakT}
\int_{r_0-\delta/2}^{r_0+\delta/2}\left[\left|U_{\rm hor}'\right|^2 + \Lambda\left|U_{\rm hor}\right|^2\right]\, dr^* \leq \frac{B\left(\omega,m,r_0\right)}{\Lambda}\left[1 + \frac{(\omega-\upomega_+m)^2\left|\mathfrak{R}\right|^2}{\left|\mathfrak{T}\right|^2}\right].
\end{equation}
Intuitively, the estimate~(\ref{amostatfrakT}) shows that that $U_{\rm hor}$ must be small near the large potential barrier.

We now want to use an energy estimate to show that if $U_{\rm hor}$ is small near the potential barrier, then $\mathfrak{T}$ must be small. We thus consider the microlocal $K$-energy
current $(\ref{orismostouKcur})$
from Proposition~\ref{microreddef}.
Now let $\chi(r)$ denote a cut-off function which is identically $1$ for $r \in [r_+,r_0-\delta/2]$ and identically $0$ for $r \in [r_0+\delta/2,\infty)$. Then, keeping~(\ref{amostatfrakT}) and~(\ref{Kconscons}) in mind,
\begin{align}\label{sosososoclose}
\left(\omega-\upomega_+m\right)^2 &= \int_{-\infty}^{\infty}\left(\chi{\rm Q}^K\right)'\, dr^*
\\ \nonumber &\leq B\int_{r_0-\delta/2}^{r_0+\delta/2}\left[\left|U_{\rm hor}'\right|^2 +\left|U_{\rm hor}\right|^2\right]\, dr^*
\\ \nonumber &\leq \frac{B}{\Lambda}\left[1 + \frac{(\omega-\upomega_+m)^2\left|\mathfrak{R}\right|^2}{\left|\mathfrak{T}\right|^2}\right].
\end{align}
Now we may multiply~(\ref{sosososoclose}) through by $\mathfrak{T}$, divide through by $\left(\omega-\upomega_+m\right)^2$, take $\ell \to \infty$ and apply Theorem~\ref{refltransBound} to conclude that
\[\lim_{\ell\to\infty}\mathfrak{T}\left(\omega,m,\ell\right) = 0.\]
\end{proof}

The following corollary follows easily from Proposition~\ref{largelT}.
\begin{corollary}\label{largelR}For each fixed value of $\omega$ and $m$ satisfying $\omega - \upomega_+m \neq 0$, we have
\[\lim_{\ell\to\infty}\mathfrak{R}\left(\omega,m,\ell\right) = 1,\]
\[\lim_{\ell\to\infty}\mathfrak{\tilde R}\left(\omega,m,\ell\right) = 1.\]
\end{corollary}
\begin{proof}This follows immediately from Corollary~\ref{superradiantAmp} and Proposition~\ref{largelR}.
\end{proof}

\subsection{Nonvanishing of $\mathfrak{R}$}\label{isovanish}
The next proposition shows that for any fixed $m$ and $\ell$, the reflection coefficient $\mathfrak{R}$ cannot be identically $0$.
\begin{proposition}\label{notid0}For each $m$ and $\ell$, there exists $\omega$ such that $\mathfrak{R}\left(\omega,m,\ell\right) \neq 0$ and $\mathfrak{\tilde R}\left(\omega,m,\ell\right) \neq 0$.
\end{proposition}
\begin{proof}We will only consider the case of $\mathfrak{R}$ since $\mathfrak{\tilde R}$ is treated in a similar fashion.

Fix a choice of $m$ and $\ell$. Then, for the sake of contradiction, assume that $\mathfrak{R}\left(\omega,m,\ell\right)$ is identically $0$ in $\omega$. We first consider the case when $\upomega_+m \neq 0$. Then Corollary~\ref{superradiantAmp} implies
\[
\frac{\omega}{\omega-\upomega_+m}\left|\mathfrak{T}\right|^2 = 1.
\]
Then we get a contradiction by considering any $\omega$ such that $\omega(\omega - \upomega_+m) < 0$.

The case when $\upomega_+m = 0$ is a bit more subtle. First of all, observe that the vanishing of $\mathfrak{R}\left(\omega,m,\ell\right)$ implies that for each $\omega$, we can construct a (non-zero!) solution $u = u(r^*,\omega,m,\ell)$ to the radial o.d.e.~such that $u \sim e^{-i\omega r^*}$ as $r^* \to -\infty$ and $u \sim e^{-i\omega r^*}$ as $r^* \to \infty$. By direct inspection, one finds that the estimates of Section 8.7.1 of (version 2! of)~\cite{partIII} go through for such a solution (see Remark 8.7.1 at the end of Section 8.7.1), and in particular prove that for $\omega$ sufficiently small, $u$ must vanish. This contradiction finishes the proof.
\end{proof}

\begin{corollary}\label{novanishR}The reflection coefficients $\mathfrak{R}$  and $\mathfrak{\tilde R}$ cannot vanish on an open set of $\omega$.
\end{corollary}
\begin{proof}Standard o.d.e.~theory implies that for each fixed $m$ and $\ell$, $\mathfrak{R}$ and $\mathfrak{\tilde R}$ are analytic in $\omega \in \mathbb{R}\setminus \{0\}$, and, because Proposition~\ref{notid0} implies that they are not identically $0$, we conclude that they can only vanish at isolated points in $\omega$.
\end{proof}

\subsection{The microlocal $r^p$ estimate}
\label{MICROLrp}
In this section we will establish an analogue of Proposition~\ref{rp} for the function $u$,
using the microlocal $r^p$ current $(\ref{microrpcur})$.

The following proposition is the microlocal analogue of Proposition~\ref{rp}.
\begin{proposition}\label{microrp}Fix parameters $(\omega,m,\ell) \in \mathbb{R}\times \mathbb{Z}\times \mathbb{Z}_{\geq |m|}$ with $\omega \neq 0$, and let $u$ be a smooth solution of the radial o.d.e.~(\ref{e3iswsntouu})
\[u'' + (\omega^2-V)u = H,\]
such that $H(r^*)$ is compactly supported and the constant $a_{\mathcal{I}^-}$ from Proposition~\ref{asymAnalysis} vanishes. Then, for all $p \in [0,2]$ and sufficiently large $R$ (independent of $(\omega,m,\ell)$!),
\begin{align*}
\int_{R+1}^{\infty}&\left[r^{p-1}\left|u' - i\omega u\right|^2 + \left[(2-p)r^{p-3}\Lambda + r^{p-4}\right]\left|u\right|^2\right]\, dr^*  \\ \nonumber &\leq B\int_R^{R+1}r^p\left(1+\omega^2+\Lambda\right)\left|u\right|^2\, dr^* + B\int_R^{\infty}\left|H\right|\left[r^p\left|u'-i\omega u\right| + \left|u'\right|\right]\, dr^*.
\end{align*}
In the case $p = 2$, then we may moreover add the term $\Lambda\left|a_{\mathcal{I}^+}\right|^2$ to the left hand side.
\end{proposition}

\begin{proof}
We observing that a further asymptotic analysis (see Appendix A of~\cite{realmodestability}) of $u$ yields
\[u = a_{\mathcal{I}^+}e^{i\omega r^*}\left(1 + \frac{C}{r} + O\left(r^{-2}\right)\right)\text{ as }r\to\infty,\]
where $C \in \mathbb{C}$ is a constant independent of $u$ but depending on $(\omega,m,\ell)$. In particular, we find that
\[u' - i\omega u = O\left(r^{-2}\right)\text{ as }r\to\infty.\]

Next, let $R < \infty$ be sufficiently large and let $z = \chi r^p$ where $p\in[0,2]$
and $\chi$ is a cut-off function which is monotonically increasing, identically $0$ for $r \leq R$, and identically $1$ for $r \geq R+1$.  Keeping in mind that
\[V = \frac{\Lambda}{r^2} + \frac{2M\left[1- (\Lambda - 2am\omega)\right]}{r^3} + O\left(r^{-4}\right)\text{ as }r\to\infty,\]
we find that
\begin{align*}
&Q^z_{r^p}[u]|_{r = \infty} = 0\qquad\ \ \ \ \ \ \ \ \ \ \, \text{ if }p\in [0,2),\\
&Q^z_{r^p}[u]|_{r = \infty} = -\Lambda\left|a_{\mathcal{I}^+}\right|^2\qquad \text{ if }p = 2.
\end{align*}

Furthermore, recalling that by $(\ref{lamBound})$ we have
\[
\Lambda \geq 2\left|am\omega\right|,
\]
one may easily check that $r$ sufficiently large and $p\in [0,2]$ imply
\[
-\left(r^pV\right)' \geq b\left[(2-p)r^{p-3}\Lambda + r^{p-4}\right] - B\frac{\Lambda}{r^{p-4}}.
\]

Thus, applying the fundamental theorem of calculus to the identity~$(\ref{derrp})$ yields
\begin{align}\label{onelastannoyingtermontheright}
\int_{R+1}^{\infty}&\left[pr^{p-1}\left|u' - i\omega u\right|^2 + \left[(2-p)r^{p-3}\Lambda + r^{p-4}\right]\left|u\right|^2\right]  \\ \nonumber &\leq B\int_R^{R+1}r^p\left(1+\omega^2+ \Lambda\right)\left|u\right|^2\ dr^* + B\int_R^{\infty}r^p\left|H\right|\left|u'-i\omega u\right| \, dr^* + B\int_{R+1}^{\infty}\frac{\Lambda}{r^{p-4}}\left|u\right|^2\ dr^*,
\end{align}
where in the case $p=2$ we may add $\Lambda\left|a_{\mathcal{I}^+}\right|^2$ to the left hand side.

It remains to estimate the last term on the right hand side of~(\ref{onelastannoyingtermontheright}). (Note that for any $p \in [0,2)$ we could take $R$ sufficiently large depending on $p$ and absorb the troublesome term onto the left hand side. However, this cannot work in the case $p=2$.) Let $\tilde\chi$ be a cut-off which is identically $0$ for $r \in [r_+,R]$ and identically $1$ on $[R+1,\infty)$. Then, taking $R$ sufficiently large and applying the fundamental theorem of calculus to the identity~(\ref{thekoppa}) with $y = \tilde\chi$ easily yields
\[\int_{R+1}^{\infty}\frac{\Lambda}{r^3}\left|u\right|^2 \ dr^*\leq B\int_R^{R+1}\left(1+\omega^2+\Lambda\right)\left|u\right|^2 \ dr^*+ B\int_R^{\infty}\left|H\right|\left|u'\right|\ dr^*,\]
and thus concludes the proof.
\end{proof}

\subsection{A quantitative estimate on the rate of convergence of the microlocal radiation field}
\label{Qestie}
The following proposition will be used  in Section~\ref{relatephy} below and also in Section~\ref{someboundedargument}.
\begin{proposition}\label{microradconverge}Fix parameters $(\omega,m,\ell) \in \mathbb{R}\times \mathbb{Z}\times \mathbb{Z}_{\geq |m|}$ with $\omega \neq 0$, and let $u$ be a smooth solution of the radial o.d.e.~(\ref{e3iswsntouu}) with a right hand side $H$ vanishing for sufficiently large $r^*$, such that the constant $a_{\mathcal{I}^-}$ from Proposition~\ref{asymAnalysis} vanishes. Then there exists a sufficiently large constant $R$, independent of the frequency parameters, such that for every $\epsilon > 0$
\[\left|\omega\left(u - e^{i\omega r^*}a_{\mathcal{I}^+}\right)\Big|_{r = r_0}\right|^2 \leq B(\epsilon)r_0^{-2+\epsilon}\int_R^{R+1}\left(1+ \Lambda^3\right)\left|u\right|^2\, dr^*,\qquad \forall r_0 \geq R.\]
\end{proposition}
\begin{remark}Note that if we allowed the constants $B$ and $R$ to depend on the frequency parameters, standard o.d.e.~theory (e.g., see \cite{olver}) would allow one to replace $-2+\epsilon$ with the sharp exponent $-2$.
\end{remark}
\begin{remark}As far as the applications of Proposition~\ref{microradconverge} are concerned the only thing important about the $\Lambda$ dependence is that it is polynomial.
\end{remark}
\begin{proof}Set
\[E \doteq u - e^{i\omega r^*}a_{\mathcal{I}^+}.\]
Recall that standard o.d.e.~theory implies that $E = O\left(r^{-1}\right)$ as $r\to\infty$ (where the implied constant may depend on $(\omega, m, \ell)$).

Next, we observe that one may find a sufficiently large $R < \infty$ not depending on the frequency parameters so that $r \geq R$ implies $\left|V\right| \leq B\left(\frac{\Lambda}{r^2} + \frac{1}{r^3}\right)$.

A simple computation gives
\[E'' + \omega^2E = VE + e^{i\omega r^*}a_{\mathcal{I}^+} V.\]
Variation of parameters\footnote{More concretely, we define a function $\tilde E$ by the formula given, note that $(E-\tilde E)'' + \omega^2(E-\tilde E) = 0$, observe the trivial fact that any solution to $g'' + \omega^2g = 0$ which satisfies $g = O(r^{-1})$ must be identically $0$, and deduce that $E = \tilde E$.} then implies
\[
E(r) = -\int_{r^*}^{\infty}\left(\frac{e^{i\omega(r^*-s^*)}-e^{-i\omega(r^*-s^*)}}{2i\omega}\right)\left(V(s)E(s)+e^{i\omega s^*}a_{\mathcal{I}^+}V(s)\right)ds^*.
\]
In particular,
\begin{equation}\label{linfl1estiamte}
\left|\omega E(r)\right|^2 \leq B\left[\left(\int_{r^*}^{\infty}\frac{\left(1+\Lambda\right)|E(s)|}{s^2}ds^*\right)^2 + \left|a_{\mathcal{I}^+}\right|^2\left(\frac{\Lambda^2}{r^2} + \frac{1}{r^4}\right)\right].
\end{equation}

Now we consider the two terms on the right hand side of~(\ref{linfl1estiamte}) separately. For the first term, we begin by observing that $e^{i\omega r^*}\left(e^{-i\omega r^*}E\right)' = u' - i\omega u$. Keeping this in mind, we have
\begin{align}\label{otherterm}
\left(\int_{r^*}^{\infty}\frac{|E(s)|}{s^2}ds^*\right)^2 &\leq B(\epsilon)\int_{r^*}^{\infty}\frac{\left|E(s)\right|^2}{s^{3-\epsilon}}ds^* \\
                                                         \nonumber&= B(\epsilon)\int_{r^*}^{\infty}\frac{\left|e^{-i\omega s^*}E(s)\right|^2}{s^{3-\epsilon}}ds^* \\
                                                         \nonumber&\leq B(\epsilon)\int_{r^*}^{\infty}\frac{\left|\left(e^{-i\omega s^*}E(s)\right)'\right|^2}{s^{1-\epsilon}}ds^*\\
                                                         \nonumber&\leq B(\epsilon)r^{-2+\epsilon}\int_{r^*}^{\infty}s\left|u'-i\omega u\right|^2\, ds\\
                                                         \nonumber&\leq (1+\Lambda) B(\epsilon)r^{-2+\epsilon}\int_R^{R+1}\left|u\right|^2\, dr^*.
\end{align}
In the third inequality we used a standard Hardy inequality, and in the final inequality we appealed to Proposition~\ref{microrp}.

For the second term in~(\ref{linfl1estiamte}), we first note that Proposition~\ref{microrp}
with $p=2$ gives
\begin{equation}\label{nondegennullinf}
\Lambda\frac{\left|a_{\mathcal{I}^+}\right|^2}{r^2} \leq B\Lambda r^{-2}\int_R^{R+1}\left|u\right|^2\, dr^*.
\end{equation}
For the lower order term we use
\begin{equation}\label{blahblah}
\frac{\left|a_{\mathcal{I}^+}\right|^2}{r^4} \leq B\int_{r^*}^{\infty}\frac{\left|a_{\mathcal{I}^+}\right|^2}{s^5}ds^* \leq B\int_{r^*}^{\infty}\left[\frac{\left|u\right|^2 + \left|E\right|^2}{s^5}\right]ds^* \leq B(1+\Lambda)r^{-3}\int_R^{R+1}\left|u\right|^2\, dr^*.
\end{equation}
In the last inequality we used the estimates done in~(\ref{otherterm}) and Proposition~\ref{microrp}.

Combining~(\ref{linfl1estiamte}),~(\ref{otherterm}),~(\ref{nondegennullinf}), and~(\ref{blahblah}) concludes the proof.
\end{proof}

\subsection{Relation to the physical space radiation fields}\label{relatephy}
Definition~\ref{microRad} is motivated by the following propositions.
\begin{proposition}\label{microEqualInf}For all smooth solutions $\psi$ to~(\ref{WAVE}) on $\mathcal{D}$ arising from smooth compactly supported data along $\overline{\Sigma}$, let $a_{\mathcal{I}^+}\left(\omega,m,\ell\right)$ be the microlocal radiation field along $\mathcal{I}^+$. Then $\omega a_{\mathcal{I}^+} \in L^2_{\omega}l^2_{m\ell}$ and
\[
\partial_{\tau}\varphi(\tau,\infty,\theta,\phi)=\frac{1}{\sqrt{2\pi}}\int_{-\infty}^\infty\sum_{m\ell} \omega e^{-i\omega \tau} a_{\mathcal{I}^+}(\omega,m,\ell)S_{m\ell}(a\omega,\cos\theta)e^{im\phi} d\omega.
\]
Recall that $\varphi(\tau,\infty,\theta,\phi)$ denotes the radiation field of $\psi$ along future null infinity $\mathcal{I}^+$.
\end{proposition}

\begin{proof}First of all, as noted in Remark~\ref{obvsuffint}, $\psi$ is sufficiently integrable in the sense of Definition~\ref{sufficient} and thus the microlocal radiation field $a_{\mathcal{I}^+}$ is a well defined measurable function.

Now, define
\[\psi_+ \doteq \chi(t^*)\psi,\]
\[\psi_- \doteq (1-\chi(t^*))\psi,\]
where $\chi(x)$ is a cutoff function which is identically $0$ for $x \leq 0$ and is identically $1$ for $x \geq 1$ (the apparent asymmetry in the use of a cutoff depending on $t^*$ will not be a problem).

We shall denote $\left(r^2+a^2\right)^{1/2}\psi$, $\left(r^2+a^2\right)^{1/2}\psi_+$, and $\left(r^2+a^2\right)^{1/2}\psi_-$ by $\varphi$, $\varphi_+$, and $\varphi_-$ respectively.

The following facts are immediate consequences of $\psi$'s compact support along $\overline{\Sigma}$ and the finite speed of propagation.
\begin{enumerate}
    \item $\varphi_+|_{\mathcal{I}^+} = \varphi|_{\mathcal{I}^+}$.
    \item $\varphi_-|_{\mathcal{I}^-} = \varphi|_{\mathcal{I}^-}$.
    \item $\psi = \psi_+ + \psi_-$.
    \item $\Box_g\psi_+$ vanishes for large $r$.
    \item $\Box_g\psi_-$ vanishes for large $r$.
\end{enumerate}
Next, we observe the following immediate consequence of Proposition~\ref{rp} with $p = 1$ and Theorem~\ref{theResult} (note that the compact support of $\psi$'s initial data implies that the norms on the right sides of the estimates of Theorem~\ref{theResult} are finite and thus the right hand side of the estimate of Proposition~\ref{rp} is uniformly bounded as $\tau_2\to \infty$):
\begin{equation}\label{rp1}
\int_{-\infty}^{\infty}\int_{r\geq R}\int_0^{2\pi}\int_0^{\pi}\left|\left(\partial_t+\partial_{r^*}\right)\left(\left(r^2+a^2\right)^{1/2}\psi_+\right)\right|^2\sin\theta\, dt\, dr\, d\theta\, d\phi < \infty,
\end{equation}
where $R$ is sufficiently large. Applying the discrete isometry $(t,\phi) \mapsto (-t,-\phi)$ and repeating the above argument implies
\begin{equation}\label{rp2}
\int_{-\infty}^{\infty}\int_{r\geq R}\int_0^{2\pi}\int_0^{\pi}\left|\left(\partial_t-\partial_{r^*}\right)\left(\left(r^2+a^2\right)^{1/2}\psi_-\right)\right|^2\sin\theta\, dt\, dr\, d\theta\, d\phi < \infty,
\end{equation}
where $R$ is sufficiently large.

Noting that $\psi_{\pm}$ are easily seen to be sufficiently integrable in the sense of Definition~\ref{sufficient}, we may apply Carter's separation to $\psi_+$ and $\psi_-$ and define $u_+$ and $u_-$. Now we observe that Plancherel and~(\ref{rp1}) are easily seen to imply the existence of a dyadic sequence $\{r_n\}$ such that
\[\lim_{n\to\infty}\int_{-\infty}^{\infty}\sum_{m\ell}\left|u_+'-i\omega u_+\right|^2|_{r = r_n} = 0.\]
In turn, upon passing to a subsequence, this implies that for almost every $\omega$ and every $(m,\ell)$ we have
\[\lim_{n\to\infty}\left|u_+'-i\omega u_+\right||_{r = r_n} = 0.\]
Finally, combining this with Proposition~\ref{asymAnalysis} clearly implies that $u_+ \sim e^{i\omega r^*}$ as $r\to\infty$. Similarly, we observe that $u_- \sim e^{-i\omega r^*}$ as $r\to\infty$. Since we clearly have $u = u_+ + u_-$, we finally conclude that for almost every $\omega$ and each $(m,\ell)$ we have
\begin{equation}\label{theSplit1}
u_+ = a_{\mathcal{I}^+}e^{i\omega r^*} + O(r^{-1})\text{ as }r\to\infty,
\end{equation}
\begin{equation}\label{theSplit2}
u_- = a_{\mathcal{I}^-}e^{-i\omega r^*} + O(r^{-1})\text{ as }r\to\infty.
\end{equation}

Observe that the Fourier transform in $\tau$ of $\partial_{\tau}\varphi_+$ is given by $\left(e^{-i\omega r^*} + O\left(\frac{\omega}{r}\right)\right)\omega u_+$ as $r \to \infty$. Furthermore, observe that Theorem~\ref{h.o.s.} and Plancherel are easily seen to imply that
\[\int_{-\infty}^{\infty}\sum_{m\ell}\int_R^{R+1}\left(1+ \Lambda^3 + \omega^2\right)\left|u_+\right|^2\, dr^*\, d\omega < \infty.\]
Thus we may apply Proposition~\ref{microradconverge} to conclude that $\omega e^{-i\omega r^*}u_+$ and $\omega^2 e^{i\omega r^*}u_+$ converge in $L^2_{\omega}l^2_{m\ell}$ as $r\to\infty$ to $\omega a_{\mathcal{I}^+}$ and $\omega^2 a_{\mathcal{I}^+}$ respectively. In particular, the Fourier transform in $\tau$ of $\partial_{\tau}\varphi_+$ converges to $\omega a_{\mathcal{I}^+}$ in $L^2_{\omega}l^2_{m\ell}$ as $r\to\infty$. Plancherel then implies that any subsequence $\{\partial_{\tau}\varphi_+\}_{r_n}$ is Cauchy in $L^2_{\mathbb{R}\times\mathbb{S}^2}$. Now, recalling that $\partial_{\tau}\varphi_+\left(\tau,r,\theta,\phi\right)$ converges to $\partial_{\tau}\varphi_+\left(\tau,\infty,\theta,\phi\right)$ in $L^{\infty}_{\mathbb{R}\times \mathbb{S}^2}$ (see Remark~\ref{linf} and keep in mind that the finite speed of propagation implies $\varphi_+$ is only supported along $\tau \geq \tau_0$ for some $\tau_0 \in \mathbb{R}$), we conclude, using the uniqueness of $L^p$ limits, that $\partial_{\tau}\varphi_+\left(\tau,r,\theta,\phi\right)$ converges to $\partial_{\tau}\varphi_+(\tau,\infty,\theta,\phi)$ in $L^2_{\mathbb{R}\times \mathbb{S}^2}$. Finally, continuity of the Fourier transform on $L^2$ implies that
\[
\partial_{\tau}\varphi_+(\tau,\infty,\theta,\phi)=\frac{1}{\sqrt{2\pi}}\int_{-\infty}^\infty\sum_{m\ell} \omega e^{-i\omega \tau} a_{\mathcal{I}^+}(\omega,m,\ell)S_{m\ell}(a\omega,\cos\theta)e^{im\phi} d\omega.
\]
To conclude the proof we simply recall that $\varphi_+|_{\mathcal{I}^+} = \varphi|_{\mathcal{I}^+}$.
\end{proof}

Now we turn to the horizon flux.
\begin{proposition}\label{microequalhor}For all solutions $\psi$ to~(\ref{WAVE}) on $\mathcal{D}$ arising from smooth compactly supported initial data along $\overline{\Sigma}$, let $a_{\mathcal{H}^+}(\omega,m,\ell)$ be the microlocal radiation field along $\mathcal{H}^+$. Then $(\omega-\upomega_+m)a_{\mathcal{H}^+} \in L^2_{\omega}l^2_{m\ell}$ and
\begin{equation}\label{theassertion}
K\psi(t^*,r_+,\theta,\phi) = \frac{1}{\sqrt{4M\pi r_+}}\int_{-\infty}^\infty\sum_{m\ell} \left(\omega-\upomega_+m\right)e^{-i\omega t^*} a_{\mathcal{H}^+}(\omega,m,\ell)S_{m\ell}(a\omega,\cos\theta)e^{im\phi^*} d\omega.
\end{equation}
\end{proposition}
\begin{proof}First of all, as noted in Remark~\ref{obvsuffint}, $\psi$ is sufficiently integrable in the sense of Definition~\ref{sufficient} and thus the microlocal radiation field $a_{\mathcal{H}^+}$ is a well defined measurable function.

We first consider the case where the initial data for $\psi$ are in fact compactly supported along $\mathring{\Sigma}$. We may then proceed in a completely analogous manner to the proof of Proposition~\ref{microEqualInf}. We note that the argument is in fact simpler since we will be able to rely directly on Theorem~\ref{theResult} instead of developing an analogue of Theorem~\ref{rp} near the horizon.

Define
\[\psi_+ \doteq \chi(t)\psi,\]
\[\psi_- \doteq (1-\chi(t))\psi,\]
where $\chi(x)$ is a cutoff function which is identically $0$ for $x \leq 0$ and is identically $1$ for $x \geq 1$.

The following facts are immediate consequences of $\psi$'s compact support away from the {bifurcate sphere} $\mathcal{B}$ and the finite speed of propagation.
\begin{enumerate}
    \item $\psi_+|_{\mathcal{H}^+} = \psi|_{\mathcal{H}^+}$.
    \item $\psi_-|_{\mathcal{H}^-} = \psi|_{\mathcal{H}^-}$.
    \item $\psi = \psi_+ + \psi_-$.
    \item $\Box_g\psi_+$ vanishes for small $r-r_+$.
    \item $\Box_g\psi_-$ vanishes for small $r-r_+$.
\end{enumerate}

Recalling that the smooth extension of $\partial_{r^*}$ to $\mathcal{H}^+\cup\mathcal{H}^-$ satisfies $\partial_{r^*}|_{\mathcal{H}^+} = K$ and $\partial_{r^*}|_{\mathcal{H}^-} = -K$, we see that Theorem~\ref{h.o.s.} immediately implies
\begin{equation}\label{boundHorizonEstplus}
\lim_{r\to r_+}\left(\partial_{r^*} - K\right)\psi_+ = 0\text{ in }L^2_{t^*,\theta^*,\phi^*}\left(\sin\theta^*\, dt^*\, d\theta^*\, d\phi^*\right),
\end{equation}
\begin{equation}\label{boundHorizonEstminus}
\lim_{r\to r_+}\left(\partial_{r^*} + K\right)\psi_- = 0\text{ in }L^2_{{}^*t,{}^*\theta,{}^*\phi}\left(\sin\text{}^*\theta\, d\text{}^*t\, d\text{}^*\theta\, d\text{}^*\phi\right).
\end{equation}

Appealing to Theorems~\ref{theResult} and~\ref{h.o.s.}, we may apply Carter's separation to $\psi_+$ and $\psi_-$ and define $u_+$ and $u_-$. Since we clearly have $u = u_+ + u_-$, Proposition~\ref{asymAnalysis},~(\ref{boundHorizonEstplus}),~(\ref{boundHorizonEstminus}) and a similar argument as we used near $\mathcal{I}^+$ (note that the convergence of $\psi_+$ to its radiation field along the horizon in both $L^{\infty}_{\mathbb{R}_{\geq 0}\times \mathbb{S}^2}$ and $L^2_{\mathbb{R}_{\geq 0} \times \mathbb{S}^2}$ follows immediately from the fundamental theorem of calculus and Theorem~\ref{h.o.s.}) imply that for almost every $\omega$ and for each $(m,\ell)$, we have
\begin{equation}\label{theSplithor1}
u_+ = a_{\mathcal{H}^+}e^{-i\left(\omega-\upomega_+m\right)r^*} + O(r-r_+)\text{ as }r\to r_+,
\end{equation}
\begin{equation}\label{theSplithor2}
u_- = a_{\mathcal{H}^-}e^{i\left(\omega-\upomega_+m\right) r^*} + O(r-r_+)\text{ as }r\to r_+.
\end{equation}

Now, we note that Theorem~\ref{theResult} is easily seen to imply that $\psi_+|_{r=s} \to \psi_+|_{r=r_+}$ as $s\to r_+$ in $L^2_{t^*,\theta^*,\phi^*}$. Arguing in a similar fashion as in the proof of Proposition~\ref{microEqualInf} we conclude that $a_{\mathcal{H}^+}$ is in $L^2_{\omega}l^2_{m\ell}$ and
\begin{equation}
\psi(t^*,r_+,\theta,\phi) = \frac{1}{\sqrt{4M\pi r_+}}\int_{-\infty}^\infty\sum_{m\ell}e^{-i\omega t^*} a_{\mathcal{H}^+}(\omega,m,\ell)S_{m\ell}(a\omega,\cos\theta)e^{im\phi^*} d\omega.
\end{equation}

Now we consider the case where the support of $\psi$ may contain the {bifurcate sphere}
$\mathcal{B}$.
We begin by commuting~(\ref{WAVE}) with $K$ and conclude that $\Box_g\left(K\psi\right) = 0$. Then we recall that $K\psi$ vanishes on the bifurcate sphere in view of~(\ref{vanish}). Now, let $\chi(x)$ be a smooth function which is identically $0$ for $x \in (-\infty,1]$ and identically $1$ for $x \in [2,\infty)$. Set $\chi_{\epsilon}(x) \doteq \chi\left(\frac{x}{\epsilon}\right)$, and, recalling the coordinate system $(U^+,V^+,\theta,\phi)$ near the bifurcate sphere which was introduced in Section~\ref{diffCoord}, let $\left(K\psi\right)_{\epsilon}$ denote the solution to the wave equation with the initial data of $\chi_{\epsilon}\left(V^+\right)K\psi$. Using that $K\psi$ is smooth and vanishes at the bifurcate sphere, one may easily verify that
\[\lim_{\epsilon \to 0}\int_{\mathring{\Sigma}}\mathbf{J}^N_{\mu}\left[K\psi - (K\psi)_{\epsilon}\right]n^{\mu}_{\mathring{\Sigma}} = 0.\]
Theorem~\ref{theResult} then implies that
\[\lim_{\epsilon\to 0}\int_{\mathcal{H}^+}\mathbf{J}^N_{\mu}\left[K\psi - (K\psi)_{\epsilon}\right]n^{\mu}_{\mathcal{H}^+} = 0.\]

Since $(K\psi)_{\epsilon}$ is compactly supported away from the bifurcate sphere,
\begin{equation}
(K\psi)_{\epsilon}(t^*,r_+,\theta,\phi) = \frac{1}{\sqrt{4 M\pi r_+}}\int_{-\infty}^\infty\sum_{m\ell}e^{-i\omega t^*} a^{(K)}_{\epsilon,\mathcal{H}^+}(\omega,m,\ell)S_{m\ell}(a\omega,\cos\theta)e^{im\phi^*} d\omega,
\end{equation}
where $a^{(K)}_{\epsilon,\mathcal{H}^+}$ is the microlocal radiation field along $\mathcal{H}^+$ for $(K\psi)_{\epsilon}$ (observe that $(K\psi)_{\epsilon}$ is easily seen to be sufficiently integrable in the sense of Definition~\ref{sufficient}).

In order to finish the proof, we just need to establish that $a^{(K)}_{\epsilon,\mathcal{H}^+} \to (\omega-\upomega_+m)a_{\mathcal{H}^+}$ in $L^2_{\omega}l^2_{m\ell}$ as $\epsilon \to 0$. We begin by noting that the convergence of $(K\psi)_{\epsilon}$ to $K\psi$ and Plancherel imply that $\{a^{(K)}_{\epsilon,\mathcal{H}^+}\}$ has an $L^2_{\omega}l^2_{m\ell}$ limit as $\epsilon \to 0$; hence, it suffices to check that $\left(\omega-\upomega_+m\right)a^{(K)}_{\epsilon,\mathcal{H}^+}$ converges to $\left(\omega-\upomega_+m\right)a_{\mathcal{H}^+}$ pointwise almost everywhere. In order to see this, we let $u^{(K)}_{\epsilon}$ denote the result of applying Carter's separation to $\left(K\psi\right)_{\epsilon}$, and observe that $a^{(K)}_{\epsilon,\mathcal{H}^+}$ is, up to an appropriate normalisation, equal to the Wronskian of $u^{(K)}_{\epsilon}$ with $U_{\rm hor}$:
\[a^{(K)}_{\epsilon,\mathcal{H^+}} = \left(-2i\left(\omega-\upomega_+m\right)\right)^{-1}\left(\left(u^{(K)}_{\epsilon}\right)'\overline{U_{\rm hor}} - u^{(K)}_{\epsilon}\overline{U'_{\rm hor}}\right).\]
Since Theorem~\ref{theResult} may be easily used to show that for each $(\omega,m,\ell)$ and $r^*$, $u^{(K)}_{\epsilon}\left(r^*,\omega,m,\ell\right)$ converges to $u_{\epsilon}\left(r^*,\omega,m,\ell\right)$ as $\epsilon \to 0$, we conclude that $\left(\omega-\upomega_+m\right)a^{(K)}_{\epsilon,\mathcal{H}^+}\left(\omega,m,\ell\right)$ converges to $\left(\omega-\upomega_+m\right)a^{(K)}_{\mathcal{H}^+}$ as $\epsilon \to 0$.
\end{proof}

\section{Boundedness revisited: A degenerate-energy boundedness statement}\label{secDegenBound}
This section is dedicated to refining our recent proof from~\cite{partIII} of boundedness for the wave equation so as to apply for finite degenerate $V$-energy solutions.

We will collect all statements which we shall need for the remainder of
the paper in Section~\ref{MTandC}. The key statement is Theorem~\ref{boundDegen}
together with one immediate corollary.
(In particular, after digesting these statements, the reader impatient to proceed to the scattering theory
constructions can skip to
Section~\ref{forward}.)

In the brief aside of Section~\ref{yetanotheraside},
we shall also state the full degenerate-energy analogue
of Theorem~\ref{theResult} in Section~\ref{yetanotheraside}
as Theorem~\ref{degentheResult}.
We shall not actually require the latter result in the paper and it in fact is more
convenient to infer it a posteriori with the help of the backwards scattering maps
which we shall construct in Section~\ref{secScatter}.
Thus, the proof of
Theorem~\ref{degentheResult}
 is in fact deferred till Section~\ref{asideforthepr}.

Section~\ref{outline} gives the proof of Theorem~\ref{boundDegen}.
We note that the proof
will crucially use Proposition~\ref{fundamental}, Proposition~\ref{theWronkIsReallyBounded} and Proposition~\ref{asymExpandGood}.

\subsection{The main theorem and corollary}
\label{MTandC}
The main result which we shall require for later sections is the following.
\begin{moderthem}\label{boundDegen}For all solutions $\psi$ to~(\ref{WAVE}) on $\mathcal{R}_{\geq 0}$ arising from smooth initial data on $\Sigma_0^*$
which are compactly supported, we have
\begin{equation}\label{theEstimate0}
 \int_{\mathcal{I}^+} {\bf J}_\mu^T[\psi]
n^\mu_{\mathcal{I}^+} + \int_{\mathcal{H}^+_{\geq 0}}{\bf J}^K_\mu[\psi]n^\mu_{\mathcal{H}^+}\le B \int_{\Sigma_0^*} {\bf J}_\mu^V[\psi]n^\mu_{\Sigma_0^*}.
\end{equation}
\end{moderthem}
\begin{remark}One can easily formulate and prove higher order versions of Theorem~\ref{boundDegen} but we will not pursue this here.
\end{remark}

Given that the restriction of the deformation tensor of $V$ to $J^-\left(\Sigma_0^*\right) \cap J^+\left(\overline{\Sigma}\right)$ is compactly supported away from $\mathcal{H}^+\cup\mathcal{H}^-\cup\mathcal{B}$, a finite in time energy estimate,~i.e.~(\ref{ingeneralform}) with $X = V$, immediately implies
\begin{modercor}\label{boundDegent0}For all solutions $\psi$ to~(\ref{WAVE}) on $J^+\left(\overline{\Sigma}\right)$ arising from smooth compactly supported initial data along $\overline{\Sigma}$, we have
\begin{equation}\label{theEstimatet0}
 \int_{\mathcal{I}^+} {\bf J}_\mu^T[\psi]
n^\mu_{\mathcal{I}^+} + \int_{\overline{\mathcal{H}^+}}{\bf J}^K_\mu[\psi]n^\mu_{\overline{\mathcal{H}^+}}\le B \int_{\overline{\Sigma}} {\bf J}_\mu^V[\psi]n^\mu_{\overline{\Sigma}}.
\end{equation}
\end{modercor}

\subsection{Aside: the full degenerate boundedness and integrated decay statements}
\label{yetanotheraside}
We note that we can in fact obtain the full analogue
of Theorem~\ref{theResult} where  energy boundedness is given with respect to
a spacelike foliation, and where integrated local energy decay is proven, both now involving the degeneate energy.  We will not require this result in the rest
of paper and it is in fact convenient to obtain it a posteriori using our scattering theory.

\begin{moderthem}\label{degentheResult}For all solutions $\psi$ to~(\ref{WAVE}) on $\mathcal{R}_{\geq 0}$ arising from smooth initial data on $\Sigma_0^*$
which are compactly supported, we have
\begin{equation}
 \int_{\Sigma_s^*} {\bf J}_\mu^V[\psi]
n^\mu_{\Sigma_s^*}\le B \int_{\Sigma_0^*} {\bf J}_\mu^V[\psi]n^\mu_{\Sigma_0^*}, \qquad
\forall s\ge 0,
\end{equation}
\begin{equation}
\int_{\mathcal{R}_{\geq 0}}\Big(r^{-1}\zeta |\nabb\psi|^2+r^{-1-\delta}\zeta \left|T\psi\right|^2+\left(r-r_+\right)^2r^{-3-\delta}\left|\tilde Z^*\psi\right|^2+ r^{-3-\delta} \left|\psi\right|^2\Big) \le B(\delta)
\int_{\Sigma_0^*} {\bf J}_\mu^V[\psi]n^\mu_{\Sigma_0^*},
\end{equation}
where $\zeta$ is defined as in the statement of Theorem~\ref{theResult}.
\end{moderthem}
The proof is defered till Section~\ref{asideforthepr}.

\begin{remark}Note the degeneration of the  bulk integral at the horizon.
One can easily formulate and prove higher order versions of Theorem~\ref{degentheResult} but we will not pursue this here.
\end{remark}

\subsection{The proof of Theorem~\ref{boundDegen}}\label{outline}

Before we begin the discussion of the proof of Theorem~\ref{boundDegen}, let us briefly indicate what would go wrong if we simply tried to repeat the proof of Theorem~\ref{theResult} as given in~\cite{partIII}.
\begin{enumerate}
    \item Anytime the redshift estimate of~\cite{lectnotes} and~\cite{redshift} is applied to $\psi$, one must put a term $\int_{\Sigma_0^*}{\bf J}_{\mu}^N[\psi]n^{\mu}_{\Sigma_0^*}$ on the right hand side of the resulting estimate.
    \item In~\cite{partIII}, when we proved the integrated energy decay statement for $\psi$ we first proved an estimate for $\chi\psi$ where $\chi(t^*)$ was a cutoff function which was identically $0$ in the past of $\Sigma_0^*$ and identically $1$ in the future of $\Sigma_1^*$. We then studied the inhomogeneous wave equation
        \[\Box_g(\chi\psi) = 2g^{\mu\nu}\nabla_{\mu}\chi\nabla_{\nu}\psi + \left(\Box_g\chi\right)\psi \doteq F.\]
        The resulting estimate in~\cite{partIII} had, in particular, a term on the right hand side proportional to
        \[\int_{\mathcal{R} \cap \{r \leq R\}}\left|F\right|^2,\]
        for some constant $R > r_+$. Note that on the horizon, $F$ will contain a term proportional to $Z^*\psi$. Unfortunately, this is exactly the derivative that the $\mathbf{J}^V$ energy loses control of as $r\to r_+$.
\end{enumerate}

In order to prove Theorem~\ref{boundDegen} we will first observe that without loss of generality, we can assume that the initial data for $\psi$ is supported near the horizon. Applying a $\mathbf{J}^K$ energy estimate for $\psi$ and Plancherel then immediately reduce the problem to estimating the microlocal radiation fields for $\psi_{\text{\Rightscissors}}$ along $\mathcal{I}^+$ in the superradiant frequency regime $\mathcal{F}^{(\epsilon)}_{\sharp}$. Next, using the fundamental solution representation of Proposition~\ref{fundamental} we will represent the microlocal radiation fields along $\mathcal{I}^+$ as an integral in $r^*$ of the Fourier transform of $F$ against $U_{\rm hor}$. Following this, in the most subtle part of the proof, we will crucially exploit the fact that we are in a superradiant frequency regime where we can afford to lose a derivative, the fact we only need to estimate the flux to $\mathcal{I}^+$, the fact that $F$ is supported near the horizon and the oscillations of $U_{\rm hor}$ in $r^*$ (as embodied in Proposition~\ref{asymExpandGood}) in order to gain some degeneration in $r-r_+$. Somewhat surprisingly, this step does not use that $F = 2g^{\mu\nu}\nabla_{\mu}\chi\nabla_{\nu}\psi + \left(\Box_g\chi\right)\psi$; it treats $F$ as an arbitrary inhomogeneity. Finally, the proof concludes with finite in time energy estimates and Hardy inequalities (of course, the fact that $F = 2g^{\mu\nu}\nabla_{\mu}\chi\nabla_{\nu}\psi + \left(\Box_g\chi\right)\psi$ is used in this step).

\begin{proof}[Proof of Theorem~\ref{boundDegen}]
We start with an easy reduction; we may split $\psi$ into $\psi_1$ and $\psi_2$ where $\psi_1$ has initial data supported near the horizon and $\psi_2$ has initial data supported away from the horizon. Of course, the estimate~(\ref{theEstimate0}) for $\psi_2$ follows from Theorem~\ref{theResult}. Thus, without loss of generality, we will assume that $\psi_2 = 0$ and that $\psi = \psi_1$ has initial data whose support is contained in $r \in [r_+,10M]$.

We now define $\psi_{\text{\Rightscissors}} \doteq \chi\psi$ where $\chi$ is a function which is identically $1$ in the future of $\Sigma_1^*$, and identically $0$ in the past of $\Sigma_0^*$. This satisfies
\[\Box_g\psi_{\text{\Rightscissors}} = 2g^{\mu\nu}\nabla_{\mu}\chi\nabla_{\nu}\psi + \left(\Box_g\chi\right)\psi \doteq F.\]

The functions $u$ and $H$ are then defined by applying Carter's separation to $\chi\psi$ and $F$ respectively. This satisfies the radial o.d.e.~(\ref{e3iswsntouu}) with a non-zero right hand side $H$. Let $a_{\mathcal{I}^+}$ denote the corresponding microlocal radiation field of $u$.

We begin by showing
\begin{equation}\label{hawkenergyestphysical}
\int_{\mathcal{H}^+_{\geq 0}}\mathbf{J}^K_{\mu}[\psi]n^{\mu}_{\mathcal{H}^+} + \int_{-\infty}^{\infty}\sum_{m\ell} \omega\left(\omega-\upomega_+m\right)\left|a_{\mathcal{I}^+}\right|^2\ d\omega \leq B\int_{\Sigma_0^*}\mathbf{J}^V_{\mu}[\psi]n^{\mu}_{\Sigma_0^*}.
\end{equation}
Note that part of the proof of this statement will be that the unsigned quantity  $\int_{-\infty}^{\infty}\sum_{m\ell} \omega\left(\omega-\upomega_+m\right)\left|a_{\mathcal{I}^+}\right|^2\ d\omega$ is absolutely convergent. (One should think of~(\ref{hawkenergyestphysical}) as corresponding to the formal statement $
\int_{\mathcal{H}^+_{\geq 0}}\mathbf{J}^K_{\mu}[\psi]n^{\mu}_{\mathcal{H}^+} + \int_{\mathcal{I}^+}\mathbf{J}^K_{\mu}[\psi]n^{\mu}_{\mathcal{I}^+} \leq B\int_{\Sigma_0^*}\mathbf{J}^V_{\mu}[\psi]n^{\mu}_{\Sigma_0^*}$. However, we will wish to avoid a discussion of the convergence of the unsigned integral $\int_{\mathcal{I}^+}\mathbf{J}^K_{\mu}[\psi]n^{\mu}_{\mathcal{I}^+}$.)

Let $s > 0$ and $r_0 > r_+$. We start with a $\mathbf{J}^K$ energy estimate in the region bounded by $\mathcal{H}^+(0,s)$, $\Sigma_s^* \cap \{r \leq r_0\}$, $\{r = r_0\} \cap J^-(\Sigma_s^*)$, and $\Sigma_0^*$. We obtain
\begin{equation}\label{finitehawk}
\int_{\mathcal{H}^+(0,s)}\mathbf{J}^K_{\mu}[\psi]n^{\mu}_{\mathcal{H}^+} + \int_{\Sigma_s^*\cap \{r \leq r_0\}}\mathbf{J}^K_{\mu}[\psi]n^{\mu}_{\Sigma_s^*} +
\int_{\{r=r_0\}\cap J^-(\Sigma_s^*) \cap J^+(\Sigma_0^*)}\mathbf{J}^K_{\mu}[\psi]n^{\mu}_{\{r=r_0\}} = \int_{\Sigma_0^*\cap \{r\leq r_0\}}\mathbf{J}^K_{\mu}[\psi]n^{\mu}_{\Sigma_0^*}.
\end{equation}
It easily follows from Theorem~\ref{h.o.s.} that for each $r_0$, there exists a dyadic sequence $\{s_i\}_{i=1}^{\infty}$ such that
\[\lim_{i\to\infty}\int_{\Sigma_{s_i}^*\cap \{r \leq r_0\}}\mathbf{J}^K_{\mu}[\psi]n^{\mu}_{\Sigma_{s_i}^*} = 0.\]
We thus obtain
\begin{equation}\label{finitehawkhighr0}
\int_{\mathcal{H}^+_{\geq 0}}\mathbf{J}^K_{\mu}[\psi]n^{\mu}_{\mathcal{H}^+} + \int_{\{r=r_0\}\cap J^+(\Sigma_0^*)}\mathbf{J}^K_{\mu}[\psi]n^{\mu}_{\{r=r_0\}} = \int_{\Sigma_0^*\cap \{r\leq r_0\}}\mathbf{J}^K_{\mu}[\psi]n^{\mu}_{\Sigma_0^*}.
\end{equation}
Observe that Theorem~\ref{h.o.s.} allows us to unambiguously assign a value to the unsigned quantity
\[\int_{\{r=r_0\}\cap J^+(\Sigma_0^*)}\mathbf{J}^K_{\mu}[\psi]n^{\mu}_{\{r=r_0\}}.\]

Next, recalling that $\psi|_{\Sigma_0^*}$ is supported with $[r_+,r_++10M]$, we observe that if $r_0$ is sufficiently large, then~(\ref{finitehawkhighr0}) becomes
\begin{equation}\label{finitehawkhighr0large}
\int_{\mathcal{H}^+_{\geq 0}}\mathbf{J}^K_{\mu}[\psi]n^{\mu}_{\mathcal{H}^+} + \int_{\{r=r_0\}}\mathbf{J}^K_{\mu}[\psi_{\text{\Rightscissors}}]n^{\mu}_{\{r=r_0\}} \leq B\int_{\Sigma_0^*}\mathbf{J}^V_{\mu}[\psi]n^{\mu}_{\Sigma_0^*}.
\end{equation}
Now we explicitly compute, apply Plancherel, and integrate by parts:
\begin{align}\label{r0energy}
\int_{\{r=r_0\}}\mathbf{J}^K_{\mu}[\psi_{\text{\Rightscissors}}]n^{\mu}_{\{r=r_0\}} &= \int_{\{r = r_0\}}\left(\left(T\psi_{\text{\Rightscissors}} + \omega_+\Phi\psi_{\text{\Rightscissors}}\right)\partial_{r^*}\psi_{\text{\Rightscissors}}\right)(r^2+a^2)\sin\theta\, dt\, d\theta\, d\phi
\\ \nonumber &= \int_{\{r = r_0\}}\left(\left(T((r^2+a^2)^{1/2}\psi_{\text{\Rightscissors}}) + \omega_+\Phi((r^2+a^2)^{1/2}\psi_{\text{\Rightscissors}})\right)\partial_{r^*}((r^2+a^2)^{1/2}\psi_{\text{\Rightscissors}})\right)\sin\theta\, dt\, d\theta\, d\phi
\\ \nonumber &= \int_{-\infty}^{\infty}\sum_{m\ell}\left(\omega-\upomega_+m\right)\text{Im}\left(u'\overline{u}\right)|_{r=r_0}d\omega.
\end{align}
Next, we consider the microlocal $K$-energy current (see Section~\ref{sepcurr}):
\[
Q^K[u] \doteq \left(\omega-\upomega_+m\right)\text{Im}\left(u'\overline{u}\right).
\]
This is conserved for $r \geq r_0$ for $r_0$ sufficiently large, i.e.
\[
\left(Q^K\right)' = 0.
\]
Noting that the proof of Proposition~\ref{microEqualInf} implies $\left(u' - i\omega u\right)|_{r=\infty} = 0$, we thus obtain
\begin{equation}\label{microenergy}
\int_{-\infty}^{\infty}\sum_{m\ell}\left(\omega-\upomega_+m\right)\text{Im}\left(u'\overline{u}\right)|_{r=r_0} d\omega= \int_{-\infty}^{\infty}\sum_{m\ell}\omega\left(\omega-\upomega_+m\right)\left|a_{\mathcal{I}^+}\right|^2d\omega.
\end{equation}
In particular, the right hand side of~(\ref{microenergy}) is absolutely convergent. Combining~(\ref{finitehawkhighr0large}),~(\ref{r0energy}), and~(\ref{microenergy}) yields~(\ref{hawkenergyestphysical}).

Next, we observe that Propositions~\ref{microEqualInf} and~\ref{efluxnullinf} together imply that
\begin{equation}\label{theymustbeequal}
\int_{\mathcal{I}^+}\mathbf{J}^T_{\mu}[\psi]n^{\mu}_{\mathcal{I}^+} = \int_{-\infty}^{\infty}\sum_{m\ell}\omega^2\left|a_{\mathcal{I}^+}\right|^2\, d\omega.
\end{equation}

Now, observing that $(\omega,m,\ell) \not\in \mathcal{F}^{(\epsilon)}_{\sharp}$ imply that $\omega\left(\omega - \upomega_+m\right) \geq b(\epsilon)\omega^2$, it is clear that in order to finish the proof we need only show
\begin{equation}\label{tofinish}
\int_{-\infty}^{\infty}\sum_{\{(m,\ell) :(\omega,m,\ell) \in \mathcal{F}^{(\epsilon)}_{\sharp}\}}\left|\upomega_+\omega m\right|\left|a_{\mathcal{I}^+}\right|^2\ d\omega \leq B(\epsilon)\int_{\Sigma_0^*}\mathbf{J}^V_{\mu}[\psi]n^{\mu}_{\Sigma_0^*},
\end{equation}
for some sufficiently small $\epsilon > 0$.

We turn thus to the proof of $(\ref{tofinish})$.

First, note that Proposition~\ref{fundamental} allows us to write
\begin{equation}
\label{graytoedw}
\left|a_{\mathcal{I}^+}\right|^2 = \left|\mathfrak{W}\right|^{-2}\left|\int_{-\infty}^{\infty}U_{\rm hor}(x^*)H(x^*)dx^*\right|^2.
\end{equation}

Keeping in mind that the set $\{(\omega,m,\ell) \in \mathcal{F}^{(\epsilon)}_{\sharp} : \Lambda \leq c\}$ is compact, standard o.d.e. theory implies
\begin{equation}\label{somestuff000}
U_{\rm hor}(r^*) = e^{-i\left(\omega-\upomega_+m\right)r^*} + E_{\rm hor}(r^*),
\end{equation}
where $(\omega,m,\ell) \in \mathcal{F}^{(\epsilon)}_{\sharp}$ and $\Lambda \leq c$ implies
\begin{equation}\label{somestuff0000}
\left|E_{\rm hor}(r^*)\right| \leq B\left(\epsilon,c\right)\left|r-r_+\right|,
\end{equation}
for $r-r_+$ sufficiently small. As $c\to \infty$, however, the dependence of $B(\epsilon, c)$ may be bad.
Fortunately Proposition~\ref{asymExpandGood} shows that if $\epsilon > 0$ is sufficiently small and $\Lambda$ is sufficiently large, then we have
\[
\left|E_{\rm hor}\right| \leq \frac{B(\epsilon)\left|\mathfrak{W}\right|}{\sqrt{\Lambda}}\sqrt{r-r_+},
\]
for sufficiently small $r-r_+$.

Applying $(\ref{graytoedw})$,~$(\ref{somestuff000})$,~$(\ref{somestuff0000})$,
Proposition~\ref{theWronkIsReallyBounded} and Proposition~\ref{asymExpandGood},
we obtain
\begin{align}\label{whatweworkedhardfor}
\int_{-\infty}^{\infty}&\sum_{\{(m,\ell) :(\omega,m,\ell) \in \mathcal{F}^{(\epsilon)}_{\sharp}\}}\left|\upomega_+ \omega m\right|\left|a_{\mathcal{I}^+}\right|^2\ d\omega
\\ \nonumber &= \int_{-\infty}^{\infty}\sum_{\{(m,\ell) :(\omega,m,\ell) \in \mathcal{F}^{(\epsilon)}_{\sharp}\}}\frac{\left|\upomega_+\omega m\right|}{\left|\mathfrak{W}\right|^2}\left|\int_{-\infty}^{\infty}U_{\rm hor}(r^*)H(r^*)dr^*\right|^2\ d\omega
\\ \nonumber &\leq B \int_{-\infty}^{\infty}\sum_{\{(m,\ell) :(\omega,m,\ell) \in \mathcal{F}^{(\epsilon)}_{\sharp}\}}\Bigg[\frac{\left|\upomega_+ \omega m\right|}{\left|\mathfrak{W}\right|^2}\left|\int_{-\infty}^{\infty}e^{-i\left(\omega-\upomega_+m\right)r^*}H(r^*)dr^*\right|^2
\\ \nonumber &\qquad \qquad\qquad\qquad \qquad\ \ \ \ \ +\frac{\left|\upomega_+ \omega m\right|}{\left|\mathfrak{W}\right|^2}\left|\int_{-\infty}^{\infty}e^{-i\left(\omega-\upomega_+m\right)r^*}E_{\rm hor}(r^*)H(r^*)dr^*\right|^2\Bigg]\ d\omega
\\ \nonumber &\leq B(\epsilon)\int_{-\infty}^{\infty}\sum_{\{(m,\ell) :(\omega,m,\ell) \in \mathcal{F}^{(\epsilon)}_{\sharp}\}}\left[\left|\int_{-\infty}^{\infty}e^{-i\left(\omega-\upomega_+m\right)r^*}H(r^*)\, dr^*\right|^2 +\left|\int_{-\infty}^{\infty}\sqrt{r-r_+}\left|H(r^*)\right|\, dr^*\right|^2\right]
\\ \nonumber &\doteq B(\epsilon)\left[I + II\right].
\end{align}

Let us now recall the explicit form of $H$:
\begin{equation}\label{explicit1}
H = \frac{\Delta}{(r^2+a^2)^{3/2}}\int_{-\infty}^{\infty}\int_{\mathbb{S}^2}e^{i\omega t}e^{-im\phi}S_{m\ell}(\theta,a\omega)\left(\rho^2F\right)\sin\theta\, dt\, d\theta\, d\phi,
\end{equation}
\begin{equation}\label{explicit2}
F = 2g^{\mu\nu}\nabla_{\mu}\chi\nabla_{\nu}\psi + \left(\Box_g\chi\right)\psi.
\end{equation}

In particular, directly applying Cauchy-Schwarz, Plancherel, a straightforward Hardy inequality, and a finite in time energy inequality, one may easily check that
\begin{equation}\label{IIOK}
\int_{-\infty}^{\infty}\sum_{\{(m,\ell) :(\omega,m,\ell) \in \mathcal{F}^{(\epsilon)}_{\sharp}\}}\left|II\right| \leq B\int_{\Sigma_0^*}\mathbf{J}^V_{\mu}[\psi]n^{\mu}_{\Sigma_0^*}.
\end{equation}

For every $\gamma > 0$, the same direct application of Cauchy-Schwarz, Plancherel, a straightforward Hardy inequality, and a finite in time energy inequality to the term $I$ will only give
\begin{equation}\label{INOTOK}
\int_{-\infty}^{\infty}\sum_{\{(m,\ell) :(\omega,m,\ell) \in \mathcal{F}^{(\epsilon)}_{\sharp}\}}\left|I\right| \leq B(\gamma)\int_{\Sigma_0^* \cap [r_+,r_++10M]}(r-r_+)^{1-\gamma}\left|\tilde Z^*\psi\right|^2 + B\int_{\Sigma_0^*}\mathbf{J}^V_{\mu}[\psi]n^{\mu}_{\Sigma_0^*}.
\end{equation}
Unfortunately, the first term on the right hand side is (barely) not controlled by $\int_{\Sigma_0^*}\mathbf{J}^V_{\mu}[\psi]n^{\mu}_{\Sigma_0^*}$.

We control the term $I$ as follows (we will not lose anything by allowing the sum in $m$ and $\ell$ to be over all of $\mathbb{Z}\times \mathbb{Z}_{\geq |m|}$):
\begin{align}
\nonumber &\int_{-\infty}^{\infty}\sum_{m\ell}\left|I\right|\, d\omega
\\ \nonumber &=\int_{-\infty}^{\infty}\sum_{m\ell}\left|\int_0^{\pi}\int_0^{2\pi}\int_{-\infty}^{\infty}\int_{-\infty}^{\infty}e^{-i\left(\omega-\upomega_+m\right)r^*}e^{i\omega t}e^{-im\phi}S_{m\ell}(\theta,a\omega)\frac{\Delta\rho^2}{(r^2+a^2)^{3/2}}F\, \sin\theta\,d\theta\, d\phi\, dt\, dr^*\right|^2\, d\omega
\\
 \label{IOK2} &\leq \int_0^{\pi}\int_{-\infty}^{\infty}\sum_m\left|\int_{-\infty}^{\infty}\int_{\infty}^{\infty}\int_0^{2\pi}e^{-i\omega(r^*-t)}e^{im\left(\upomega_+r^*-\phi\right)}\frac{\Delta\rho^2}{(r^2+a^2)^{3/2}}F\, d\phi\, dt\, dr^*\right|^2\sin\theta\, d\omega\, d\theta.
\end{align}
For each fixed $m$ we have used the orthogonality of the $S_{m\ell}$ in the last inequality.

Now we introduce the variables $\tilde v \doteq t + r^*$ and $\tilde u \doteq t - r^*$ and keep in mind that $F$ is only supported in a compact range of $\tilde v$. Then~(\ref{IOK2}) becomes
\begin{align}\label{IOK3}
&\int_0^{\pi}\sum_m\int_{-\infty}^{\infty}\left|\int_{-\infty}^{\infty}\int_{\infty}^{\infty}\int_0^{2\pi}e^{i\omega \tilde u}e^{im\left(\upomega_+\frac{\tilde v-\tilde u}{2}-\phi\right)}\frac{\Delta\rho^2}{(r^2+a^2)^{3/2}}F\, d\phi \, d\tilde u\, d\tilde v\right|^2\sin\theta\, d\omega\, d\theta
\\ \nonumber &\qquad\leq B\int_0^{\pi}\int_{-\infty}^{\infty}\sum_m\int_{-\infty}^{\infty}\left|\int_{-\infty}^{\infty}\int_0^{2\pi}e^{i\omega \tilde u}e^{im\left(\upomega_+\frac{\tilde v-\tilde u}{2} - \phi\right)}\frac{\Delta\rho^2}{(r^2+a^2)^{3/2}}F\, d\phi\, d\tilde u\right|^2\sin\theta\, d\omega\, d\tilde v\, d\theta.
\\ \nonumber &\qquad= (2\pi)B\int_0^{\pi}\int_{-\infty}^{\infty}\int_{-\infty}^{\infty}\sum_m\left|\int_0^{2\pi}e^{im\left(\upomega_+\frac{\tilde v-\tilde u}{2} - \phi\right)}\frac{\Delta\rho^2}{(r^2+a^2)^{3/2}}F\, d\phi\right|^2\sin\theta\, d\tilde u\, d\tilde v\, d\theta
\\ \nonumber &\qquad= (2\pi)B\int_0^{\pi}\int_{-\infty}^{\infty}\int_{-\infty}^{\infty}\sum_m\left|\int_0^{2\pi}e^{-im \phi}\frac{\Delta\rho^2}{(r^2+a^2)^{3/2}}F\, d\phi\right|^2\sin\theta\, d\tilde u\, d\tilde v\, d\theta
\\ \nonumber &\qquad= (2\pi)^2B\int_{-\infty}^{\infty}\int_{-\infty}^{\infty}\int_{\mathbb{S}^2}\left|\frac{\Delta\rho^2}{(r^2+a^2)^{3/2}}F\right|^2\sin\theta \, d\tilde u\, d\tilde v\, d\theta\, d\phi
\\ \nonumber &\qquad\leq B\int_{\Sigma_0^*}\mathbf{J}^V_{\mu}[\psi]n^{\mu}_{\Sigma_0^*}.
\end{align}
We have used Plancherel in the $\omega$ variable and the orthogonality of the $e^{im\phi}$. In the last line we used finite in time energy estimates and the Hardy inequality $\int_{r_+}^{\infty}f^2\ dr \leq B\int_{r_+}^{\infty}(r-r_+)^2(\partial_rf)^2\ dr$, which holds for smooth functions $f$ which vanish for large $r$.

Putting together $(\ref{whatweworkedhardfor})$, $(\ref{IIOK})$, $(\ref{IOK2})$
and $(\ref{IOK3})$, we
 have indeed obtained $(\ref{tofinish})$. The theorem is thus proven.
\end{proof}

\section{The forward maps}\label{forward}
We now turn to our scattering theory proper.

The first order of business is to carefully set up the relevant spaces described in
Section~\ref{FROMTHEIN}
of the introduction. This will be accomplished in Section~\ref{FSsec} below.

We will then define in Section~\ref{DefANDbound}
the various forward maps $\mathscr{F}_+$
and infer their boundedness.
The boundedness of the map with domain $\mathcal{E}_{\Sigma^*_0}^N$
(Theorem~\ref{forwarddegen}) is independent of Section~\ref{secDegenBound}.
This will give {\bf Theorem~\ref{THEOREM1}} of Section~\ref{Nenergyforsec}.

The boundedness of  the degenerate-energy theory maps
with domain $\mathcal{E}^V_{\mathring\Sigma}$ and $\mathcal{E}^V_{\overline{\Sigma}}$
(Theorems~\ref{forwardt00} and~\ref{forwardt00bif}, respectively) indeed requires
the statement of Theorem~\ref{boundDegen} just proven.
This will give
{\bf Theorems~\ref{toinvert}} of Section~\ref{venergyforwardsec}.

\subsection{Function spaces}
\label{FSsec}
In this section we will define the function spaces for which we will formulate our scattering theory.

\subsubsection{Initial data on $\Sigma_0^*$, $\mathring{\Sigma}$ and $\overline{\Sigma}$}
Let us denote by
${}^2\mathcal{C}^{\infty}_{cp}(\Sigma_0^*)$, ${}^2\mathcal{C}^{\infty}_{cp}(\mathring\Sigma)$,
${}^2\mathcal{C}^{\infty}_{cp}({\overline\Sigma})$ the vector space of
smooth compactly supported pairs of functions $(\uppsi,\uppsi')$ defined
on $\Sigma_0^*$, $\mathring{\Sigma}$,
$\overline{\Sigma}$, respectively.
We will complete these vector spaces with respect
to appropriate norms to define the Hilbert spaces of
our scattering theory.

We start with the non-degenerate $N$-energy space. We shall only in fact consider this
for initial data on $\Sigma_0^*$.
\begin{definition}\label{nonDegNorm}For $(\uppsi,\uppsi')\in
{}^2\mathcal{C}^{\infty}_{cp}(\Sigma_0^*)$
we set
\[
\left\vert\left\vert (\uppsi,\uppsi')\right\vert\right\vert_{\mathcal{E}^N_{\Sigma_0^*}} \doteq \sqrt{\int_{\Sigma_0^*}\mathbf{J}^N_{\mu}[\Psi]n^{\mu}_{\Sigma_0^*}},
\]
where $\Psi$ is any extension of $\uppsi$ to $\mathcal{R}$ such that
$n_{\Sigma_0^*}\Psi = \uppsi'$.

The above expression gives a norm on the vector space ${}^2\mathcal{C}^{\infty}_{cp}(\Sigma_0^*)$,
and we  define the space
\[
(\mathcal{E}^N_{\Sigma_0^*}, \|\cdot \|_{\mathcal{E}^N_{\Sigma_0^*}})
\]
to be its completion.
\end{definition}

Next, we define the degenerate $V$-energy spaces
along $\Sigma_0^*$, $\mathring\Sigma$,
and $\overline\Sigma$, respectively.
\begin{definition}\label{degNorm}For $(\uppsi,\uppsi')\in
{}^2\mathcal{C}^{\infty}_{cp}(\Sigma_0^*)$,
${}^2\mathcal{C}^{\infty}_{cp}(\mathring\Sigma)$, and
${}^2\mathcal{C}^{\infty}_{cp}(\overline\Sigma)$,
respectively, we
 set
\[
\left\vert\left\vert (\uppsi,\uppsi')\right\vert\right\vert_{\mathcal{E}^V_{\Sigma_0^*}} \doteq \sqrt{\int_{\Sigma_0^*}\mathbf{J}^V_{\mu}[\Psi]n^{\mu}_{\Sigma_0^*}},\qquad
\left\vert\left\vert (\uppsi,\uppsi')\right\vert\right\vert_{\mathcal{E}^V_{\mathring{\Sigma}}} \doteq \sqrt{\int_{\mathring{\Sigma}}\mathbf{J}^V_{\mu}[\Psi]n^{\mu}_{\mathring{\Sigma}}},\qquad
\left\vert\left\vert (\uppsi,\uppsi')\right\vert\right\vert_{\mathcal{E}^V_{\overline{\Sigma}}} \doteq \sqrt{\int_{\overline{\Sigma}}\mathbf{J}^V_{\mu}[\Psi]n^{\mu}_{\overline{\Sigma}}},
\]
where $\Psi$ is any extension of $\uppsi$ to $\mathcal{D}$ such that
$n_{\Sigma_0^*}\Psi = \uppsi'$, $n_{\mathring\Sigma}\Psi=0$, $n_{\overline\Sigma}\Psi=0$ respectively.

The above expression give norms on the vector spaces
${}^2\mathcal{C}^{\infty}_{cp}(\Sigma_0^*)$,
${}^2\mathcal{C}^{\infty}_{cp}(\mathring\Sigma)$, and
${}^2\mathcal{C}^{\infty}_{cp}(\overline\Sigma)$, respectively, and we define
the spaces
\[
(\mathcal{E}^V_{\Sigma_0^*}, \|\cdot \|_{\mathcal{E}^V_{\Sigma_0^*}}),
\qquad
(\mathcal{E}^V_{\mathring\Sigma}, \|\cdot \|_{\mathcal{E}^V_{\mathring\Sigma}}),
\qquad
(\mathcal{E}^V_{\overline\Sigma}, \|\cdot \|_{\mathcal{E}^V_{\overline\Sigma}}),
\]
to be their respective completions.
\end{definition}

\begin{remark}\label{degen}Note that the energy density is pointwise degenerate because as $r \to r_+$ the vector field $V$ becomes null. An explicit calculation gives
\[
\mathbf{J}^V_{\mu}[\tilde \uppsi]n^{\mu}_{\Sigma_0^*} \sim \left|\partial_{t^*}\tilde \uppsi\right|^2 + \left(r - r_+\right)\left|Z^*\tilde \uppsi\right|^2 + \left|\nabb \tilde \uppsi\right|^2\text{ as }r\to r_+.
\]
This does not however affect the positive definitivity of the above norms, which
moreover are easily to seen to arise from a positive definite inner product.
Thus, $\mathcal{E}^N_{\Sigma_0^*}$, $\mathcal{E}^V_{\Sigma_0^*}$, $\mathcal{E}_{\mathring{\Sigma}}^V$, $\mathcal{E}^V_{\underline{\Sigma}}$ are all in fact Hilbert spaces.
Note moreover that both $\mathcal{E}_{\Sigma_0^*}^N$ and $\mathcal{E}^V_{\Sigma_0^*}$ may be identified with subsets of $L^2_{\rm loc}(\Sigma_0^*)\times L^2_{\rm loc}(\Sigma_0^*)$ and, after this identification is made, $\mathcal{E}_{\Sigma_0^*}^N$ is a proper subset of $\mathcal{E}^V_{\Sigma_0^*}$.

Finally, we note that one may easily check that a sufficient condition for a pair of smooth functions $(\uppsi,\uppsi')$ to lie in $\mathcal{E}^V_{\mathring{\Sigma}}$ is that $\| (\uppsi,\uppsi')\|_{\mathcal{E}^V_{\mathring{\Sigma}}} < \infty$ and $\lim_{r\to r_+}(\uppsi,\uppsi') = \lim_{r\to\infty}(\uppsi,\uppsi') = 0$.
\end{remark}

\subsubsection{Scattering data along $\mathcal{H}_{\geq 0}^+$, $\mathcal{H}^+$ and $\overline{\mathcal{H}^+}$}
We now carry out similar constructions for data along $\mathcal{H}^+_{\geq 0}$, $\mathcal{H}^+$ and $\overline{\mathcal{H}^+}$. Let us denote by
$\mathcal{C}^{\infty}_{cp}(\mathcal{H}^+_{\geq 0})$, $\mathcal{C}^{\infty}_{cp}(\mathcal{H}^+)$,
$\mathcal{C}^{\infty}_{cp}(\overline{\mathcal{H}^+})$ the vector space of
smooth compactly supported functions $\uppsi$ defined
on $\mathcal{H}^+_{\geq 0}$, $\mathcal{H}^+$,
$\overline{\mathcal{H}^+}$, respectively.

We start with the case of finite non-degenerate energy data along $\mathcal{H}^+_{\geq 0}$.
\begin{definition}For $\uppsi \in
\mathcal{C}^{\infty}_{cp}(\mathcal{H}^+_{\geq 0})$
we set
\[
\left\vert\left\vert \uppsi\right\vert\right\vert_{\mathcal{E}^N_{\mathcal{H}^+_{\geq 0}}} \doteq \sqrt{\int_{\mathcal{H}^+_{\geq 0}}\mathbf{J}^N_{\mu}[\uppsi]n^{\mu}_{\mathcal{H}^+}}.
\]

The above expression gives a norm on the vector space $\mathcal{C}^{\infty}_{cp}(\mathcal{H}^+_{\geq 0})$,
and we  define the space
\[
(\mathcal{E}^N_{\mathcal{H}^+_{\geq 0}}, \|\cdot \|_{\mathcal{E}^N_{\mathcal{H}^+_{\geq 0}}})
\]
to be its completion.
\end{definition}

Next, we define the $K$-energy spaces
along $\mathcal{H}^+_{\geq 0}$, $\mathcal{H}^+$,
and $\overline{\mathcal{H}^+}$, respectively.
\begin{definition}For $\uppsi\in
\mathcal{C}^{\infty}_{cp}(\mathcal{H}^+_{\geq 0})$,
$\mathcal{C}^{\infty}_{cp}(\mathcal{H}^+)$, and
$\mathcal{C}^{\infty}_{cp}(\overline{\mathcal{H}^+})$,
respectively, we
 set
\[
\left\vert\left\vert \uppsi \right\vert\right\vert_{\mathcal{E}^K_{\mathcal{H}^+_{\geq 0}}} \doteq \sqrt{\int_{\mathcal{\mathcal{H}^+_{\geq 0}}}\mathbf{J}^K_{\mu}[\uppsi]n^{\mu}_{\mathcal{H}^+_{\geq 0}}},\qquad
\left\vert\left\vert \uppsi \right\vert\right\vert_{\mathcal{E}^K_{\mathcal{H}^+}} \doteq \sqrt{\int_{\mathcal{H}^+}\mathbf{J}^K_{\mu}[\uppsi]n^{\mu}_{\mathcal{H}^+}},\qquad
\left\vert\left\vert \uppsi \right\vert\right\vert_{\mathcal{E}^K_{\overline{\mathcal{H}^+}}} \doteq \sqrt{\int_{\overline{\mathcal{H}^+}}\mathbf{J}^K_{\mu}[\uppsi]n^{\mu}_{\overline{\Sigma}}}.
\]

The above expression give norms on the vector spaces
$\mathcal{C}^{\infty}_{cp}(\mathcal{H}^+_{\geq 0})$,
$\mathcal{C}^{\infty}_{cp}(\mathcal{H}^+)$, and
$\mathcal{C}^{\infty}_{cp}(\overline{\mathcal{H}^+})$, respectively, and we define
the spaces
\[
(\mathcal{E}^K_{\mathcal{H}^+_{\geq 0}}, \|\cdot \|_{\mathcal{E}^K_{\mathcal{H}^+_{\geq 0}}}),
\qquad
(\mathcal{E}^K_{\mathcal{H}^+}, \|\cdot \|_{\mathcal{E}^K_{\mathcal{H}^+}}),
\qquad
(\mathcal{E}^K_{\overline{\mathcal{H}^+}}, \|\cdot \|_{\mathcal{E}^K_{\overline{\mathcal{H}^+}}}),
\]
to be their respective completions.
\end{definition}

\begin{remark}Note that the $K$-based energy densities are pointwise degenerate in that the norms do not control $\partial_{\theta*}\psi$ and $\partial_{\phi^*}\psi$. An explicit calculation gives
\[
\mathbf{J}^K_{\mu}[\uppsi]n^{\mu}_{\mathcal{H}^+} \sim \left|K\psi\right|^2.
\]
Again, this degeneration
does not however affect the positive definitivity of the above norms, which
moreover are again easily to seen to arise from a positive definite inner product. Thus, $\mathcal{E}^N_{\mathcal{H}^+_{\geq 0}}$, $\mathcal{E}^K_{\mathcal{H}^+_{\geq 0}}$, $\mathcal{E}_{\mathcal{H}^+}^K$, $\mathcal{E}^K_{\overline{\mathcal{H}^+}}$ are all in fact Hilbert spaces.
Note moreover that both $\mathcal{E}_{\mathcal{H}^+_{\geq 0}}^N$ and $\mathcal{E}^K_{\mathcal{H}^+_{\geq 0}}$ may be identified with subsets of $L^2_{\rm loc}(\mathcal{H}^+)$ and, after this identification is made, $\mathcal{E}_{\mathcal{H}^+_{\geq 0}}^N$ is a proper subset of $\mathcal{E}^K_{\mathcal{H}^+_{\geq 0}}$.
\end{remark}

\subsubsection{Scattering data along $\mathcal{I}^+$}
Finally, we turn to null infinity. Let us denote by $\mathcal{C}^{\infty}_{cp}(\mathcal{I}^+)$ the vector space of
smooth compactly supported functions $\upphi$ defined
on $\mathcal{I}^+$.

The space of finite energy data along $\mathcal{I}^+$ is then defined as follows.
\begin{definition}\label{nonDegNormHor}For $\upphi \in
\mathcal{C}^{\infty}_{cp}(\mathcal{I}^+)$
we set
\[
\left\vert\left\vert \upphi\right\vert\right\vert_{\mathcal{E}^T_{\mathcal{I}^+}} \doteq \sqrt{\int_{\mathcal{I}^+}\left|\partial_{\tau}\upphi\right|^2}.
\]

The above expression gives a norm on the vector space $\mathcal{C}^{\infty}_{cp}(\mathcal{I}^+)$,
and we  define the space
\[
(\mathcal{E}^T_{\mathcal{I}^+}, \|\cdot \|_{\mathcal{E}^T_{\mathcal{I}^+}})
\]
to be its completion.
\end{definition}

\begin{remark}Note that this energy density is pointwise degenerate in that it does not control $\partial_{\theta}\upphi$ and $\partial_{\phi}\upphi$.
As before, this does not however affect the positive definitivity of the above norms, which
moreover are easily to seen to arise from a positive definite inner product.
Thus, $\mathcal{E}^T_{\mathcal{I}^+}$ is in fact a Hilbert space.
\end{remark}

\subsection{Definition and boundedness of the forward maps}
\label{DefANDbound}
In this section we will define the  various forward maps from Cauchy data to scattering
data and infer their boundedness.
However, we first need the following corollary of Theorems~\ref{h.o.s.} and~\ref{theResultinf}.

\begin{corollary}\label{weakDecay}For all
solutions $\psi$ to~(\ref{WAVE}) on $\mathcal{R}_{\geq 0}$ arising from initial data in
${}^2\mathcal{C}^{\infty}_{cp}(\Sigma_0^*)$, we have that the radiation fields to $\mathcal{H}^+_{\geq 0}$ and $\mathcal{I}^+$ lie in the spaces $\mathcal{E}_{\mathcal{H}^+_{\geq 0}}^N$ and $\mathcal{E}_{\mathcal{I}^+}^T$ respectively.
\end{corollary}
\begin{proof}Given Theorems~\ref{theResult} and~\ref{theResultinf}, the only statement we need to check is that the radiation fields lie in the closure of compactly supported smooth functions.

In order to prove this, we start by giving a (standard) argument which upgrades Theorem~\ref{h.o.s.} to the statement that
\begin{equation}\label{thatsweakindeed}
\lim_{s\to\infty}\int_{\Sigma_s^*\cap \{r \in [r_+,R]\}}\mathbf{J}^N_{\mu}[\psi]n^{\mu}_{\Sigma_s^*} = 0\qquad \forall R > r_+.
\end{equation}

First we observe that the fundamental theorem of calculus and Theorem~\ref{h.o.s.} immediately imply the following Lipschitz property:
\begin{equation}\label{lip}
\left|\int_{\Sigma_{s_2}^*\cap \{r \in [r_+,R]\}}\mathbf{J}^N_{\mu}[\psi]n^{\mu}_{\Sigma_{s_2}^*} -
\int_{\Sigma_{s_1}^*\cap \{r \in [r_+,R]\}}\mathbf{J}^N_{\mu}[\psi]n^{\mu}_{\Sigma_{s_1}^*}\right| \leq B(\psi)\left|s_2-s_1\right|.
\end{equation}

Let $\epsilon > 0$. Using~(\ref{lip}) we may obtain
\begin{equation}\label{weakest}
\int_{\Sigma_s^*\cap \{r \in [r_+,R]\}}\mathbf{J}^N_{\mu}[\psi]n^{\mu}_{\Sigma_s^*} \leq B(\psi)\epsilon + \inf_{s'\in[s-\epsilon,s+\epsilon]}\int_{\Sigma_{s'}^*\cap \{r \in [r_+,R]\}}\mathbf{J}^N_{\mu}[\psi]n^{\mu}_{\Sigma_{s'}^*}.
\end{equation}
Of course, Theorem~\ref{h.o.s.} implies that
\[\lim_{s\to\infty}\inf_{s'\in[s-\epsilon,s+\epsilon]}\int_{\Sigma_{s'}^*\cap \{r \in [r_+,R]\}}\mathbf{J}^N_{\mu}[\psi]n^{\mu}_{\Sigma_{s'}^*} = 0.\]
Thus~(\ref{weakest}) implies that
\[\limsup_{s\to\infty}
\int_{\Sigma_s^*\cap \{r \in [r_+,R]\}}\mathbf{J}^N_{\mu}[\psi]n^{\mu}_{\Sigma_s^*} \leq B(\psi)\epsilon.\]
Since $\epsilon$ was arbitrary,~(\ref{thatsweakindeed}) follows.

Now, using Theorem~\ref{h.o.s.} we immediately obtain higher order versions of~(\ref{thatsweakindeed}). Sobolev inequalities then imply that
\begin{equation}\label{pointweak}
\lim_{s\to\infty}\sup_{\Sigma_s^*\cap\{r \in [r_+,R]\}}\left|\psi\right| = 0.
\end{equation}
In particular, we may conclude that the radiation field along the horizon $\mathcal{H}^+$ lies in the closure of compactly supported functions.

For the radiation field along null infinity, we recall that in the proof of Proposition~\ref{microEqualInf}, we proved that $\partial_{\tau}\varphi|_{r = r_0}$ converges as $r_0 \to \infty$ to the $\partial_{\tau}$ derivative of the radiation field in $L^2_{\mathbb{R}_{\geq \tau_0}\times \mathbb{S}^2}$ for some sufficiently negative $\tau_0$. For each $r_0$,~(\ref{pointweak}) implies that $\varphi|_{r = r_0}$ lies in the closure of smooth compactly supported functions; completeness thus implies that the radiation field along null infinity also lies in this closure.
\end{proof}

Similarly, we have the following two corollaries.
\begin{corollary}\label{weakDecay2}For all
solutions $\psi$ to~(\ref{WAVE}) on $\mathcal{R}$ arising from initial data in ${}^2\mathcal{C}^{\infty}_{cp}(\mathring{\Sigma})$, we have that the radiation fields to $\mathcal{H}^+$ and $\mathcal{I}^+$ lie in the spaces $\mathcal{E}_{\mathcal{H}^+}^V$ and $\mathcal{E}_{\mathcal{I}^+}^T$ respectively.
\end{corollary}
\begin{proof}It follows immediately from a $K$ energy estimate near the bifurcate sphere that the radiation field of $\psi$ along $\mathcal{H}^+$ vanishes for sufficiently negative $t^*$. Since a finite-in-time energy estimate implies that $(\psi|_{\Sigma_0^*},n_{\Sigma_0^*}\psi|_{\Sigma_0^*}) \in {}^2\mathcal{C}^{\infty}_{cp}(\Sigma_0^*)$, the rest of the proof may be concluded with an appeal to Corollary~\ref{weakDecay}.
\end{proof}
\begin{corollary}\label{weakDecay3}For all
solutions $\psi$ to~(\ref{WAVE}) on $\mathcal{R}$ arising from initial data in ${}^2\mathcal{C}^{\infty}_{cp}(\overline{\Sigma})$, we have that the radiation fields to $\overline{\mathcal{H}^+}$ and $\mathcal{I}^+$ lie in the spaces $\mathcal{E}_{\overline{\mathcal{H}^+}}^V$ and $\mathcal{E}_{\mathcal{I}^+}^T$ respectively.
\end{corollary}
\begin{proof}The proof is the same as the proof of Corollary~\ref{weakDecay2}.
\end{proof}

The above three corollaries allow us to make the following definition.
\begin{definition}We define the ``forward maps''
\[
\mathscr{F}_+ : {}^2\mathcal{C}^{\infty}_{cp}(\Sigma_0^*)\to \mathcal{E}_{\mathcal{H}^+_{\geq 0}}^N\oplus\mathcal{E}_{\mathcal{I}^+}^T,
\qquad
\mathscr{F}_+ : {}^2\mathcal{C}^{\infty}_{cp}(\mathring{\Sigma})\to \mathcal{E}_{\mathcal{H}^+}^K\oplus\mathcal{E}_{\mathcal{I}^+}^T,
\qquad
\mathscr{F}_+ : {}^2\mathcal{C}^{\infty}_{cp}(\overline{\Sigma})\to \mathcal{E}_{\overline{\mathcal{H}^+}}^K\oplus\mathcal{E}_{\mathcal{I}^+}^T,
\]
to be the maps
\begin{equation}
(\uppsi|_{\Sigma_0^*, \, \mathring\Sigma, {\rm\ or\ }\overline{\Sigma}},\, \uppsi'|_{\Sigma_0^*,\, \mathring\Sigma, {\rm\ or\ }\overline{\Sigma}})\mapsto \psi \mapsto (\uppsi|_{\mathcal{H}^+_{\ge 0}, \, \mathcal{H}^+,{\rm\ or\ } \overline{\mathcal{H}^+}} \doteq \psi|_{\mathcal{H}^+_{\ge 0}, \, \mathcal{H}^+,{\rm\ or\ } \overline{\mathcal{H}^+}},\,\upphi|_{\mathcal{I}^+}\doteq  r\psi|_{\mathcal{I}^+})
\end{equation}
which take smooth initial data in ${}^2\mathcal{C}^{\infty}_{cp}(\Sigma_0^*)$, ${}^2\mathcal{C}^{\infty}_{cp}(\mathring{\Sigma})$ or ${}^2\mathcal{C}^{\infty}_{cp}(\overline{\Sigma})$, solve the wave equation
to
the future and then compute the radiation fields along $\mathcal{H}^+_{\geq 0}$, $\mathcal{H}^+$ or $\overline{\mathcal{H}^+}$,
respectively, and $\mathcal{I}^+$.
\end{definition}

Theorem~\ref{theResultinf} now implies
\begin{theorem}
\label{boundNONDeg}
The forward map $\mathscr{F}_+$ uniquely extends by density to a bounded map
\[
\mathscr{F}_+ : \mathcal{E}_{\Sigma_0^*}^N \to \mathcal{E}_{\mathcal{H}^+_{\geq 0}}^N\oplus\mathcal{E}_{\mathcal{I}^+}^T.
\]
\end{theorem}
This gives {\bf Theorem~\ref{THEOREM1}} of Section~\ref{Nenergyforsec}.

Similarly, Theorem~\ref{boundDegen} implies the following theorem.
\begin{theorem}\label{forwarddegen}
The forward map $\mathscr{F}_+$ uniquely extends by density to a bounded map
\[\mathscr{F}_+ : \mathcal{E}^V_{\Sigma_0^*} \to \mathcal{E}^K_{\mathcal{H}^+_{\geq 0}}\oplus\mathcal{E}^T_{\mathcal{I}^+}.\]
\end{theorem}

Lastly, Corollaries~\ref{boundDegent0},~\ref{weakDecay2} and~\ref{weakDecay3} now imply the following two theorems.
\begin{theorem}\label{forwardt00}The forward map $\mathscr{F}_+$ uniquely extends by density to a bounded map
\[\mathscr{F}_+ : \mathcal{E}^V_{\mathring{\Sigma}} \to \mathcal{E}^K_{\mathcal{H}^+} \oplus \mathcal{E}^T_{\mathcal{I}^+}.\]
\end{theorem}
\begin{theorem}\label{forwardt00bif}The forward map $\mathscr{F}_+$ uniquely extends by density to a bounded map
\[\mathscr{F}_+ : \mathcal{E}^V_{\overline{\Sigma}} \to \mathcal{E}^K_{\overline{\mathcal{H}^+}} \oplus \mathcal{E}^T_{\mathcal{I}^+}.\]
\end{theorem}
We have obtained thus {\bf Theorem~\ref{toinvert}} of the Section~\ref{venergyforwardsec}.

\section{The backwards maps and the scattering matrix}\label{secScatter}
This section represents the heart of the paper. We will here
construct bounded maps
\begin{equation}
\label{three3inverses}
\mathscr{B}_- : \mathcal{E}^K_{\mathcal{H}^+_{\geq 0}} \oplus \mathcal{E}^T_{\mathcal{I}^+} \to \mathcal{E}^V_{\Sigma_0^*},
\qquad
\mathscr{B}_- : \mathcal{E}^K_{\mathcal{H}^+} \oplus \mathcal{E}^T_{\mathcal{I}^+} \to \mathcal{E}^V_{\mathring{\Sigma}},
\qquad
\mathscr{B}_- : \mathcal{E}^K_{\overline{\mathcal{H}^+}} \oplus \mathcal{E}^T_{\mathcal{I}^+} \to \mathcal{E}^V_{\overline{\Sigma}},
\end{equation}
which will invert the maps $\mathscr{F}_+$ from
Theorem~\ref{forwarddegen},~\ref{forwardt00} and~\ref{forwardt00bif} and then we shall construct the scattering maps
\begin{equation}
\label{HEREsca}
\mathscr{S} : \mathcal{E}^V_{\mathcal{H}^-} \oplus \mathcal{E}^T_{\mathcal{I}^-} \to \mathcal{E}^V_{\mathcal{H}^+} \oplus \mathcal{E}^T_{\mathcal{I}^+},
\qquad
\mathscr{S} : \mathcal{E}^V_{\overline{\mathcal{H}^-}} \oplus \mathcal{E}^T_{\mathcal{I}^-} \to \mathcal{E}^V_{\overline{\mathcal{H}^+}} \oplus \mathcal{E}^T_{\mathcal{I}^+}.
\end{equation}

It turns out that for technical reasons, it is easiest to first construct the middle
map of $(\ref{three3inverses})$ and show  that it is a two-sided inverse
of the corresponding forward map on $\mathcal{E}^V_{\mathring\Sigma}$.
This will be the content of Section~\ref{backt2oring} where
the main result is stated as Theorem~\ref{isot0}.
The remaining two backwards maps to $\Sigma_0^*$ and $\overline{\Sigma}$ will then
be easily constructed in Sections~\ref{MOREBACK1} and~\ref{MOREBACK2},
and these will be shown in Theorems~\ref{defScat} and~\ref{torevisitornottorevisit}
to be two-sided  inverses of the corresponding maps $\mathscr{F}_+$.
The above three theorems will give {\bf Theorem~\ref{existofsc}} of Section~\ref{vbackwardssec}.

The scattering maps $(\ref{HEREsca})$
and their boundedness will be deduced as Theorem~\ref{scatAll}
in Section~\ref{scatteringfrom} after
introducing the past-analogues $\mathscr{F}_-$ and $\mathscr{B}_+$
and inferring their boundedness (Theorem~\ref{scatPast}).
This will give {\bf Theorem~\ref{SMatrix}} of Section~\ref{EBSmat}.
We shall also represent $\mathscr{S}$ in the frequency domain by
Theorem~\ref{constructaway}, giving
the relationship between the fixed-frequency and physical space theories.
This will imply in particular {\bf Theorem~\ref{PSpREPint}} of Section~\ref{connectPST}.

Finally, this section contains two separate ``asides'', Sections~\ref{asideforthepr} and~\ref{veryselfcontained}, either of which can be skipped, but both
of which could have interest independent of the rest of the paper.
In Section~\ref{asideforthepr},
we will use the maps $(\ref{three3inverses})$
to complete the theory of boundedness and integrated
 decay for the degenerate $V$-energy by giving
 the proof of Theorem~\ref{degentheResult} from Section~\ref{yetanotheraside}.
In Section~\ref{veryselfcontained},
we will give an alternative,
self-contained discussion of the Schwarzschild $a=0$ case using
exclusively physical-space (i.e.~``time-dependent'') methods.

\subsection{The backwards map to $\mathring{\Sigma}$}
\label{backt2oring}
We begin by constructing the map $\mathscr{B}_- : \mathcal{E}^K_{\mathcal{H}^+} \oplus \mathcal{E}^T_{\mathcal{I}^+} \to \mathcal{E}^V_{\mathring{\Sigma}}$.

\subsubsection{A frequency-space definition of $\mathscr{B}_-$}
First, we define what will turn out to be essentially the Fourier transform of our backwards map. We begin by recalling the coefficients $\mathfrak{A}_{\mathcal{I}^{\pm}}$, $\mathfrak{A}_{\mathcal{H}^{\pm}}$ and the Wronskian $\mathfrak{W}$ from Definition~\ref{wronk} and~\ref{reflectransmit},
as well as Theorem~\ref{modestability} which states that $\mathfrak{W}$ never vanishes.

\begin{definition}\label{hatS}
For all smooth functions $a_{\mathcal{I}^+}(\omega,m,\ell)$ and $a_{\mathcal{H}^-}(\omega,m,\ell)$ which are only supported on a compact set of $(\omega,m,\ell)$, for all $(\omega,m,\ell)$ with $\omega \neq 0$ and $\omega \neq \upomega_+ m$, we define
\[\hat{\mathscr{B}}_-\left(a_{\mathcal{H}^+},a_{\mathcal{I}^+}\right)|_{(r,\omega,m,\ell)} \doteq \left(a_{\mathcal{I}^+}\overline{\mathfrak{W}^{-1}(2i\omega)}\right) \overline{U_{\rm hor}} + \left(a_{\mathcal{H}^+}\overline{\mathfrak{W}^{-1}(2i(\omega-\upomega_+m))}\right) \overline{U_{\rm inf}}.\]
\end{definition}

The next proposition explains the definition of $\hat{\mathscr{B}}_-$.
\begin{proposition}\label{hatSwellDef}For $\omega \neq 0$ and $\omega \neq \upomega_+m$, $\hat{\mathscr{B}}_-(a_{\mathcal{H}^+},a_{\mathcal{I}^+})$ is the unique solution $u$ to the radial o.d.e.~(\ref{e3iswsntouu}) with vanishing right hand side $H=0$ such that there exist complex
numbers $\alpha(\omega,m,\ell)$ and $\beta(\omega,m,\ell)$ satisfying
\begin{equation}\label{require1}
u = a_{\mathcal{H}^+}U_{\rm hor} + \alpha(\omega,m,\ell)\overline{U_{\rm hor}},
\end{equation}
\begin{equation}\label{require2}
u = a_{\mathcal{I}^+}U_{\rm inf} + \beta(\omega,m,\ell)\overline{U_{\rm inf}}.
\end{equation}
\end{proposition}
\begin{proof}We start with uniqueness. Suppose that we have two solutions $u$ and $\tilde u$ to the radial o.d.e.~(\ref{e3iswsntouu}) with a vanishing right hand side $H$ such that
\[
u = a_{\mathcal{H}^+}U_{\rm hor} + \alpha(\omega,m,\ell)\overline{U_{\rm hor}},
\qquad \tilde u = a_{\mathcal{H}^+}U_{\rm hor} + \tilde \alpha(\omega,m,\ell)\overline{U_{\rm hor}},
\]
\[
u = a_{\mathcal{I}^+}U_{\rm inf} + \beta(\omega,m,\ell)\overline{U_{\rm inf}},
\qquad
\tilde u = a_{\mathcal{I}^+}U_{\rm inf} + \tilde \beta(\omega,m,\ell)\overline{U_{\rm inf}}.
\]

Then, for each $(\omega,m,\ell)$ with $\omega \neq 0$ and $\omega \neq \upomega_+m$, $\overline{u-\tilde u}$ would be a solution the radial o.d.e.~(\ref{e3iswsntouu}) with a vanishing right hand side such that
\[\overline{u-\tilde u} \sim e^{i\omega r^*} \text{ as }r\to\infty,\]
\[\overline{u-\tilde u} \sim e^{-i\left(\omega-\upomega_+m\right)r^*}\text{ as }r \to r_+.\]

These asymptotic conditions imply that $\overline{u-\tilde u}$ corresponds to a ``mode solution'' (see Definition 1.1 of~\cite{realmodestability}), and Theorem 1.6 of~\cite{realmodestability} proves that there are no non-zero mode solutions.

To see that $\hat{\mathscr{B}}_-$ verifies~(\ref{require1}) and~(\ref{require2}), it suffices to recall the relations
\[\overline{U_{\rm hor}} = \overline{\mathfrak{A}_{\mathcal{I}^+}}\overline{U_{\rm inf}} + \overline{(2i\omega)^{-1}\mathfrak{W}}U_{\rm inf},\]
\[\overline{U_{\rm inf}} = \overline{\mathfrak{A}_{\mathcal{H}^+}}\overline{U_{\rm hor}} + \overline{(2i(\omega-\upomega_+m))^{-1}\mathfrak{W}}U_{\rm hor}.\]
\end{proof}

We now introduce a useful function space.
\begin{definition}\label{nohatSdef}Let $\check{\mathcal{C}}^{\infty}_{cp}$ denote the set of functions $f : \mathbb{R} \times \mathbb{S}^2 \to \mathbb{C}$ such that
\[\hat{f}\left(\omega,m,\ell\right) \doteq \frac{1}{\sqrt{2\pi}}\int_{-\infty}^{\infty}e^{i\omega t}e^{-im\phi}S_{m\ell}\left(a\omega,\theta\right)f\sin\theta\, dt\, d\theta\, d\phi\]
is smooth in $\omega$ and vanishes for $\left(\omega,m,\ell\right)$ outside a compact set of $\mathbb{R}\times \mathbb{Z} \times \mathbb{Z}_{\geq \left|m\right|}$.

Next, observing that $\check{\mathcal{C}}^{\infty}_{cp}$ may be naturally identified as a subset of either $L^2_{\rm loc}\left(\mathcal{H}^+\right)$ or $L^2_{\rm loc}\left(\mathcal{I}^+\right)$, we let $\check{\mathcal{C}}^{\infty}_{cp}(\mathcal{H}^+)$ be the result of identifying $\check{\mathcal{C}}^{\infty}_{cp}$ with a subset of $\mathcal{E}^K_{\mathcal{H}^+}$, and let $\check{\mathcal{C}}^{\infty}_{cp}(\mathcal{I}^+)$ be the result of identifying $\check{\mathcal{C}}^{\infty}_{cp}$ with a subset of $\mathcal{E}^T_{\mathcal{H}^+}$.
\end{definition}

\begin{remark}\label{theyaredense}
One may easily check that $\check{\mathcal{C}}^{\infty}_{cp}(\mathcal{H}^+)$ is dense in $\mathcal{E}^K_{\mathcal{H}^+}$ and that $\check{\mathcal{C}}^{\infty}_{cp}(\mathcal{I}^+)$ is dense in $\mathcal{E}^T_{\mathcal{H}^+}$.
\end{remark}

We now define the map $\mathscr{B}_-$ on the space $\check{\mathcal{C}}^{\infty}_{cp}(\mathcal{H}^+)\oplus \check{\mathcal{C}}^{\infty}_{cp}(\mathcal{I}^+)$.
\begin{definition}\label{truedefB}For all $(\uppsi,\upphi) \in \check{\mathcal{C}}^{\infty}_{cp}(\mathcal{H}^+)\oplus \check{\mathcal{C}}^{\infty}_{cp}(\mathcal{I}^+)$, we define the function $\mathscr{B}_-\left(\uppsi,\upphi\right) : \mathring{\mathcal{R}} \to \mathbb{C}$ by
\[\mathscr{B}_-\left(\uppsi,\upphi\right)|_{(t,r,\theta,\phi)} \doteq \frac{1}{(r^2+a^2)^{1/2}\sqrt{2\pi}}\int_{-\infty}^{\infty}\sum_{m\ell}e^{-it\omega}e^{im\phi}S_{m\ell}\left(a\omega,\theta\right)\hat{\mathscr{B}}_-\left(\hat{\uppsi},\hat{\upphi}\right)|_{(\omega,r,m,\ell)}\, d\omega.\]
Note that $\hat{\mathscr{B}}_-\left(\hat{\uppsi},\hat{\upphi}\right)$ vanishes
for all $(\omega,m,\ell)$ outside a compact set; it immediately follows that $\mathscr{B}_-\left(\uppsi,\upphi\right)$ is a smooth function of $(t,r,\theta,\phi)$.
\end{definition}

\subsubsection{Boundedness}\label{someboundedargument}
The following proposition will be used to show that the map $(\uppsi,\upphi) \mapsto \left(\mathscr{B}_-(\uppsi,\upphi)|_{\mathring{\Sigma}},n_{\mathring{\Sigma}}\mathscr{B}_-(\uppsi,\upphi)|_{\mathring{\Sigma}}\right)$ is bounded.
\begin{proposition}\label{basicallybounded}For all $(\uppsi,\upphi) \in \check{\mathcal{C}}^{\infty}_{cp}(\mathcal{H}^+)\oplus \check{\mathcal{C}}^{\infty}_{cp}(\mathcal{I}^+)$, we have
\[\int_{\mathring{\Sigma}}\mathbf{J}^V_{\mu}\left[\mathscr{B}_-(\uppsi,\upphi)\right]n^{\mu}_{\mathring{\Sigma}} \leq B\int_{-\infty}^{\infty}\int_{\mathbb{S}^2}\left[\left|\left(\partial_{\tau}+\upomega_+\partial_{\phi}\right)\uppsi\right|^2 + \left|\partial_{\tau}\upphi\right|^2\right]\sin\theta\, d\tau\, d\theta\, d\phi.\]
\end{proposition}

\begin{proof}
Set
\[
u \doteq \hat{\mathscr{B}}_-\left(\hat{\uppsi},\hat{\upphi}\right)|_{(\omega,r,m,\ell)},
\]
\[
\psi \doteq \mathscr{B}_-\left(\uppsi,\upphi\right)|_{(t,r,\theta,\phi)}.
\]

First of all, we observe that $\psi : \mathring{\mathcal{R}} \to \mathbb{C}$ is a smooth solution to $\Box_g\psi = 0$, is easily seen to be sufficiently integrable in the sense of Definition~\ref{sufficient}, and that applying Carter's separation to $\psi$ yields $u$.

Keeping the explicit formula for $\hat{\mathscr{B}}_-$ in mind, applying Theorem~\ref{scatterEst2} to $u$ implies that for each $-\infty < R^*_- < R^*_+ < \infty$ we have
\begin{align}\label{microILED}
(\omega-\upomega_+m)^2|a_{\mathcal{H}^-}|^2 + \omega^2|a_{\mathcal{I}^-}|^2 + &\int_{R^*_-}^{R_+^*}\left[\left| u'\right|^2 + \left(\left(1-r_{\rm trap}r^{-1}\right)^2(\omega^2 + \Lambda) + 1\right)\left|u\right|^2\right]\, dr^*  \\ \nonumber &\leq B(R^*_-,R^*_+)\left[(\omega-\upomega_+m)^2|\hat{\uppsi}|^2 + \omega^2|\hat{\upphi}|^2\right].
\end{align}

The rest of the proof will borrow some ideas from Section 13 of~\cite{partIII}. In order to work around the presence of the $\left(1-r_{\rm trap}r^{-1}\right)^2$ term in~(\ref{microILED}), it will be useful to decompose $\psi$ in pieces, each of which experience trapping near a specific value of $r$. We first define the following ranges of $(\omega,m,\ell)$:
\begin{definition}\label{decomp}Let $\epsilon > 0$ be a sufficiently small parameter to fixed later. We define
\[\mathscr{F}_0 \doteq \left\{\left(\omega,m,\ell\right) : r_{trap}=0
\right\},\]
\[\mathscr{F}_i\doteq \left\{\left(\omega,m,\ell\right) : r_{trap} \in \left[3M - s^- + \epsilon\left(i-1\right),3M-s^- +\epsilon i\right)\right\}\forall \quad i = 1,\ldots,\lceil \epsilon^{-1}\left(s^++s^-\right)\rceil.\]
\end{definition}

Observe that each value of $\left(\omega,m,\ell\right)$ lies in exactly one of the $\mathscr{F}_i$.

\begin{definition}We define $\psi_i$ by a phase space multiplication of $\psi$ by $1_{\mathscr{F}_i}$, the indicator function of $\mathscr{F}_i$:
\[\psi_i \doteq \frac{1}{(r^2+a^2)^{1/2}\sqrt{2\pi}}\int_{-\infty}^\infty\sum_{m\ell} e^{-i\omega t} 1_{\mathscr{F}_i}S_{m\ell}(a\omega,\cos\theta)e^{im\phi}\hat{\mathscr{B}}_-\left(\hat{\uppsi},\hat{\upphi}\right) d\omega.\]
\end{definition}
Note that each $\psi_i$ is a smooth function from $\mathring{\mathcal{R}}$ to $\mathbb{C}$, satisfies $\Box_g\psi_i = 0$ and is sufficiently integrable in the sense of Definition~\ref{sufficient}.

Next, keeping in mind that each $\hat{\psi}_i$ is compactly supported in $(\omega,m,\ell)$, Plancherel immediately implies that for each $r_+ < r_0 < r_1 < \infty$ we have
\begin{equation}\label{gettingreadyfordyadic}
\int_{-\infty}^{\infty}\int_{\{t = s\}\cap \{r \in [r_0,r_1]\}}\mathbf{J}^V_{\mu}[\psi_i]n^{\mu}_{\{t=s\}}\, ds < \infty.
\end{equation}

In particular, for each $r_+ < r_0 < r_1 < \infty$ and $i = 0,\ldots,\lceil \epsilon^{-1}\left(s^++s^-\right)\rceil$
there exists a constant $C_i(r_0,r_1)$ and a dyadic sequence $\{s^{(i)}_n\}_{n=1}^{\infty}$
such that $s^{(i)}_n \to \infty$ as $n\to\infty$ and
\begin{equation}\label{dyadicdecay}
\int_{\{t = s^{(i)}_n\} \cap \{r \in [r_0,r_1]\}}\mathbf{J}^N_{\mu}\left[\psi_i\right]n^{\mu}_{\{t = s^{(i)}_n\}} \leq \frac{C_i(r_0,r_1)}{s^{(i)}_n}.
\end{equation}

Next, taking $\epsilon$ from Definition~\ref{decomp} sufficiently small (and then fixing $\epsilon$), for each $r_i$ we appeal to Corollary~\ref{tempKill} and construct a $T$-invariant
timelike vector field $V_i$ on $\mathring{\mathcal{R}}$
which is Killing in the region
\[
r \in\left[3M - s^- + \left(i-1\right)\epsilon,3M-s^- + i\epsilon\right),
\]
and is equal to $V$ for $r$ sufficiently close to $r_+$ and $r$ sufficiently large.

Finally, we are ready for our main estimate. For each $r_+ < r_0 < r_1 < \infty$ such that $r_0 - r_+$ is sufficiently small and $r_1$ is sufficiently large, we apply the energy identity associated to $V_i$ in between the hypersurfaces $\mathring{\Sigma} \cap \{r \in [r_0,r_1]\}$, $\{r = r_0\} \cap J^+\left(\mathring{\Sigma}\right) \cap J^-\left(\left\{t=s^{(i)}_n\right\}\right)$, $\{r = r_1\} \cap J^+\left(\mathring{\Sigma}\right) \cap J^-\left(\left\{t=s^{(i)}_n\right\}\right)$, and $\{t = s^{(i)}_n\} \cap \{r \in [r_0,r_1]\}$. We obtain
\begin{align}\label{whatwegot2}
\int_{\mathring{\Sigma}\cap \{r \in [r_0,r_1]\}}&\mathbf{J}^{V_i}_{\mu}\left[\psi_i\right]n^{\mu}_{\mathring{\Sigma}}
\\ \nonumber &\leq B\int_0^{s^{(i)}_n}\int_{\{t=s\} \cap\, {\rm supp}\left(\mathbf{K}^{V_i}\right)}\mathbf{J}^{V_i}_{\mu}[\psi_i]n^{\mu}_{\{t=s\}}\, ds + \int_{\{t = s^{(i)}_n\}\cap \{r \in [r_0,r_1]\}}\mathbf{J}^{V_i}_{\mu}\left[\psi_i\right]n^{\mu}_{\{t = s^{(i)}_n\}}
\\ \nonumber &+B\left(\int_{\{r=r_0\}} + \int_{\{r=r_1\}}\right)\left|V\left(\left(r^2+a^2\right)^{1/2}\psi_i\right)\partial_{r^*}\left(\left(r^2+a^2\right)^{1/2}\psi_i\right)\right|\, \sin\theta\, dt\, d\theta\, d\phi,
\end{align}
where we have used the calculation~(\ref{r0energy}) and that fact that
\[\left|\mathbf{K}^{V_i}\right| \leq B\mathbf{J}^{V_i},\]
where we recall that $\mathbf{K}^{V_i} = \text{}^{(V_i)}\pi_{\alpha\beta}\mathbf{T}^{\alpha\beta}$ and $\text{}^{(V_i)}\pi_{\alpha\beta}$ denotes the deformation tensor of $V_i$.

Taking $n\to \infty$ and appealing to~(\ref{dyadicdecay}) then yields
\begin{align}\label{whatwegot3}
\int_{\mathring{\Sigma}\cap \{r \in [r_0,r_1]\}}&\mathbf{J}^{V_i}_{\mu}\left[\psi_i\right]n^{\mu}_{\mathring{\Sigma}}
\\ \nonumber &\leq B\int_0^{\infty}\int_{\{t=s\} \cap\, {\rm supp}\left(\mathbf{K}^{V_i}\right)}\mathbf{J}^{V_i}_{\mu}[\psi_i]n^{\mu}_{\{t=s\}}\, ds
\\ \nonumber &+ B\left(\int_{\{r=r_0\}} + \int_{\{r=r_1\}}\right)\left|V\left(\left(r^2+a^2\right)^{1/2}\psi_i\right)\partial_{r^*}\left(\left(r^2+a^2\right)^{1/2}\psi_i\right)\right|\, \sin\theta\, dt\, d\theta\, d\phi,
\end{align}

Next, since $\left(1-r_{\rm trap}r^{-1}\right)^2\left|\hat{\psi}_i\right|^2 \sim \left|\hat{\psi}_i\right|^2$ for values of $r$ in the support of $\mathbf{K}^{(V_i)}$, we observe that applying Plancherel and~(\ref{microILED}) yields
\begin{align}\label{anEst15}
\int_0^{\infty}&\int_{\{t=s\} \cap\, {\rm supp}\left(\mathbf{K}^{V_i}\right)}\mathbf{J}^{V_i}_{\mu}[\psi_i]n^{\mu}_{\{t=s\}}\, ds
\\ \nonumber &\leq \int_{-\infty}^{\infty}\int_{\{t=s\} \cap\, {\rm supp}\left(\mathbf{K}^{V_i}\right)}\left|\mathbf{J}^{V_i}_{\mu}[\psi_i]n^{\mu}_{\{t=s\}}\right|\, ds
\\ \nonumber &\leq B\int_{-\infty}^{\infty}\sum_{(\omega,m,\ell)\in \mathscr{F}_i}\int_{r_{\rm min}^*}^{r_{\rm max}^*}\left[\left|u'\right|^2 + \left(\left(1-r_{\rm trap}r^{-1}\right)^2(\omega^2 + \Lambda) + 1\right)\left|u\right|^2\right]\, dr^*\, d\omega
\\ \nonumber &\leq B\int_{-\infty}^{\infty}\sum_{m\ell}\left[(\omega-\upomega_+m)^2|\hat{\uppsi}|^2 + \omega^2|\hat{\upphi}|^2\right]\, d\omega
\\ \nonumber &\leq B\int_{-\infty}^{\infty}\int_{\mathbb{S}^2}\left[\left|\left(\partial_{\tau}+\upomega_+\partial_{\phi}\right)\uppsi\right|^2 + \left|\partial_{\tau}\upphi\right|^2\right]\sin\theta\, d\tau\, d\theta\, d\phi.
\end{align}

We conclude that
\begin{align}\label{whatwegot4}
\int_{\mathring{\Sigma}}&\mathbf{J}^{V_i}_{\mu}\left[\psi_i\right]n^{\mu}_{\mathring{\Sigma}} \leq
\\ \nonumber &B\int_{-\infty}^{\infty}\int_{\mathbb{S}^2}\left[\left|\left(\partial_{\tau}+\upomega_+\partial_{\phi}\right)\uppsi\right|^2 + \left|\partial_{\tau}\upphi\right|^2\right]\sin\theta\, d\tau\, d\theta\, d\phi
\\ \nonumber &+B\liminf_{r_0 \to r_+}\liminf_{r_1\to\infty}\left(\int_{\{r=r_0\}} + \int_{\{r=r_1\}}\right)\left|V\left(\left(r^2+a^2\right)^{1/2}\psi_i\right)\partial_{r^*}\left(\left(r^2+a^2\right)^{1/2}\psi_i\right)\right|\, \sin\theta\, dt\, d\theta\, d\phi.
\end{align}

It immediately follows from Proposition~\ref{smoothenough}, the compact support of $\hat{\psi}_i$ in $(\omega,m,\ell)$, and~(\ref{microILED}) that
\begin{align}\label{anEst16}
\liminf_{r_0\to r_+}\int_{r = r_0}&\left|V\left(\left(r^2+a^2\right)^{1/2}\psi_i\right)\partial_{r^*}\left(\left(r^2+a^2\right)^{1/2}\psi_i\right)\right|\, \sin\theta\, dt\, d\theta\, d\phi
\\ \nonumber &= \liminf_{r_0\to r_+}\int_{r = r_0}\left|K\left(\left(r^2+a^2\right)^{1/2}\psi_i\right)\partial_{r^*}\left(\left(r^2+a^2\right)^{1/2}\psi_i\right)\right|\, \sin\theta\, dt\, d\theta\, d\phi
\\ \nonumber &\leq \liminf_{r_0\to r_+}B\int_{-\infty}^{\infty}\sum_{m\ell}\left|\left(\omega-\upomega_+m\right)uu'\right|\, d\omega
\\ \nonumber &\leq B\int_{-\infty}^{\infty}\sum_{m\ell}(\omega-\upomega_+m)^2\left[|a_{\mathcal{H}^+}|^2+\left|a_{\mathcal{H}^-}\right|^2\right]\, d\omega
\\ \nonumber &\leq B\int_{-\infty}^{\infty}\sum_{m\ell}\left[(\omega-\upomega_+m)^2|\hat{\uppsi}|^2 + \omega^2|\hat{\upphi}|^2\right]\, d\omega
\\ \nonumber &\leq B\int_{-\infty}^{\infty}\int_{\mathbb{S}^2}\left[\left|\left(\partial_{\tau}+\upomega_+\partial_{\phi}\right)\uppsi\right|^2 + \left|\partial_{\tau}\upphi\right|^2\right]\sin\theta\, d\tau\,
d\theta\, d\phi.
\end{align}
\begin{remark}Note that the passing of the limit through the integral and sum that implicitly occurs between lines $3$ and $4$ is justified by Proposition~\ref{smoothenough} and the compact support of $\hat{\psi}_i$ in $(\omega,m,\ell)$.
\end{remark}

Similarly, it immediately follows from Proposition~\ref{microradconverge}, Proposition~\ref{microrp}, the compact support of $\hat{\psi}_i$, and~(\ref{microILED}) that
\begin{align}\label{anEst17}
\liminf_{r_1\to \infty}\int_{r = r_1}&\left|V\left(\left(r^2+a^2\right)^{1/2}\psi_i\right)\partial_{r^*}\left(\left(r^2+a^2\right)^{1/2}\psi_i\right)\right|\, \sin\theta\, dt\, d\theta\, d\phi
\\ \nonumber &= \liminf_{r_1\to \infty}\int_{r = r_1}\left|T\left(\left(r^2+a^2\right)^{1/2}\psi_i\right)\partial_{r^*}\left(\left(r^2+a^2\right)^{1/2}\psi_i\right)\right|\, \sin\theta\, dt\, d\theta\, d\phi
\\ \nonumber &\leq \liminf_{r_1\to \infty}B\int_{-\infty}^{\infty}\sum_{m\ell}\left|\omega uu'\right|
\\ \nonumber &\leq B\int_{-\infty}^{\infty}\sum_{m\ell}\omega^2\left[|a_{\mathcal{I}^+}|^2+\left|a_{\mathcal{I}^-}\right|^2\right]\, d\omega
\\ \nonumber &\leq B\int_{-\infty}^{\infty}\sum_{m\ell}\left[(\omega-\upomega_+m)^2|\hat{\uppsi}|^2 + \omega^2|\hat{\upphi}|^2\right]\, d\omega
\\ \nonumber &\leq B\int_{-\infty}^{\infty}\int_{\mathbb{S}^2}\left[\left|\left(\partial_{\tau}+\upomega_+\partial_{\phi}\right)\uppsi\right|^2 + \left|\partial_{\tau}\upphi\right|^2\right]\sin\theta\, d\tau\, d\theta\, d\phi.
\end{align}
Combining~(\ref{whatwegot4}),~(\ref{anEst16}), and~(\ref{anEst17}) yields
\begin{equation}\label{whatwegot5}
\int_{\mathring{\Sigma}}\mathbf{J}^{V_i}_{\mu}\left[\psi_i\right]n^{\mu}_{\mathring{\Sigma}} \leq B\int_{-\infty}^{\infty}\int_{\mathbb{S}^2}\left[\left|\left(\partial_{\tau}+\upomega_+\partial_{\phi}\right)\uppsi\right|^2 + \left|\partial_{\tau}\upphi\right|^2\right]\sin\theta\, d\tau\, d\theta\, d\phi.
\end{equation}
We conclude the proof with the (trivial) observation that
\[\int_{\mathring{\Sigma}}\mathbf{J}^V_{\mu}\left[\psi\right]n^{\mu}_{\mathring{\Sigma}} \leq B\sum_{i=1}^{\lceil \epsilon^{-1}\left(s^++s^-\right)\rceil}
\int_{\{t=0\}}\mathbf{J}^{V_i}_{\mu}\left[\psi_i\right]n^{\mu}_{\{t=0\}}.\]
\end{proof}

The following proposition will be used to show that the range of $\left(\mathscr{B}_-|_{\mathring{\Sigma}},n_{\mathring{\Sigma}}\mathscr{B}_-|_{\mathring{\Sigma}}\right)$ lies in $\mathcal{E}^V_{\mathring{\Sigma}}$.
\begin{proposition}\label{vanishbifurcate}For all $(\uppsi,\upphi) \in \check{\mathcal{C}}^{\infty}_{cp}(\mathcal{H}^+)\oplus \check{\mathcal{C}}^{\infty}_{cp}(\mathcal{I}^+)$, we have
\[\lim_{r\to r_+}\sup_{\mathbb{S}^2}\left|\mathscr{B}_-\left(\uppsi,\upphi\right)|_{t=0}\right| = 0,\]
\[\lim_{r\to \infty}\sup_{\mathbb{S}^2}\left|\mathscr{B}_-\left(\uppsi,\upphi\right)|_{t=0}\right| = 0.\]
\end{proposition}
\begin{proof}We start with the limit as $r \to r_+$. Since $\hat{\mathscr{B}}_-\left(\hat{\uppsi},\hat{\upphi}\right)$ is compactly supported in $(\omega,m,\ell)$, and $U_{\rm hor}$ and $\mathfrak{W}$ are smooth for $\omega \in \mathbb{R}\setminus \{0\}$, one may easily establish that for every $\delta > 0$
\[{\rm min}\left(\left|\omega\right|,\left|\omega-\upomega_+m\right|\right) \geq \delta \Rightarrow \hat{\mathscr{B}}_-\left(\hat{\uppsi},\hat{\upphi}\right) = a_{\mathcal{H}^+}e^{-i\left(\omega-\upomega_+m\right)r^*} + a_{\mathcal{H}^-}e^{i\left(\omega-\upomega_+m\right)r^*} + {\rm Error},\]
where
\[\left|{\rm Error}\right| \leq B(\delta,\uppsi,\upphi)(r-r_+).\]

Let $\chi(x)$ be a cutoff function which is identically $1$ in a neighborhood of $0$ and identically $0$ for $\left|x\right| > 1$. For every $\delta > 0$ we have
\begin{align}\label{0est}
\limsup_{r\to r_+}&\left|\left(r^2+a^2\right)^{1/2}\sqrt{2\pi}\mathscr{B}_-|_{(0,r,\theta,\phi)}\right|
\\ \nonumber &= \limsup_{r\to r_+}\left|\int_{-\infty}^{\infty}\sum_{m\ell}e^{im\phi}S_{m\ell}\left(a\omega,\theta\right)\hat{\mathscr{B}}_-\, d\omega\right|.
\\ \nonumber &\leq \limsup_{r\to r_+}\left|\int_{-\infty}^{\infty}\sum_{m\ell}\chi\left(\omega\delta^{-1}\right)\chi\left((\omega-\upomega_+m)\delta^{-1}\right)e^{im\phi}S_{m\ell}\left(a\omega,\theta\right)\hat{\mathscr{B}}_-\, d\omega\right|
\\ \nonumber &\qquad +\limsup_{r\to r_+}\left|\int_{-\infty}^{\infty}\left(1-\chi\left(\omega\delta^{-1}\right)\right)\left(1-\chi\left((\omega-\upomega_+m)\delta^{-1}\right)\right)\sum_{m\ell}e^{im\phi}S_{m\ell}\left(a\omega,\theta\right)\hat{\mathscr{B}}_-\, d\omega\right|.
\end{align}

We estimate the first term simply with Cauchy-Schwarz and~(\ref{microILED}):
\begin{align}\label{1est}
\limsup_{r\to r_+}&\left|\int_{-\infty}^{\infty}\sum_{m\ell}\chi\left(\omega\delta^{-1}\right)\chi\left((\omega-\upomega_+m)\delta^{-1}\right)e^{im\phi}S_{m\ell}\left(a\omega,\theta\right)\hat{\mathscr{B}}_-\, d\omega\right| \\ \nonumber &\leq B\left(\uppsi,\upphi\right)\limsup_{r\to r_+}\delta^{1/2}\sqrt{\int_{-\infty}^{\infty}\sum_{m\ell}\left|\hat{\mathscr{B}}_-\right|^2\, d\omega}
\\ \nonumber &\leq B\left(\uppsi,\upphi\right)\delta^{1/2}\sqrt{\int_{-\infty}^{\infty}\sum_{m\ell}\left[\left|a_{\mathcal{H}^+}\right|^2 + \left|a_{\mathcal{H}^-}\right|^2\right]}
\\ \nonumber & \leq B\left(\uppsi,\upphi\right)\delta^{1/2}.
\end{align}

Set $\tilde\chi_{\delta} \doteq \left(1-\chi\left(\omega\delta^{-1}\right)\right)\left(1-\chi\left((\omega-\upomega_+m)\delta^{-1}\right)\right)$. For the second term we use the oscillation in $r^*$:
\begin{align}\label{2est}
\limsup_{r\to r_+}&\left|\int_{-\infty}^{\infty}\sum_{m\ell}e^{im\phi}\tilde\chi_{\delta}S_{m\ell}\left(a\omega,\theta\right)\hat{\mathscr{B}}_-\, d\omega\right|
\\ \nonumber &\leq B\left(\uppsi,\upphi,\delta\right)\limsup_{r\to r_+}\left|\int_{-\infty}^{\infty}\sum_{m\ell}\tilde\chi_{\delta}e^{im\phi}S_{m\ell}\left(a_{\mathcal{H}^+}e^{-i\left(\omega-\upomega_+m\right)r^*} + a_{\mathcal{H}^-}e^{i\left(\omega-\upomega_+m\right)r^*}\right)\, d\omega\right|
\\ \nonumber &\leq B\left(\uppsi,\upphi,\delta\right)\limsup_{r\to r_+}\delta^{-1}\left|r^*\right|^{-1}
\\ \nonumber &= 0.
\end{align}
In the second to last line, the decay in $r^*$ came from an integration by parts in $\omega$.

Since $\delta$ may be taken arbitrary small, combining~(\ref{0est}),~(\ref{1est}), and~(\ref{2est}) concludes the proof for the limit when $r \to r_+$. Moreover, it is easy to see that essentially the same proof works for the limit as $r \to \infty$.
\end{proof}

The previous two propositions and Remark~\ref{theyaredense} immediately imply the following corollary.
\begin{corollary}\label{allScatBound}The map $\left(\mathscr{B}_-|_{\mathring{\Sigma}},n_{\mathring{\Sigma}}\mathscr{B}_-|_{\mathring{\Sigma}}\right)$, which we shall, by a mild abuse of notation, now denote by $\mathscr{B}_-$, extends by density to a bounded map
\[\mathscr{B}_- : \mathcal{E}^K_{\mathcal{H}^+} \oplus \mathcal{E}^T_{\mathcal{I}^+} \to \mathcal{E}^V_{\mathring{\Sigma}}.\]
\end{corollary}
\begin{proof}The key point is that a straightforward calculation (remember that $V$ vanishes at the bifurcate sphere!) shows that $\lim_{r\to r_+}\left(\mathscr{B}_-|_{\mathring{\Sigma}},n_{\mathring{\Sigma}}\mathscr{B}_-|_{\mathring{\Sigma}}\right) = 0$ and $\lim_{r\to \infty}\left(\mathscr{B}_-|_{\mathring{\Sigma}},n_{\mathring{\Sigma}}\mathscr{B}_-|_{\mathring{\Sigma}}\right) = 0$ imply that $\left(\mathscr{B}_-|_{\mathring{\Sigma}},n_{\mathring{\Sigma}}\mathscr{B}_-|_{\mathring{\Sigma}}\right)$ lies in $\mathcal{E}^V_{\mathring{\Sigma}}$.
\end{proof}

\subsubsection{Inverting the forward map}
Finally, we are ready for the key result of the section.
\begin{theorem}\label{isot0}
Let $\mathscr{B}_-$ and $\mathscr{F}_+$ be as in Corollary~\ref{allScatBound} and Theorem~\ref{forwardt00}. Then $\mathscr{B}_-$ and $\mathscr{F}_+$ are both bounded isomorphisms and satisfy $\mathscr{B}_-\circ \mathscr{F}_+ = {\rm Id}$ and $\mathscr{F}_+ \circ\mathscr{B}_- = {\rm Id}$.
\end{theorem}
\begin{proof}Of course, it suffices to prove the assertions $\mathscr{B}_-\circ \mathscr{F}_+ = {\rm Id}$ and $\mathscr{F}_+ \circ\mathscr{B}_- = {\rm Id}$.

We start with establishing $\mathscr{F}_+ \circ\mathscr{B}_- = {\rm Id}$. By density
(Remark~\ref{theyaredense}), it suffices to check that
\[
\left(\mathscr{F}_+ \circ\mathscr{B}_-\right)|_{\check{\mathcal{C}}^{\infty}_{cp}(\mathcal{H}^+)\oplus\check{\mathcal{C}}^{\infty}_{cp}(\mathcal{I}^+)} = {\rm Id}.
\]
Let $(\uppsi,\upphi) \in \check{\mathcal{C}}^{\infty}_{cp}(\mathcal{H}^+) \oplus \check{\mathcal{C}}^{\infty}_{cp}(\mathcal{I}^+)$. Proposition~\ref{hatSwellDef} implies that there exists functions $\alpha(\omega,m,\ell)$ and $\beta(\omega,m,\ell)$ such that
\[\hat{\mathscr{B}}_-\left(\hat{\uppsi},\hat{\upphi}\right) = \hat{\uppsi}U_{\rm hor} + \alpha \overline{U_{\rm hor}},\]
\[\hat{\mathscr{B}}_-\left(\hat{\uppsi},\hat{\upphi}\right) = \hat{\upphi}U_{\rm inf} + \beta \overline{U_{\rm inf}}.\]
Now, using that $(\mathscr{B}_-(\uppsi,\upphi)|_{\mathring{\Sigma}},n_{\mathring{\Sigma}}\mathscr{B}_-(\uppsi,\upphi)|_{\mathring{\Sigma}})$ lies $\mathcal{E}^V_{\mathring{\Sigma}}$, and Theorem~\ref{boundDegent0}, one may easily check that the same arguments used in the proofs of Propositions~\ref{microEqualInf} and ~\ref{microequalhor} immediately imply
\[
\left(\mathscr{F}_+\circ \mathscr{B}_-\right)(\uppsi,\upphi)  = (\uppsi,\upphi) .
\]

We now turn to establishing $\mathscr{B}_- \circ\mathscr{F}_+ = {\rm Id}$. By density, it suffices to study solutions arising from initial data $(\uppsi,\uppsi')\in {}^2\mathcal{C}^{\infty}_{cp}(\mathring{\Sigma})$. Let $a_{\mathcal{H}^+}$ and $a_{\mathcal{I}^+}$ denote the microlocal radiation fields. Then Proposition~\ref{microEqualInf} and~\ref{microequalhor} yield
\[\mathscr{F}_+\left(\uppsi,\uppsi'\right) = \left(\frac{1}{\sqrt{4M\pi r_+ }}\int_{-\infty}^{\infty}\sum_{m\ell}e^{-it^*\omega}e^{im\phi}S_{m\ell}a_{\mathcal{H}^+}\, d\omega,\ \frac{1}{\sqrt{2\pi}}\int_{-\infty}^{\infty}\sum_{m\ell}e^{-i\tau\omega}e^{im\phi}S_{m\ell}a_{\mathcal{I}^+}\, d\omega\right).\]
It immediately follows from Proposition~\ref{hatSwellDef} that $\left(\mathscr{B}_-\circ\mathscr{F}_+\right)\left(\uppsi,\uppsi'\right) = \left(\uppsi,\uppsi'\right)$.
\end{proof}

\subsubsection{A physical-space characterization of $\mathscr{B}_-$}\label{constbackphys}
Before we close the section it will be conceptually clarifying and technically useful to observe that the backwards map may also be characterized in physical space.
\begin{proposition}\label{backwardinthephysicalspace}Let $\left(\uppsi_{\mathcal{H}^+},\upphi_{\mathcal{I}^+}\right) \in \mathcal{C}^{\infty}_{cp}(\mathcal{H}^+)\oplus \mathcal{C}^{\infty}_{cp}(\mathcal{I}^+)$. Pick $\tau_0 < \infty$ such that $\uppsi_{\mathcal{H}^+}$ is compactly supported in $\mathcal{H}^+(-\infty,\tau_0)$ and $\upphi_{\mathcal{I}^+}$ is compactly supported in $\mathcal{I}^+_{\tau_0}$, and then let $\Phi_{\mathcal{I}^+}$ be any smooth extension of $\upphi_{\mathcal{I}^+}$ to the manifold with boundary $\tilde{\mathcal{R}}$ (see Definition~\ref{defNullInf}) such that $\Phi_{\mathcal{I}^+}$ vanishes in neighborhood of $S_{\tau_0}$.

Next, using Proposition~\ref{phybackwp}, for each $s > 0$ sufficiently large we may uniquely define a smooth solution $\psi_s$ to~(\ref{WAVE}) in the past of $\overline{\mathcal{H}^+_{\leq \tau_0}} \cup \left(S_{\tau_0} \cap \{r \leq r(\tau_0,s)\}\right) \cup \left(\{t = s\} \cap \{r \geq r(\tau_0,s)\}\right)$ by requiring
\[\psi_s|_{\overline{\mathcal{H}^+_{\leq \tau_0}}} = \uppsi_{\mathcal{H}^+},\]
\[(\psi_s|_{S_{\tau} \cap \{r \leq r(\tau,s)\}},n_{S_{\tau}}\psi_s|_{S_{\tau} \cap \{r \leq r(\tau,s)\}}) =(0,0),\]
\[r\psi_s|_{\{t = s\} \cap \{r \geq r(\tau,s)\}} = \Phi_{\mathcal{I}^+}|_{\{t = s\} \cap \{r \geq r(\tau,s)\}}.\]

Let $\{s_i\}_{i=1}^{\infty}$ be a sequence satisfying $s_i \to \infty$ as $i \to \infty$. It follows
 that $\psi_{s_i}|_{J^+(\mathring{\Sigma})}$  and any finite number of derivatives form a bounded equicontinuous sequence. In particular, we may extract a smooth limit $\psi$ which will be a solution to~(\ref{WAVE}) in the region $D^-\left(S_{\tau_0}\right) \cap J^+(\mathring{\Sigma})$. Finally, we have
\begin{equation}\label{theyareequalandthatsconvenient}
\mathscr{B}_-\left(\uppsi_{\mathcal{H}^+},\upphi_{\mathcal{I}^+}\right) = (\psi|_{\mathring{\Sigma}},n_{\mathring{\Sigma}}\psi_{\mathring{\Sigma}}).
\end{equation}
\end{proposition}

\begin{proof}The boundedness and equicontinuity of any finite number of derivatives of $\{\psi_{s_i}\}$ follows immediately from (higher order) $\mathbf{J}^V$ energy estimates (it may be useful for the reader to note that the intersection of $D^-\left(S_{\tau_0}\right) \cap J^+(\mathring{\Sigma})$ and the support of ${}^{(V)}\pi$ is compact and contained in $\cup_{s \in [0,\tau_0]}S_s$).

Next, using Theorem~\ref{isot0}, we note that~(\ref{theyareequalandthatsconvenient}) would follow from
\begin{equation}\label{theyareequalandthatsconvenient2}
\mathscr{F}_+(\psi|_{\mathring{\Sigma}},n_{\mathring{\Sigma}}\psi_{\mathring{\Sigma}}) = \left(\uppsi_{\mathcal{H}^+},\upphi_{\mathcal{I}^+}\right).
\end{equation}

Now, the equality $\mathscr{F}_+(\psi|_{\mathring{\Sigma}},n_{\mathring{\Sigma}}\psi_{\mathring{\Sigma}})|_{\mathcal{H}^+} = \uppsi_{\mathcal{H}^+}$ is a trivial consequence of the definition of the radiation field and Proposition~\ref{t0wellposed}.

Finally, the equality $\mathscr{F}_+(\psi|_{\mathring{\Sigma}},n_{\mathring{\Sigma}}\psi_{\mathring{\Sigma}})|_{\mathcal{I}^+} = \upphi_{\mathcal{I}^+}$ follows from Proposition~\ref{rp} and a straightforward modification of the arguments given in the proof of Proposition~\ref{radinfwelldef}.
\end{proof}

\subsection{The backwards map to $\Sigma_0^*$}\label{MOREBACK1}
In this section we will define the  backwards
map $\mathscr{B}_-$ on $\mathcal{E}^K_{\mathcal{H}^+_{\geq 0}}\oplus \mathcal{E}^T_{\mathcal{I}^+}$.

\begin{definition}\label{truedefScat}
Let $\mathscr{E} : \mathcal{C}^{\infty}_{cp}(\mathcal{H}^+_{\geq 0}) \to \mathcal{C}^{\infty}_{cp}(\mathcal{H}^+)$ be any map satisfying
\begin{enumerate}
    \item $\mathscr{E}\left(f\right)|_{\mathcal{H}^+_{\geq 0}} = f$.
    \item $\int_{\mathcal{H}^+}\mathbf{J}^K_{\mu}\left[\mathscr{E}\left(f\right)\right]n^{\mu}_{\mathcal{H}^+} \leq B\int_{\mathcal{H}^+_{\geq 0}}\mathbf{J}^K_{\mu}\left[f\right]n^{\mu}_{\mathcal{H}^+}$.
\end{enumerate}
Note that such a map is easily constructed.

Then we define the backwards map
\[\mathscr{B}_- : \mathcal{C}^{\infty}_{cp}(\mathcal{H}^+_{\geq 0}) \oplus \mathcal{C}^{\infty}_{cp}(\mathcal{I}^+) \to \mathcal{E}^V_{\Sigma_0^*},\]
by
\begin{equation}\label{Bisonbothsides}
\mathscr{B}_-\left(\uppsi_{\mathcal{H}^+_{\geq 0}},\upphi_{\mathcal{I}^+}\right) \doteq \left(\mathscr{B}_-\left(\mathscr{E}\left(\uppsi_{\mathcal{H}^+_{\geq 0}}\right),\upphi_{\mathcal{I}^+}\right)|_{\Sigma_0^*}, n_{\Sigma_0^*}\mathscr{B}_-\left(\mathscr{E}\left(\uppsi_{\mathcal{H}^+_{\geq 0}}\right),\upphi_{\mathcal{I}^+}\right)|_{\Sigma_0^*}\right).
\end{equation}
The reader should keep in mind our standard recycling of the notation concerning the symbol $\mathscr{B}_-$. In particular, $\mathscr{B}_-$ on the right hand side of~(\ref{Bisonbothsides}) is as in Definition~\ref{truedefB}.
\end{definition}

The next theorem establishes that the backwards map extends to $\mathcal{E}^K_{\mathcal{H}^+_{\geq 0}} \oplus \mathcal{E}^T_{\mathcal{I}^+}$ and inverts the forward map $\mathscr{F}_+$.
\begin{theorem}\label{defScat}
The map $\mathscr{B}_-$ defined above is a bounded map and thus uniquely extends to a map
\[
\mathscr{B}_- : \mathcal{E}^K_{\mathcal{H}^+_{\geq 0}} \oplus \mathcal{E}^T_{\mathcal{I}^+} \to \mathcal{E}^V_{\Sigma_0^*}.
\]
Let $\mathscr{F}_+$ denote the forward map $\mathscr{F}_+ : \mathcal{E}^V_{\Sigma_0^*} \to \mathcal{E}^K_{\overline{\mathcal{H}^+}} \oplus \mathcal{E}^T_{\mathcal{I}^+}$. Then, $\mathscr{B}_-\circ \mathscr{F}_+ = {\rm Id}$ and $\mathscr{F}_+\circ \mathscr{B}_- = {\rm Id}$ and thus $\mathscr{B}_-$ and $\mathscr{F}_+$ are bounded isomorphisms.
\end{theorem}
\begin{remark}Observe that one corollary of Theorem~\ref{defScat} is that $\mathscr{B}_-$ does not depend on the choice of extension $\mathscr{E}$.
\end{remark}
\begin{proof}
First of all, we observe that the boundedness of $\mathscr{B}_-$ and the statement $\mathscr{F}_+\circ \mathscr{B}_- = {\rm Id}$ follow immediately from Theorem~\ref{isot0}, Proposition~\ref{mixedwellposed} and finite in time energy estimates (cf. the proof of Corollary~\ref{boundDegent0}).

The equality $\mathscr{B}_-\circ \mathscr{F}_+ = {\rm Id}$ is a bit more subtle. The key observation is that it suffices to check this on a dense subject and it pays to expend a little effort in creating a convenient one. We thus turn to the construction of a useful dense subset. First of all, $\mathcal{C}^{\infty}_{cp}(\mathcal{H}^+) \oplus \mathcal{C}^{\infty}_{cp}(\mathcal{I}^+)$ is a dense subset of $\mathcal{E}^K_{\mathcal{H}^+} \oplus \mathcal{E}^T_{\mathcal{I}^+}$, and thus Theorem~\ref{isot0} implies that $\mathscr{B}_-\left(\mathcal{C}^{\infty}_{cp}(\mathcal{H}^+) \oplus \mathcal{C}^{\infty}_{cp}(\mathcal{I}^+)\right)$ is a dense subset of $\mathcal{E}^V_{\mathring{\Sigma}}$. Now, considering the elements of $\mathscr{B}_-\left(\mathcal{C}^{\infty}_{cp}(\mathcal{H}^+) \oplus \mathcal{C}^{\infty}_{cp}(\mathcal{I}^+)\right)$ as Cauchy data along $\mathring{\Sigma}$, we may solve the wave equation to the future of $\mathring{\Sigma}$ with Proposition~\ref{t0wellposed} and restrict the solutions to $\Sigma_0^*$. This defines a subset $\tilde{\mathcal{C}}_{\Sigma_0^*}$ of $\mathcal{E}^V_{\Sigma_0^*}$. It follows from Proposition~\ref{phybackwp} and finite in time energy estimates (cf. the proof of Corollary~\ref{boundDegent0}) that $\tilde{\mathcal{C}}_{\Sigma_0^*}$ is in fact a dense subset of $\mathcal{E}^V_{\Sigma_0^*}$.

We now turn to proving that $\mathscr{B}_-\circ \mathscr{F}_+|_{\tilde{\mathcal{C}}_{\Sigma_0^*}} = {\rm Id}$. Let $\psi$ be a solution to~(\ref{WAVE}) in $\mathcal{R}_{\geq 0}$ whose
initial data along $\Sigma_0^*$ lie in $\tilde{\mathcal{C}}_{\Sigma_0^*}$. We then define a solution $\tilde\psi$ to~(\ref{WAVE}) in $\mathcal{R}_{\geq 0}$ by applying Proposition~\ref{WPProp} to solve the wave equation with initial data $\left(\mathscr{B}_-\circ\mathscr{F}_+\right)\left(\psi|_{\Sigma_0^*},n_{\Sigma_0^*}\psi|_{\Sigma_0^*}\right)$ along $\Sigma_0^*$.
We need to prove that $\psi - \tilde\psi = 0$. Now, the key advantage to considering initial data in $\tilde{\mathcal{C}}_{\Sigma_0^*}$ is that it immediately follows from Proposition~\ref{backwardinthephysicalspace} that $\psi|_{\Sigma_0^*}$, $n_{\Sigma_0^*}\psi$, $\tilde\psi|_{\Sigma_0^*}$ and $n_{\Sigma_0^*}\tilde\psi|_{\Sigma_0^*}$ are smooth functions and hence that $\psi$ and $\tilde\psi$ extend smoothly to $\mathcal{H}^+_{\geq 0}$. Since $\mathscr{F}_+\circ \mathscr{B}_- = {\rm Id}$ we conclude in particular that $K\left(\psi-\tilde\psi\right)|_{\mathcal{H}^+_{\geq 0}} = 0$.

Set $\psi^{\dagger} \doteq K\left(\psi-\tilde\psi\right)$. Since the Cauchy data for $\psi^{\dagger}$ along $\Sigma_0^*$ vanishes at $\mathcal{H}^+$, we may easily construct a sequence $\{\psi^{\dagger}_i\}_{i=1}^{\infty}$ of solutions to~(\ref{WAVE}) whose initial data along $\Sigma_0^*$ are smooth and compactly supported away from $\mathcal{H}^+ \cap \Sigma_0^*$ and spacelike infinity and which satisfy
\[\lim_{i\to\infty}\int_{\Sigma_0^*}\mathbf{J}^V_{\mu}\left[\tilde\psi - \tilde\psi_i\right]n^{\mu}_{\Sigma_0^*} = 0.\]
Since the $\psi^{\dagger}_i$ are compactly supported away from $\mathcal{H}^+\cap \Sigma_0^*$, they may easily be extended as solutions to~(\ref{WAVE}) to all of $\mathcal{R}$ by applying Proposition~\ref{mixedwellposed} with vanishing initial data along $\overline{\mathcal{H}^+_{\leq 0}}$. We will also denote the extension by $\psi^{\dagger}_i$. Since $\psi^{\dagger}_i$ is easily seen to be sufficiently integrable in the sense of Definition~\ref{sufficient}, we may apply Carter's separation to $\psi^{\dagger}_i$ to define $u^{\dagger}_i$ and the corresponding microlocal fluxes $a^{\dagger}_{i,\mathcal{H}^+}$ and $a^{\dagger}_{i,\mathcal{I}^+}$. It follows immediately from the construction of $\psi^{\dagger}_i$, Corollary~\ref{boundDegent0}, Proposition~\ref{microEqualInf} and Proposition~\ref{microequalhor} that
\begin{equation}\label{gettingthere}
\lim_{i\to\infty}\int_{-\infty}^{\infty}\sum_{m\ell}\left[\left|a^{\dagger}_{i,\mathcal{H}^+}\right|^2 +\left|a^{\dagger}_{i,\mathcal{I}^+}\right|^2\right] = 0.
\end{equation}
Proposition~\ref{basicallybounded} (and an easy density argument) then imply that
\[\lim_{i\to\infty}\int\mathbf{J}^V_{\mu}[\psi^{\dagger}_i]n^{\mu}_{\{t=0\}} = 0.\]
Then, finite in time energy estimates show that $\psi^{\dagger}$ vanishes. Finally, using also~(\ref{thatsweakindeed}) from the proof of Corollary~\ref{weakDecay}, we conclude that $(\psi-\tilde\psi)|_{\mathcal{R}_{\geq 0}}$ vanishes.

\end{proof}

\subsection{The backwards map to $\overline{\Sigma}$}\label{MOREBACK2}

With Theorem~\ref{defScat} proven, we can now revisit scattering to $\overline{\Sigma}$ and prove a version of Theorem~\ref{isot0} where $\mathcal{E}^V_{\mathring{\Sigma}}$ is replaced by $\mathcal{E}^V_{\overline{\Sigma}}$ and $\mathcal{E}^K_{\mathcal{H}^+}$ is replaced with $\mathcal{E}^K_{\overline{\mathcal{H}^+}}$.

\begin{theorem}\label{torevisitornottorevisit}
Let $\mathscr{F}_+$ be the forward map $\mathscr{F}_+ : \mathcal{E}^V_{\overline{\Sigma}} \to \mathcal{E}^K_{\overline{\mathcal{H}^+}} \oplus \mathcal{E}^T_{\mathcal{I}^+}$.
Then there exists a backwards map
\[\mathscr{B}_- : \mathcal{E}^K_{\overline{\mathcal{H}^+}} \oplus \mathcal{E}^T_{\mathcal{I}^+} \to \mathcal{E}^V_{\overline{\Sigma}}\]
such that $\mathscr{B}_-$ is a bounded map, $\mathscr{B}_-\circ \mathscr{F}_+ = {\rm Id}$ and $\mathscr{F}_+\circ \mathscr{B}_- = {\rm Id}$. Thus $\mathscr{B}_-$ and $\mathscr{F}_+$ are both bounded isomorphisms.

\end{theorem}

\begin{proof}
We begin by introducing the notation
\[
\overline{\mathcal{H}^+_{\leq 0}} \doteq \overline{\mathcal{H}^+} \cap J^-(\Sigma_0^*).
\]
Then, we define the function space ${}^3\mathcal{C}^{\infty}_{cp}(\overline{\mathcal{H}^+_{\leq 0}} \cup \Sigma_0^*)$ to consist of triples $(\uppsi_{\overline{\mathcal{H}^+_{\leq 0}}},\uppsi_{\Sigma_0^*},\uppsi'_{\Sigma_0^*})$ such that $\uppsi_{\overline{\mathcal{H}^+_{\leq 0}}}$ is a smooth function on $\overline{\mathcal{H}^+_{\leq 0}}$, $\uppsi_{\Sigma_0^*}$ and $\uppsi'_{\Sigma_0^*}$ are smooth functions of compact support on $\Sigma_0^*$ and there exists a smooth function $\tilde\Psi$ on $\mathcal{D}$ such that $\tilde\Psi|_{\overline{\mathcal{H}^+_{\leq 0}}} = \uppsi_{\overline{\mathcal{H}^+_{\leq 0}}}$, $\tilde\Psi|_{\Sigma_0^*} = \uppsi_{\Sigma_0^*}$ and $n_{\Sigma_0^*}\tilde\Psi|_{\Sigma_0^*} = \uppsi'|_{\Sigma_0^*}$.

Proposition~\ref{mixedwellposed} states that to each $(\uppsi_{\overline{\mathcal{H}^+_{\leq 0}}},\uppsi_{\Sigma_0^*},\uppsi'_{\Sigma_0^*}) \in {}^3\mathcal{C}^{\infty}_{cp}(\overline{\mathcal{H}^+_{\leq 0}} \cup \Sigma_0^*)$ there exists a unique smooth solutions $\psi$ to~(\ref{WAVE}) in $J^-(\Sigma_0^*)$. Restricting these solutions to $\overline{\Sigma}$ thus defines a map
\begin{equation}\label{finitemixed}
{}^3\mathcal{C}^{\infty}_{cp}(\overline{\mathcal{H}^+_{\leq 0}} \cup \Sigma_0^*) \mapsto {}^2\mathcal{C}^{\infty}_{cp}(\overline{\Sigma}).
\end{equation}

Conversely, given any element of $(\uppsi,\uppsi') \in {}^2\mathcal{C}^{\infty}_{cp}(\overline{\Sigma})$, Proposition~\ref{t0wellposed} yields a unique solution to~(\ref{WAVE}) whose Cauchy data along $\overline{\Sigma}$ are given by $(\uppsi,\uppsi')$. Restricting these solutions to $\overline{\mathcal{H}^+_{\leq 0}} \cup \Sigma_0^*$ defines a map
\begin{equation}\label{forwardmixed}
{}^2\mathcal{C}^{\infty}_{cp}(\overline{\Sigma}) \mapsto {}^3\mathcal{C}^{\infty}_{cp}(\overline{\mathcal{H}^+_{\leq 0}} \cup \Sigma_0^*).
\end{equation}
It immediately follows that the maps~(\ref{finitemixed}) and~(\ref{forwardmixed}) are inverses of each other and hence that both are bijections.

Next, we let $\mathcal{E}^V_{\overline{\mathcal{H}^+_{\leq 0}}\cup\Sigma_0^*}$ denote the completion of $\mathcal{C}^{\infty}_{cp}(\overline{\mathcal{H}^+_{\leq 0}}\cup\Sigma_0^*)$ under the norm
\[\left\vert\left\vert (\uppsi_{\overline{\mathcal{H}^+_{\leq 0}}},\uppsi_{\Sigma_0^*},\uppsi'_{\Sigma_0^*})\right\vert\right\vert_{\mathcal{E}^V_{\overline{\mathcal{H}^+_{\leq 0}}\cup\Sigma_0^*}} \doteq \sqrt{\int_{\overline{\mathcal{H}^+_{\leq 0}}}\mathbf{J}^K_{\mu}[\tilde\Psi]n^{\mu}_{\mathcal{H}^+} + \int_{\Sigma_0^*}\mathbf{J}^V_{\mu}[\tilde\Psi]n^{\mu}_{\Sigma_0^*}},\]
where $\tilde\Psi$ is the smooth extension mentioned in the definition of the space $\mathcal{C}^{\infty}_{cp}(\overline{\mathcal{H}^+_{\leq 0}}\cup\Sigma_0^*)$.
Finite in time $\mathbf{J}^V$ energy estimates and the bijection~(\ref{forwardmixed}) immediately yield a bounded isomorphism
\begin{equation}\label{boundisomixed}
\mathcal{E}^V_{\overline{\Sigma}} \mapsto \mathcal{E}^V_{\overline{\mathcal{H}^+_{\leq 0}} \cup \Sigma_0^*}.
\end{equation}

We conclude the proof by combining~(\ref{boundisomixed}) with the easily observed fact that Theorem~\ref{defScat} implies that forward evolution yields a bounded isomorphism
\[
\mathcal{E}^V_{\overline{\mathcal{H}^+_{\leq 0}} \cup \Sigma_0^*} \mapsto \mathcal{E}^K_{\overline{\mathcal{H}^+}} \oplus \mathcal{E}^T_{\mathcal{I}^+}.
\]
\end{proof}

\subsection{Aside: Proof of Theorem~\ref{degentheResult}}
\label{asideforthepr}

At this point, using the properties of the backwards map $\mathscr{B}_-$,
we can now complete our study of boundedness
and integrated local energy decay for the degenerate $V$-energy theory
by giving the proof of Theorem~\ref{degentheResult} of Section~\ref{yetanotheraside}.

\begin{proof}[Proof of Theorem~\ref{degentheResult}]
Observe that for any $s \geq 0$, we could have defined a forward map $\mathscr{F}^{(s)}_+ : \mathcal{E}^V_{\Sigma_s^*} \to \mathcal{E}^K_{\mathcal{H}^+_{\geq s}} \oplus \mathcal{E}^T_{\mathcal{I}^+}$ which, in the case of smooth compactly supported data, computes the radiation field of Cauchy data along $\Sigma_s^*$ and, similarly, we could have defined a backwards map $\mathscr{B}^{(s)}_- : \mathcal{E}^K_{\mathcal{H}^+_{\geq s}} \oplus \mathcal{E}^T_{\mathcal{I}^+} \to  \mathcal{E}^V_{\Sigma_s^*}$.  Just as before, we would obtain that $\mathscr{F}^{(s)}_+$ and $\mathscr{B}^{(s)}_-$ are both bounded (with a constant \underline{independent of $s$})
and inverses of each other. In particular, since $\mathscr{F}^{(s)}_+(\psi|_{\Sigma_s^*},n_{\Sigma_s^*}\psi|_{\Sigma_s^*}) = \mathscr{F}_+(\psi|_{\Sigma_0^*},n_{\Sigma_0^*}\psi|_{\Sigma_0^*})$, we obtain
\begin{align}\label{blahjblahg}
\int_{\Sigma_s^*}\mathbf{J}^V_{\mu}[\psi]n^{\mu}_{\Sigma_s^*} &\leq B\left\vert\left\vert\mathscr{F}^{(s)}_+(\psi|_{\Sigma_s^*},n_{\Sigma_s^*}\psi|_{\Sigma_s^*})\right\vert\right\vert_{\mathcal{E}^K_{\mathcal{H}^+_{\geq s}\oplus\mathcal{E}^T_{\mathcal{I}^+}}}
\\ \nonumber &\leq B\left\vert\left\vert\mathscr{F}_+(\psi|_{\Sigma_0^*},n_{\Sigma_0^*}\psi|_{\Sigma_0^*})\right\vert\right\vert_{\mathcal{E}^K_{\mathcal{H}^+_{\geq 0}}\oplus\mathcal{E}^T_{\mathcal{I}^+}}
\\ \nonumber &\leq  B\int_{\Sigma_0^*}\mathbf{J}^V_{\mu}[\psi]n^{\mu}_{\Sigma_0^*}
\end{align}

Next, we observe that during the proof Proposition~\ref{basicallybounded} an integrated estimate in $r$ is in fact established. Using this, an easy density argument, a finite in time energy estimate and an application of Plancherel easily show that for any compact set $K \subset (r_+,\infty)$ we have
\begin{align}\label{blahjjj}
\int_{\mathcal{R}_{\geq 0} \cap \{r \in K\}}\Big(\zeta |\nabb\psi|^2+\zeta \left|T\psi\right|^2+\left|\tilde Z^*\psi\right|^2+ \left|\psi\right|^2\Big) &\leq B(K)\left\vert\left\vert\mathscr{F}_+(\psi|_{\Sigma_0^*},n_{\Sigma_0^*}\psi|_{\Sigma_0^*})\right\vert\right\vert_{\mathcal{E}^K_{\mathcal{H}^+_{\geq 0}}\oplus\mathcal{E}^T_{\mathcal{I}^+}}
\\ \nonumber &\leq B(K)\int_{\Sigma_0^*}\mathbf{J}^V_{\mu}[\psi]n^{\mu}_{\Sigma_0^*}.
\end{align}

In order to finish the proof we need to exchange the restriction $\{r \in K\}$ in~(\ref{blahjjj}) for the appropriate weights in $r$ and $r-r_+$. For $r$ large, the desired estimate is a trivial consequence of the ``large-$r$ estimate'' of Proposition 4.6.1 in \cite{partIII} and the arguments of Section 9.4 in \cite{partIII}. For $r$ close to the horizon it is possible to apply a degenerate version of the redshift effect~\cite{degenred} to achieve the desired estimate.
\end{proof}

\subsection{The scattering matrix $\mathscr{S}=\mathscr{F}_+\circ \mathscr{B}_+$}
\label{scatteringfrom}
For notational convenience, we have so far restricted our attention to scattering data along $\mathcal{H}^+$ and $\mathcal{I}^+$. However, in view of the discrete isometry
$(\ref{discreteIso})$
of $\mathcal{D}$, all of our theorems have exact analogues where $\mathcal{H}^+$ is replaced by $\mathcal{H}^-$ and $\mathcal{I}^+$ is replaced by $\mathcal{I}^-$. In particular, we have the following version of Theorems~\ref{isot0} and~\ref{torevisitornottorevisit}.

\begin{theorem}\label{scatPast}
Forward evolution (towards the past)
uniquely extends to the bounded maps
\[
\mathscr{F}_- : \mathcal{E}^V_{\mathring{\Sigma}} \to \mathcal{E}^K_{\mathcal{H}^-} \oplus \mathcal{E}^T_{\mathcal{I}^-}, \qquad
\mathscr{F}_- : \mathcal{E}^V_{\overline{\Sigma}} \to \mathcal{E}^K_{\overline{\mathcal{H}^-}} \oplus \mathcal{E}^T_{\mathcal{I}^-}.
\]
There exist bounded maps $\mathscr{B}_+$
\[
\mathscr{B}_+ : \mathcal{E}^K_{\mathcal{H}^-} \oplus \mathcal{E}^T_{\mathcal{I}^-} \to  \mathcal{E}^V_{\mathring{\Sigma}} ,
\qquad
\mathscr{B}_+ : \mathcal{E}^K_{\overline{\mathcal{H}^-}} \oplus \mathcal{E}^T_{\mathcal{I}^-} \to  \mathcal{E}^V_{\overline{\Sigma}} ,
\]
such that $\mathscr{F}_-\circ \mathscr{B}_+ = {\rm Id}$ and $\mathscr{B}_+\circ\mathscr{F}_- = {\rm Id}$.
\end{theorem}

Combining Theorems~\ref{isot0} and~\ref{scatPast} allows
us to define the maps between scattering data along $\mathcal{H}^-\cup\mathcal{I}^-$ and $\mathcal{H}^+ \cup \mathcal{I}^+$. We immediately obtain the following theorem.

\begin{theorem}\label{scatAll}
We define the {\bf scattering map} (or {\bf $S$-matrix})
\begin{equation*}
\mathscr{S} : \mathcal{E}^K_{\mathcal{H}^-} \oplus \mathcal{E}^T_{\mathcal{I}^-} \to \mathcal{E}^K_{\mathcal{H}^+} \oplus \mathcal{E}^T_{\mathcal{I}^+},
\qquad
\mathscr{S} : \mathcal{E}^K_{\overline{\mathcal{H}^-}} \oplus \mathcal{E}^T_{\mathcal{I}^-} \to \mathcal{E}^K_{\overline{\mathcal{H}^+}} \oplus \mathcal{E}^T_{\mathcal{I}^+},
\end{equation*}
by
\begin{equation}\label{themapwhichscattersreallywell}
\mathscr{S} \doteq \mathscr{F}_+\circ \mathscr{B}_+.
\end{equation}
The map $\mathscr{S}$ is then a bounded isomorphism from
$\mathcal{E}^K_{\mathcal{H}^-} \oplus \mathcal{E}^T_{\mathcal{I}^-}$ to
$\mathcal{E}^K_{\mathcal{H}^+} \oplus \mathcal{E}^T_{\mathcal{I}^+}$ and $\mathcal{E}^K_{\overline{\mathcal{H}^-}} \oplus \mathcal{E}^T_{\mathcal{I}^-}$ to
$\mathcal{E}^K_{\overline{\mathcal{H}^+}} \oplus \mathcal{E}^T_{\mathcal{I}^+}$

Furthermore, for every $(\uppsi_{\mathcal{H}^-},\upphi_{\mathcal{I}^-}) \in \mathcal{E}^K_{\mathcal{H}^-} \oplus \mathcal{E}^T_{\mathcal{I}^-}$, there exists a unique set of initial data $(\uppsi,\uppsi') \in \mathcal{E}^V_{\mathring{\Sigma}}$ such that $\mathscr{F}_-\left(\uppsi,\uppsi'\right) = (\uppsi_{\mathcal{H}^-},\upphi_{\mathcal{H}^-})$ and $\mathscr{F}_+\left(\uppsi,\uppsi'\right) = \mathscr{S}\left(\uppsi_{\mathcal{H}^-},\upphi_{\mathcal{I}^-}\right)$. An analogous statement holds for $(\uppsi_{\overline{\mathcal{H}^-}},\upphi_{\mathcal{I}^-}) \in \mathcal{E}^K_{\overline{\mathcal{H}^-}} \oplus \mathcal{E}^T_{\mathcal{I}^-}$
\end{theorem}
This is the precise statement of~{\bf Theorem~\ref{SMatrix}} of Section~\ref{EBSmat}.

Next, we observe that our scattering map $\mathscr{S}$ may be given by an explicit formula involving involving the reflection and transmission coefficients.
\begin{theorem}\label{constructaway}Let $\mathscr{S}$ denote the scattering map from Theorem~\ref{scatAll}. Then, for $\left(\uppsi_{\mathcal{H}^-},\upphi_{\mathcal{I}^-}\right)$ lying in either domain $\mathcal{E}^K_{\mathcal{H}^-} \oplus \mathcal{E}^T_{\mathcal{I}^-}$ or $\mathcal{E}^K_{\overline{\mathcal{H}^-}} \oplus \mathcal{E}^T_{\mathcal{I}^-}$, we have
\begin{align}\label{anexpressionofscattering}
\mathscr{S}&\left(\uppsi_{\mathcal{H}^-},\upphi_{\mathcal{I}^-}\right)
\\ \nonumber &= \Bigg(\frac{1}{\sqrt{4M\pi r_+}}\int_{-\infty}^{\infty}\sum_{m\ell}\left(-\left(\frac{\omega}{\omega-\upomega_+}\right)a_{\mathcal{I}^-}\mathfrak{T} + a_{\mathcal{H}^-}\tilde{\mathfrak{R}} \right)e^{-it\omega}e^{im\phi}S_{m\ell}\ d\omega,
\\ \nonumber &\ \ \ \ \ \frac{1}{\sqrt{2\pi}}\int_{-\infty}^{\infty}\sum_{m\ell}\left(-\left(\frac{\omega-\upomega_+m}{\omega}\right)a_{\mathcal{H}^-}\tilde{\mathfrak{T}} + a_{\mathcal{I}^-}\mathfrak{R} \right)e^{-it\omega}e^{im\phi}S_{m\ell}\ d\omega \Bigg).
\end{align}
Here
\[-i\left(\omega-\upomega_+m\right)a_{\mathcal{H}^-} \doteq \sqrt{\frac{Mr_+}{\pi}}\int_{-\infty}^{\infty}\int_{\mathbb{S}^2}\left(\partial_t+\upomega_+\partial_{\phi}\right)\uppsi_{\mathcal{H}^-}e^{it^*\omega}e^{-im\phi}S_{m\ell}\sin\theta\ dt^*\, d\theta\, d\phi,\]
\[-i\omega a_{\mathcal{I}^-} \doteq \frac{1}{\sqrt{2\pi}}\int_{-\infty}^{\infty}\int_{\mathbb{S}^2}\partial_t\upphi_{\mathcal{I}^-}e^{it\omega}e^{-im\phi}S_{m\ell}\sin\theta\ dt\, d\theta\, d\phi,\]
and we emphasize that in interpreting the formula~(\ref{anexpressionofscattering}), one must keep in mind that the $a_{\mathcal{I}^{\pm}}\left(\omega,m,\ell\right)$ are only defined as functions such that $\omega a_{\mathcal{I}^{\pm}} \in L^2_{\omega}l^2_{m\ell}$ and that the $a_{\mathcal{H}^{\pm}}\left(\omega,m,\ell\right)$ are only defined as functions such that $(\omega-\upomega_+m)a_{\mathcal{H}^{\pm}} \in L^2_{\omega}l^2_{m\ell}$.
\end{theorem}
\begin{proof}This follows immediately from the construction of $\mathscr{S}$, Propositions~\ref{microEqualInf} and~\ref{microequalhor} and an easy density argument.
\end{proof}
In particular, specialising to the case where $\uppsi_{\mathcal{H}^-} = 0$, this establishes {\bf Theorem~\ref{PSpREPint}} of Section~\ref{connectPST}.

\begin{remark}
We note that one can \underline{define} the map $\mathscr{S}$ by the expression~(\ref{anexpressionofscattering}) and prove directly by Theorem~\ref{refltransBound} that the map is a bounded isomorphism without relying on Theorem~\ref{isot0}; in fact, the proof is a good deal easier because one need never establish the boundedness of the map $\mathscr{B}_-$. However, one would still have to prove the decomposition~(\ref{themapwhichscattersreallywell}) so as to identify elements of $\mathcal{E}^K_{\mathcal{H}^-} \oplus \mathcal{E}^T_{\mathcal{I}^-}$ and $\mathcal{E}^K_{\mathcal{H}^+} \oplus \mathcal{E}^T_{\mathcal{I}^+}$ as radiation fields of solutions to the wave equation arising from finite energy Cauchy data.
\end{remark}

\subsection{Aside: A self-contained physical-space treatment of the Schwarzschild case}\label{veryselfcontained}

In the Schwarzschild case ($a = 0$) there is no superradiance and it is much easier to establish that $\mathscr{F}$ is invertible using the unitarity property.
Furthermore, the proofs may all be carried out in physical space, i.e.~using ``time-dependent methods''.
In this section we will give a self-contained treatment of how this can be done in our set-up (cf.~the related \cite{nicolas2}).

The ease of the Schwarzschild scattering theory is all associated with the following unitarity property.
\begin{proposition}\label{sounitary}Let $a = 0$, and observe that in this case $K = T=V$.
Then the forward maps $\mathscr{F}_+$ of Theorems~\ref{forwarddegen},~\ref{forwardt00}
and~\ref{forwardt00bif} are unitary:
\begin{equation}\label{unit1}
\left\vert\left\vert\mathscr{F}_+\left(\uppsi,\uppsi'\right)\right\vert\right\vert_{\mathcal{E}^T_{\mathcal{H}^+_{\geq 0}}}^2 + \left\vert\left\vert\mathscr{F}_+\left(\uppsi,\uppsi'\right)\right\vert\right\vert_{\mathcal{E}^T_{\mathcal{I}^+}}^2 = \left\vert\left\vert\left(\uppsi,\uppsi'\right)\right\vert\right\vert_{\mathcal{E}^T_{\Sigma_0^*}}^2,
\end{equation}
\begin{equation}\label{unit2}
\left\vert\left\vert\mathscr{F}_+\left(\uppsi,\uppsi'\right)\right\vert\right\vert_{\mathcal{E}^T_{\mathcal{H}^+}}^2 + \left\vert\left\vert\mathscr{F}_+\left(\uppsi,\uppsi'\right)\right\vert\right\vert_{\mathcal{E}^T_{\mathcal{I}^+}}^2 = \left\vert\left\vert\left(\uppsi,\uppsi'\right)\right\vert\right\vert_{\mathcal{E}^T_{\mathring{\Sigma}}}^2,
\end{equation}
\begin{equation}\label{unit3}
\left\vert\left\vert\mathscr{F}_+\left(\uppsi,\uppsi'\right)\right\vert\right\vert_{\mathcal{E}^T_{\overline{\mathcal{H}^+}}}^2 + \left\vert\left\vert\mathscr{F}_+\left(\uppsi,\uppsi'\right)\right\vert\right\vert_{\mathcal{E}^T_{\mathcal{I}^+}}^2 = \left\vert\left\vert\left(\uppsi,\uppsi'\right)\right\vert\right\vert_{\mathcal{E}^T_{\overline{\Sigma}}}^2.
\end{equation}
\end{proposition}
\begin{proof}We will only prove~(\ref{unit1}) as the proof of~(\ref{unit2}) and~(\ref{unit3}) is exactly the same. By density it suffices to prove~(\ref{unit1}) in the case when
$\left(\uppsi,\uppsi'\right) \in {}^2\mathcal{C}^{\infty}_{cp}(\Sigma_0^*)$.
Let us assume this for the remainder of the proof.
It follows now from Proposition~\ref{rp} and Theorem~\ref{h.o.s.} that we can find a dyadic sequence $\{\tau_i\}_{i=1}^{\infty}$ such that
\[\lim_{i\to\infty}\int_{S_{\tau_i}}\mathbf{J}^T_{\mu}\left[\mathscr{F}\left(\uppsi,\uppsi'\right)\right]n^{\mu}_{S_{\tau_i}} = 0.\]
Next, for each $\tau_i$ a $\mathbf{J}^T$ energy estimate yields
\[\int_{\mathcal{H}^+(0,\tau_i)}\mathbf{J}^T_{\mu}\left[\mathscr{F}_+\left(\uppsi,\uppsi'\right)\right]n^{\mu}_{\mathcal{H}^+} + \int_{\mathcal{I}^+_{\leq \tau_i}}\mathbf{J}^T_{\mu}\left[\mathscr{F}_+\left(\uppsi,\uppsi'\right)\right]n^{\mu}_{\mathcal{I}^+} + \int_{S_{\tau_i}}\mathbf{J}^T_{\mu}\left[\mathscr{F}_+\left(\uppsi,\uppsi'\right)\right]n^{\mu}_{S_{\tau_i}} = \left\vert\left\vert\left(\uppsi,\uppsi'\right)\right\vert\right\vert_{\mathcal{E}^T_{\Sigma_0^*}}^2.\]
We conclude the proof by taking $i \to \infty$.
\end{proof}

\begin{remark}
Let us remark that in the Schwarzschild $a=0$ case suitable versions of the statements
of Theorems~\ref{theResult} and~\ref{h.o.s.}
can be obtained without phase space analysis with respect
to either time or angular frequency decompositions. See~\cite{lectnotes}
and~\cite{anote}. Thus, not only the construction but also all relavent properties
of $\mathscr{F}$ are obtained purely with physical space
(i.e.~``time-dependent'') methods. Cf.~with the Kerr $a\ne 0$ case where the construction
of $\mathscr{F}$ is still formulated in the time domain
but requires the result of Theorem~\ref{theResult} which is itself based
on frequency-analysis.
\end{remark}

The injectivity of the forward map is an immediate corollary.
\begin{corollary}\label{injSchw}Let $a = 0$. Then the forward map $\mathscr{F}_+$ is injective.
\end{corollary}

We now construct the backwards map.
\begin{theorem}\label{isoSchw}Let $a = 0$. Then the forward map $\mathscr{F}_+$ is a unitary isomorphism (with either domain $\mathcal{E}^V_{\mathring{\Sigma}}$ or $\mathcal{E}^V_{\Sigma_0^*}$ or $\mathcal{E}^V_{\overline{\Sigma}}$) with two-sided
unitary inverse $\mathscr{B}_-$ satisfying $\mathscr{B}_-\circ \mathscr{F}_+ =Id$,
$\mathscr{F}_+\circ \mathscr{B}_-=Id$.
\end{theorem}
\begin{proof}We consider the case where the domain is $\mathcal{E}^V_{\mathring{\Sigma}}$, the cases $\mathcal{E}^V_{\overline{\Sigma}}$ or $\mathcal{E}^V_{\Sigma_0^*}$ are handled in an analogous fashion.

First of all, using the physical space construction from the proof of Proposition~\ref{backwardinthephysicalspace} we may \underline{define} the backwards map on a dense set:
\[\mathscr{B}_- : \mathcal{C}^{\infty}_{cp}(\mathcal{H}^+_{\geq 0})\oplus \mathcal{C}^{\infty}_{cp}(\mathcal{I}^+) \to \mathcal{E}^T_{\Sigma_0^*}.\]
Furthermore, the proof of Proposition~\ref{backwardinthephysicalspace} shows that $\left(\uppsi_{\mathcal{H}^+},\upphi_{\mathcal{I}^+}\right) \in \mathcal{C}^{\infty}_{cp}(\mathcal{H}^+_{\geq 0})\oplus \mathcal{C}^{\infty}_{cp}(\mathcal{I}^+)$ implies
\[
\mathscr{F}_+\left(\mathscr{B}_-\left(\uppsi_{\mathcal{H}^+},\upphi_{\mathcal{I}^+}\right)\right) = \left(\uppsi_{\mathcal{H}^+},\upphi_{\mathcal{I}^+}\right).
\]

We thus conclude that the forward map $\mathscr{F}_+$ has a dense image. Since the unitarity of $\mathscr{F}_+$ implies that the backwards map $\mathscr{B}_-$ is bounded on its domain, it follows immediately that $\mathscr{F}_+$ is in fact surjective. The rest of theorem follows immediately.
\end{proof}

\begin{remark}
It is instructive to compare the above ``time-dependent method'' construction
of $\mathscr{B}_-$ to the stationary-method construction of Defintion~\ref{hatS}.
Of course, one could have defined $\mathscr{B}_-$ on a dense
subset in the general
Kerr case with Proposition~\ref{backwardinthephysicalspace}, but one would still need to have used the representation
of Definition~\ref{hatS} to estimate it so as to take the completion.
\end{remark}

Applying the discrete isometry $t \mapsto -t$ of Schwarzschild yields the analogues of the above statements for $\mathscr{F}_-$ and $\mathscr{B}_+$. As before, we then define the scattering map $\mathscr{S} = \mathscr{F}_+ \circ \mathscr{B}_+$. We immediately obtain the following corollary.
\begin{theorem}\label{schscatunit}Let $a = 0$. Then
the scattering maps $\mathscr{S} : \mathcal{E}^T_{\mathcal{H}^-} \oplus \mathcal{E}^T_{\mathcal{I}^-} \mapsto \mathcal{E}^T_{\mathcal{H}^+} \oplus \mathcal{E}^T_{\mathcal{I}^+}$
are
$\mathscr{S} : \mathcal{E}^T_{\overline{\mathcal{H}^-}} \oplus \mathcal{E}^T_{\mathcal{I}^-} \mapsto \mathcal{E}^T_{\overline{\mathcal{H}^+}} \oplus \mathcal{E}^T_{\mathcal{I}^+}$
 unitary isomorphisms.
\end{theorem}

\section{Further applications}\label{somanyapplications}

We collect  here some further applications of our scattering theory.

In Section~\ref{PSTSR},
we will construct a physical-space  (time-domain)
theory of superradiant reflection.
Theorem~\ref{notreallysparta}  will give the results of {\bf Theorem~\ref{isbound}}
and {\bf Theorem~\ref{lowbn}} of Section~\ref{pST}.
We will also formulate and prove an analogous amplification statement
in terms of compactly supported smooth Cauchy data (Theorem~\ref{makeitsuperr}).

We will then show in Section~\ref{PSu}
a ``pseudo-unitary'' property (Theorem~\ref{aunitscatter})
of our scattering map $\mathscr{S}$ restricted to
past scattering data supported only on $\mathcal{I}^-$,
as well as a genuine unitarity property of $\mathscr{S}$ restricted to an appopriate
Hilbert space of
non-superradiant data (Theorem~\ref{unitNonsup}).
This will give {\bf Theorem~\ref{bb}} and {\bf Theorem~\ref{bbb}} of Section~\ref{pseunitsec}.

Finally, in Section~\ref{uniqueconti}
we will establish  the injectivity result Theorem~\ref{uniquecont},
which corresponds to  ``uniqueness of scattering states''  for improperly posed
scattering problems (for which there is no existence).
This will give {\bf Theorem~\ref{introuC}} of Section~\ref{theillpos}.

\subsection{A physical space theory of superradiant reflection}
\label{PSTSR}
First we define the physical-space reflection and transmission maps
referred to already in Section~\ref{pST}.

\begin{definition}Define the {\bf reflection map} $\mathscr{R}$ and  the {\bf transmission map}
$\mathscr{T}$  by
\[
\mathscr{R}\doteq \pi_{\mathcal{E}^T_{\mathcal{I}^+}}\circ
\mathscr{S}|_{\{0\}\oplus \mathcal{E}^T_{\mathcal{I}^-}}, \qquad
\mathscr{T}\doteq \pi_{\mathcal{E}^K_{\mathcal{H}^+}}\circ
\mathscr{S}|_{\{0\}\oplus \mathcal{E}^T_{\mathcal{I}^-}}
\]
where
\[
\pi_{\mathcal{E}^T_{\mathcal{I}^+}} : \mathcal{E}^K_{\mathcal{H}^+} \oplus
\mathcal{E}^T_{\mathcal{I}^+} \to \mathcal{E}^T_{\mathcal{I}^+},
\qquad
\pi_{\mathcal{E}^K_{\mathcal{H}^+}} : \mathcal{E}^K_{\mathcal{H}^+} \oplus
\mathcal{E}^T_{\mathcal{I}^+} \to \mathcal{E}^K_{\mathcal{H}^+}
\]
are the natural projections.
\end{definition}
We can view
\[
\mathscr{S}|_{\{0\}\oplus \mathcal{E}^T_{\mathcal{I}^-}}=\mathscr{R}\oplus \mathscr{T}.
\]
We are now ready for the following theorem.
\begin{theorem}
\label{notreallysparta}
The operator norms of $\mathscr{T}$ and $\mathscr{R}$ are bounded
\[
\|\mathscr{T}\| \le B, \qquad
\|\mathscr{R}\| \le B
\]
If $a=0$, then $\|\mathscr{R}\|=1$, whereas if $a\ne 0$ then
\[
\|\mathscr{R}\| >1.
\]
\end{theorem}
\begin{proof}The maps $\mathscr{T}$ and $\mathscr{R}$ are compositions of the bounded maps $\pi_{\mathcal{E}^T_{\mathcal{I}^+}}$, $\pi_{\mathcal{E}^K_{\mathcal{H}^+}}$ and $\mathscr{S}$ and hence are bounded.

Next, it follows immediately from the formula~(\ref{anexpressionofscattering}) that
\[
\left\vert\left\vert \mathscr{R}\right\vert\right\vert = \sup_{(\omega,m,\ell)}\mathfrak{R}(\omega,m,\ell).
\]
Thus, when $a \neq 0$, Corollary~\ref{superradiantAmp} shows that $\|\mathscr{R}\| > 1$, and when $a = 0$, Corollary~\ref{largelR} shows that $\|\mathscr{R}\| = 1$.
\end{proof}

We have now established {\bf Theorems~\ref{isbound} and~\ref{lowbn}}
 from Section~\ref{pST}.

With a little more work, we can upgrade the above result to the following statement.
\begin{theorem}\label{makeitsuperr}
Let $a\ne 0$.
There exists a smooth solution $\psi$ on $\mathcal{D}$ such that the initial data for $\psi$ along $\mathring{\Sigma}$ is supported away from the bifurcate sphere $\mathcal{B}$ (though not necessarily of compact support), $\psi$ has finite $V$-energy along $\mathring{\Sigma}$ and we have
\[
\int_{\mathcal{H}^-}\mathbf{J}^K_{\mu}[\psi]n^{\mu}_{\mathcal{H}^-} = 0,
\]
\[
\int_{\mathcal{I}^+}\mathbf{J}^T_{\mu}[\psi]n^{\mu}_{\mathcal{I}^+} > \int_{\mathcal{I}^-}\mathbf{J}^T_{\mu}[\psi]n^{\mu}_{\mathcal{I}^+}.
\]

Also, for all $R < \infty$ there exists a solution $\psi_R$ to~(\ref{WAVE}) on $\mathcal{R}$ such that the initial data for $\psi_R$ along $\mathring{\Sigma}$ are compactly supported within $r \in [R,\infty)$, and $\psi_R$ exhibits superradiance in the sense that
\[
\int_{\mathcal{I}^+}\mathbf{J}^T_{\mu}[\psi_R]n^{\mu}_{\mathcal{I}^+} > \int_{\mathring{\Sigma}}\mathbf{J}^T_{\mu}[\psi_R]n^{\mu}_{\mathring{\Sigma}}.
\]

\end{theorem}
\begin{proof}
We start by letting $a_{\mathcal{I}^-}(\omega,m,\ell)$ be a non-zero smooth function which is compactly supported in the set of $(\omega,m,\ell)$ which satisfy
\[\omega > 0,\]
\[\left(\omega-\upomega_+m\right) < 0.\]

We define
\[u \doteq \frac{\mathfrak{T}}{-i\left(\omega-\upomega_+m\right)}a_{\mathcal{I}^-}U_{\rm hor} = \frac{\mathfrak{R}}{i\omega}a_{\mathcal{I}^-}U_{\rm inf} + \frac{1}{i\omega}a_{\mathcal{I}^-}\overline{U_{\rm inf}},\]
\[
\psi_0 \doteq \frac{1}{(r^2+a^2)^{1/2}\sqrt{2\pi}}\int_{-\infty}^{\infty}\sum_{m\ell}e^{-i\omega t}e^{im\phi}S_{m\ell}\left(a\omega,\theta\right)u\, d\omega.
\]

Note that Proposition~\ref{microEqualInf}, Theorem~\ref{boundDegent0}, and Proposition~\ref{basicallybounded} imply
\[\int_{\mathcal{I}^-}\mathbf{J}^T_{\mu}[\psi_0]n^{\mu}_{\mathcal{I}^-} = \sqrt{\int_{\infty}^{\infty}\sum_{m\ell}\left|a_{\mathcal{I}^-}\right|^2}.\]

Now, Corollary~\ref{superradiantAmp} implies
\begin{align*}
\left|\mathfrak{R}\right|^2\left|a_{\mathcal{I}^-}\right|^2 \geq \left|a_{\mathcal{I}^-}\right|^2 + \epsilon,
\end{align*}
for some sufficiently small $\epsilon > 0$ on a compact set of frequencies. Integrating and summing and applying Proposition~\ref{microEqualInf}, Theorem~\ref{boundDegent0}, and Proposition~\ref{basicallybounded} yields

\begin{align}\label{thefirstineq}
  \int_{\mathcal{I}^+}\mathbf{J}^T_{\mu}[\psi_0]n^{\mu}_{\mathcal{I}^+} = \sqrt{\int_{-\infty}^{\infty}\sum_{m\ell}\left|\mathfrak{R}\right|^2\left|a_{\mathcal{I}^-}\right|^2\, d\omega}> \sqrt{\int_{-\infty}^{\infty}\sum_{m\ell}\left|a_{\mathcal{I}^-}\right|^2\, d\omega} = \int_{\mathcal{I}^-}\mathbf{J}^T_{\mu}[\psi_0]n^{\mu}_{\mathcal{I}^-}.
\end{align}

Finally, applying Proposition~\ref{microequalhor}, Theorem~\ref{boundDegent0}, and Proposition~\ref{basicallybounded} yields
\[\int_{\mathcal{H}^-}\mathbf{J}^K_{\mu}[\psi_0]n^{\mu}_{\mathcal{H}^-} = 0.\]
Thus, we may multiply $\psi_0$ be an appropriate constant to define a solution $\psi_1$ which will satisfy
\begin{enumerate}
    \item $\int_{\mathcal{I}^+}\mathbf{J}^T_{\mu}\left[\psi_1\right]n^{\mu}_{\mathcal{I}^+} > 1$.
    \item $\int_{\mathcal{I}^-}\mathbf{J}^T_{\mu}\left[\psi_1\right]n^{\mu}_{\mathcal{I}^-} = 1$.
    \item $\int_{\mathcal{H}^-}\mathbf{J}^K_{\mu}\left[\psi_1\right]n^{\mu}_{\mathcal{H}^-} = 0$.
\end{enumerate}

Let $\tilde\varphi$ denote the radiation field for $\psi_1$ along $\mathcal{I}^-$. Let $\chi(\tau)$ be a bump function, $\epsilon > 0$ be sufficiently small, and let $\psi_2$ be the unique solution to~(\ref{WAVE}) whose radiation field vanishes along $\mathcal{H}^-$ and has the radiation field \[\frac{\chi\left(\tau\epsilon^{-1}\right)\tilde\varphi}{\sqrt{\int_{\mathcal{I}^-}\mathbf{J}^T_{\mu}\left[\chi\left(\tau\epsilon^{-1}\right)\tilde\varphi\right]n^{\mu}_{\mathcal{I}^-}}}\] along $\mathcal{I}^-$. Using the boundedness of the map $\mathscr{S}$ from Definition~\ref{scatAll}, it is clear that taking $\epsilon$ sufficiently small (and then fixing $\epsilon$) will imply that
\begin{enumerate}
    \item $\int_{\mathcal{I}^+}\mathbf{J}^T_{\mu}\left[\psi_2\right]n^{\mu}_{\mathcal{I}^+} > 1$.
    \item $\int_{\mathcal{I}^-}\mathbf{J}^T_{\mu}\left[\psi_2\right]n^{\mu}_{\mathcal{I}^-} = 1$.
    \item $\int_{\mathcal{H}^-}\mathbf{J}^K_{\mu}\left[\psi_2\right]n^{\mu}_{\mathcal{H}^-} = 0$.
\end{enumerate}

Note that an easy domain of dependence argument shows that the initial data for $\psi_2$ along $\mathring{\Sigma}$ is compactly supported;
thus we may set $\psi \doteq \psi_2$.

In order to construct $\psi_R$ we need to do a little more work. We begin by recalling the estimate~(\ref{thatsweakindeed}), the proof of which (being invariant under time reversal) implies
\begin{equation}\label{soweakinthepast}
\lim_{s \to -\infty}\int_{\{t=s\}\cap [r_++\epsilon',r_+ + A]}\mathbf{J}^N_{\mu}[\psi_2]n^{\mu}_{\{t=s\}} = 0\qquad \forall\ 0 < \epsilon' < A< \infty.
\end{equation}

Let $\chi(x)$ be cut-off which is $0$ for $x \in [0,1]$ and identically $1$ for $x \in [2,\infty)$. Letting $\epsilon'$ be small enough so that $K$ is timelike for $r \in [r_+,r_++2\epsilon']$, applying a $\mathbf{J}^K$ energy estimate to $\left(1-\chi\left(r\left(\epsilon'\right)^{-1}\right)\right)\psi_{\epsilon}$ easily implies
\begin{align}\label{anEst45678}
\int_{\{t = s\}\cap [r_+,r_++\epsilon']}&\mathbf{J}^K_{\mu}[\psi_2]n^{\mu}_{\{t= s\}}  \\ \nonumber &
\leq\int_{\mathcal{H}^-}\mathbf{J}^K_{\mu}\left[\psi_2\right]n^{\mu}_{\mathcal{H}^-} + B(\epsilon')\int_{s}^{\infty}\int_{\{t = s\}\cap [r_++\epsilon',r_++2\epsilon']}\left[\mathbf{J}^K_{\mu}[\psi_2]n^{\mu}_{\{t= s\}} + \left|\psi_2\right|^2\right]\, ds.
\end{align}
Theorem~\ref{theResult} implies the the second term on the right hand side of this estimate converges to $0$ as $s \to \infty$. Since the first term on the right hand side vanishes, we conclude that
\begin{equation}\label{anEst12345}
\limsup_{s\to -\infty}\int_{\{t = s\}\cap [r_+,r_++\epsilon']}\mathbf{J}^K_{\mu}[\psi_2]n^{\mu}_{\{t= s\}} = 0.
\end{equation}

Taking $R$ suitably large, and applying a similar argument in the region $r \geq R$,  one may easily deduce that
\begin{equation}\label{anEst1111111}
\limsup_{s\to -\infty}\int_{\{t = s\}\cap \{r \in [R/10,\infty)\}}\mathbf{J}^T_{\mu}[\psi_2]n^{\mu}_{\{t= s\}} \leq 1.
\end{equation}

Let $\epsilon'' > 0$ be a small constant to be fixed later. Now we choose $R$ sufficiently large and $s = s(\epsilon'',R)$ sufficiently large and negative so that
\begin{equation}
\int_{\{t=s\}}\mathbf{J}^V_{\mu}\left[\psi_2-\chi\left(rR^{-1}\right)\psi_2\right]n^{\mu}_{\{t=s\}} < \epsilon'',
\end{equation}
\begin{equation}
\int_{\{t=s\}}\mathbf{J}^V_{\mu}\left[\chi\left(rR^{-1}\right)\psi_2\right]n^{\mu}_{\{t=s\}} \leq 1+\epsilon'',
\end{equation}

Let $\psi_3$ be the solution to~(\ref{WAVE}) whose initial data along $\{t = s\}$ are given by $\chi\left(rR^{-1}\right)\psi_2$. Now set
$\psi_4\left(t,r,\theta,\phi\right) \doteq \tilde\psi_3\left(t-s,r,\theta,\phi\right)$. It is clear that if we choose $\epsilon''$ small enough, then Theorem~\ref{boundDegent0} will imply that
\[\int_{\mathcal{I}^+}\mathbf{J}^T_{\mu}\left[\psi_4\right]n^{\mu}_{\mathcal{I}^+} > \int_{\{t=0\}}\mathbf{J}^T_{\mu}\left[\psi_4\right]n^{\mu}_{\{t=0\}}.\]

Finally, appealing to Theorem~\ref{boundDegent0} one more time, we may define $\psi_R$ to be the unique solution to~(\ref{WAVE}) whose initial data along $\{t = 0\}$ is given by $\left(1-\chi\left(rS^{-1}\right)\right)\psi_4$ for some sufficiently large $S$.
\end{proof}

\subsection{Pseudo-unitarity and non-superradiant unitarity}
\label{PSu}
The next sequence of results expresses the conservation of the $\mathbf{J}^T$ flux. Since this flux is unsigned along $\mathcal{H}^+$ we may interpret this as a statement of ``pseudo-unitarity''.
\begin{proposition}\label{easyunit}
Let $\psi$ be a solution to~(\ref{WAVE}) whose initial data lines in $\mathcal{E}^N_{\Sigma_0^*}$. Observe that Theorem~\ref{theResult} implies that
\[
\int_{\mathcal{H}^+_0}\left|\mathbf{J}^T_{\mu}[\psi]n^{\mu}_{\mathcal{H}^+}\right| \leq B\int_{\mathcal{H}^+_0}\mathbf{J}^N_{\mu}[\psi]n^{\mu}_{\mathcal{H}^+} \leq B\int_{\Sigma_0^*}\mathbf{J}^N_{\mu}[\psi]n^{\mu}_{\Sigma_0^*}.
\]
In particular, even though the integrand is unsigned, the integral
\[
\int_{\mathcal{H}^+_0}\mathbf{J}^T_{\mu}[\psi]n^{\mu}_{\mathcal{H}^+}
\]
is well defined and finite.

We then have
\[\int_{\mathcal{H}^+_0}\mathbf{J}^T_{\mu}[\psi]n^{\mu}_{\mathcal{H}^+} = \int_{\Sigma^*_0}\mathbf{J}^T_{\mu}[\psi]n^{\mu}_{\Sigma^*_0} - \int_{\mathcal{I}^+}\mathbf{J}^T_{\mu}[\psi]n^{\mu}_{\mathcal{I}^+}.\]
\end{proposition}
\begin{proof}By density considerations, we may assume that $\psi$ lies in $\mathcal{C}^{\infty}_{cp}(\Sigma^*_0)$. As we have already argued a few times before, Proposition~\ref{rp} and Theorem~\ref{h.o.s.} then allow us to find a dyadic sequence $\{\tau_i\}$ such that $\int_{S_{\tau_i}}\mathbf{J}^N_{\mu}[\psi]n^{\mu}_{S_{\tau_i}} \to 0$ as $i\to \infty$. For each $\tau_i$, a $\mathbf{J}^T$ energy estimate yields
\[\int_{\mathcal{H}^+(0,\tau_i)}\mathbf{J}^T_{\mu}[\psi]n^{\mu}_{\mathcal{H}^+} + \int_{S_{\tau_i}}\mathbf{J}^T_{\mu}[\psi]n^{\mu}_{S_{\tau_i}} + \int_{\mathcal{I}^+_{\leq \tau_i}}\mathbf{J}^T_{\mu}[\psi]n^{\mu}_{\mathcal{I}^+} = \int_{\Sigma_0^*}\mathbf{J}^T_{\mu}[\psi]n^{\mu}_{\Sigma_0^*}.\]
Now we simply take $\tau_i \to \infty$ and observe that $\left|\mathbf{J}^T_{\mu}[\psi]n^{\mu}_{S_{\tau_i}}\right| \leq B\mathbf{J}^N_{\mu}[\psi]n^{\mu}_{S_{\tau_i}}$.
\end{proof}
\begin{remark}\label{aneverysoslightmodification}Of course, one may prove a version of Proposition~\ref{easyunit} where the hypersurface $\Sigma_0^*$ is replaced by $\overline{\Sigma}$.
\end{remark}

\begin{theorem}\label{aunitscatter}
For any $\upphi \in \mathcal{C}^{\infty}_{cp}(\mathcal{I}^-)$ we have
\begin{equation}\label{thisiswhatunitaritylookslike}
\int_{\mathcal{H}^+}\mathbf{J}^T_{\mu}\left[\mathscr{T}\upphi\right]n^{\mu}_{\mathcal{H^+}} + \int_{\mathcal{I}^+}\mathbf{J}^T_{\mu}\left[\mathscr{R}\upphi\right]n^{\mu}_{\mathcal{I^+}} = \int_{\mathcal{I}^-}\mathbf{J}^T_{\mu}\left[\upphi\right]n^{\mu}_{\mathcal{I^-}},
\end{equation}
\begin{equation}\label{Tintegrable}
\int_{\mathcal{H}^+}\left|\mathbf{J}^T_{\mu}\left[\mathscr{T}\upphi\right]n^{\mu}_{\mathcal{H^+}}\right| \leq B\int_{\mathcal{I}^-}\mathbf{J}^T_{\mu}\left[\upphi\right]n^{\mu}_{\mathcal{I^-}}.
\end{equation}
Then, an easy density argument shows that~(\ref{thisiswhatunitaritylookslike}) and~(\ref{Tintegrable}) hold for arbitrary $\upphi \in \mathcal{E}^T_{\mathcal{I}^-}$.

\end{theorem}
\begin{remark}As is immediately clear from the proof below, the inequality~(\ref{Tintegrable}) holds without the non-superradiant assumption.
\end{remark}
\begin{proof}The equality~(\ref{thisiswhatunitaritylookslike}) follows immediately from Remark~\ref{aneverysoslightmodification} and the fact that $\upphi \in \mathcal{C}^{\infty}_{cp}(\mathcal{I}^-)$ implies that
\[\int_{\overline{\Sigma}}\mathbf{J}^N_{\mu}\left[\mathscr{B}_+\left(0,\upphi\right)\right]n^{\mu}_{\overline{\Sigma}} < \infty.\]

The inequality~(\ref{Tintegrable}) follows immediately from Plancherel, Theorem~\ref{constructaway}, Theorem~\ref{notreallysparta} and the fact that combining Theorem~\ref{notreallysparta} and Corollary~\ref{superradiantAmp} implies that $\left|\frac{\omega}{\omega-\upomega_+}\mathfrak{T}\right|$ is uniformly bounded.
\end{proof}
This gives {\bf Theorem~\ref{bb}} of Section~\ref{pseunitsec}.
\begin{remark}
Note that we cannot consider the case of general
initial data in $\mathcal{E}^K_{\mathcal{H}^-}$ as
$\uppsi \in \mathcal{E}^K_{\mathcal{H}^-}$ does \underline{not} imply that
\[
\int_{\mathcal{H}^-}\mathbf{J}^T_{\mu}[\uppsi]n^{\mu}_{\mathcal{H}^-} < \infty.
\]
\end{remark}

Finally, we observe that if we restrict the initial data along $\mathcal{H}^-$ and $\mathcal{I}^-$ to be non-superradiant, then the map $\mathscr{S}$ will be unitary in the standard sense. First we introduce the relevant function spaces.
\begin{definition}\label{nonsupfunc}We define $\mathcal{E}^{T,\natural}_{\mathcal{I}^{\pm}}$ to be the Hilbert space consisting of functions $f(\tau,\theta,\phi) : \mathcal{I}^{\pm} \to \mathbb{C}$ such that \[\hat{f}\left(\omega,m,\ell\right) = \frac{1}{\sqrt{2\pi}}\int_{-\infty}^{\infty}\int_{\mathbb{S}^2}e^{i\omega t}e^{-im\phi}S_{m\ell} f\, \sin\theta\, dt\, d\theta\, d\phi,\]
lies in the closure of functions compactly supported in $\{(\omega,m,\ell) : \omega\left(\omega-\upomega_+m\right) > 0\}$ under the inner product
\[\int_{-\infty}^{\infty}\sum_{m\ell}\omega^2\text{Re}\left(f_1\overline{f_2}\right).\]
\end{definition}
\begin{definition}\label{nonsupfunc2}We define $\mathcal{E}^{T,\natural}_{\mathcal{H}^{\pm}}$ to be the Hilbert space consisting of functions $f(\tau,\theta,\phi) : \mathcal{H}^{\pm} \to \mathbb{C}$ such that \[\hat{f}\left(\omega,m,\ell\right) = \frac{1}{\sqrt{2\pi}}\int_{-\infty}^{\infty}\int_{\mathbb{S}^2}e^{i\omega t}e^{-im\phi}S_{m\ell} f\, \sin\theta\, dt\, d\theta\, d\phi,\]
lies in the closure of functions compactly supported in
\begin{equation}\label{someweirdrange}
\{(\omega,m,\ell) : \omega\left(\omega-\upomega_+m\right) > 0\}
\end{equation}
under the inner product
\[\int_{-\infty}^{\infty}\sum_{m\ell}\omega\left(\omega-\upomega_+m\right)\text{Re}\left(f_1\overline{f_2}\right).\]
\end{definition}
The following theorem is an immediate consequence of the microlocal energy identity Proposition~\ref{microEnergyEst}.
\begin{theorem}\label{unitNonsup}The restriction of the map $\mathscr{S}$ to functions whose Fourier transforms are compactly supported in~(\ref{someweirdrange}) extends by density to a map $\mathscr{S} : \mathcal{E}^{T,\natural}_{\mathcal{H}^-} \oplus \mathcal{E}^{T,\natural}_{\mathcal{I}^-} \to \mathcal{E}^{T,\natural}_{\mathcal{H}^-} \oplus \mathcal{E}^{T,\natural}_{\mathcal{I}^-}$ which is a unitary isomorphism with respect to the positive definite inner product
\[\langle \left(\uppsi_1,\upphi_1\right),\left(\uppsi_2,\upphi_2\right)\rangle = \int_{-\infty}^{\infty}\sum_{m\ell}\left[\omega\left(\omega-\upomega_+m\right)\text{Re}\left(\hat{\uppsi}_1\overline{\hat{\uppsi}}_2\right) + \omega^2\text{Re}\left(\hat{\upphi}_1\overline{\hat{\upphi}_2}\right)\right].\]
\end{theorem}
This gives {\bf Theorem~\ref{bbb}} of Section~\ref{pseunitsec}.
Note that the above reduces again to Theorem~\ref{schscatunit} in the case $a=0$, where
$\mathcal{E}^{T,\natural}_{\mathcal{H}^\pm} \oplus \mathcal{E}^{T,\natural}_{\mathcal{I}^\pm}$
coincide with
$\mathcal{E}^T_{\mathcal{H}^\pm}\oplus \mathcal{E}^T_{\mathcal{I}^\pm}$.
It also yields in particular that $\mathscr{S}$ restricted to axisymmetric scattering
data is unitary.

\subsection{Uniqueness of ill-posed scattering states}
\label{uniqueconti}

We turn finally to the ``ill-posed case'', where one attempts to pose
scattering data on $\mathcal{H}^+\cup\mathcal{H}^-$, $\mathcal{I}^+\cup
\mathcal{I}^-$, $\mathcal{H}^-\cup\mathcal{I}^+$ or $\mathcal{H}^+\cup\mathcal{I}^-$.

To state our theorems, let us note first that we may define the forward maps
\begin{equation}
\label{IMP1}
\mathscr{F}:\mathcal{E}_{\overline{\Sigma}}^{V}\to \mathcal{E}^T_{\mathcal{I}^+}\oplus\mathcal{E}^T_{\mathcal{I}^-},
\qquad
\mathscr{F}:\mathcal{E}_{\overline{\Sigma}}^{V}\to\mathcal{E}^K_{\overline{\mathcal{H}^+}}\oplus\mathcal{E}^K_{\mathcal{H}^-},
\qquad
\mathscr{F}:\mathcal{E}_{\overline{\Sigma}}^{V}\to \mathcal{E}^K_{\overline{\mathcal{H}^-}}\oplus \mathcal{E}^T_{\mathcal{I}^+},
\qquad
\mathscr{F}:\mathcal{E}_{\overline{\Sigma}}^{V}\to
\mathcal{E}^K_{\overline{\mathcal{H}^+}}\oplus
\mathcal{E}^T_{\mathcal{I}^-},
\end{equation}
by completion of
\begin{equation}
\label{IMPcomplete}
(\uppsi, \uppsi')\mapsto\psi\mapsto (\upphi|_{\mathcal{I}^+},\upphi|_{\mathcal{I}^-})\text{ or }(\uppsi|_{\mathcal{H}^+},\uppsi|_{\mathcal{H}^-})\text{ or }(\uppsi|_{\mathcal{H}^-},\upphi|_{\mathcal{I}^+})\text{ or }(\uppsi_{\mathcal{H}^+},\upphi|_{\mathcal{I}^-}),
\end{equation}
and these are again bounded maps by our previous results.
We have the following statement of uniqueness (but not existence!) of
``improper'' scattering states:
\begin{theorem}
\label{uniquecont}
The maps $\mathscr{F}$ of $(\ref{IMP1})$ are all injective.
\end{theorem}

\begin{proof}We start with the case of the first two maps of $(\ref{IMP1})$.

First of all, the proof is conceptually clearer in the case of smooth compactly supported initial data, and we thus begin with this case. Consider
$(\uppsi, \uppsi')\in {}^2\mathcal{C}^{\infty}_{cp}({\overline\Sigma})$,
let $\psi$ be solution of $(\ref{IMPcomplete})$
and assume $(\uppsi, \uppsi')$ is in the kernel of the first or second map of $(\ref{IMP1})$.
Then, upon an application of Carter's separation to $\psi$ we have that
for almost every $(\omega,
m, \ell)$, the resulting $u$ is a smooth solution to
the radial o.d.e.~(\ref{e3iswsntouu}) such that when $\omega \neq 0$, $\left|a_{\mathcal{I}^+}\right|^2 + \left|a_{\mathcal{I}^-}\right|^2 = 0$ or $\left|a_{\mathcal{H}^+}\right|^2 +\left|a_{\mathcal{H}^-}\right|^2 = 0$, respectively.
It follows immediately from the local existence theory for these o.d.e.'s that $u$ is identically $0$ (see~\cite{olver}) whenever $\omega \neq 0$, and thus $\psi$ is $0$.  It follows
that $(\uppsi, \uppsi')=(0,0)$.

For general $(\uppsi, \uppsi')\in \mathcal{E}^V_{\overline\Sigma}$, let
us first consider the case only of the first map of $(\ref{IMP1})$,
i.e., let $(\uppsi, \uppsi')\in \ker \mathscr{F}:\mathcal{E}_{\overline{\Sigma}}^{V}\to \mathcal{E}^T_{\mathcal{I}^+}\oplus\mathcal{E}^T_{\mathcal{I}^-}$. Let $\psi$ denote the solution
of the wave equation $(\ref{WAVE})$ arising from $(\uppsi, \uppsi')$.
Theorem~\ref{theResult} implies that $\psi$ lies in $L^2_{{\rm loc}, r}L^2_{t,r,\theta,\phi}$. In particular, we can take the Fourier transform of $\psi$ and define the Carter separated function $u\left(r,\omega,m,\ell\right)$ which will lie in $L^2_{{\rm loc}, r}L^2_{\omega}l^2_{m,\ell}$.

Let $\epsilon > 0$, let $\mathcal{F}_\flat$ denote an arbitrary compact set of $(\omega,m,\ell)$ and let $\mathcal{K}$ denote an arbitrary compact set in $(r_+,\infty)$. Now, by regularizing the initial data for $\psi$, we can produce a solution $\psi_{\epsilon}$ to~(\ref{WAVE}) with smooth compactly supported initial data such that
\[
\int_{\overline{\Sigma}}\mathbf{J}^V_{\mu}\left[\psi - \psi_{\epsilon}\right]n^{\mu}_{\overline{\Sigma}} \leq \epsilon.
\]
It follows immediately from the fact that the forward map is well defined, that
\[
\int_{\mathcal{I}^{\pm}}\mathbf{J}^T_{\mu}\left[\psi_{\epsilon}\right]n^{\mu}_{\mathcal{I}^{\pm}} \leq B\epsilon.
\]
In particular, if we let $a_{\epsilon,\mathcal{I}^{\pm}}$ denote the microlocal radiation fields for $\psi_{\epsilon}$, Propositions~\ref{microEqualInf} and~\ref{efluxnullinf} imply that
\[
\int_{-\infty}^{\infty}\sum_{m\ell}\left[\left|a_{\epsilon,\mathcal{I}^+}\right|^2 +  \left|a_{\epsilon,\mathcal{I}^-}\right|^2\right]\ d\omega \leq B\epsilon.
\]
Letting $u_{\epsilon}$ denote the result of applying Carter's separation to $\psi_{\epsilon}$, it now follows immediately from standard o.d.e.~theory that
\[
\int_{\mathcal{K}}
\int_{(\omega,m,\ell) \in \mathcal{F}_{\flat}}\left|u_{\epsilon}\right|^2 \leq B(\mathcal{K},
\mathcal{F}_{\flat})\epsilon.
\]
Finally, an application of Theorem~\ref{theResult} to the $\psi-\psi_{\epsilon}$ followed by an application of Plancherel implies
\[
\int_{\mathcal{K}}\int_{(\omega,m,\ell) \in \mathcal{F}_{\flat}}\left|u\right|^2 \leq
B(\mathcal{K},
\mathcal{F}_{\flat})\epsilon,
\]
where $u\left(r,\omega,m,\ell\right)$ is the result of applying Carter's separation to $\psi$. Since $\epsilon$, $\mathcal{K}$, and $\mathcal{F}_{\flat}$ were arbitrary, we conclude that $u$ and hence $\psi$ vanishes.

The case where $(\uppsi,\uppsi')$ lies in the kernel of the second map of~$(\ref{IMP1})$
is treated in exactly the same way.

We turn now to the case when $(\uppsi,\uppsi')$ lies in the kernel of the third and fourth map of~$(\ref{IMP1})$. Since $\psi$ is not necessarily sufficiently integrable, we cannot use Definition~\ref{microRad} to define the microlocal radiation fields; instead we define $a_{\mathcal{I}^{\pm}}(\omega,m,\ell)$ and $a_{\mathcal{H}^{\pm}}(\omega,m,\ell)$ by applying Carter's separation to the functions $\mathscr{F}_{\pm}(\uppsi,\uppsi')|_{\mathcal{I}^{\pm}}$ and $\mathscr{F}_{\pm}(\uppsi,\uppsi')|_{\mathcal{H}^{\pm}}$:
\[a_{\mathcal{I}^{\pm}} \doteq \int_{-\infty}^{\infty}\int_{\mathbb{S}^2}e^{it\omega}e^{-im\phi}S_{m\ell}\mathscr{F}_{\pm}(\uppsi,\uppsi')|_{\mathcal{I}^{\pm}}\sin\theta\ dt\ d\theta\ d\phi,\]
\[a_{\mathcal{H}^{\pm}} \doteq \sqrt{2Mr_+}\int_{-\infty}^{\infty}\int_{\mathbb{S}^2}e^{it^*\omega}e^{-im\phi}S_{m\ell}\mathscr{F}_{\pm}(\uppsi,\uppsi')|_{\mathcal{H}^{\pm}}\sin\theta\ dt^*\ d\theta^*\ d\phi^*.\]
Now, we may apply Theorem~\ref{constructaway} (and its complex conjugated version) to conclude that
\begin{equation}\label{scatYay1}
\mathscr{F}_+(\uppsi,\uppsi')|_{\mathcal{I}^+} = \frac{1}{\sqrt{2\pi}}\int_{-\infty}^{\infty}\sum_{m\ell}\left(-\left(\frac{\omega-\upomega_+m}{\omega}\right)a_{\mathcal{H}^-}\tilde{\mathfrak{T}} + a_{\mathcal{I}^-}\mathfrak{R} \right)e^{-it\omega}e^{im\phi}S_{m\ell}\ d\omega,
\end{equation}
\begin{equation}\label{scatYay2}
\mathscr{F}_-(\uppsi,\uppsi')|_{\mathcal{I}^-} = \frac{1}{\sqrt{2\pi}}\int_{-\infty}^{\infty}\sum_{m\ell}\left(-\left(\frac{\omega-\upomega_+m}{\omega}\right)a_{\mathcal{H}^+}\overline{\tilde{\mathfrak{T}}} + a_{\mathcal{I}^+}\overline{\mathfrak{R}} \right)e^{-it\omega}e^{im\phi}S_{m\ell}\ d\omega,
\end{equation}
\begin{equation}\label{scatYay3}
\mathscr{F}_+(\uppsi,\uppsi')|_{\mathcal{H}^+} = \frac{1}{\sqrt{4M\pi r_+}}\int_{-\infty}^{\infty}\sum_{m\ell}\left(-\left(\frac{\omega}{\omega-\upomega_+}\right)a_{\mathcal{I}^-}\mathfrak{T} + a_{\mathcal{H}^-}\tilde{\mathfrak{R}} \right)e^{-it\omega}e^{im\phi}S_{m\ell}\ d\omega,
\end{equation}
\begin{equation}\label{scatYay4}
\mathscr{F}_-(\uppsi,\uppsi')|_{\mathcal{H}^-} = \frac{1}{\sqrt{4M\pi r_+}}\int_{-\infty}^{\infty}\sum_{m\ell}\left(-\left(\frac{\omega}{\omega-\upomega_+}\right)a_{\mathcal{I}^+}\overline{\mathfrak{T}} + a_{\mathcal{H}^+}\overline{\tilde{\mathfrak{R}}} \right)e^{-it\omega}e^{im\phi}S_{m\ell}\ d\omega.
\end{equation}

 Observe that if $(\uppsi,\uppsi')$ lies in the kernel of the third map of~$(\ref{IMP1})$, then $a_{\mathcal{H}^-}$ and $a_{\mathcal{I}^+}$ will vanish almost everywhere. Then~(\ref{scatYay1}) and~(\ref{scatYay4}) imply that $a_{\mathcal{I}^-}\mathfrak{R}$ and $a_{\mathcal{H}^+}\overline{\tilde{\mathfrak{R}}}$ both vanish almost everywhere. However, Corollary~\ref{novanishR} implies that $\mathfrak{R}$ and $\tilde{\mathfrak{R}}$ can only vanish at isolated points in $\omega$. We conclude that $a_{\mathcal{H}^+}$ and $a_{\mathcal{I}^-}$ can only be non-zero at isolated points and hence that $a_{\mathcal{H}^+}$ and $a_{\mathcal{I}^-}$ vanish almost everywhere. We conclude that $\mathscr{F}_{\pm}\left(\uppsi,\uppsi'\right) = (0,0)$ and thus that $\psi$ vanishes.

The case where $(\uppsi,\uppsi')$ lies in the kernel of the fourth map of~$(\ref{IMP1})$
is treated in a similar fashion.
\end{proof}
We have thus obtained now {\bf Theorem~\ref{introuC}} of Section~\ref{theillpos}.
\begin{remark}
In regard to the first two maps of $(\ref{IMP1})$, we note that is possible to prove localised
versions of the above  via the techniques of ``unique continuation'',
where $\psi$ is only assumed to vanish on certain portions of $\mathcal{H}^+\cup\mathcal{H}^-$ or portions of
$\mathcal{I}^+\cup\mathcal{H}^+$, but with stronger regularity assumptions and decay at infinity. See~\cite{baskinwang} for such results in the Schwarzschild case,~\cite{alexakisschlueshao} for such results on general asymptotically flat spacetimes and~\cite{alexakisshao} for such results for (among other things) certain non-linear wave equations on Minkowski space.
\end{remark}

\section{The backwards blue-shift instability and horizon-singular solutions}\label{correlate}

In this final section, we shall show that any solution of the wave equation $(\ref{WAVE})$ on
Schwarzschild assumed to have a particular choice of
radiation field necessarily would have infinite
$N$-energy on the hypersurface $\Sigma_0^*$.
Our theorem can be stated as follows:

\begin{moderthem}\label{noScatter}
Let $a=0$ and let $\psi$ be a smooth spherically symmetric
solution of the wave equation in the region $\mathring{\mathcal{R}}_{\geq 0}$ such that
\begin{enumerate}
    \item The initial data for $\psi$ lies in the closure of compactly supported initial data under the norm \[\int_{\Sigma_0^*}\left[\mathbf{J}^T_{\mu}[\psi] + \mathbf{J}^T_{\mu}[T\psi]\right]n^{\mu}_{\Sigma_0^*}.\]
    \item $\partial_{\tilde v}\psi$ extends continuously to the function $(t^*+1)^{-p}$ on $\mathcal{H}^+_{\geq 0}$ for some $p>2$.
    \item $\partial_{\tilde v}\left(T\psi\right)$ extends continuously to the function $-p(t^*+1)^{-p-1}$ on $\mathcal{H}^+_{\geq 0}$ for the same $p$ as above.
    \item There exists $\tau_0$ such that $\tau_1 > \tau_0$ implies \[\lim_{r\to\infty}r\psi|_{\tau = \tau_1} = 0.\]
\end{enumerate}
Then
\[
\int_{\Sigma_0^*}\mathbf{J}^N_{\mu}[\psi]n^{\mu}_{\Sigma_0^*} = \infty.
\]
\end{moderthem}

We will prove Theorem~\ref{noScatter} in Sections~\ref{metricandwave}--\ref{proofofithere} below.
We have stated our theorem in the above form so as to be independent
of the existence of the scattering theory maps $\mathscr{F}_+$, $\mathscr{B}_-$, etc.,
proven in this paper. Thus,
the proof of Theorem~\ref{noScatter} can be read independently of the rest of our paper.
The argument exploits the blue-shift factor of the horizon together with
a simple monotonicity property of the spherically symmetric
wave equation.

In combination with the results of our paper, Theorem~\ref{noScatter} can be reinterpreted
in the context of both our $N$-energy and our $T$-energy theories.
First, applying Theorem~\ref{defScat}, we shall
construct solutions $\psi$ satisfying the assumptions of Theorem~\ref{noScatter}
such that their induced data lie in $\mathcal{E}^T_{\Sigma_0^*}$ and give a short discussion of the
significance of the existence of such solutions. Finally, in Section~\ref{nonsurji}, we shall reinterpret Theorem~\ref{constructi} as a statement of
the non-surjectivity  of the map $\mathscr{F}_+:\mathcal{E}^N_{\Sigma_0^*}\to
\mathcal{E}^N_{\mathcal{H}^+}\oplus \mathcal{E}^N_{\mathcal{I}^+}$ of
Theorem~\ref{boundNONDeg}.
This will thus give {\bf Theorem~\ref{failsurj}} of Section~\ref{failsurjdisc}.

\subsection{Schwarzschild computations}
\label{metricandwave}
Setting $a= 0$ in~(\ref{eleme}), the Schwarzschild metric in Boyer-Lindquist coordinates
takes the form
\begin{equation}\label{schw}
g_{\rm Schw} = -\left(1-\frac{2M}{r}\right)dt^2 + \left(1-\frac{2M}{r}\right)^{-1}dr^2 + r^2\left(d\theta^2+\sin^2\theta\ d\phi^2\right).
\end{equation}

To get an explicitly regular expression for the metric near the event horizon $\mathcal{H}^+$ we introduce the $(v,r,\theta,\phi)$ coordinate system defined by
\[
\frac{dr^*}{dr} \doteq \left(1-\frac{2M}{r}\right)^{-1},
\qquad \tilde v \doteq t + r^*.
\]
The metric then takes the form
\begin{equation}\label{ingoing}
g_{\rm Schw} = -\left(1-\frac{2M}{r}\right)d\tilde v^2 + 2d\tilde vdr + r^2\left(d\theta^2+\sin^2\theta\ d\phi^2\right).
\end{equation}
Note that we have $T = \partial_{\tilde v}$ in the $(\tilde v,r,\theta,\phi)$ coordinate system. Let us also agree to set $Y \doteq \partial_r$.

It will also be useful to introduce a $(\tilde t, r,\theta,\phi)$ coordinate system in the following fashion. Let $\chi(r)$ be a cut-off which is identically $0$ for $r \in [2M,3M]$ and identically $1$ for $r \in [4M,\infty)$. We then set
\[\tilde t(t,r) \doteq t + r^* - 2\chi(r)r^*.\]
Note that $\partial_{\tilde t} = T$ is Killing.

Finally, it turns that it often convenient to work in the null coordinate system $(\tilde u,\tilde v,\theta,\phi)$ where $\tilde v$ is defined as before and
\[\tilde u \doteq t - r^*.\]
Then metric then takes the form
\begin{equation}\label{null}
g_{\rm Schw} = -2\left(1-\frac{2M}{r}\right)d\tilde ud\tilde v + r^2\left(d\theta^2 +\sin^2\theta\ d\phi^2\right).
\end{equation}
\begin{remark}
These coordinates break down at the horizon, where $\tilde u = \infty$. Nevertheless we can still use these coordinates in an effective manner near $\mathcal{H}^+$ as long we remember that the $\left(1-\frac{2M}{r}\right)^{-1}\partial_{\tilde u} = Y$ is a regular vector field on $\mathcal{H}^+$.
\end{remark}

If not explicitly noted otherwise, $\partial_{\tilde v}$ and $\partial_{\tilde u}$ will also be defined in the $(\tilde u,\tilde v,\theta,\phi)$ coordinate system. Let us agree to set $L \doteq \partial_{\tilde v}$.

In null coordinates, the wave equation~(\ref{WAVE}) applied to a spherically symmetric function $\psi$ takes the form
\[\partial_{\tilde v}\left(r^2\partial_{\tilde u}\psi\right) + \partial_{\tilde u}\left(r^2\partial_{\tilde v}\psi\right) = 0 \Leftrightarrow \]
\begin{equation}\label{waveNull}
\partial_{\tilde v,\tilde u}^2\psi + \frac{(\partial_{\tilde u}r)\partial_{\tilde v}\psi}{r} + \frac{(\partial_{\tilde v}r)\partial_{\tilde u}\psi}{r} = 0.
\end{equation}
The equation~(\ref{waveNull}) is equivalent to the following coupled transport equations for $r\partial_{\tilde u}\psi$ and $r\partial_{\tilde v}\psi$:
\begin{equation}\label{transport1}
\partial_{\tilde u}\left(r\partial_{\tilde v}\psi\right) = -\frac{(\partial_{\tilde v}r)(r\partial_{\tilde u}\psi)}{r} = -\frac{\left(1-\frac{2M}{r}\right)r\partial_{\tilde u}\psi}{r},
\end{equation}
\begin{equation}\label{transport2}
\partial_{\tilde v}\left(r\partial_{\tilde u}\psi\right) = -\frac{(\partial_{\tilde u}r)(r\partial_{\tilde v}\psi)}{r} = \frac{\left(1-\frac{2M}{r}\right)r\partial_{\tilde v}\psi}{r}.
\end{equation}
Near the event horizon it will be useful to work with transport equations for $r\left(1-\frac{2M}{r}\right)^{-1}\partial_{\tilde u}\psi$ and $r\partial_{\tilde v}\psi$:
\begin{equation}\label{transport1reg}
\left(1-\frac{2M}{r}\right)^{-1}\partial_{\tilde u}\left(r\partial_{\tilde v}\psi\right) = -\frac{\left(1-\frac{2M}{r}\right)\left[r\left(1-\frac{2M}{r}\right)^{-1}\partial_{\tilde u}\psi\right]}{r},
\end{equation}

\begin{equation}\label{transport2reg}
\partial_{\tilde v}\left(r\left(1-\frac{2M}{r}\right)^{-1}\partial_{\tilde u}\psi\right) + \frac{2M}{r^2}\left(r\left(1-\frac{2M}{r}\right)^{-1}\partial_{\tilde u}\psi\right) = \frac{r\partial_{\tilde v}\psi}{r}.
\end{equation}
\begin{remark}The fact that $\frac{2M}{r^2}|_{\mathcal{H}^+} = \frac{1}{2M} > 0$ represents the positivity of surface gravity and is intimately tied to the (local)
redshift effect. See the discussion in~\cite{lectnotes}.
\end{remark}

\subsection{Proof of Theorem~\ref{noScatter}}
\label{proofofithere}
We are now ready for the proof of Theorem~\ref{noScatter}. We will proceed in four steps.
\begin{enumerate}
\item Letting $\psi$ be as in Theorem~\ref{noScatter}, we begin by establishing a local energy decay statement with a sharp rate:
        \[\int_{S_{\tau}\cap \{r \leq R\}}\left[\left(T\psi\right)^2 + \left(\partial_{r^*}\psi\right)^2\right] \leq B(R)(1+\tau)^{-2p}\qquad \forall R > r_+.\]

\item Using the decay from the previous step, for all sufficiently small $\epsilon > 0$ we will propagate the $(1+\tilde v)^{-p}$ lower bound for $\partial_{\tilde v}\psi$ along $\mathcal{H}^+$ to a $(1+v)^{-p}$ lower bound on the hypersurfaces $\{r = 2M + \epsilon\}$. Cf.~\cite{cbh, price'slaw}.

\item Using the $(\tilde t,r,\theta,\phi)$ coordinate system, define $\tilde \psi(\tilde t,r) \doteq -\int_{\tilde t}^{\infty}\psi(\tilde s,r)\, d\tilde s$. Using the equation~(\ref{transport2reg}) and the previous steps, we will prove that unless $\int_{\Sigma_0^*}\mathbf{J}^N_{\mu}[\psi]n^{\mu}_{\Sigma_0^*} = \infty$, then $\left(1-\frac{2M}{r}\right)^{-1}\partial_{\tilde u}\tilde\psi$ and $\partial_{\tilde v}\tilde\psi$ both are positive along one of the hypersurfaces $\{r = 2M + \epsilon\}$.

\item Under the assumption that $\int_{\Sigma_0^*}\mathbf{J}^N_{\mu}[\psi]n^{\mu}_{\Sigma_0^*} < \infty$ we will use some monotonicity hidden in the system~(\ref{transport1}) and~(\ref{transport2}), and show that the positivity of $\partial_{\tilde u}\tilde\psi$ and $\partial_{\tilde v}\tilde \psi$ along $\{r = 2M + \epsilon\}$ propagates along outgoing null curves. Finally, we will see that this positivity of $r\partial_{\tilde u}\tilde\psi$ and $r\partial_{\tilde v}\tilde\psi$ implies that $r\psi$ cannot vanish along $\mathcal{I}^+$, yielding a contradiction to the assumption $\int_{\Sigma_0^*}\mathbf{J}^N_{\mu}[\psi]n^{\mu}_{\Sigma_0^*} < \infty$.
\end{enumerate}

\subsubsection{Local energy decay}
\label{forcontrahere}

We begin with the following proposition.
\begin{proposition}
\label{usefulpropo}
Let $\psi$ and $p$ be as in the statement of Theorem~\ref{noScatter}
and satisfy $(\ref{forcontrahere})$. Then, for all $R < \infty$ we have
\[
\int_{S_{\tau}\cap \{r \leq R\}}\left[\left(T\psi\right)^2 + \left(\partial_{r^*}\psi\right)^2\right] \leq B(R)(1+\tau)^{-2p}.
\]
\end{proposition}
\begin{proof}We begin by arguing that
\begin{equation}\label{energyDecay}
\int_{S_{\tau}}\mathbf{J}^T_{\mu}[\psi]n^{\mu}_{S_{\tau}} \leq B\left[\int_{\mathcal{H}^+_{\geq\tau}}\mathbf{J}^T_{\mu}[\psi]n^{\mu}_{\mathcal{H}^+} + \int_{\mathcal{I}^+_{\geq\tau}}\mathbf{J}^T_{\mu}[\psi]n^{\mu}_{\mathcal{I}^+}\right] \leq B\left(1+\tau\right)^{-2p+1}.
\end{equation}

Let $\tau < \infty$. Since $\int_{\Sigma_0^*}\mathbf{J}^T_{\mu}[\psi]n^{\mu}_{\Sigma_0^*} < \infty$ we may find a sequence of solutions $\{\psi_i\}$ to~(\ref{WAVE}) whose initial data lies in $\tilde{\mathcal{E}}_{\Sigma_0^*}$ and which satisfy $\lim_{i\to\infty}\int_{S_{\tau}}\mathbf{J}^T_{\mu}[\psi_i]n^{\mu}_{S_{\tau}} = \int_{S_{\tau}}\mathbf{J}^T_{\mu}[\psi]n^{\mu}_{S_{\tau}}$. As we have already observed multiple times, Theorem~\ref{h.o.s.} and Proposition~\ref{rp} imply that we may find a dyadic sequence $\{\tau_j^{(i)}\}_{j=1}^{\infty}$ such that $\lim_{j\to\infty}\int_{S_{\tau_j^{(i)}}}\mathbf{J}^T_{\mu}[\psi_i]n^{\mu}_{S_{\tau_j^{(i)}}} = 0$. Then we may apply a $\mathbf{J}^T$ energy estimate to each $\psi_i$ and conclude
\begin{equation*}
\int_{S_{\tau}}\mathbf{J}^T_{\mu}[\psi_i]n^{\mu}_{S_{\tau}} = \int_{\mathcal{H}^+(\tau,\tau_j^{(i)})}\mathbf{J}^T_{\mu}[\psi_i]n^{\mu}_{\mathcal{H}^+} + \int_{\mathcal{I}^+(\tau,\tau_j^{(i)})}\mathbf{J}^T_{\mu}[\psi_i]n^{\mu}_{\mathcal{I}^+} + \int_{S_{\tau_j^{(i)}}}\mathbf{J}^T_{\mu}[\psi_i]n^{\mu}_{S_{\tau_j^{(i)}}}.
\end{equation*}
Taking $j$ to infinity and then $i$ to infinity yields
\begin{equation*}
\int_{S_{\tau}}\mathbf{J}^T_{\mu}[\psi]n^{\mu}_{S_{\tau}} = \int_{\mathcal{H}^+_{\geq \tau}}\mathbf{J}^T_{\mu}[\psi]n^{\mu}_{\mathcal{H}^+} + \int_{\mathcal{I}^+_{\geq \tau}}\mathbf{J}^T_{\mu}[\psi]n^{\mu}_{\mathcal{I}^+}.
\end{equation*}
Finally, using that $\partial_{\tilde v}\psi$ extends continuously to $\left(1+t^*\right)^{-p}$ one may easily show that $\int_{\mathcal{H}^+_{\geq \tau}}\mathbf{J}^T_{\mu}[\psi]n^{\mu}_{\mathcal{H}^+} \leq B\int_{\tau}^{\infty}\left(1+t^*\right)^{-2p}\, dt^*$, and hence establish~(\ref{energyDecay}).

Next, we commute with the Killing vector field $T$ and consider the solution $T\psi$. Repeating the above procedure (using in particular that $T\psi$ is assumed that have a finite $\mathbf{J}^T$ energy along $\Sigma_0^*$ and the assumption on the limit of $\partial_{\tilde v}(T\psi)$ to $\mathcal{H}^+_{\ge 0}$) another $\mathbf{J}^T$ energy estimate implies
\begin{equation}\label{energyDecayT}
\int_{S_{\tau}}\mathbf{J}^T_{\mu}[T\psi]n^{\mu}_{S_{\tau}} \leq B\left[\int_{\mathcal{H}^+_{\geq\tau}}\mathbf{J}^T_{\mu}[T\psi]n^{\mu}_{\mathcal{H}^+} + \int_{\mathcal{I}^+_{\geq\tau}}\mathbf{J}^T_{\mu}[T\psi]n^{\mu}_{\mathcal{I}^+}\right] \leq B\left(1+\tau\right)^{-2p-1}.
\end{equation}

The final ingredient is an integrated local energy decay estimate. Setting $X \doteq f(r^*)\partial_{r^*}$ for a function $f$ to be fixed later, a straightforward calculation yields the following general formula:
\begin{align}\label{divCalc}
\nabla^{\mu}\mathbf{J}_{\mu}^X[\psi] &= \left(\frac{f'}{2}\left(1-\frac{2M}{r}\right)^{-1} + fr^{-1}\right)\left(T\psi\right)^2 +\left(\frac{f'}{2}\left(1-\frac{2M}{r}\right)^{-1} - fr^{-1}\right)\left(\partial_{r^*}\psi\right)^2.
\end{align}
We set $f \doteq - r^{-3}$ and obtain
\begin{align}\label{ILEDsph1}
\nabla^{\mu}J_{\mu}^X[\psi] = \frac{1}{2}r^{-4}\left(T\psi\right)^2 + \frac{5}{2}r^{-4}\left(\partial_{r^*}\psi\right)^2.
\end{align}

Keeping in mind that $X|_{\mathcal{H}^+} = -(2M)^{-3}T$ and $\left|\int_{S_{\tau}}\mathbf{J}^X_{\mu}[\psi]n^{\mu}_{S_{\tau}}\right| \leq B\int_{S_{\tau}}\mathbf{J}^T_{\mu}[\psi]n^{\mu}_{S_{\tau}}$, combining~(\ref{ILEDsph1}) with~(\ref{energyDecay}) and~(\ref{energyDecayT}) yields the following two estimates:
\begin{align}\label{spacetimeEst}
\int_{\tau}^{\infty}\int_{S_{\tau} \cap \{r \leq R\}}\left[\left(T\psi\right)^2 + \left(\partial_{r^*}\psi\right)^2\right] \leq B(R)\left(1+\tau\right)^{-2p+1}\qquad \forall R > r_+,
\end{align}
\begin{align}\label{spacetimeEstT}
\int_{\tau}^{\infty}\int_{S_{\tau} \cap \{r \leq R\}}\left[\left(T^2\psi\right)^2 + \left(\partial_{r^*}T\psi\right)^2\right] \leq B(R)\left(1+\tau\right)^{-2p-1}\qquad \forall R > r_+.
\end{align}

We will now interpolate between these four estimates in a straightforward fashion.\footnote{For example, see~\cite{schlue} for an application of such an interpolation argument to interior decay for the wave equation.}

For every $k \geq 1$, using the fact that
\begin{align*}
\int_{2^k}^{2^{k+1}}\int_{S_{\tau}\cap\{r \leq R\}}\left[\left(T\psi\right)^2 + \left(\partial_{r^*}\psi\right)^2\right]\leq B(R)\left(2^k\right)^{-2p+1},
\end{align*}
we may find a $\tau_k \in [2^k,2^{k+1}]$ such that
\begin{align}\label{pigeonhole}
\int_{S_{\tau_k}\cap\{r \leq R\}}\left[\left(T\psi\right)^2 + \left(\partial_{r^*}\psi\right)^2\right] \leq B(R)\tau_k^{-2p}.
\end{align}
Now consider $\tau \in [\tau_k,\tau_{k+1}]$. The fundamental theorem of calculus and the estimates~(\ref{spacetimeEst}),~(\ref{spacetimeEstT}), and~(\ref{pigeonhole}) imply
\begin{align}\label{degenDecay}
\int_{S_{\tau}\cap\{r \leq R\}}&\left[\left(T\psi\right)^2 + \left(\partial_{r^*}\psi\right)^2\right]
\\ \nonumber &\leq \int_{S_{\tau_{k+1}}\cap \{r \leq R\}}\left[\left(T\psi\right)^2 + \left(\partial_{r^*}\psi\right)^2\right]  \\ \nonumber &\qquad +B\int_{\tau_{k}}^{\tau_{k+1}}\int_{S_s\cap\{r \leq R\}}\left[\tau_k^{-1}\left(T\psi\right)^2 + \tau_k^{-1}\left(\partial_{r^*}\psi\right)^2 + \tau_k\left(T^2\psi\right)^2 + \tau_k\left(\partial_{r^*}T\psi\right)^2\right]
\\ \nonumber &\leq B(R)\left[\tau_{k+1}^{-2p} + \tau_k^{-2p} + \tau_k^{-2p}\right]
\leq B(R)\tau^{-2p}.
\end{align}
\end{proof}
\begin{remark}If one adds the assumption that $\int_{\Sigma_0^*}\mathbf{J}^N_{\mu}[\psi]n^{\mu}_{\Sigma_0^*} < \infty$, then one could establish the local energy decay statement using only transport equations in the region $\{r \leq R\}$.
\end{remark}
\begin{corollary}\label{vconstDec}Of course, the use of the particular foliation $\{S_{\tau}\}$ is not important for the above proposition. By modifying $S_{\tau}$ to equal $\{\tilde v = \tau\}$ in the region $\{r \leq R\}$ and repeating the above proof, one immediately obtains
\[\int_{\{\tilde v = \tau\} \cap \{r \leq R\}}\left[\left(T\psi\right)^2 + \left(\partial_{r^*}\psi\right)^2\right] \leq B(R)(1+\tau)^{-2p}.\]
Analogously, for $(r_0,r_1) \subset (r_+,\infty)$ one may show
\[\int_{\{\tilde u = \tau\} \cap \{r \in [r_0,r_1]\}}\left[\left(T\psi\right)^2 + \left(\partial_{r^*}\psi\right)^2\right] \leq B(r_0,r_1)(1+\tau)^{-2p}.\]
\end{corollary}

\subsubsection{Pushing the tail off the horizon}

We now turn to the proof of
\begin{proposition}\label{anicetail}Let $\psi$ be as in Theorem~\ref{noScatter}. Then, for all $\epsilon > 0$ sufficiently small,
\[b(1+\tilde v)^{-p} \leq \left(r\partial_{\tilde v}\psi\right)|_{\{r \leq 2M + \epsilon\}} \leq B(1+\tilde v)^{-p}.\]
\end{proposition}
\begin{proof}Keeping in mind that $\left(1-\frac{2M}{r}\right)^{-1}\partial_{\tilde u} = Y$ is equal to $\partial_r$ in $(\tilde v,r,\theta,\phi)$ coordinates, we integrate the transport equation~(\ref{transport1}) and obtain
\begin{equation}\label{intthetras}
\left(r\partial_{\tilde v}\psi\right)|_{(\tilde v,r) = (\tau,2M+\epsilon)} = \left(r\partial_{ v}\psi\right)|_{(\tilde v,r) = (\tau,2M)} + \int_{\{\tilde v=\tau\}\cap \{r \leq 2M+\epsilon\}}\partial_{\tilde u}\psi\, dr.
\end{equation}
Cauchy-Schwarz and Corollary~\ref{vconstDec} then yield
\begin{align}\label{gotthetail}
\left(\left(r\partial_{\tilde v}\psi\right)|_{(\tilde v,r) = (\tau,2M+\epsilon)} - \left(1+\tau\right)^{-p}\right)^2 &\leq \epsilon\int_{\{\tilde v=\tau\}\cap \{r \leq 2M+\epsilon\}}\left[\left(T\psi\right)^2 + \left(\partial_{r^*}\psi\right)^2\right]
\\ \nonumber &\leq B\epsilon(1+\tau)^{-2p}.
\end{align}
\end{proof}

\subsubsection{Positivity of $\tilde\psi$}

As we have already indicated in the outline, it will be useful to introduce the function
\begin{equation}\label{psitilde}
\tilde\psi(\tilde t,r) \doteq \int_{\tilde t}^{\infty}\psi(\tilde s,r)\, d\tilde s.
\end{equation}

Using the fact that $T$ is Killing, and the fact that $\left|\psi(\tau,r)\right| \leq B(r)(1+\tau)^{-p+1}$, one may easily check that $\tilde\psi$ is a smooth solution to~(\ref{WAVE}) in $\mathring{\mathcal{R}}_0$ and that $T\tilde\psi = -\psi$. The goal of this section is to use the transport equation~(\ref{transport2reg}) to show that $r\left(1-\frac{2M}{r}\right)^{-1}\partial_{\tilde u}\tilde\psi$ inherits some of $r\partial_{\tilde v}\psi$'s positivity.

We begin by studying $r\left(1-\frac{2M}{r}\right)^{-1}\partial_{\tilde u}\psi$.
\begin{proposition}\label{almosttail}Let $\psi$ be as in the statement of Theorem~\ref{noScatter}, $\tilde v_0$ be a fixed sufficiently large constant, and $(\tilde u,\tilde v) \in \{r = 2M +\epsilon\}$ for $\tilde v \geq v_0$ and $\epsilon > 0$ sufficiently small. Then
\[\left(r\left(1-\frac{2M}{r}\right)^{-1}\partial_{\tilde u}\psi\right)\Big|_{(\tilde u,\tilde v)} \geq B\exp\left(-\int_{\tilde v_0}^{\tilde v}\frac{2M}{r^2}d\tilde v'\right)\left|\left(r\left(1-\frac{2M}{r}\right)^{-1}\partial_{\tilde u}\psi\right)\Big|_{(\tilde u,\tilde v_0)}\right| + b\tilde v^{-p}.\]
\end{proposition}
\begin{proof}

We may write equation~(\ref{transport2reg}) as
\begin{equation}\label{expdamp}
\partial_{\tilde v}\left[\exp\left(\int_{\tilde v_0}^{\tilde v}\frac{2M}{r^2}d \tilde v'\right)r\left(1-\frac{2M}{r}\right)^{-1}\partial_{\tilde u}\psi\right] = \exp\left(\int_{\tilde v_0}^{\tilde v}\frac{2M}{r^2}d\tilde v'\right)\partial_{\tilde v}\psi.
\end{equation}
We conclude that
\begin{align}\label{expdamp2}
&\left(r\left(1-\frac{2M}{r}\right)^{-1}\partial_{\tilde u}\psi\right)\Big|_{(\tilde u,\tilde v)}  \\ \nonumber &\qquad=\exp\left(-\int_{\tilde v_0}^{\tilde v}\frac{2M}{r^2}d\tilde v'\right)\left[\left(r\left(1-\frac{2M}{r}\right)^{-1}\partial_{\tilde u}\psi\right)\Big|_{(\tilde u,\tilde v_0)} + \int_{\tilde v_0}^{\tilde v}\left(\exp\left(\int_{\tilde v_0}^{\tilde v'}\frac{2M}{r^2}d\tilde v''\right)\partial_{\tilde v'}\psi d\tilde v'\right)\right].
\end{align}

Next, using Proposition~\ref{anicetail}, we observe that
\begin{align}
&\exp\left(-\int_{\tilde v_0}^{\tilde v}\frac{2M}{r^2}d\tilde v'\right)\int_{\tilde v_0}^{\tilde v}\left(\exp\left(\int_{\tilde v_0}^{\tilde v'}\frac{2M}{r^2}d\tilde v''\right)\partial_{\tilde v'}\psi d\tilde v'\right)  \\ \nonumber &\qquad \geq b
\exp\left(-\int_{\tilde v_0}^{\tilde v}\frac{2M}{r^2}d\tilde v'\right)\int_{\tilde v_0}^{\tilde v}\left(\exp\left(\int_{\tilde v_0}^{\tilde v'}\frac{2M}{r^2}d\tilde v''\right)\left(\tilde v'\right)^{-p} d\tilde v'\right).
\end{align}

Using that $\left|\partial_{\tilde v}r\right| \leq B\epsilon$, a straightforward series of integration by parts yields
\begin{align}
\exp\left(-\int_{\tilde v_0}^{\tilde v}\frac{2M}{r^2}d\tilde v'\right)&\int_{\tilde v_0}^{\tilde v}\left(\exp\left(\int_{\tilde v_0}^{\tilde v'}\frac{2M}{r^2}d\tilde v''\right)\left(\tilde v'\right)^{-p} d\tilde v'\right)
\\ \nonumber & \geq \left(\frac{r^2}{2M}\right)\Big|_{(\tilde v,\tilde u)}\tilde v^{-p} - \left(\frac{r^2}{2M}\right)\Big|_{(\tilde v_0,\tilde u)}(\tilde v_0)^{-p}\exp\left(-\left(1-B\epsilon\right)(\tilde v-\tilde v_0)\right)
\\ \nonumber &\qquad- B\epsilon\exp\left(-\int_{\tilde v_0}^{\tilde v}\frac{2M}{r^2}d\tilde v'\right)\int_{\tilde v_0}^{\tilde v}\left(\exp\left(\int_{\tilde v_0}^{\tilde v'}\frac{2M}{r^2}d\tilde v''\right)\left(\tilde v'\right)^{-p} d\tilde v'\right)
\\ \nonumber &\qquad -B\left(\frac{r^2}{2M}\right)\Big|_{(\tilde v,\tilde u)}\tilde v^{-p-1} + b\left(\frac{r^2}{2M}\right)\Big|_{(\tilde v_0,\tilde u)}(\tilde v_0)^{-p-1}\exp\left(-\left(1-B\epsilon\right)(\tilde v-\tilde v_0)\right) - B\tilde v^{-p-1}.
\end{align}
We conclude that
\begin{align}\label{lowerboundforint}
\exp\left(-\int_{\tilde v_0}^{\tilde v}\frac{2M}{r^2}d\tilde v'\right)&\int_{\tilde v_0}^{\tilde v}\left(\exp\left(\int_{\tilde v_0}^{\tilde v'}\frac{2M}{r^2}d\tilde v''\right)\left(\tilde v'\right)^{-p} d\tilde v'\right) \geq b\tilde v^{-p}.
\end{align}
Combining~(\ref{lowerboundforint}) with~(\ref{expdamp2}) finishes the proof.

\end{proof}

\begin{remark}If we added the assumption that $\left(1-\frac{2M}{r}\right)^{-1}\partial_{\tilde u}\psi$ was uniformly bounded, then for sufficiently large $\tilde v$ this proposition would prove that $\left(1-\frac{2M}{r}\right)^{-1}\partial_{\tilde u}\psi \geq b\tilde v^{-p}$. Cf.~\cite{cbh}.
\end{remark}

We now have
\begin{proposition}\label{positivetildepsi}Let $\psi$ be as in Theorem~\ref{noScatter}, $\tilde\psi$ defined by~(\ref{psitilde}), and $\tilde v_0$ be a sufficiently large constant. Then $(\tilde u,\tilde v) \in \{r = 2M+\epsilon\}$ , $\epsilon > 0$ sufficiently small, and $\tilde v \geq \tilde v_0$ imply
\[r\left(1-\frac{2M}{r}\right)^{-1}\partial_{\tilde u}\tilde\psi\Big|_{(\tilde u,\tilde v)} \geq -Be^{-\frac{(1-B\epsilon)\tilde v}{4M}}\sqrt{\int_{\{\tilde v = \tilde v_0\}\cap \{r \leq 2M+\epsilon\}}\mathbf{J}^N_{\mu}[\psi]n^{\mu}_{\{\tilde v = \tilde v_0\}}} + b\tilde v^{-p+1}.\]
\end{proposition}
\begin{proof}Using that $(\tilde u,\tilde v) \in \{r = 2M + \epsilon\}$ implies that $\tilde u  = \tilde v - (4M + 2\epsilon)^*$, applying Proposition~\ref{almosttail} to $\psi$ and integrating implies that
\begin{equation}
r\left(1-\frac{2M}{r}\right)^{-1}\partial_{\tilde u}\tilde\psi \geq -B\int_{\tilde v}^{\infty}\left|\exp\left(-\int_{\tilde v_0}^{\tilde v}\frac{2M}{r^2}d\tilde v'\right)\left(1-\frac{2M}{r}\right)^{-1}\partial_{\tilde u}\psi\Big|_{(\tilde v-(4M+2\epsilon)^*,\tilde v_0)} \right|d\tilde v + b\tilde v^{-p+1}.
\end{equation}

Now, we observe that a change of variables yields
\begin{equation}\label{changethatvariable}
\int_{\tilde v}^{\infty}\left|\exp\left(-\int_{\tilde v_0}^{\tilde v}\frac{2M}{r^2}d\tilde v'\right)\left(1-\frac{2M}{r}\right)^{-1}\partial_{\tilde u}\psi\Big|_{(\tilde v-2(2M+\epsilon),\tilde v_0) }\right| d\tilde v \leq \int_{2M}^{2M + Be^{-\frac{\tilde v}{2M}}}\left(r-2M\right)^{-B\epsilon}\left|Y\psi\right||_{\tilde v = \tilde v_0} dr
\end{equation}

Cauchy-Schwarz then gives us
\begin{equation}\label{totheN}
\int_{2M}^{2M + Be^{-\frac{\tilde v}{2M}}}\left(r-2M\right)^{-B\epsilon}\left|Y\psi\right||_{\tilde v = \tilde v_0} dr \leq Be^{-\frac{(1-B\epsilon)\tilde v}{4M}}\sqrt{\int_{\{\tilde v = \tilde v_0\}\cap \{r \leq 2M+\epsilon\}}\mathbf{J}^N_{\mu}[\psi]n^{\mu}_{\{\tilde v = \tilde v_0\}}}.
\end{equation}
\end{proof}

\subsubsection{Positivity on $\mathcal{I}^+$ and the contradiction}
\label{positivityonI+}
Finally, we will show that if $\partial_{\tilde v}\tilde\psi$ and $\partial_{\tilde u}\tilde\psi$ are eventually positive along $\{r = 2M+\epsilon\} $, then the null derivatives of $\tilde\psi$ must eventually be positive in a
neigbourhood of $\mathcal{I}^+$.
\begin{proposition}\label{theyarequitepositive}Let $\psi$ be as in Theorem~\ref{noScatter} and define $\tilde\psi$ by~(\ref{psitilde}). Additionally, let us assume that $\int_{\{\tilde v = \tilde v_0\}\cap\{r \leq 2M+\epsilon\}}\mathbf{J}^N_{\mu}[\psi]n^{\mu}_{\{\tilde v= \tilde v_0\}} < \infty$. Then, there exists a constant $c$ such that
\begin{equation}
\label{Igave}
r\partial_{\tilde v}\tilde\psi \geq c\tilde v^{-p+1},
\qquad
r\partial_{\tilde u}\tilde\psi \geq c\tilde u^{-p+1},
\end{equation}
for all sufficiently large $r$ and $\tilde v$.
\end{proposition}
\begin{proof}Propositions~\ref{anicetail} and~\ref{positivetildepsi} imply that we may find $r_0 > 2M$ and $\tilde v_0 < \infty$ such that $\tilde v \geq \tilde v_0$ and $(\tilde u,\tilde v) \in \{r^* = r^*_0\}$ implies that there exists a constant $c>0$ such that
\begin{equation}\label{tail1}
r\partial_{\tilde v}\tilde\psi \geq c\tilde v^{-p+1},\qquad
r\partial_{\tilde u}\tilde\psi \geq c\tilde u^{-p+1}.
\end{equation}

Now, we define
\[\mathcal{A} \doteq \left\{s \in [0,\infty] : \tilde v \geq \tilde v_0\text{ and }r^*(\tilde u,\tilde v) \in [r^*_0,r^*_0+s) \Rightarrow r\partial_{\tilde v}\tilde\psi \geq c\tilde v^{-p+1}\text{ and }r\partial_{\tilde u}\tilde\psi \geq c\tilde u^{-p+1}\right\},\]
where we emphasize that $c$ is the constant from~(\ref{tail1}).
The proof will be finished if we can prove that $\mathcal{A} = [0,\infty)$.

It is clear that $\mathcal{A}$ is a closed and non-empty subset of $[0,\infty)$, so it suffices to prove that $\mathcal{A}$ is open. Suppose that $s_0 \in \mathcal{A}$. We show that $s_0 + \epsilon \in \mathcal{A}$ for $\epsilon > 0$ sufficiently small.

It immediately follows from the transport equations~(\ref{transport1}) and~(\ref{transport2}) and Corollary~\ref{vconstDec} that $\epsilon$ sufficiently small implies that for all $r^* \in [r_0^* + s_0,r^*_0 + s_0+\epsilon]$ and $\tilde v \geq \tilde v_0$ we have
\begin{equation}\label{tail1a}
r\partial_{\tilde v}\tilde\psi \geq \frac{1}{2}c\tilde v^{-p+1}, \qquad
r\partial_{\tilde u}\tilde\psi \geq \frac{1}{2}c\tilde u^{-p+1}.
\end{equation}

Given these estimates, we integrate again the transport equations~(\ref{transport1}) and~(\ref{transport2}) and now use~(\ref{tail1a}) to determine that in the region $r^* \in [r_0^* + s_0,r^*_0 + s_0+\epsilon]$ and $\tilde v \geq \tilde v_0$, $r\partial_{\tilde v}\tilde\psi$ is monotonically increasing in $-\tilde u$ and $r\partial_{\tilde u}\tilde\psi$ is monotonically increasing in $\tilde v$. We conclude that $r^* \in [r_0^* + s_0,r^*_0 + s_0+\epsilon]$ and $\tilde v \geq \tilde v_0$ imply
$(\ref{Igave})$.
\end{proof}

The next corollary establishes the desired contradiction and thus concludes the proof of Theorem~\ref{noScatter}.
\begin{corollary}Let $\psi$ be as in Theorem~\ref{noScatter} and define $\tilde\psi$ by~(\ref{psitilde}). Then, for each sufficiently large $\tau_1$, \[\lim_{r\to \infty}r\psi\left(r,\tau_1\right) < 0.\]
\end{corollary}
\begin{proof}
Using Proposition~\ref{theyarequitepositive} and the facts $\partial_{\tilde u} + \partial_{\tilde v} = T$ and $T\tilde\psi = -\psi$ we find that
\[
-r\psi = rT\tilde\psi = r\partial_{\tilde u}\tilde\psi + r\partial_{\tilde v}\tilde\psi \geq c\left(\tilde u^{-p+1}+\tilde v^{-p+1}\right).
\]
The result follows since $\lim _{r\to \infty} \tilde u(r,\tau_1)$ is bounded above.
\end{proof}

\subsection{Construction of $\psi$ using the degenerate $T$-scattering theory}
\label{constructi}

We now apply our degenerate scattering theory of Theorem~\ref{defScat} (see also Section~\ref{veryselfcontained}) to
indeed construct  solutions $\psi$ as in the statement of Theorem~\ref{noScatter}. Let
$\uppsi_{\mathcal{H}^+_{\geq 0}}:\mathcal{H}^+_{\geq 0}\to \mathbb R$ denote
the function
\[
\uppsi_{\mathcal{H}^+_{\geq 0}} (t^*) = \frac{(t^*+1)^{-p+1}}{-p+1}
\]
for $p>2$.
\begin{proposition}Let $a=0$.
For $\uppsi_{\mathcal{H}^+_{\geq 0}}\oplus 0\in \mathcal{E}^N_{\mathcal{H}^+_{\geq 0}}\oplus
\mathcal{E}^T_{\mathcal{I}^+}
\subset \mathcal{E}^T_{\mathcal{H}^+_{\geq 0}}\oplus \mathcal{E}_{\mathcal{I}^+}^T$
above, the solution $\mathscr{B}_-\left(\uppsi_{\mathcal{H}^+_{\geq 0}},0\right)$ satisfies all of the hypothesis of Theorem~\ref{noScatter}.
\end{proposition}
\begin{proof}Let us set $\psi \doteq \mathscr{B}_-\left(\uppsi_{\mathcal{H}^+_{\geq 0}},0\right)$. We first note that the spherical symmetry of Schwarzschild, Theorem~\ref{defScat}, and commutations with $T$ and $\Omega^{\alpha}$ 
are easily seen to imply that $\psi$ is a smooth  spherically symmetric solution in $\mathring{\mathcal{R}}$ and that
\[\int_{\Sigma_0^*}\left(\mathbf{J}^T_{\mu}\left[\psi\right] + \mathbf{J}^T_{\mu}\left[T\psi\right]\right)n^{\mu}_{\Sigma_0^*} < \infty.\]

Next, we observe that~(\ref{transport1}), the fundamental theorem of calculus, Cauchy Schwarz and an easy density argument show that $\partial_{\tilde v}\psi$ and $\partial_{\tilde v}\left(T\psi\right)$ extend continuously to the functions $(1+t^*)^{-p}$ and $-p(1+t^*)^{-p-1}$ respectively along $\mathcal{H}^+_{\geq 0}$.

In order to establish that $\lim_{r\to\infty}r\psi|_{S_{\tau}} = 0$ we first observe the unitarity property of Theorem~\ref{sounitary} yields
\begin{equation}\label{whatanestimate}
\int_{\Sigma_t^*}\mathbf{J}^T_{\mu}\left[\psi\right]n^{\mu}_{\Sigma_t^*} = \int_{\mathcal{H}^+_{\geq t}}\mathbf{J}^T_{\mu}[\psi]n^{\mu}_{\mathcal{H}^+_{\geq t}} \leq B\left(1+t\right)^{-2p+1}.
\end{equation}

Next, for $r$ sufficiently large, the fundamental theorem of calculus implies
\begin{equation}\label{decaysquickly}
\left|\psi(t,r)\right| \leq \int_r^{\infty}\left|\partial_r\psi\right|\, dr \leq r^{-1/2}\sqrt{\int_r^{\infty}(\partial_r\psi)^2r^2\ dr} \leq r^{-1/2}\sqrt{\int_{\Sigma_t^*}\mathbf{J}^T_{\mu}\left[\psi\right]} \leq Br^{-1/2}(1+t)^{-p+1/2}.
\end{equation}

Since $r$ is comparable to $t$ along any fixed hypersurface $S_{\tau}$, and $p > 2$, the estimate~(\ref{decaysquickly}) immediately implies that
\[\lim_{r\to\infty}r\psi|_{S_{\tau}} = 0.\]
\end{proof}

We immediately obtain the following corollary.
\begin{modercor}\label{awonderfulcor}
Let $a=0$.
For $\uppsi_{\mathcal{H}^+_{\geq 0}}\oplus 0\in \mathcal{E}^N_{\mathcal{H}^+_{\geq 0}}\oplus
\mathcal{E}^T_{\mathcal{I}^+}
\subset \mathcal{E}^T_{\mathcal{H}^+_{\geq 0}}\oplus \mathcal{E}_{\mathcal{I}^+}^T$
above,
then the map $\mathscr{B}_-$ of Theorem~\ref{defScat} maps
\[
\mathscr{B}_- ( \uppsi_{\mathcal{H}^+_{\geq 0}},0)  \in  \mathcal{E}^T_{\Sigma_0^*}\setminus \mathcal{E}^N_{\Sigma_0^*}.
\]
\end{modercor}

More pedestrianly,
\begin{modercor}
There indeed exist $\psi$ as in Theorem~\ref{noScatter}.
\end{modercor}

We note that by what we have shown in Proposition~\ref{usefulpropo},
$\psi$ has several nice additional properties. In particular, we have the following decay result.

\begin{modercor}
Let $\uppsi_{\mathcal{H}^+_{\geq 0}}$ be as in Corollary~\ref{awonderfulcor}.
Then, for every $R < \infty$ we have
\[
\int_{S_{\tau}\cap \{r \leq R\}}\left[\left|T\mathscr{B}_-(\uppsi_{\mathcal{H}^+_{\geq 0}},0)\right|^2
  + \left|\partial_{r^*}\mathscr{B}_-(\uppsi_{\mathcal{H}^+_{\geq 0}},0)\right|^2\right]
\leq B(R)(1+\tau)^{-2p}\qquad \forall \tau \geq 0.
\]
\end{modercor}
These strong decay properties lend further support to Conjecture~\ref{BHscatterconji}.

\subsection{Non-surjectivity of the $N$-energy forward map}
\label{nonsurji}

Lastly, we can immediately reinterpret
Corollary~\ref{awonderfulcor} as a non-surjectivity result (cf.~the discussion in
Section~\ref{failsurjdisc}).

\begin{modercor}
\label{thenonsurcor}
Let $a=0$. The asymptotic state $\uppsi_{\mathcal{H}^+_{\geq 0}}\oplus 0$ is not in
the image of the map $\mathcal{F}_+: \mathcal{E}_{\Sigma_0^*}^N \to \mathcal{E}_{\mathcal{H}^+_{\geq 0}}^N\oplus\mathcal{E}_{\mathcal{I}^+}^T$. Thus,
the map
of Theorem~\ref{boundNONDeg}
is not surjective, in fact,
the image
$\mathscr{F}_+\left(\mathcal{E}_{\Sigma_0^*}^N\right)$ has infinite co-dimension in $\mathcal{E}_{\mathcal{H}^+_{\geq 0}}^N\oplus\mathcal{E}_{\mathcal{I}^+}^T$ and infinite
codimension when intersected with $\mathcal{E}^{t_*^n N}_{\mathcal{H}_{\geq 0}^+}\oplus 0$
for any $n\ge 0$.
\end{modercor}
\begin{proof}If $\uppsi_{\mathcal{H}^+_{\geq 0}}\oplus 0$ was in
the image of the map $\mathcal{F}_+: \mathcal{E}_{\Sigma_0^*}^N \to \mathcal{E}_{\mathcal{H}^+_{\geq 0}}^N\oplus\mathcal{E}_{\mathcal{I}^+}^T$, then Theorem~\ref{defScat} and Corollary~\ref{awonderfulcor} would immediately yield a contradiction.
\end{proof}
We have thus obtained the final remaining~{\bf Theorem~\ref{failsurj}} of Section~\ref{failsurjdisc}.

\end{document}